\documentclass[useAMS,usegraphicx,usenatbib]{mn2e}
\usepackage{url}
\usepackage{amssymb}
\usepackage{graphicx}
\usepackage{dcolumn}
\usepackage{bm}
\usepackage{epsf}
\usepackage{psfig}
\usepackage{verbatim}
\usepackage{amsmath}
\usepackage{amsfonts}
\usepackage[usenames]{color}
\usepackage{fix-cm}
\usepackage{rotating}
\usepackage{stfloats} 
\usepackage{fixltx2e} 

\def\myfarfig#1{#1}

\numberwithin{equation}{section}

\newcommand\pasj{{PASJ}}   
\newcommand\araa{{ARA\&A}} 
\newcommand\apj{{ApJ}}     
\newcommand\apjl{{ApJ}}    
\newcommand\aj{{AJ}}       
\newcommand\mnras{{MNRAS}} 
\newcommand\aap{{A\&A}}    
\newcommand\aaps{{A\&AS}}  
\newcommand\physrep{{Phys.~Rep.}}  
\newcommand\nat{{Nature}}  
\newcommand\apss{{APSS}}   
\newcommand\na{{NA}}       

\newcounter{CommentNumber}
\newcommand{\TeXComment}[1]{{\textcolor{blue}{\tiny$^{[\theCommentNumber]}$}\stepcounter{CommentNumber}}}
\newcommand{\cm}{\mbox{ cm}}
\newcommand{\km}{\mbox{ km}}
\newcommand{\pc}{\mbox{ pc}}
\newcommand{\kpc}{\mbox{ kpc}}
\newcommand{\Mpc}{\mbox{ Mpc}}
\newcommand{\se}{\mbox{ s}}

\newcommand{\Gyr}{\mbox{ Gyr}}
\newcommand{\Myr}{\mbox{ Myr}}

\newcommand{\MHz}{\mbox{ MHz}}
\newcommand{\GHz}{\mbox{ GHz}}
\newcommand{\erg}{\mbox{ erg}}
\newcommand{\eV}{\mbox{ eV}}
\newcommand{\keV}{\mbox{ keV}}

\newcommand{\GeV}{\mbox{ GeV}}

\newcommand{\muG}{\mbox{ $\mu$G}}
\newcommand{\mb}{\mbox{ mb}}
\newcommand{\asec}{\mbox{ arcsec}}

\newcommand{\muJy}{\mbox{ $\mu$Jy}}
\newcommand{\sr}{\mbox{ ster}}
\newcommand{\pr}{\partial}
\newcommand{\grad}{\bm{\nabla}}

\newcommand{\vectwo}[1]{\vec{#1}}
\newcommand{\vecthree}[1]{\mathbf{#1}}
\newcommand{\fin}{\mbox{ .}}
\newcommand{\coma}{\mbox{ ,}}
\newcommand{\ie}{\emph{i.e.} }
\newcommand{\eg}{\emph{e.g.,} }
\newcommand{\cf}{\emph{cf.} }

\newcommand{\myp}{{\kappa}}

\newcommand{\myeta}{{\eta}}

\newcommand{\mach}{{\mathcal{M}}}
\newcommand{\constant}{\mbox{constant}}

\newcommand{\const}{\mbox{const.}}
\newcommand{\myetaL}{\bar{\eta}}
\newcommand{\CRP}{{CRP}}
\newcommand{\CRE}{{CRE}}
\newcommand{\CRI}{{CRI}}
\newcommand{\CR}{{CR}}
\newcommand{\CRPs}{{CRPs}}
\newcommand{\CREs}{{CREs}}
\newcommand{\CRIs}{{CRIs}}
\newcommand{\CRs}{{CRs}}
\newcommand{\mypsi}{{\psi}}
\newcommand{\myPsi}{{\Psi}}
\newcommand{\myz}{{\zeta}}
\newcommand{\myzz}{{\varsigma}}
\newcommand{\myr}{{\mathcal{R}}}
\newcommand{\myQ}{{\mathcal{Q}}}
\newcommand{\myQQ}{{\mathcal{\widetilde{Q}}}}
\newcommand{\myPalpha}{{\widetilde{\alpha}}}
\newcommand{\avalpha}{{\bar{\alpha}}} 
\newcommand{\edalpha}{{\hat{\alpha}}} 
\newcommand{\KL}{{KL10}}
\newcommand{\Fig}{{Fig.}}
\newcommand{\Tab}{{Table}}
\newcommand{\Figs}{{Figs.}}
\newcommand{\Eq}[1]{{Equation~(#1)}}
\newcommand{\SEq}[1]{{Eq.~(#1)}}
\newcommand{\Eqs}[1]{{Equations~(#1)}}
\newcommand{\eq}[1]{{equation~(#1)}}
\newcommand{\eqs}[1]{{equations~(#1)}}
\newcommand{\EqO}{{equation}}
\newcommand{\EqsO}{{equations}}
\newcommand{\myX}{{x}}

\newcommand{\myrg}{{r_g}}
\newcommand{\myrp}{{r_p}}
\newcommand{\myrcr}{{r_{cr}}}
\newcommand{\myrcre}{{r_{cre}}}
\newcommand{\myrcrp}{{r_{crp}}}
\newcommand{\myrB}{{r_{B}}}
\newcommand{\myn}{{N}}
\newcommand{\mykappa}{{\kappa_\phi}}
\newcommand{\myQpsi}{{\mathcal{F}}}
\newcommand{\myQQpsi}{{\mathcal{\widetilde{F}}}}
\newcommand{\Dltcr}{{\Delta_{cr}}}
\newcommand{\Dltg}{{\Delta_{g}}}
\newcommand{\Dltcool}{{\Delta_{cool}}}
\newcommand{\DltB}{{\Delta_{B}}}
\newcommand{\myx}{{X}}
\newcommand{\myy}{{Y}}
\newcommand{\myh}{{h}}
\newcommand{\mysigma}{{\sigma}}
\newcommand{\myk}{{k}}
\newcommand{\myCp}{{C_p}}
\newcommand{\nosymb}{{---}}
\newcommand{\till}{\text{--}}
\newcommand{\range}[2]{{[#1, #2]}}

\begin{document}

\title[Relics and Halos: homogeneous CRIs, evolving magnetic fields]
{Common origin for radio relics and halos: \\galaxy cluster-wide, homogeneous cosmic-ray distribution, and evolving magnetic fields}

\author[Uri Keshet]
       {\parbox[]{6.0in}
        {Uri Keshet\thanks{E-mail: ukeshet@cfa.harvard.edu}\thanks{Einstein fellow}\\
        \footnotesize
        Harvard-Smithsonian Center for Astrophysics, 60 Garden Street,  Cambridge, MA 02138, USA\\}}

\pagerange{\pageref{firstpage}--\pageref{lastpage}} \pubyear{2010} 
\maketitle 

\label{firstpage}

\begin{abstract}
Some galaxy clusters show diffuse radio emission in the form of peripheral relics (so far attributed to primary, shock-(re)accelerated electrons) or central halos.
Analysing radio and X-ray data from the literature, we find new connections between halos and relics, such as
a universal linear relation between their peak radio brightness and the gas column density.
Our results indicate that halos, relics, and halo--relic bridges in a cluster, all arise from the same, homogeneous cosmic ray ({\CR}) ion ({\CRI}) distribution.
We analytically derive the signature of synchrotron emission from secondary electrons and positrons ({\CREs}) produced in hadronic {\CRI} collisions, for an arbitrary magnetic field evolution.
In our model, flat spectrum halos (both giant and minihalos) arise from steady-state magnetic fields, whereas relics and steep halos reflect recent or irregular magnetic growth.
This naturally explains the properties of halos, relics, and the connections between them, without invoking particle (re)acceleration in weak shocks or turbulence.
We find {\CRI} energy densities in the range $u_p \simeq 10^{-\range{12.4}{13.3}} \erg\cm^{-3}$, with a spectral index $s_p=-2.20\pm0.05$, and identify an $\epsilon_B\sim 0.1$ magnetic fraction in some halos and behind relics, as far as $2\Mpc$ from the cluster's centre.
The {\CRI} homogeneity suggests strong {\CR} diffusion, $D(100\GeV)\gtrsim 10^{32}\cm^2\se^{-1}$.
The strong magnetisation imposes strict upper limits on $>10\GeV$ {\CRE} (re)acceleration in weak shocks (efficiency $\epsilon_e<10^{-4}$) and turbulence; indeed, each weak shock slightly lowers the energy fraction of flat {\CRs}.
\end{abstract}
\begin{keywords}
galaxies: clusters: general --- galaxies: clusters: intracluster medium --- X-rays: galaxies: clusters --- radio continuum: general --- magnetic fields
\end{keywords}


\section{Introduction}
\label{sec:Introduction}

As the largest gravitationally bound structures in the Universe, galaxy clusters are the focus of intense cosmological and astrophysical research.
Nonthermal radiation from clusters was observed in the radio band \citep[for review, see][]{Feretti05} and in hard X-rays \citep[for review, see ][]{RephaeliEtAl08}, and is expected to be observed in $\gamma$-rays in the near future \citep[by the 5-year Fermi mission; see][]{KeshetEtAl03}.

Such nonthermal signals trace the cosmic rays ({\CRs}) and magnetic fields permeating the intracluster medium (ICM).
These nonthermal components play an important role in the evolution of clusters on multiple scales, affecting their dynamical and thermal structure, for example by modifying the transport and dissipation processes.
The distributions of {\CRs} and magnetic fields in the ICM hold a unique record of past dynamical processes, such as the history of merger-induced shocks and turbulence in the cluster.
Modeling these component also constrains the poorly understood processes of particle acceleration and plasma magnetisation.

A fair fraction of the hot galaxy clusters \citep[$\sim35\%$ of the clusters with X-ray luminosty $L_X>10^{45}\erg\se^{-1}$;][]{GiovanniniEtAl02} show extended, nonthermal radio emission with low surface brightness, which is not associated with any particular member galaxy.
This is believed to be synchrotron radiation emitted by CR electrons or positrons ({\CREs}), injected locally into the ICM and gyrating in its pervasive magnetic fields.
Arguably, such radio observations hold more information regarding the nonthermal components of the ICM than presently available in any other band.

\subsection{Source classification: halos and relics}

ICM radio sources are broadly classified, according to their location, morphology, and polarisation, as giant halos (GHs; also known as a cluster-wide halos), minihalos (MHs; or core halos), or relics \citep{FerettiGiovannini96}.
In general, halos (both GHs and MHs) are regular, unpolarised emission around the cluster's centre, whereas relics are peripheral, polarised, typically elongated, and thought to be associated with shocks.
For a recent review, see \citet{FerrariEtAl08}.
\citep[We use the conventional term ``relic'', although the recently suggested terms ``flotsam'' or ``gischt'' may be more appropriate; see][]{KempnerEtAl04}.

GHs are found in the centres of merger, non-cool core clusters.
They are typically unpolarised, and show a regular morphology which follows the thermal plasma.
Their spectral indices lie in the range $\alpha_\nu \equiv d\log(P_\nu)/d\log\nu=-\range{1.0}{1.5}$ (flat halos) or $\alpha_\nu=-\range{1.5}{2.0}$ (steep halos), where $P_\nu$ is the specific radio power and $\nu$ is the frequency.
GHs extend over large, $\sim\text{Mpc}$ scales, farther than the distance a {\CRE} can cross before cooling.
Therefore, {\CREs} must be injected locally and continuously into the ICM. Two types of models have been proposed for {\CRE} injection in GHs: {\it (i)} secondary production by hadronic collisions between CR ions ({\CRI}) and the ambient plasma \citep{Dennison80, BlasiColafrancesco99}; and {\it (ii)} in-situ turbulent acceleration or reacceleration of primary {\CREs} \citep{EnsslinEtAl99, BrunettiEtAl01, Petrosian01}.
It was recently shown \citep{KushnirEtAl09, KeshetLoeb10} that the radio--X-ray correlations in GH luminosity \citep{BrunettiEtAl07} and in surface brightness \citep{GovoniEtAl01,KeshetLoeb10} strongly support the first, secondary {\CRE} model, and imply that the defining property of GHs is a strongly magnetised, $B\gtrsim 3\muG$ ICM. (For a different view, see \citet{BrunettiEtAl09}.)
This model reproduces the spectral, morphological and energetic properties of flat GHs \citep[][henceforth {\KL}]{KeshetLoeb10}.
Independent measurements of $B$ within halos are presently not sufficiently precise to test this connection; low-significance evidence for higher magnetisation in halo clusters was reviewed in {\KL}.

MHs are found in the centres of more relaxed, cool-core clusters (CCs).
They extend roughly over the cooling region \citep{GittiEtAl02}, encompassing up to a few percent of the typical GH volume, and often overlap the radio emission from an active galactic nucleus (AGN).
They resemble miniature versions of flat GHs, typically being unpolarised, regular, and spectrally flat with $\alpha=-\range{1.0}{1.5}$.
They show radio--X-ray correlations consistent with those of GHs, and a similar ratio $\eta\equiv \nu I_\nu/F_X$ between the radio and X-ray surface brightness ({\KL}).
This indicates that they arise from the same mechanism as GHs: secondary {\CREs} losing most of their energy to synchrotron radiation in highly magnetised cores ({\KL}).
This conclusion is supported by the morphological association between MH edges and cold fronts (CFs), reported by \citet{MazzottaGiacintucci08}.
Such CFs, present in most CCs \citep{MarkevitchVikhlinin07}, were identified as tangential discontinuities lying above (\ie at larger distances $r$ from the cluster's centre) regions magnetised by bulk shear flow \citep{KeshetEtAl10}.
We do not focus on MHs here; for a discussion of their properties, see {\KL}.

As opposed to halos, radio relics are typically polarised (at $\gtrsim 10\%$ levels), irregular (often elongated), and far from the cluster's centre (up to $r\sim$ a few Mpc).
Different classification schemes have been proposed for relics \citep{GiovanniniFeretti04, KempnerEtAl04}; here we focus on relics which are true ICM emission, not associated with any galaxy or AGN.
All relic clusters which have been carefully analysed in the optical or X-ray bands show evidence of a recent merger \citep{GiovanniniFeretti04}.
These properties, and the absence of nearby {\CRE} sources, suggest that relics are associated with merger shocks propagating through the ICM.
Indeed, in some cases \citep[\eg in A521; ][]{GiacintucciEtAl08}, a relic was found to coincide with an X-ray brightness edge consistent with a shock front.
Thus far, it has been thought that {\CREs} in relics must be primary particles, injected by the weak collisionless shock.
Two mechanisms were proposed for such injection:
\emph{(i)} diffusive shock acceleration (DSA) in merger or accretion shocks \citep{EnsslinEtAl98}; and
\emph{(ii)} adiabatic compression of fossil radio plasma caught by a shock \citep[a ``radio phoenix''; ][]{EnsslinGopalKrishna01}.
However, a secondary {\CRE} model for relics was never ruled out (D. Kushnir 2008, private communications).

\subsection{Motivation: observational inconsistencies and unconstrained assumptions}

Accumulating evidence indicates that the present modeling of these ICM radio sources is, at best, incomplete.
Observationally, there are several results that are peculiar or inconsistent with the present models.
This includes multiple similarities between halos and relics, radio bridges sometimes observed to connect a halo and a relic, halos with a steep spectrum, the remarkably similar spectrum at the edges of all relics, and the selective appearance of spectral steepening inward of relics.
Interpreting these results under the assumption that halos are produced by secondary {\CREs} while relics arise from primary {\CREs} requires fine tuning and implausible assumptions, as we show in \S\ref{sec:ModelProblems}.
The problem becomes worse if one assumes that halos arise from turbulent-accelerated primary {\CREs}, as this leads to additional inconsistencies and unnatural assumptions, as discussed in {\KL} and below.

On the theoretical side, the present models rely on some questionable, unconstrained  assumptions.
Primary {\CRE} models are sensitive to the poorly understood processes of particle acceleration and magnetisation in weak collisionless shocks and in turbulence.
Halo primary {\CRE} models thus make multiple assumptions which are not independently tested or constrained \citep[see, for example][]{BrunettiLazarian07}.
Models assuming pristine particle acceleration at merger shocks typically compute the spectrum from diffusive shock acceleration (DSA) theory, but the predicted spectrum was not confirmed for weak shocks and the acceleration efficiency is unknown; only loose upper limits on the efficiency are available.
Analogously, due to the poor understanding of {\CRI} diffusion through the ICM, secondary {\CRI} models typically make some simplifying assumptions regarding the distribution of {\CRIs} or magnetic fields within the cluster.

In addition, previous secondary {\CRE} models assume a steady-state magnetic field, in the sense that the magnetic energy density $u_B=B^2/8\pi$ evolves on a timescale much longer than the {\CRE} cooling time, $t_{cool}\sim 0.1\Gyr$.
These models assume, in addition, a steady-state {\CRE} injection rate.
These assumptions are violated for example in the vicinity of shocks and during the onset of turbulence.
Halo models are sensitive to these assumptions, in particular when the halos are young, near the edge of halos, and near shocks embedded in the halo or at its edge, as observed in several cases.

\subsection{A unified halo--relic, secondary {\CRE} model}

We find that a single secondary {\CRE} model simultaneously accounts for all types of diffusive radio emission from the ICM, including GHs, MHs, relics, and halo--relic bridges.
We begin in \S\ref{sec:HaloAndRelicEta} by studying radio data, extracted from the literature, for all known relic clusters and for a sample of halo clusters.
We parameterise the distributions of {\CREs} and magnetic fields as unknown power-law functions of the bulk plasma (for brevity: gas) density, thus avoiding  unnecessary assumptions regarding the nature of the {\CREs} and the distributions of {\CREs}, magnetic fields, and gas.

A useful diagnostic of diffuse radio sources is the ratio between radio and X-ray emission, as it relates the nonthermal and thermal components of the plasma.
The X-ray emissivity, dominated by thermal bremsstrahlung, is proportional to the gas density squared, $j_X\propto n^2$, where $n$ is the electron number density.
The radio emissivity is roughly proportional to the product of the {\CRE} energy density $u_e$ and the magnetic energy density, $j_\nu\propto u_e u_B$.
As clusters are approximately isothermal, the ratio $j_\nu/j_X\propto \epsilon_e \epsilon_B$ gauges the energy fractions $\epsilon_e\equiv u_e/u_{th}$ of the {\CREs} and $\epsilon_B\equiv u_B/u_{th}$ of the magnetic field, measured with respect to the thermal energy density $u_{th}=(3/2)\mu^{-1}n k_B T$.
Here, $T$ is the temperature, $k_B$ is Boltzmann's constant, and the $\mu\equiv\bar{m}/m_p\simeq 0.6$ factor accounts for the thermal contribution of ions, with $\bar{m}$ being the average particle mass.

The radio to X-ray ratio is particularly useful in regions where $u_B$ exceeds the energy density $u_{cmb}$ of the cosmic microwave background (CMB), $B>B_{cmb}\equiv (8\pi u_{cmb})^{1/2}$.
Here, {\CRE} cooling is regulated by the magnetic field with little Compton losses, so in a steady state $u_e\propto u_B^{-1}$, and $j_\nu/j_X\propto \epsilon_e/n$ is independent of the magnetic field.
This appears to be the case near the centres of halos, according to the $P_\nu\till L_X$ correlation between radio power and X-ray luminosity in GHs \citep{KushnirEtAl09}, and the linear $I_\nu\till F_X$ correlation between radio and X-ray surface brightness near the centres of both GHs and MHs ({\KL}).

Therefore, we begin by modeling the radio to X-ray brightness ratio $\eta\equiv \nu I_\nu/F_X$ near the centres of well studied halos.
We find that $\eta\propto F_X^{-2/3}$ in these clusters, valid out to distances $r\gtrsim 400\kpc$.
Equivalently, this may be written as $I_\nu\propto F_X^{1/3}\sim \lambda_n$, where $\lambda_n$ is the gas column density.
As $j_\nu\propto u_e\propto u_p n$ in strongly magnetised regions, this is our first direct indication that the {\CRI} energy density $u_p$ is homogeneous.
We identify a radial break in $\eta$ in some halos, which we interpret as the transition from strong to weak fields (see {\KL}), complicating any estimate of $u_p$ at larger distances based on halo data alone.

Next, we examine $\eta$ in relics, using $\beta$ models of the clusters based on ASCA data \citep{FukazawaEtAl04}.
In particular, we choose the location along the relic where radio brightness is maximal, presumably corresponding to the highest magnetic field and a favourable projection.
We find that these relics $\eta$ values lie close to, but slightly above, the $\eta = \eta_0(n/n_0)^{-1} \simeq \eta_0(F_X/F_{X,0})^{-2/3}$ curve normalised by halos, where subscripts $0$ denote a quantity measured at the centre of the cluster, $r=0$, and $n$ here is the projected density in the plane of the cluster.
Therefore, relics too approximately satisfy $I_\nu\propto \lambda_n$, and with a proportionality coefficient similar but slightly higher than in halos.

This surprising result is very unnatural in the context of present models, because all model variants attribute halos and relics to different {\CRE} populations.
It strongly suggests that relics, like halos, arise from secondary {\CREs}, produced by the same population of {\CRIs}.
This would imply that the distribution of {\CRIs} remains homogeneous out to $r\sim 2\Mpc$, close to the virial shock of the cluster.

\subsection{Incorporating deviations from a {\CRE} steady state: essential for  relics and spectral analyses}

In order to test the applicability of one secondary {\CRE} model for both halos and relics, we examine the morphological and spectral properties of relics, and the halo--relic regions in clusters that harbour both.
Recognising that the rapid changes in the magnetic field and in {\CRE} injection at the relic discontinuity are responsible for the elevated (with respect to $\eta=\eta_0(n/n_0)^{-1}$) relic brightness and the spectral steepening observed inward of several relics, indicates that the space-time evolution of the {\CRE} population must be incorporated in the model.

Therefore, in \S\ref{sec:TimeDependentTheory} we compute the {\CRE} evolution and the resulting synchrotron signal, for an arbitrary temporal evolution of the magnetic field and of the {\CRE} injection rate, and examine the effects of {\CRE} diffusion across magnetic irregularities. 
In particular, we study the structure of a weak shock, and derive the properties of radio emission arising from weak shocks and turbulence.
We show that a weak shock of Mach number $\mach\lesssim 5$ raises the pressure of flat spectrum {\CRIs} and {\CREs} by a factor $\leq\mach^2$, thus lowering the {\CR} energy fraction with respect to the (shocked) gas.

Quite generally, synchrotron emission brightens, and subsequently steepens, in regions that experienced strong recent magnetic growth which exceeds the gas compression.
This explains the steepening observed downstream of several (but not all) relics, provided that these relics are strongly magnetised, as confirmed in some cases by independent estimates of the relic magnetic field. 
This effect also provides an alternative explanation for the spectral steepening observed near the edges of some halos, interpreted by {\KL} as evidence for a steep cosmic ray proton ({\CRP}) spectrum, and explains the very steep spectrum of a subset of GHs.

\subsection{Model calibration, tests, and implications}

Our time-dependent model is applied to halo and relic observations in \S\ref{sec:ModelApplications}, in order to test the model and calibrate its parameters.
We find that the model reproduces the observations, provided that the {\CRI} distribution is homogeneous, and that a fraction $\epsilon_B\simeq 0.1$ of the thermal energy density downstream is deposited in magnetic fields.
We then show that the model naturally explains the multiple connections inferred between halos and relics, such as the halo--relic bridges, which all arise because the same {\CRIs} are involved.
The model also explains the universally flat spectrum observed at the edges of relics, provided that the {\CRP} spectrum is flat out to $\sim 2\Mpc$ from the centre.

The brightening (dimming) and spectral steepening (flattening) of the radio signal reflect recent magnetic growth (decay), gauging the dynamical state of the cluster. 
The possibility that steep GHs are young mergers associated with recent or irregular magnetic growth is tested, by showing that they are preferentially associated with nearby relics.

The homogeneous {\CRI} distribution we infer on cluster scales, and the strong, evolving magnetic fields required to explain the various radio sources, bear several implications for the energy budget in clusters, their nonthermal emission in radio and other bands, and the physics of magnetisation and particle acceleration and evolution.
After briefly reviewing the model in \S\ref{sec:Discussion}, we estimate the {\CRI} energy density $u_p$ and spectrum $s_p$, and show that the magnetic fields in several halos are consistent with $\epsilon_B\simeq \constant$, on the order of $10\%$.
We show that assuming that $u_p$ and $\epsilon_B$ are universal constants in GHs, approximately reproduces the $P_\nu\till L_X$ and $P_\nu\till R_\nu$ correlations observed, where $R_\nu$ is the halo size.

We show that the {\CRI} distribution can be explained by particle acceleration in strong shocks: SNe shocks, the virial shock, or a combination of both.
The homogeneous {\CRI} distribution then requires that either the diffusion of {\CRIs} is sufficiently strong and their escape from the cluster is quenched, or that gas mixing is highly efficient.
Various implications of our results are discussed, in particular the connections between the different radio sources, and additional hadronic signals.
As the magnetic fields are found to be strong in both halos and relics, we impose strict upper limits on the efficiency of particle acceleration in weak shocks and turbulence.

\subsection{Central argument}

The diffuse (not associated with any local source) radio emission observed from the ICM, in its difference forms (flat and steep GHs, MHs, relics, and halo--relic bridges), can be explained as synchrotron emission from secondary {\CREs}, produced by hadronic collisions involving {\CRIs} with a flat spectrum ($s_p=-2.20\pm0.05$) and with a homogeneous distribution, with energy density in the range $u_p\simeq 10^{-\range{12.4}{13.3}}\erg\cm^{-3}$, provided that the spatial and temporal variations of the magnetic field are taken into account.

This model resolves the present puzzles outlined in \S\ref{sec:ModelProblems}, reproduces the morphology, spectra and energetics of flat spectrum halos (for example in \Figs~\ref{fig:ProfilesA2163}--\ref{fig:ProfilesA665B}), and explains the spectral and morphological properties of relics and halo--relics bridges (see \S\ref{sec:ModelApplications}).
Taking into account the magnetic amplification by shocks, the model also reproduces the brightness of relics (compare \Figs~\ref{fig:SourcesEtaN} and \ref{fig:SourcesEtaNModelM2}, before and after accounting for $\epsilon=4\%$ shock magnetisation).
Interpreting steep spectrum GHs as young mergers, based on their association with nearby relics (see \Fig~\ref{fig:HaloAlphaVsRelicR}), explains their spectral steepening as arising from an increasing level of magnetic turbulence, in particular if {\CRE} diffusion is strong (see \S\ref{sec:ModelTest_RelicGHConnection}).

\subsection{Paper layout and definitions}

The paper is organised as follows.
In \S\ref{sec:ModelProblems} we discuss some peculiarities and inconsistencies of present halo and relic models.
In \S\ref{sec:HaloAndRelicEta} we analyse halo and relic observations, present the phenomenological evidence for a homogeneous {\CRI} distribution, and show that a time-dependent model is required in order to explain the morphologies and spectra observed.
A time-dependent model is derived in \S\ref{sec:TimeDependentTheory}, generalising secondary {\CRE} emission for the case where the magnetic field and {\CRE} injection evolve rapidly, in particular near a weak shock and under turbulent conditions.
The model is then tested against observations in \S\ref{sec:ModelApplications}, both among different sources and within well-studied halo clusters that harbour shocks or relics. Here we derive $\epsilon_B$ for relics, discuss the spectral steepening and curvature, and show the association between steep GHs and nearby relics.
In \S\ref{sec:Discussion}, we discuss the model and its implications.
In particular, we compute the energy density $u_p$ and the spectral index $s_p$ of the {\CRIs}, and outline the implications of their homogeneous distribution.
Also discussed are the implied constraints on {\CRI} diffusion, primary {\CRE} acceleration in shocks and turbulence, and additional hadronic signals.
Finally, our analysis and results are briefly summarised in \S\ref{sec:Summary}.
Supporting computations for the model are provided in Appendices \ref{sec:transient_B_evolution}--\ref{sec:FiniteSpectrumAndBeam}.

Considering the observationally-driven structure of the paper, the reader may wish to skip the observational motivation for the model in \S\ref{sec:ModelProblems} and \S\ref{sec:HaloAndRelicEta}, and the analysis of variable fields and injection in \S\ref{sec:TimeDependentTheory}, and proceed directly to the description of the model and its application to observations in \S\ref{sec:ModelApplications}, or to its review and discussion in \S\ref{sec:Discussion}.
Alternatively, the summary in \S\ref{sec:Summary} provides references to all the results derived in the text.

We assume a concordance $\Lambda$CDM model with dark matter fraction $\Omega_M=0.26$, baryon fraction $\Omega_b=0.04$, and a Hubble constant $H=70\km\se^{-1}\Mpc^{-1}$.
Error bars are $1\sigma$ confidence intervals, unless otherwise stated.
The main parameters used in the study are defined in Table \ref{tab:Parameters}.
We use the term cosmic-ray proton ({\CRP}) instead of {\CRI} when discussing processes in which a proton plays an individual role, for example when describing the high energy spectrum.


\section{Present halo and relic models: puzzles and inconsistencies}
\label{sec:ModelProblems}

Several observations are inconsistent with, or unexplained by, the present halo and relic models.
These discrepancies and coincidences --- manifestations of the same physical process, as we shall see --- indicate that the present models are, at best, incomplete.
In order to better understand the present models and their limitations, and to motivate the search for a more successful model, we now review these observational clues.

\subsection{Giant halos with a steep spectrum}
\label{sec:SteepHalos}

In the secondary {\CRE} model for halos (both GHs and MHs), the {\CREs} are produced through hadronic collisions, and their energy spectrum closely reflects the spectrum of their parent {\CRPs}.
This primary {\CRP} spectrum is uncertain, but thought to be well approximated by a power-law of index $s_p\equiv d\log n_p/d\log E_p =-\range{2.0}{3.0}$ in the relevant, $E_p\sim$few--$100\GeV$ proton energy range (see {\KL} for a discussion).
For slowly evolving magnetic fields, the corresponding radio spectrum is roughly $\alpha\sim s_p/2=-(1.0$--$1.5)$, as observed in flat halos.
Here, $\alpha=-1$ ($s_p=-2$) implies equal energy per logarithmic interval in photon (proton) energy.

Recent years saw increasing evidence for the existence of GHs with a steep spectrum, where $\alpha<-1.5$.
An extreme example is the halo in A521, where the spectral index in the frequency range $330\MHz$--$1.4\GHz$ was recently shown to be $\alpha_{0.3}^{1.4}=-1.86\pm0.08$ \citep{DallacasaEtAl09}.
(We shall henceforth use this $\alpha_{\nu_1}^{\nu_2}$ notation to represent the spectrum between frequencies $\nu_1$ and $\nu_2$ measured in GHz.)
There are currently six known steep halos --- in A521, A697, A754, A1300, A1914, and A2256; their parameters are summarised in Table \ref{tab:ClusterData}.
This sample of steep GHs constitutes a minority  --- less than $20\%$ --- of all GHs, as at least $31$ GHs \citep[see][]{GiovanniniEtAl09} are currently known.
However, halo observations are usually selected based on high frequencies maps, so steep halos could in principle be much more common than revealed by the present data.
Future low frequency studies with MWA\footnote{http://www.mwatelescope.org}, LOFAR\footnote{http://www.lofar.org}, and SKA\footnote{http://www.skatelescope.org} are expected to discover many more halos, and would better estimate the steep fraction.

It was recently claimed that such steep GHs cannot arise from secondary {\CREs}, because this would require a primary {\CRI} population with unrealistically large energy and steep spectrum \citep{BrunettiEtAl08, DallacasaEtAl09, Brunetti09}.
Indeed, present secondary {\CRE} models must be revised if they are to explain halos with $\alpha\lesssim -1.5$.

A closely related phenomenon, which challenges secondary {\CRE} models, involves the strong spectral steepening observed as a function of frequency in some GHs.
Examples include Coma \citep[A1656; $\alpha_{0.3}^{1.4}\simeq -1.16$ and $\alpha_{1.4}^{4.8}\simeq -2.3$;][]{GiovanniniEtAl09}, A2319 \citep[$\alpha_{0.4}^{0.6}\sim -0.92$ and $\alpha_{0.6}^{1.4}\sim -2.2$;][]{FerettiEtAl97}, and A3562 \citep[$\alpha_{0.3}^{0.8}=-1.3\pm0.2$ and $\alpha_{0.8}^{1.4}=-(1.9$--$2.3)$;][]{VenturiEtAl03, GiacintucciEtAl05}.
Such steepening exceeds the weak spectral variations expected in secondary {\CRE} models due to the energy dependence of the cross section for secondary production, unless $s_p<-3$ (see {\KL}).
The thermal Sunyaev-Zeldovich (SZ) effect \citep{ZeldovichSunyaev69, SunyaevZeldovich80} contributes to such steepening, and was suggested as a possible explanation for the spectrum of halos such as Coma \citep{Ensslin02}.
However, it was recently argued that the SZ effect is not sufficiently strong to account for the substantial steepening observed \citep{Brunetti04,DonnertEtAl10A}.

\subsection{Spectral peculiarities of relics}
\label{sec:PeculiarRelics}

Detailed spectral maps obtained recently pose a challenge for relic models, as they do for halos.
One challenge here is to explain why some relics show substantial spectral steepening inward of the relic, while others do not.
Another difficulty involves several relics, found in different clusters and radii (\ie distance from the cluster centre), all showing a nearly identical, flat spectrum, which is unnatural in a primary {\CRE} model.

All relic models so far agree that due to {\CRE} cooling, the radio spectrum should gradually steepen away from the shock \citep[\eg][]{GiacintucciEtAl08}; in most cases this implies steepening with decreasing radius.
Indeed, in several cases (A521, A3667, A1240, A2256, A2345), the radio spectrum was found to vary significantly across the relic, being flatter ($\alpha\simeq -1$) at the outer rim and steeper ($\alpha< -1.5$) towards the cluster's centre.
If the spectrum at the outer rim is a pure power-law with index $\edalpha$, one expects the spectrum of the integrated radio signal to be steeper, $\avalpha=\edalpha-1/2$, due to {\CRE} cooling.
This appears to be the case for example in A521, where $\edalpha_{0.3}^{0.6}=-1.05\pm0.05$, but inward of the rim the spectrum dramatically steepens to values $<-2.5$, leading to an average $\avalpha_{0.2}^{4.9}=-1.48\pm0.01$ spectrum \citep{GiacintucciEtAl08}.

However, some relics show only a modest \citep[\eg A2744; ][]{OrruEtAl07} or no \citep[\eg A2163; ][]{FerettiEtAl04a} steepening.
For example, the textbook relic A1253+275 in Coma shows mild steepening, with $\edalpha_{0.6}^{1.4}\simeq -1.0$ at the outer rim and $\alpha_{0.6}^{1.4}=-(1.0$--$1.4)$ at smaller radii \citep{GiovanniniEtAl91}.
This leads to an integrated, pure power-law spectrum with index $\avalpha_{0.15}^{4.7}=-1.18\pm0.02$ \citep{ThierbachEtAl03}, flatter than the $\avalpha\simeq-1.5$ spectrum anticipated from the above arguments.
A clue to the difference between these spectral behaviours is the distance of the relics from the centre of their clusters; all steepening relics are found at radii $r<1\Mpc$, whereas little or no steepening is found in relics beyond $r=1\Mpc$.

Considering the observational difficulty of integrating the entire diffuse, weak radio emission, a useful measure of the relic spectra is the (usually) flatter spectrum $\edalpha$ at the outer edge.
All models associate relics with shocks, so $\edalpha$ is the most pristine measure of the spectrum just behind (\ie downstream of) the shock.
It was measured for a handful of relics, by binning the radio maps radially.
The spectra of the outermost bins are summarised in Table \ref{tab:RelicSpec}; all the relic agree with the range $\edalpha=-\range{0.95}{1.10}$.
These measurement can be supplemented by the edge spectrum in relics that show little spatial variations, such that $\edalpha$ can be estimated from the uniform spectral map or from the computed $\avalpha$.
Examples include Coma \citep[$\edalpha_{0.6}^{1.4}\simeq -1.0$ typically found by][]{GiovanniniEtAl91}, A2163, and A2744 (see Table \ref{tab:RelicSpec}).

The narrow distribution of the edge spectral indices $\edalpha\simeq -1$ among the different relics shown in Table \ref{tab:RelicSpec} is unnatural or inconsistent with the present relic models, because they either cannot explain the pure power-law spectrum observed (radio Phoenix model), or attribute no special significance to the value $\edalpha=-1$ (DSA models), as we now show.

\begin{table*}
 \caption{Spectrum of radio emission from the outer rim of relics.}
 \label{tab:RelicSpec}
 \begin{tabular}{@{}|lcccc|}
  \hline
  Cluster & Relic position & Outer spectrum & Reference & Shock Mach number \\
  name & (or name) &    &  & according to DSA \\
  \hline
    A521 & Southeast & $\edalpha_{0.3}^{0.6}=-1.05\pm0.05$ & \citet[][figure 7]{GiacintucciEtAl08} & $\mach=2.15_{-0.07}^{+0.08}$ \\
    A1240 & North (A1240-1) & $\edalpha_{0.3}^{1.4}=-1.09\pm0.17$ & \citet[][figure 8]{BonafedeEtAl09Relics}  & $\mach=2.10_{-0.19}^{+0.31}$ \\
    A1240 & South (A1240-2) & $\edalpha_{0.3}^{1.4}=-1.10\pm0.15$ & \citet[][figure 8]{BonafedeEtAl09Relics}  & $\mach=2.08_{-0.17}^{+0.25}$ \\
    A2345 & East (A2345-2) & $\edalpha_{0.3}^{1.4}=-1.06\pm0.08$ & \citet[][figure 3]{BonafedeEtAl09Relics}  & $\mach=2.14_{-0.11}^{+0.13}$ \\
    A2256 & North & $\edalpha_{1.4}^{1.7} = -0.95\pm0.15$ & \citet[][figure 5]{ClarkeEnsslin06}  & $\mach=2.33_{-0.25}^{+0.44}$ \\
  \hline
    A1656 & West & $\edalpha_{0.6}^{1.4}\simeq -1.0$ & \citet{GiovanniniEtAl91}  & $\mach\simeq 2.24$ \\
    A2163 & Northeast & $\avalpha_{0.3}^{1.4}=-1.02\pm 0.04$ & \citet{FerettiEtAl04a}  & $\mach=2.20_{-0.06}^{+0.07}$ \\
    A2744 & Northeast & $\edalpha_{0.3}^{1.4}=-0.9\pm0.1$ & \citet{OrruEtAl07}  & $\mach=2.45_{-0.21}^{+0.32}$ \\
  \hline
 \end{tabular} 

 \medskip
    \raggedright
    The estimated outer spectral index $\edalpha$ is shown for each relic, along with the corresponding shock Mach number $\mach$ according to DSA theory.
    Quoted values of $\edalpha$ give the spectrum of the outermost, flattest bin extracted from published spectral profiles (upper rows), or otherwise estimated for relics with little spectral variability (bottom rows).
    We exclude the exceptionally flat, non-power-law spectrum of the relic in A2255 \citep{PizzoDeBruyn09}, which cannot be modeled in a DSA context, and the West relic in A2345 \citep[A2345-1;][]{BonafedeEtAl09Relics}, which has an unclear projected geometry and shows inward flattening, rather than steepening.
\end{table*}

In some relics, $\avalpha$ was found to be very well fit by a pure power-law, spanning a wide frequency range, with no evidence for spectral curvature.
The best examples are Coma \citep[$\avalpha_{0.15}^{4.7}=-1.18\pm0.02$;][]{ThierbachEtAl03} and A521 \citep[$\avalpha_{0.2}^{4.9}=-1.48\pm0.01$;][]{GiacintucciEtAl08}.
The lack of spectral curvature rules out the radio Phoenix model as a plausible explanation for such relics \citep[see][]{GiacintucciEtAl08}.
Henceforth we focus on the shock acceleration model.

In the DSA model, the outer rim of the relic is identified with a shock, where {\CREs} with a power-law energy distribution, $N_e\propto E_e^{s_e}$, are injected into the downstream plasma.
The injection spectrum is determined solely by the shock compression ratio, or --- henceforth assuming an adiabatic index $\Gamma=5/3$ --- by the Mach number $\mach$ of the shock \citep{Krymskii77, AxfordEtAl77, Bell78, BlandfordOstriker78},
\begin{equation}
s_e = -2 \frac{\mach^2+1}{\mach^2-1} \fin
\end{equation}
This fixes the radio spectrum at the outer rim of the relic, where {\CREs} had no time to cool, through $\edalpha=(s_e+1)/2$ \citep[\eg][]{RybickiLightman86}.
Consequently,
\begin{equation} \label{eq:edalpha}
\edalpha = -\frac{3+\mach^2}{2(\mach^2-1)} \fin
\end{equation}
{\CREs} farther inward of the rim had progressively more time to cool, so the spectrum steepens.
If all the radio emission is accounted for, the integrated spectrum $\avalpha$ becomes
\begin{equation}
\avalpha= \edalpha-\frac{1}{2} = -\frac{\mach^2+1}{\mach^2-1} \fin
\end{equation}

One can invert \eq{\ref{eq:edalpha}} and find the shock Mach numbers corresponding to the $\edalpha\simeq -1$ measurements.
As shown in Table \ref{tab:RelicSpec}, this yields Mach numbers in the range $\range{2.1}{2.3}$ for the five relics with measured $\edalpha$ (and similarly for A2163, A2744, and Coma).
This range is alarmingly narrower than expected.
Numerical simulations show a wide distribution of shock Mach numbers, in the range $1$--$3$ at the radii characteristic of relics, with rare shocks as strong as $\mach=3.5$ \citep[][and references therein]{VazzaEtAl10}.
Moreover, simulations predict a radial dependence of the Mach number distribution \citep[\eg][]{VazzaEtAl10}.
The above relics span a wide range of radii, $\sim \range{0.5}{2}\Mpc$, and yet show very similar $\mach$ according to the DSA model.
Finally, X-ray data suggests that some of these relics are associated with shocks stronger \citep[\eg A521, see ][]{GiacintucciEtAl08} or weaker \citep[\eg Coma, see ][]{FerettiNeumann06} than the $\mach=2.2$ values inferred under the DSA assumption.

Finally, we stress that in addition to the observational discrepancies demonstrated above, the DSA relic model is of limited use here because particle acceleration in \emph{weak} shocks is neither theoretically understood, nor observationally constrained.
DSA theory, while successfully reproducing the spectrum in strong shocks, does not self-consistently compute the acceleration efficiency.
The efficiency is well constrained from observations only for strong shocks, for example in supernova remnants \citep[for a discussion, see][]{KeshetEtAl03}.
Only loose upper limits on the acceleration efficiency of weak shocks are available, on the order of $\epsilon_e\sim 10\%$ of the thermal downstream energy \citep{NakarEtAl08}.
In the absence of a calibrated model for weak shocks, DSA relic models utilise ad-hoc prescriptions for the efficiency, which probably depends non-trivially on the plasma parameters and the shock strength.
Note that similar comments can be made regarding particle acceleration in turbulence, for example as the putative source of {\CREs} in radio halos.


\subsection{Halo--relic connections}
\label{sec:HaloRelicConnection}

There is still no widely accepted agreement regarding the origin of {\CREs} in neither halos (secondary vs. primary {\CREs}) nor relics (acceleration vs. reacceleration of primary {\CREs}, as so far believed).
Nevertheless, so far all models agree that the two phenomena involve {\CREs} generated in \emph{different} physical processes.
Here we point out several connections between GHs and relics, indicating that the two phenomena are more intimately related than is plausible in a framework with different {\CRE} injection mechanisms.
These connections suggest, as we confirm in \S\ref{sec:HaloAndRelicEta}, that the {\CREs} in halos and relics in fact share a common origin.

\begin{enumerate}
\vspace{-2mm}

\item
In clusters that harbour both a halo and a relic, a faint radio bridge is sometimes observed to connects the two, for example in Coma, A2255, and A2744 \citep{GiovanniniFeretti04}.
The spectrum of the bridge is similar or somewhat steeper than in the halo and in the relic \citep{KimEtAl89, PizzoDeBruyn09, OrruEtAl07}.
The low surface brightness of the radio sources combined with anecdotal evidence, such as an alignment between the relic position and the halo elongation axis in A2163 \citep[see ][]{FerettiEtAl04a}, suggest that observational limits and projection effects hide many such bridges.
The origin of the bridges is unknown; in present models they would require a fine-tuned spatial interpolation between the halo and relic {\CRE} populations.

\item
Although relics --- and not halos --- are thought to be generated by shocks, weak shocks were discovered at the edges of several GHs.
(In fact, so far no relic shock has been confirmed, due to the low ambient density).
This includes confirmed shocks in 1E 0657--56 \citep[the bullet cluster;][]{MarkevitchEtAl02}, A520 \citep{MarkevitchEtAl05} and A754 \citep{KrivonosEtAl03}, and suspected shocks in A665 \citep{MarkevitchVikhlinin01}, A2219 \citep{MillionAllen09}, and Coma \citep{BrownRudnick10}.

Assuming that as these weak shocks travel outward and disconnect from the halo they become relics, further exacerbates the problem in interpreting the two phenomena with different {\CRE} injection mechanisms.
With the present data, this would also require a fine-tuned temporal interpolation between the two phenomena.

Incidentally, notice that while a weak outgoing shock at the very edge of a GH is consistent with the secondary {\CRE} model (where shock magnetisation is sufficient to generate the halo), it is inconsistent with primary {\CRE} models: strong turbulence is not expected in the shock region \citep{GovoniEtAl04}, and even if it was, turbulent acceleration would not be effective immediately behind the shock.

\item
A telling and somewhat surprising halo-relic connection is the apparent coincidence between the presence of a steep spectrum GH in a cluster, and the presence of a nearby relic.
All six steep GHs discussed in \S\ref{sec:SteepHalos} either have a nearby relic \citep[A521, A754, A1300, A2256; see][]{GiovanniniEtAl09}, or have an irregular morphology containing a relic-like filamentary protrusion \citep[filaments West of A697 and Southwest of A1914;][]{VenturiEtAl08, BacchiEtAl03}.
This incidence rate of relics is significantly higher than found amongst flat spectrum GHs: only five ($20\%$) out of the 25 flat halo clusters reviewed in \citet{GiovanniniEtAl09} harbour a relic.
As we show in \S\ref{sec:ModelApplications}, there appears to be a bimodality in the distance of halo-cluster relics: relics in steep GH clusters are found at $r\lesssim 1\Mpc$, whereas relics in flat GH clusters are more distant.
This behaviour is reminiscent --- and indirectly related --- to the excessive steepening behind central relics mentioned in \S\ref{sec:PeculiarRelics}.

\item
Halos and relics share many traits. To some extent, the observationally-driven classification of diffuse radio emission into halos and relics is artificial and blurred, with some halos showing relic features and vice versa.

\begin{enumerate}
\vspace{-2mm}

\item
Halos and relics have a similar spectrum, in particular if we restrict ourselves to the central part of the halo and the external edge of the relic.
This can be seen, for example, in the spectral maps of A2744 \citep{OrruEtAl07} and A2163 \citep{FerettiEtAl04a}.
In the context of present halo and relic models, this is a highly peculiar coincidence.

\item
There are telling exceptions to the polarisation classification scheme described in \S\ref{sec:Introduction}.
So far, strong polarisation was detected in one GH (at a $20\%$--$40\%$ level, in A2255; see \citet{GovoniEtAl05}; intermediate polarisation, $2-7\%$ on average, was found in MACS J0717.5 +3745; see \citet{BonafedeEtAl09}), and in one MH \citep[at a $10\%-20\%$ level, in A2390; ][]{BacchiEtAl03}.
Some relics show a low polarisation level, for example $\sim 2\%$ was reported in A133 \citep{SleeEtAl01}; however this may be an AGN relic \citep{KempnerEtAl04}.
These exceptions suggest a continuous distribution of polarisation properties among halos and relics.

\item
Some GHs are irregular, showing a clumpy or filamentary morphology. Examples include RXC\,J2003.5–-2323 \citep{GiacintucciEtAl09}, A2255, and A2319 \citep{MurgiaEtAl09}.
Morphologically, they could be interpreted as an ensemble of relics, as recently suggested for A2255 \citep{PizzoEtAl10}.

\item
Some GHs show spectral steepening with increasing radius, resembling that observed in relics.
This can be seen, for example, in the spectral maps of A665, A2163 \citep{FerettiEtAl04a}, A2219, and A2744 \citep{OrruEtAl07}.

\end{enumerate}

\item
A correlation has been reported \citep{GiovanniniFeretti04} between the total $1.4\GHz$ radio power of relics and the bolometric X-ray luminosity of their host clusters, somewhat reminiscent of the radio--X-ray correlation found in GHs.
Recall that the strong correlation observed in halos provides a strong evidence that they arise from secondary {\CREs} \citep[][{\KL}]{KushnirEtAl09}.

\end{enumerate}

The above connections between halos and relics suggest that the two phenomena arise from the same population of {\CREs}.
However, no unified model has been proposed thus far for halos and relics.
There is evidence, preliminary in relics but strong in GHs and in MHs (and so, by proxy, also in relics), that secondary {\CREs} are involved.

In spite of the halo-relic connection, we do not find significant evidence for an enhanced incidence rate of halo/relic detection in relic/halo clusters.
For example, 11 out of the 31 ($\sim 35\%$) GHs summarised in \citet{GiovanniniEtAl09} also harbour at least one relic, whereas about 7 out of the 30 ($23\%$) relics reported in \citet{GiovanniniFeretti04} are found in halo clusters.
These incidence rates are similar to the unconditional halo and relic detection rates in X-ray bright clusters \citep{GiovanniniEtAl02}.
This behaviour could arise, for example, if halos are much more long-lived than the time scale during which a relic is detectable.
A careful analysis of the selection effects involved is necessary in order to quantify a correlation between the presence of halos and relics in a cluster, or the lack thereof.


\section{Relic and halo phenomenology: both arise from the same, homogeneous {\CRI} distribution}
\label{sec:HaloAndRelicEta}

\begin{figure*}
\centerline{\epsfxsize=16cm \epsfbox{\myfarfig{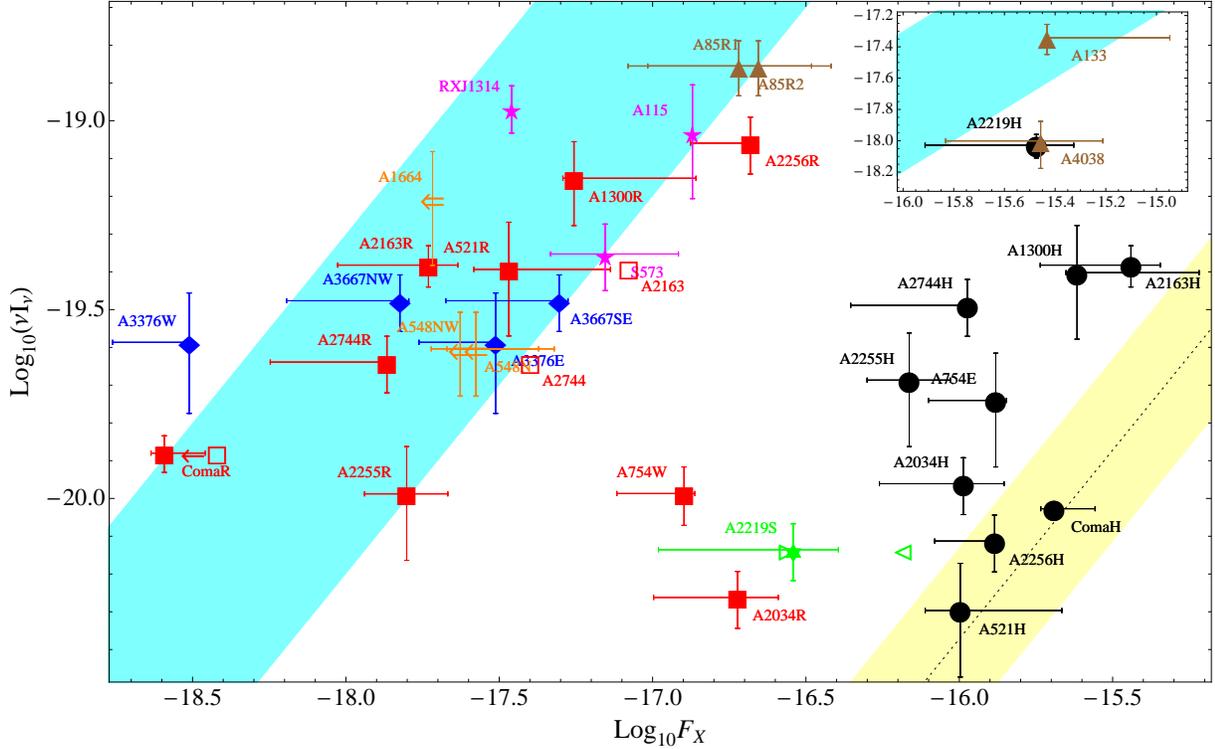}}}
\caption{
Peak $\nu\simeq 1.4\GHz$ radio brightness $\nu I_\nu$ vs. coincident $0.1$--$2.4\keV$ X-ray brightness $F_X$ of radio halos (black disks)
and relics.
\newline
The radio-to-X-ray brightness ratio $\eta\equiv \nu I_\nu/F_x$ of most halos (all but A2219 found in relic clusters) is somewhat larger than the best fit derived for halos by {\KL} (in a partly overlapping sample of GHs, mostly residing in non-relic clusters; dashed line with yellow shaded band showing the $1\sigma$ dispersion).
Relics show much higher $\eta$; $75\%$ of them lie within $-2.2<\log_{10}\eta<-1.3$ (cyan shaded region).
Note that the classification of the relics in A754 and A2034 is highly uncertain (see \S\ref{sec:SourceSelection}), and A2219 is strongly contaminated by point sources (see \S\ref{sec:DataPreparation}); the bimodality in $\eta$ between halos and relics is significant without these three sources.
\newline
Following \citet{GiovanniniFeretti04}, relics are classified as found in halo clusters (red filled squares), found in double relic systems (blue diamonds), found near the first-rank galaxy (brown up-triangles), circular peripheral relics (orange double arrows), and classical elongated relics (magenta five-pointed stars).
\newline
The radio data (vertical error bars) show the values of the two brightness contours bracketing the radio peak in published, $\sim 1.4\GHz$ contour maps, and their mean.
The X-ray brightness is based on $\beta$-models from the literature.
In order to avoid excessive error propagation, the $F_X$ (horizontal) error bars reflect the $1\sigma$ standard deviation of $n_0^2$, but not of $\beta$ or $r_c$ (see \S\ref{sec:DataPreparation}).
The $\beta$-model is supplemented by some measurements (red empty squares) of $F_X$ (in A2163 and A2744; see \S\ref{sec:EtaInWellStudiedGHs}), or an $F_X$ upper limit \citep[in Coma;][]{FerettiNeumann06}.
The data used to produce the figure are summarised in Table \ref{tab:ClusterData}.
\newline
Also shown is a possible shock region in A2219, reported by \citet{MillionAllen09}, with $F_X$ measured upstream and downstream (green right and left triangles), and as inferred from the $\beta$-model (green six-pointed star).
\newline
Inset: particularly bright radio sources.
\label{fig:SourcesInuIx}
\vspace{2mm}}
\end{figure*}

The preceding discussion, in particular the connections between radio halos and relics outlined in \S\ref{sec:HaloRelicConnection}, motivates a unified exposition of cluster radio sources.
We thus begin by showing in Figure \ref{fig:SourcesInuIx} a sample of various types of halos and relics, presented in the phase space of maximal radio brightness $\nu I_\nu$ vs. coincident X-ray brightness $F_X$.

Roughly speaking, $\nu I_\nu$ is proportional to the product of the energy densities of {\CREs} and magnetic fields,
whereas $F_X$ is proportional to the square of the plasma density.
The data used to produce this and the following figures were extracted from the literature, as summarised in Table \ref{tab:ClusterData}.
The data preparation and source selection and classification are described below in \S\ref{sec:DataPreparation} and \S\ref{sec:SourceSelection}.

Notice the similarity --- to order of magnitude, at least --- in the peak $I_\nu$ values of the relics and halos shown in \Fig~\ref{fig:SourcesInuIx}.
In particular, the peak brightness of halos and relics found in the same cluster appear to be somewhat correlated.
Could these subtle connections reflect a unified {\CR} origin in these clusters?

We begin answering this question by showing in \S\ref{sec:EtaDiagnostic} that the dimensionless ratio $\eta$ between $\nu I_\nu$ and $F_X$ provides a useful diagnostic of the nonthermal plasma.
In GHs, where the radio emission is associated with secondary {\CREs} and strong magnetic fields, $\eta$ is a direct measure of the local {\CRI} fraction (in regions of slowly varying magnetic fields).

Next, we present in \S\ref{sec:HaloRisingEta} evidence for a universal, radially rising $\eta(r)$ profile in GHs, and reconcile it with previous reports of uniform $\eta(r)$ in some clusters.
The scaling $\eta(r)\propto (n/n_0)^{-1}$ we derive indicates that the {\CRIs} are uniformly distributed in GH clusters, even beyond the $\sim 0.5\Mpc$ scales illuminated by strong magnetic fields.

Finally, in \S\ref{sec:eta_in_relics} we combine the data of halos and relics, and derive a universal $\eta(n/n_0)$ profile extending out to large, $r\gtrsim 2\Mpc$ radii.
The data motivate a unified model, in which halos and relics both arise from secondary {\CREs}, produced from the same homogeneous distribution of primary {\CRIs}.
This model resolves several of the previous model discrepancies outlined in \S\ref{sec:ModelProblems}.
Addressing the remaining discrepancies and the spectral properties of relics requires a generalised model, involving time-dependent {\CRE} injection and dynamic magnetic fields, derived in \S\ref{sec:TimeDependentTheory}.

\subsection{Data reduction}
\label{sec:DataPreparation}

For a handful of halo and relic clusters, detailed radio and X-ray brightness maps can be found in the literature.
In a small subset of these clusters, detailed maps of the radio spectrum are also available.
This includes the GH in A665, and the clusters A2163 and A2744 which harbour both a halo and a relic.
The combination of surface brightness and spectral maps provides strong constraints on the radio model and on the plasma parameters.
We examine and simultaneously model the brightness and spectral profiles in these clusters;
here we focus on the brightness maps of the GHs, and defer modeling the relics and the spectral variations to \S\ref{sec:ModelApplications}, where the description of time-dependent {\CRs} and magnetic fields developed in \S\ref{sec:TimeDependentTheory} is incorporated.
We do not discuss the published radio maps of A2219 \citep{OrruEtAl07} and A2255 \citep{PizzoDeBruyn09}, because
the halo in A2219 is strongly contaminated in the central $300\kpc$ by a blend of radio galaxies \citep{OrruEtAl07},
and the GH in A2255 is highly irregular and filamentary, and could be identified as an ensemble of relics rather than a halo \citep{PizzoEtAl10}.

Unfortunately, in most clusters, only low resolution contour maps and the integrated properties of the radio sources (power, spectrum, polarisation) are available.
We choose to examine the surface brightness, rather than the integrated luminosity of each source, in order to obtain a local measure of the plasma which is less sensitive to background and resolution effects.
However, the radio sources are diffuse and extend over large, sometimes $>1\Mpc$ scales, thus spanning a wide range of surface brightness.
In order to have a simple, yet well-defined prescription for assigning a brightness to each source, we identify it with the position (line of sight) of \emph{maximal radio brightness}.
This choice has additional advantages.
For example, in relics, it is likely that the magnetic field is maximal or the projection is most favourable at peak radio brightness, simplifying the analysis considerably, as discussed in \S\ref{sec:SecondaryCREsInRelics} and \S\ref{sec:Discussion}.

For each radio source, we define an uncertainty range $\{I_\nu^{min},I_\nu^{max}\}$ of \emph{peak} radio brightness (vertical error bars in \Fig~\ref{fig:SourcesInuIx}) such that $I_\nu^{min}$ is the value of the brightest contour found in the radio map of the source, and $I_\nu^{max}$ is the value of the next, putative contour (not found in the map).
The $I_\nu$ value assigned to each source (shown by symbols) is the arithmetic mean of these lower and upper limits.
This introduces some nonphysical error; in all cases it is less than a factor of $1.5$.
We use $\nu I_\nu$ as a measure of the radio brightness (as the ordinate in \Fig~\ref{fig:SourcesInuIx}) because it varies weakly with frequency when the spectrum is nearly flat, $\alpha\simeq -1$.
Nevertheless, due to some frequency dependence and for consistency with previous work, we only use data in frequencies around $1.4\GHz$ (within $15\%$ in $\nu$); see Table \ref{tab:ClusterData} for details.

Radio relics are often found at large distances from the cluster's centre, where the X-ray emission of the cluster becomes too faint to be distinguished from the background.
Therefore, we compute $F_X$ by extrapolating the bright, central X-ray emission out to the radio source position,
using an isothermal $\beta$-model \citep{CavaliereFusco76} for each cluster taken from the literature.
In these models, the electron number density varies as a function of radius $\vecthree{r}$ according to
\begin{equation}
n(\vecthree{r})=n_0 \left(1+\frac{r^2}{r_c^2}\right)^{-\frac{3}{2}\beta} \coma
\end{equation}
where $\beta$, $r_c$ and $n_0$ are the free parameters of the model (see Table \ref{tab:ClusterData} for individual model references), and $r\equiv|\vecthree{r}|$.
We identify the X-ray peak as the centre of the cluster.
A distance range $\{r_{min},r_{max}\}$ is assigned to each radio source according to the minimal, maximal distance of the $I_\nu^{min}$ contour from the cluster's centre.
The value $r$ associated with each symbol in \Fig~\ref{fig:SourcesInuIx} is taken as the arithmetic mean of these lower and upper limits.

The $\beta$-model extrapolation introduces inevitable errors, as illustrated in \Fig~\ref{fig:SourcesInuIx} by comparing the extrapolated and the measured values of $F_X$ in a few relics.
In nonspherical clusters where the $\beta$-model fits poorly, it may produce significant errors in $F_X$.
Another source of error is an enhanced brightness observed in several relics, thought to be caused by shock compression of the plasma.
In such cases, the $\beta$-model tends to underestimate the X-ray signal, for example in A2163 and A2744 (see \Fig~\ref{fig:SourcesInuIx}).
To illustrate this effect, \Fig~\ref{fig:SourcesInuIx} also shows (as empty triangles) $F_X$ both upstream and downstream of a suspected shock front in A2219 \citep[within the GH and outside the central, contaminated region; see][]{MillionAllen09}.
Nevertheless, the $\beta$-model extrapolation errors are much smaller than the three orders of magnitude in $F_X$ spanned by the data; in most cases where $F_X$ is known, the error is less than a factor of 3.

We estimate the confidence intervals arising from the $\beta$-model uncertainties by adopting the largest propagated uncertainty due to any single one of the three parameters of the model, $n_0$, $r_c$, or $\beta$.
As the correlations between the uncertainties of these parameter are highly correlated, this appears to be the most reliable estimator of the error in the absence of covariance matrices.
Notice that our estimate could be either larger or smaller than the true error.
Propagating all three parameter uncertainties by assuming that they are, for example, uncorrelated, would probably spuriously increase the confidence intervals.

For consistency with previous work, we use $F_X$ exclusively in the energy range $0.1$--$2.4\keV$.
Emission in this band is dominated by thermal bremsstrahlung, and has the advantage (for our present purpose) of being weakly dependent upon temperature $T$ and metallicity $Z$ in the relevant parameter range.
The X-ray emissivity in these energies, calculated using the MEKAL model
\citep{MeweEtAl85, MeweEtAl86, Kaastra92, LiedahlEtAl95} in XSPEC v.12.5 \citep{Arnaud96}, is well-fit by
\begin{equation} \label{eq:FxMEKAL}
j_{X[0.1-2.4]}(n,T,Z)\simeq 8.6\times 10^{-28}\frac{n_{-2}^2 Z_{0.3}^{0.04}}{T_{10}^{0.1}}\erg\se^{-1}\cm^{-3} \coma
\end{equation}
where $n_{-x}$ is the electron number density $n$ in units of $10^{-x}\cm^{-3}$, $T_{10}\equiv (k_BT/10\keV)$ with $k_B$ being Boltzmann's constant, and $Z_{0.3}$ is the metallicity in units of $0.3 Z_\odot$.
We use \eq{\ref{eq:FxMEKAL}} to compute $F_X$ from the $\beta$-models, assuming uniform temperature and metallicity in each cluster.
In this approximation
\begin{equation}
F_X(\vectwo{r}) = F_{X,0} \left(1+\frac{r^2}{r_c^2}\right)^{\frac{1}{2}-3\beta} \coma
\end{equation}
where the argument $\vectwo{r}$ is a two-dimensional angular vector with the centre of the cluster at $\vectwo{r}=0$, $r\equiv |\vectwo{r}|$, and
\begin{equation}
F_{X,0} = \frac{\sqrt{\pi}\Gamma(3\beta-1/2)}{\Gamma(3\beta)}r_c \, j_X(n_0,T,Z) \fin
\end{equation}

\begin{table*}
{
\begin{tabular}{|cccccccccccc|}
\hline
(1) & (2) & (3) & (4) & (5) & (6) & (7) & (8) & (9) & (10) & (11) & (12) \\ 
Name & Type & $z$ & $k_BT$ & $Z$ & $\beta$-ref & $\nu$ & $r_{min}$ & $r_{max}$ & $I_\nu^{min}$ & $I_\nu^{max}$ & $\nu$-ref \\ 
\hline
A85R1 & RG & 0.056 (F04a) & $5.9_{-0.1}^{+0.1}$  {(F04a)} & $0.36_{-0.04}^{+0.00}$  {(F04a)} & F04a & 1.4 & 444 & 451 & 8.14 & 11.40 & S01\\
A85R2 & RG & 0.056 (F04a) & $5.9_{-0.1}^{+0.1}$  {(F04a)} & $0.36_{-0.04}^{+0.00}$  {(F04a)} & F04a & 1.4 & 483 & 488 & 8.14 & 11.40 & S01\\
A115 & RI & 0.197 (G01) & $5.3_{-0.4}^{+0.4}$  {(F04a)} & $0.37_{-0.11}^{+0.10}$  {(F04a)} & G01 & 1.4 & 401 & 512 & 4.44 & 8.89 & G01\\
A133 & RG & 0.060 (F04a) & $4.1_{-0.2}^{+0.2}$  {(F04a)} & $0.43_{-0.08}^{+0.10}$  {(F04a)} & F04a & 1.4 & 42.0 & 45.0 & 254 & 396 & S01\\
A521H & H & 0.247 (B08) & $4.9_{-0.5}^{+0.5}$  {(F04a)} & $0.34_{-0.18}^{+0.20}$  {(F04a)} & F04a & 1.4 & 0 & 0 & 0.24 & 0.48 & D09\\
A521R & RH & 0.247 (B08) & $4.9_{-0.5}^{+0.5}$  {(F04a)} & $0.34_{-0.18}^{+0.20}$  {(F04a)} & F04a & 1.4 & 846 & 910 & 1.92 & 3.85 & D09\\
A548N & RC & 0.042 (F04a) & $2.9_{-0.1}^{+0.1}$  {(F04a)} & $0.30_{-0.11}^{+0.10}$  {(F04a)} & F04a & 1.4 & 416 & 549 & 1.33 & 2.22 & F06\\
A548NW & RC & 0.042 (F04a) & $2.9_{-0.1}^{+0.1}$  {(F04a)} & $0.30_{-0.11}^{+0.10}$  {(F04a)} & F04a & 1.4 & 426 & 597 & 1.33 & 2.22 & F06\\
A754E & H & 0.056 (F04a) & $9.4_{-0.3}^{+0.3}$  {(F04a)} & $0.20_{-0.05}^{+0.00}$  {(F04a)} & F04a & 1.5 & 61.0 & 167 & 0.81 & 1.62 & B03\\
A754W & RH & 0.056 (F04a) & $9.4_{-0.3}^{+0.3}$  {(F04a)} & $0.20_{-0.05}^{+0.00}$  {(F04a)} & F04a & 1.5 & 578 & 709 & 0.57 & 0.81 & B03\\
A1300H & H & 0.306 (R99) & $6.3_{-0.8}^{+0.8}$  {(F04a)} & --- & F04a & 1.4 & 0 & 1.0 & 1.89 & 3.77 & R99\\
A1300R & RH & 0.306 (R99) & $6.3_{-0.8}^{+0.8}$  {(F04a)} & --- & F04a & 1.4 & 458 & 663 & 3.77 & 6.29 & R99\\
A1664 & RC & 0.128 (G01) & $6.8_{-1.8}^{+1.2}$  {(G01)} & --- & G01 & 1.4 & 956 & 1037 & 2.96 & 5.92 & G01\\
A2034H & H & 0.151 (F04a) & $6.9_{-0.4}^{+0.4}$  {(F04a)} & $0.19_{-0.07}^{+0.10}$  {(F04a)} & F04a & 1.4 & 76.0 & 240 & 0.65 & 0.92 & G09\\
A2034R & RH & 0.151 (F04a) & $6.9_{-0.4}^{+0.4}$  {(F04a)} & $0.19_{-0.07}^{+0.10}$  {(F04a)} & F04a & 1.4 & 404 & 518 & 0.32 & 0.46 & G09\\
A2163H & H & 0.170 (F04a) & $10.3_{-0.8}^{+0.8}$  {(F04a)} & $0.15_{-0.09}^{+0.10}$  {(F04a)} & F04a & 1.4 & 0 & 231 & 2.59 & 3.33 & F04b\\
A2163R & RH & 0.170 (F04a) & $10.3_{-0.8}^{+0.8}$  {(F04a)} & $0.15_{-0.09}^{+0.10}$  {(F04a)} & F04a & 1.4 & 1258 & 1345 & 2.59 & 3.33 & F04b\\
A2219H & H & 0.226 (O07) & $9.5_{-0.4}^{+0.6}$  {(C06)} & $0.25_{-0.06}^{+0.10}$  {(W00)} & F04a & 1.4 & 0 & 10.0 & 55.40 & 78.40 & O07\\
A2219S & S & 0.226 (O07) & $9.5_{-0.4}^{+0.6}$  {(C06)} & $0.25_{-0.06}^{+0.10}$  {(W00)} & F04a & 1.4 & 430 & 430 & 0.43 & 0.61 & O07\\
A2255H & H & 0.080 (F04a) & $5.5_{-0.2}^{+0.2}$  {(F04a)} & $0.23_{-0.05}^{+0.00}$  {(F04a)} & F04a & 1.2 & 0 & 0 & 1.14 & 2.29 & P09\\
A2255R & RH & 0.080 (F04a) & $5.5_{-0.2}^{+0.2}$  {(F04a)} & $0.23_{-0.05}^{+0.00}$  {(F04a)} & F04a & 1.2 & 907 & 1110 & 0.57 & 1.14 & P09\\
A2256H & H & 0.046 (F04a) & $6.4_{-0.2}^{+0.2}$  {(F04a)} & $0.22_{-0.03}^{+0.00}$  {(F04a)} & F04a & 1.4 & 0 & 219 & 0.47 & 0.66 & CE06\\
A2256R & RH & 0.046 (F04a) & $6.4_{-0.2}^{+0.2}$  {(F04a)} & $0.22_{-0.03}^{+0.00}$  {(F04a)} & F04a & 1.4 & 438 & 543 & 5.28 & 7.47 & CE06\\
A2744H & H & 0.308 (O07) & $8.6_{-0.3}^{+0.4}$  {(C06)} & $0.24_{-0.05}^{+0.00}$  {(A09)} & F04a & 1.4 & 0 & 140 & 1.92 & 2.72 & O07\\
A2744R & RH & 0.308 (O07) & $8.6_{-0.3}^{+0.4}$  {(C06)} & $0.24_{-0.05}^{+0.00}$  {(A09)} & F04a & 1.4 & 1479 & 1641 & 1.36 & 1.92 & O07\\
A3376E & RD & 0.046 (B06) & $4.0_{-0.2}^{+0.2}$  {(F04a)} & $0.24_{-0.08}^{+0.10}$  {(F04a)} & F04a & 1.4 & 471 & 592 & 1.20 & 2.50 & B06\\
A3376W & RD & 0.046 (B06) & $4.0_{-0.2}^{+0.2}$  {(F04a)} & $0.24_{-0.08}^{+0.10}$  {(F04a)} & F04a & 1.4 & 1352 & 1447 & 1.20 & 2.50 & B06\\
A3667NW & RD & 0.055 (F10) & $5.9_{-0.2}^{+0.2}$  {(F04a)} & $0.24_{-0.05}^{+0.00}$  {(F04a)} & F04a & 1.4 & 1986 & 2068 & 1.98 & 2.79 & J04\\
A3667SE & RD & 0.055 (F10) & $5.9_{-0.2}^{+0.2}$  {(F04a)} & $0.24_{-0.05}^{+0.00}$  {(F04a)} & F04a & 1.4 & 999 & 1090 & 1.98 & 2.79 & J04\\
A4038 & RG & 0.026 (F04a) & $2.9_{0.0}^{+0.0}$  {(F04a)} & $0.35_{-0.05}^{+0.00}$  {(F04a)} & F04a & 1.4 & 29.0 & 39.0 & 46.50 & 93.00 & S01\\
ComaH & H & 0.023 (F04a) & $8.6_{-0.3}^{+0.3}$  {(F04a)} & $0.27_{-0.08}^{+0.10}$  {(F04a)} & V02 & 1.4 & 0 & 10.0 & 0.66 & 0.70 & K00\\
ComaR & RH & 0.023 (F04a) & $8.6_{-0.3}^{+0.3}$  {(F04a)} & $0.27_{-0.08}^{+0.10}$  {(F04a)} & V02 & 1.5 & 2080 & 2090 & 0.80 & 1.00 & G91; B10\\
RXJ1314 & RI & 0.247 (G01) & $8.7_{-0.6}^{+0.7}$  {(V02)} & --- & V02 & 1.4 & 1072 & 1121 & 6.70 & 8.94 & V02\\
S573 & RI & 0.014 (F04a) & $1.7_{-0.3}^{+0.3}$  {(F04a)} & $0.19_{-0.11}^{+0.10}$  {(F04a)} & F04a & 1.4 & 121 & 194 & 2.54 & 3.81 & S03\\
\hline
\end{tabular}
\caption{\label{tab:ClusterData} Parameters of the radio sources in our sample.}
\hfill{}
\\
\begin{flushleft}
{\bf Columns}: (1) source name (composed of the host cluster's name and optional suffix letters designating the source type or position);
(2) source type: \emph{H}--halo; \emph{RI}--isolated, elongated relic; \emph{RH}--relic in a halo cluster; \emph{RD}--relic in a double relic cluster; \emph{RC}-circular peripheral relic; and \emph{RG}--relic near the first-rank galaxy;
(3) redshift $z$ (with reference); (4) cluster temperature $k_BT$ in keV (with reference); (5) cluster metallicity $Z$ in solar units (with reference); (6) reference for the $\beta$-model of the cluster; (7) radio frequency $\nu$ in GHz; (8) and (9) possible distance range $r_{min}<r<r_{max}$ between the X-ray peak of the cluster and the radio peak of the source, in kpc; (10) and (11) possible range of the peak radio brightness, $I_\nu^{min}<I_\nu<I_\nu^{max}$, in units of $\mu\mbox{Jy arcsec}^{-2}$; and (12) reference for the radio data.
\\
{\bf Reference abbreviations}:
A09 -- \citet{AnderssonEtAl09};
B03 -- \citet{BacchiEtAl03};
B06 -- \citet{BagchiEtAl06};
B08 -- \citet{BrunettiEtAl08};
C06 -- \citet{CassanoEtAl06};
CE06 -- \citet{ClarkeEnsslin06};
D09 -- \citet{DallacasaEtAl09};
F04a -- \citet{FukazawaEtAl04};
F04b -- \citet{FerettiEtAl04a};
F06 -- \citet{FerettiEtAl06};
F10 -- \citet{FinoguenovEtAl10};
G91 -- \citet{GiovanniniEtAl91};
G01 -- \citet{GovoniEtAl01B};
G09 -- \citet{GiovanniniEtAl09};
J04 -- \citet{Johnston-Hollitt04};
K00 -- \citet{KimEtAl90};
O07 -- \citet{OrruEtAl07};
P09 -- \citet{PizzoDeBruyn09};
R99 -- \citet{ReidEtAl99};
S01 -- \citet{SleeEtAl01};
S03 -- \citet{SubrahmanyanEtAl03};
V02 -- \citet{ValtchanovEtAl02};
W00 -- \citet{White00}.
\\
\end{flushleft}
}
\vspace{0.5cm}
\end{table*}

\subsection{Source selection and classification}
\label{sec:SourceSelection}

Our sample consists of all the radio relics reported in the literature in clusters that satisfy the criteria outlined in \S\ref{sec:DataPreparation}, and all the halos found in those relic clusters.
This corresponds to all the known diffuse radio sources in clusters that \emph{(i)} harbour a relic; \emph{(ii)} have a published radio map at some frequency in the range $1.2$--$1.5\GHz$; and \emph{(iii)} have a published X-ray-based $\beta$-model of the cluster.
In total, our sample consists of 23 relics and 9 halos.
These halos are all GHs in relic clusters; GHs in non-relic clusters and MHs were analysed separately in {\KL}.
In addition, we include here the non-relic cluster A2219, where a suspected shock is embedded within the GH \citep[source denoted A2219S;][]{MillionAllen09}, as mentioned in \S\ref{sec:DataPreparation}.

Different classification schemes have been proposed for relics.
We follow the classification of \citet{GiovanniniFeretti04}, which is based solely on the position and morphology of the relic.
The relics in our sample are thus divided into the following five categories:
\begin{enumerate}
\vspace{-2mm}
\item
Isolated, classical elongated relics --- shown as magenta five-stars in \Fig~\ref{fig:SourcesInuIx};
\item
Relics in halo-relic systems --- red squares;
\item
Relics in double relic systems --- blue diamonds;
\item
Circular peripheral relics --- ostensibly, face-on relics; shown as orange double arrows because the coincident X-ray emission may have been overestimated due to projection;
\item
Relics near, but not coincident with, the first ranked galaxy --- brown, filled triangles.
\end{enumerate}
Due to the lack of coincident X-ray data, our sample does not include two other classes of relics proposed by \citet{GiovanniniFeretti04}: relics at a large distance ($\gtrsim3\Mpc$) from the centre of a cluster, and large-scale filaments.

Most of the known relics are found in rich, merger Abell clusters.
However, S0573 is a poor cluster that harbours a relic \citep{SubrahmanyanEtAl03}, and some of the relic clusters have a cool core: A115 (classical relic), A85, and A133 (both relics near the first ranked galaxy); see \citet{GiovanniniFeretti04}.

The classification scheme we adopt, unlike the scheme proposed for example by \citet{KempnerEtAl04}, avoids committing to any specific relic model.
This is advantageous, as the models have not yet converged.
For example, the relics in A85, A133, and A4038 are classified here as relics near the first rank galaxy, whereas \citet{vanWeerenEtAl09} classify them as radio phoenixes, and \citet{KempnerEtAl04} classify A133 as an AGN relic.

Nevertheless, the present classification scheme is by no means unambiguous.
For example, the northern emission in A2034, coincident with a CF \citep{KempnerSarazin01}, was originally identified as a relic \citep{KempnerSarazin01, KempnerEtAl03}, and we adopt this classification in \Fig~\ref{fig:SourcesInuIx}.
It was later suggested that the source be identified as an irregular GH, because its centre coincides with the X-ray peak \citep{RudnickLemmerman09, GiovanniniEtAl09}.
However, it seems most reasonable to identify the source as a (possibly disrupted) MH, considering its central position, its relatively small scale \citep[$\sim 600\kpc$; ][]{GiovanniniEtAl09}, the evidence for a cool core \citep[see discussion in][]{KempnerEtAl03}, and the association with a CF.
Such an association has not been reported so far for GHs or for relics.
(One exception is the CF near the GH in A2319, but this cluster appears to be in an intermediate state between a MH and a GH; see {\KL}.)
Note that a putative relic-CF association would imply the existence of a new class of non-shock relics.

Another example of uncertain source classification is A754.
The diffuse radio emission here is quite irregular, showing two main components with a West-East orientation \citep{BacchiEtAl03}, and a confirmed shock at the Eastern edge of the East component \citep{KrivonosEtAl03}.
Following \citet{KaleDwarakanath09}, we identify the East component (labeled A754E) as a halo and the West component (A754W) as a relic, based mostly on the location of the X-ray peak.
An opposite interpretation has been given by \citet{BacchiEtAl03}, who also suggest that the emission may be classified as either two halos or two relics.

Source names are abbreviated as the cluster name with suffix letters identifying the source type or location.
For halos, we use suffix \emph{H}.
For relics, we use suffix \emph{R} if they are found in halo clusters, the source's cardinal position initials (\emph{W} for West, \emph{NW} for Northwest, etc.) in multiple relic clusters, and no suffix otherwise.
In A85, the (Southwest) relic is composed of two disconnected regions, radially separated by $\sim 50\kpc$; we denote them by A85R1 and A85R2.
In A2219, emission from the region suspected as a shock \citep{MillionAllen09} is denoted by A2219S.

\subsection{ICM diagnostic: the ratio $\eta$ between radio and X-ray surface brightness}
\label{sec:EtaDiagnostic}

A useful dimensionless quantity in the study of diffuse emission from galaxy clusters is the ratio between the radio and the X-ray surface brightness,
\begin{equation}
\eta(\vectwo{r}) \equiv \frac{\nu I_\nu}{F_X} \fin
\end{equation}
We use (henceforth) radio frequencies $\nu\simeq 1.4\GHz$ and the X-ray energy range $0.1$--$2.4\keV$.
One advantage of using the brightness ratio is that it is not sensitive to errors in estimating the redshift of the cluster.
In most models, it is less sensitive than $I_\nu$ is to local variations in density.

The brightness ratio is particularly useful if the radio emissivity $j_\nu$ and the X-ray emissivity $j_X$ are either proportional to each other, or are strongly peaked along the line of sight.
In such cases, the brightness ratio can be approximated by the emissivity ratio
\begin{equation}
\eta(\vectwo{r}) \sim \eta_j(\vecthree{r}) \equiv \frac{\nu j_\nu}{j_X}
\end{equation}
in the densest position $\vecthree{r}$ along the line of sight $\vectwo{r}$.
The observed $\eta(\vectwo{r})$ distribution around a cluster thus translates, approximately, to the emissivity ratio in the plane of the cluster's centre, approximately perpendicular to the line of sight.

Note that the arguments $\vectwo{r}$, $\vecthree{r}$ of the functions $\eta$, $\eta_j$ are in general two-, three-dimensional vectors.
Vector notations are omitted where spherical symmetry is assumed.

The relation between $\eta$, $\eta_j$ and $n$ can be stated more precisely in the context of a density model.
The X-ray emissivity is proportional to the plasma density squared, and is approximately independent of other plasma parameters (see \Eq{\ref{eq:FxMEKAL}}), $j_X(\vecthree{r})\propto n^2$.
We shall parameterise the synchrotron emissivity as
\begin{equation} \label{eq:j_nu_vs_n_scaling}
j_\nu(\vecthree{r}) \propto n^{\mysigma+1} \coma
\end{equation}
corresponding to an $N_p\propto n^\mysigma$ scaling of {\CRIs} in the framework of a secondary {\CRE} model.
The radio--X-ray emissivity ratio then scales as
\begin{equation} \label{eq:eta_j_param}
\eta_j(\vecthree{r}) \propto n^\gamma \coma
\end{equation}
where $\gamma=\mysigma-1$.

In the $\beta$-model, if we assume that $\eta_j(\vecthree{r})\propto n(|\vecthree{r}|)^\gamma$ applies locally with some constant $\gamma>-2+(3\beta)^{-1}$, then necessarily also $\eta(\vectwo{r})\propto n(|\vectwo{r}|)^\gamma$.
Note that this property does not occur, in general, for more complicated density distributions.
It is useful because it allows us to relate the line-of-sight integrated, observed $\eta$ behaviour, to the local $\eta_j$--$n$ relation.
Another benefit of adopting a $\beta$ model is that the relevant properties can be expressed as power-law functions of the observed X-ray brightness $F_X(\vectwo{r})$.

Adopting the parametrisation \eq{\ref{eq:eta_j_param}} in the context of a $\beta$-model, we find that
\begin{align} \label{eq:EtaIxBetaModel}
& \eta(\vectwo{r}) \propto F_X^{\gamma\delta}  \,; \nonumber \\
& \delta \equiv \frac{3\beta}{6\beta-1} \fin
\end{align}
The index $\delta$ depends weakly on $\beta$ in the relevant, $1/2<\beta<1$ range, for which $3/5<\delta<3/4$.
For clusters well fit by $\beta\simeq 2/3$, \eq{\ref{eq:EtaIxBetaModel}} becomes
\begin{equation} \label{eq:EtaIxBeta23Model}
\eta(\vectwo{r})\propto F_X^{2\gamma/3} \fin
\end{equation}
In \S\ref{sec:HaloRisingEta} below we use \eq{\ref{eq:EtaIxBeta23Model}} as a proxy of $\gamma$ in $\beta\simeq 2/3$ clusters even where the $\beta$-model fails, for example in cases where $F_X(\vectwo{r})$ reveals an underlying nonspherical gas distribution.

Notice that in the $\beta$-model, the column density and the projected density are simply related to $F_X$, through
\begin{align} \label{eq:beta_n_Ix}
\lambda_n(\vectwo{r}) & \propto F_X^{\frac{3\beta-1}{6\beta-1}} \to F_X^{1/3} \, ;\nonumber \\
n(|\vectwo{r}|) & \propto F_X^{\delta} \to F_X^{2/3} \coma
\end{align}
where the expressions to the right of the arrows correspond to the case $\beta=2/3$.

For steady-state {\CREs} and static magnetic fields, $j_\nu$ is proportional to the product of the {\CRE} energy density injection rate $\dot{u}_e$, and the magnetic energy density $u_B=B^2/8\pi$, weighted by the {\CRE} cooling parameter $\psi\equiv -E_e^{-2}dE_e/dt$.
{\CRE} cooling is dominated by inverse-Compton scattering off CMB photons or by synchrotron emission due to the magnetic field.
Therefore, $\psi \propto B^2+B_{cmb}^2$, where
\begin{equation}
B_{cmb}\equiv \sqrt{8\pi u_{cmb}} \simeq 3.24(1+z)^2\muG
\end{equation}
is the putative magnetic field amplitude for which $u_B=u_{cmb}$.
Combining these relations, we obtain
\begin{equation} \label{eq:eta_prop}
\eta_j(\vecthree{r}) \sim \frac{\dot{u}_e}{n^2} \cdot \frac{B^2}{B^2+B_{cmb}^2} \fin
\end{equation}

\citet{KushnirEtAl09} have argued that the tight correlation observed between the radio power $P_\nu$ and the X-ray luminosity $L_X$ of GH clusters, and the bimodality of the GH distribution \citep[][and references therein]{BrunettiEtAl07}, strongly suggest that the magnetic fields within GHs are strong, $B\gtrsim B_{cmb}$ \citep[for a different opinion, see][]{BrunettiEtAl09}.
If so, the second factor in \eq{\ref{eq:eta_prop}} is approximately unity, such that the varying magnetisation levels among different GHs introduce only little dispersion in the $P_\nu$--$L_X$ correlation.
Furthermore, the tight GH correlation observed can be reproduced if the {\CREs} are secondary particles, produced in hadronic collisions between primary {\CRIs} and the ambient plasma.
This is most evident if the {\CRI} distribution follows the bulk plasma, $N_i(E_i)\propto n$.
In such a case, $\dot{u}_e(E_e)\propto N_i n \propto n^2$, resulting in a nearly constant $\eta$ in $B\gtrsim B_{cmb}$ regions.
Its value provides a direct measure of the {\CRI} fraction in the ICM, $\eta\sim N_i(E_i)/n$.

\begin{figure}
\centerline{\epsfxsize=9cm \epsfbox{\myfarfig{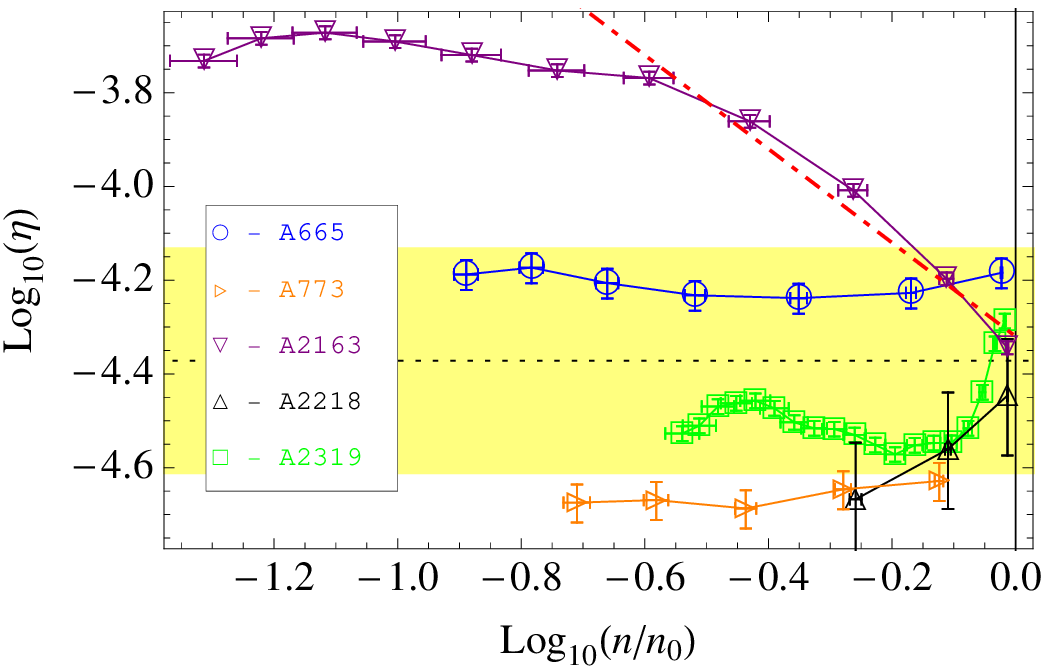}}}
\caption{
Radial profiles of the radio-to-X-ray ratio $\myeta$ in GHs with overlapping profiles from VLA \citep{MurgiaEtAl09} and from \emph{XMM-Newton} \citep[][k-corrected]{SnowdenEtAl08}.
The data is the same as is \Fig~2 of {\KL}, but here $\eta$ is plotted as a function of $n/n_0$ (so the distance from the centre of the cluster increases to the left).
We determine $n/n_0$ using individual cluster $\beta$-models, detailed in Table 1 of {\KL}.
Vertical error bars are the $1\sigma$ confidence intervals of the radio normalisation $I_0$ \citep{MurgiaEtAl09}, horizontal error bars are the propagated $r_c$ uncertainties, and the solid lines serve to guide the eye.
The best fit $\eta_0$ of {\KL} is shown as a horizontal dashed line, embedded in a shaded region showing the $1\sigma$ confidence level.
Note that in A2163, $\eta$ scales roughly as $n^{-1}$ (dot dashed red line) down to $n_0/4$ densities.
\label{fig:HalosEtaN}
}
\end{figure}

An equivalent, but stronger argument can be made regarding the radio--X-ray correlation in central surface brightness.
The $I_\nu$--$F_X$ correlation is remarkably tight near the centres of halos --- both GHs and MHs (see {\KL}).
We illustrate the GH correlation in \Fig~\ref{fig:HalosEtaN} by reproducing \Fig~2 of {\KL}, showing $\eta(r)$ for the halos with published radial profiles from both Very Large Array \citep[VLA;][]{MurgiaEtAl09} and \emph{XMM-Newton} \citep{SnowdenEtAl08}.
Here we plot $\eta$ not as a function of $r$, but rather as a function of $n(r)/n_0=(1+r^2/r_c^2)^{-3\beta/2}$, determined using the cluster $\beta$-models summarised in Table 1 of {\KL}.
The nearly identical values of $\eta$ found towards the centres ($n/n_0\to 1$) of these clusters, in regions spanning more than an order of magnitude in density, confirm both the presence of strong magnetic fields ($B\gtrsim B_{cmb}$) and suggest that $\dot{u}_e(r=0)\propto n_0^2$.

It is important to notice that the radio--X-ray correlations observed in luminosity and in central brightness do \emph{not} imply that an $N_i(E_i)\propto n$ scaling must hold throughout each halo.
A linear correlation between the \emph{central} {\CRI} density $N_{i,0}$ and the central plasma density $n_0$ must exist among different halo clusters in order to reproduce the tight $I_{\nu,0}$--$I_{X,0}$ correlation observed.
However, {\CRI} distributions that are not proportional to $n$ away from the centre are possible, and in fact better reproduce the observed correlations and morphologies.
In particular, one must take into account the weakly magnetised region, and the location of the transition between the two regimes.
Different models for the {\CRI} distribution are discussed in \S\ref{sec:etaModels}, and the luminosity correlations are revisited in \S\ref{sec:Discussion}.

\subsection{Rising $\eta(r)$ distribution in GHs: homogeneous {\CRI} distribution}
\label{sec:HaloRisingEta}

Before comparing (in \S\ref{sec:eta_in_relics}) the values of $\eta$ in relics and in halos, it is useful to examine the $\eta(\vectwo{r})$ distribution within GHs, where the X-ray emission is more constrained and the radio model is better understood.
Distinct $\eta(r)$ profiles are predicted by different variants of the secondary {\CRE} model, depending mainly on the properties of {\CRI} diffusion.
Observational evidence is now sufficient to distinguish between these models.

After briefly reviewing different {\CRI} distribution models (in \S\ref{sec:etaModels}) and previous evidence for a rising $\eta(r)$ profile in halos (in \S\ref{sec:PreviousEtaConstraints}), we analyse the radio and X-ray maps of a few well-studied GHs.
The method of analysis is presented in \S\ref{sec:EtaInWellStudiedGHs}, applied to the GHs in A2163, A665 and A2744 in \S\ref{sec:A2163}--\ref{sec:A2744}, and summarised in \S\ref{sec:HomogeneousCRIsInHalos}.
In \S\ref{sec:ReconcilingHomogeneousCRIsInHalos} we show that our results are consistent with previous studies that found linear or mildly sublinear radio--X-ray correlations in GHs.

\subsubsection{Different $\eta(r)$ distributions in a secondary {\CRE} model}
\label{sec:etaModels}

The distribution of {\CRIs} in a cluster depends on the nature of the {\CR} sources, the properties of {\CR} diffusion through the magnetised ICM, the escape of {\CRIs} beyond the virial shock, and the mixing of the gas. The most plausible sources of cluster {\CRIs} are either supernovae (SNe; see KL) or the virial shock of the cluster \citep[see][]{KushnirWaxman09, KushnirEtAl09}; weak shocks are unable to produce the flat {\CRI} spectra necessary to explain the radio observations.
This leads to the following main possibilities:
\begin{enumerate}
\vspace{-2mm}

\item
Negligible diffusion and mixing: {\CRIs} are mostly due to the virial shock and subsequent adiabatic compression.
Compression leads to $N_i(E_i)\propto n^{2/3}$ and $\eta_j\propto n^{-1/3}$, assuming an isothermal distribution.
Equivalently (see \S\ref{sec:EtaDiagnostic}), $\eta\propto n^{-1/3}$ and $\gamma=-1/3$.
Here, $N_i(E_i,\vecthree{r})$ is probably proportional to $T(\vecthree{r})$ \citep{KushnirEtAl09}.

\item
Significant escape, unsaturated diffusion and mixing: {\CRIs} are mostly produced by SNe, and are distributed roughly as the gas, $N_i(E_i)\propto n$, so $\eta\propto \constant$ ($\gamma=0$); $N_i$ could depend on the local temperature $T(\vecthree{r})$ (KL).

\item
Saturated diffusion or mixing: {\CRIs} in the halo are produced by some combination of SNe and the virial shock, and are uniformly distributed throughout the cluster, $N_i(E_i)\propto \constant$, so $\eta\propto n^{-1}$ ($\gamma=-1$).

\end{enumerate}

For brevity, we refer to the these models below as different diffusion models.
The different outcomes of strong diffusion and gas mixing are discussed in \S\ref{sec:HomogeneousCRIsImplications_Diffusion}.

As mentioned in \S\ref{sec:EtaDiagnostic}, the tight radio--X-ray correlations observed in the luminosity ($P_\nu$--$L_X$) and in the central brightness ($I_{\nu,0}$--$F_{X,0}$) of halos imply that the central density $N_{i,0}$ of the {\CRIs} is linearly correlated with the central plasma density $n_0$.
These correlations --- in particular in luminosity --- are also sensitive to the {\CRI} distribution away from the centre.
However, the interpretation of the data is complicated by effects such as a possible temperature dependence of $\eta_j$ \citep[][although such a dependence cannot be strong, see {\KL}]{KushnirEtAl09} and different scalings of the radio and X-ray bright volumes (see {\KL}).
Therefore, these correlations, by themselves, do not clearly distinguish between the possible {\CRI} distributions outlined above.

\begin{figure}
\centerline{\epsfxsize=9cm \epsfbox{\myfarfig{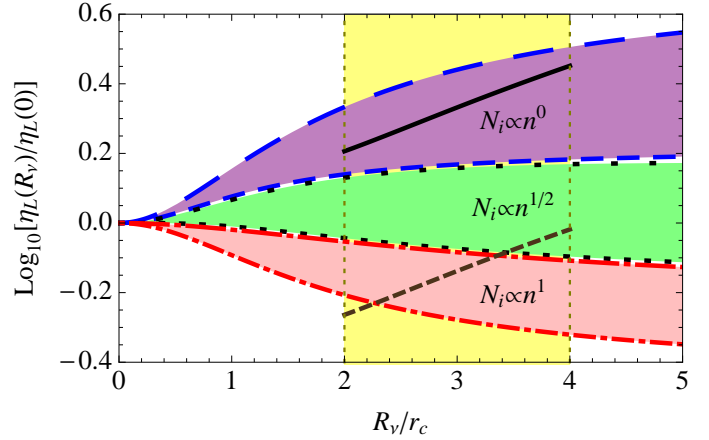}}}
\caption{
The ratio $\myetaL(R_\nu)$ between radio power and X-ray luminosity from the central, $r<R_\nu$ halo region, assuming an isothermal $\beta$-model with $\beta=2/3$, and a $B=B_0(n/n_0)^{1/2}$ magnetic scaling.
To highlight the dispersion due to $R_\nu$ and $B_0$ (which together determine the magnetic break radius $R_B$), we normalise $\myetaL$ to unity as $R_\nu\to0$.
We assume a power-law {\CRI} distribution $N_i\propto n^\sigma$, with $\sigma=0$ (homogeneous {\CRIs}; dashed curves), $1/2$ (dotted curves), and $1$ ({\CRI} density proportional to the plasma density $n$; dot-dashed curves).
Shaded, labeled regions correspond to a range of central magnetic fields between $1B_{cmb}$ (lower $\myetaL$; short dashing) and $10B_{cmb}$ (long dashing).
The shaded yellow region between vertical dotted lines shows the typical halo size range.
Also shown is the $\myetaL(R_\nu)$ relation inferred from the observed $P_\nu$--$L_X$, $L_X$--$R_{vir}$, and $R_\nu$--$R_{vir}$ relations (solid black and short-dashed brown curves; \eq{\ref{eq:etaL}}), with two arbitrary choices of normalisation (note that the full curve must begin at the origin).
\label{fig:RBDispersion}
}
\end{figure}

The small, factor $\sim 2$ dispersion in the correlations also does not, by itself, fix the {\CRI} distribution, because the different models outlined above entail a similar dispersion, as we show below and in \Fig~\ref{fig:RBDispersion}.
In particular, the small dispersion does not necessitate a spatial linear correlation between the {\CRIs} and the ambient plasma, $N_i(\vecthree{r})\propto  n(\vecthree{r})$, as assumed to hold in the models of \citet{KushnirEtAl09} and {\KL}.

To see this, consider first the $P_\nu$--$L_X$ correlation.
Assuming spherical symmetry, let $R_\nu$ denote the radius out to which the halo is observed, and $R_B$ denote the break radius where the magnetic field becomes weak ($B<B_{cmb}$; see {\KL} for a discussion of the magnetic break).
For a {\CRI} distribution that does not satisfy $N_i\propto n$, the radio emissivity $j_\nu$ does not scale as $n^2$ inside $R_B$, so $\eta_j=\nu j_\nu/j_X \sim N_i/n$ is not uniform. Consequently, the luminosity ratio $\eta_L\equiv \nu P_\nu/L_X$ depends on the halo size, introducing some spurious dispersion in the $P_\nu$--$L_X$ correlation.
However, this dispersion is not large for the typical outer radii of halos, $R_\nu\simeq (2$--$4)r_c$ \citep[\eg][]{MurgiaEtAl09}, even if the {\CRI} distribution is flat (\ie homogeneous).
Moreover, beyond $R_B$, the rapid radial decline of $j_\nu$ introduces an additional, $B$-dependent dispersion in the $P_\nu$--$L_X$ relation for all $N_i(r)$ models, in particular for the steeper, $N_i\propto n$ distributions.
As a result, the $P_\nu$--$L_X$ dispersion introduced by the varying halo sizes and magnetisation levels among different clusters is similar in the different $N_i(r)$ models outlined above, and is comparable in all cases to the (factor $\sim 2$) dispersion observed.
Analogous arguments can be made regarding the dispersion in the central brightness correlation.

To illustrate the $P_\nu$--$L_X$ dispersion corresponding to the different $N_i(r)$ models, we show in \Fig~\ref{fig:RBDispersion} the ratio $\myetaL(R_\nu)\equiv P_\nu/L_X(<R_\nu)$ as a function of halo size and magnetisation level, for different {\CRI} distributions.
For simplicity, we use the same cutoff radius $R_\nu$ for radio and X-ray emission, adopt an isothermal $\beta$-model with $\beta=2/3$, and assume that the magnetic energy density scales linearly with that of the plasma, $B^2\propto n$, as often inferred from observations \citep[\eg][]{MurgiaEtAl09, BonafedeEtAl10}.
The figure shows that for the typical ranges of halo size ($R/r_c\simeq 2$--$4$) and central halo magnetic field ($B_0/B_{cmb}\simeq 1$--$10$), the dispersion is rather similar for the different {\CRI} distributions outlined above.

\subsubsection{Previous constraints on $\eta(r)$}
\label{sec:PreviousEtaConstraints}

Inspection of the azimuthally averaged radial profiles of $\eta$ in the halos shown in \Fig~\ref{fig:HalosEtaN} reveals a substantially different behaviour in each cluster.
We ignore the dispersion seen in the central value, $\eta_0$; in an SNe model this could be attributed to different star formation histories ({\KL}).
The radial profile is quite flat ($\eta \sim n^0$) in A665 and in A773, scales roughly as $n^{-1}$ in the centre of A2163 and as $n^{+1}$ in the centre of A2218, and is irregular in A2319.
No universal {\CRI} spatial distribution can be identified based solely on these radial profiles.
The dependence of $\eta$ on the local temperature within these clusters was shown to be weak, $\eta\propto (T/T_0)^{0.2\pm0.5}$ (KL).

Notice that only one of the GH clusters in the {\KL} sample --- A2163 --- harbours a relic.
This is also the most extended and the most regular amongst the halos \citep[see][]{MurgiaEtAl09} in the figure, suggesting that it may be the most representative of the universal {\CRI} distribution, if such exists.
The profiles in the other halos could be contaminated by asymmetry and irregularities induced by substructure or regions of low magnetic field (see KL).

Evidence for a radially rising (\ie increasing with declining $n/n_0$) $\eta$ profile in some other clusters is inferred, for example, from the comparison of the radio and X-ray morphologies in four GHs studied by \citet{GovoniEtAl01}.
They find that two of these GHs (A2255 and A2744) show a linear correlation between radio and X-ray brightness (so $\eta\propto n^0$), whereas two other GHs (Coma and A2319) show a sublinear relation between $I_\nu$ and $F_X$.
Using the large radii scaling $F_X\propto r^{1-6\beta} \propto n^{2-1/(3\beta)}$ of the ASCA-based $\beta$-models of \citet{FukazawaEtAl04}, the results of \citet{GovoniEtAl01} crudely translate to $\eta\sim n^{-0.6}$ in Coma, and $\eta \sim n^{-0.2}$ in A2319.
In comparison, the sublinear $I_\nu$--$F_X$ relation found in A2163 \citep{FerettiEtAl01} translates to $\eta\sim n^{-0.5}$.

Indirect evidence for a radially rising $\eta$ profile stems from the different scalings of the integrated luminosities $P_\nu$ and $L_X$ with the respective radiating volumes.
\citet{CassanoEtAl07} found that the radio power increases rapidly with the halo size, $P_{1.4}\propto R_\nu^{4.2\pm 0.7}$.
A weaker radial dependence is found in X-rays, $L_X\sim R_X^{3.3}$, according to the relations $R_{vir}\sim T^{0.6}$ \citep{ZhangEtAl08} and $L_{X}\sim T^2$ \citep[][with $X$ referring, as usual, to $0.1$--$2.4\keV$]{Markevitch98}.
(Here we assumed that the X-rays are integrated within a radius $R_X$ proportional to the virial radius $R_{vir}$.)
These different scalings could arise, in part, from some dependence of $j_\nu$ upon the global parameters of the cluster.
However, in the absence of evidence for such a global dependence (see {\KL}), the results suggest that $\eta_j(r)$ is monotonically rising.

We may combine the above scaling $L_X\propto R_{vir}^{\myp_{L,R}\simeq 3.3}$ with the relations $R_\nu\propto R_{vir}^{\myp_{R,R}\simeq2.6}$ \citep{CassanoEtAl07} and $P_\nu\propto L_X^{\myp_{P,L}\simeq 1.7}$ \citep[][{\KL}]{KushnirEtAl09}, in order to compare the $\myetaL(R_\nu)$ profile of the different models in \Fig~\ref{fig:RBDispersion} to the observed correlations.
The $\myetaL(R_\nu)$ profile based on these phenomenological relations,
\begin{align} \label{eq:etaL}
& \myetaL(R_\nu) = \eta_L \frac{L_X}{L_X(R_\nu)} \propto \left( \frac{R_\nu}{r_c} \right)^{\myp} \frac{L_X}{L_X(R_\nu)} \, ; \nonumber \\
& \myp = \myp_{L,R} \frac{\myp_{P,L}-1}{\myp_{R,R}-1} \simeq 1.4 \coma
\end{align}
is illustrated in \Fig~\ref{fig:RBDispersion} with two different choices of normalisation (as solid black and dashed brown curves).
These curves are quite crude, as the normalisation is arbitrary (it depends on the unspecified or uncertain normalisation of the above relations), we have not incorporated the substantial dispersions of the underlying phenomenological relations, and no dependence of $\eta_j$ upon global cluster parameters was allowed.
The combined uncertainty in $\myetaL(R_\nu)$ is thus sufficiently large to allow even the $N_i\propto n$ scaling (see {\KL}).
Nevertheless, the agreement with a homogeneous {\CRI} distribution is better, as illustrated by the figure, and favours strong magnetic fields.

More work is necessary in order to resolve the processes leading to the statistical brightness and luminosity correlations observed, if they are to be used to measured the {\CRI} distribution.
A more direct approach is to study the radio and X-ray maps of individual halo clusters.

\subsubsection{Evaluating $\eta(r)$ in well-studied GHs: method}
\label{sec:EtaInWellStudiedGHs}

In order to investigate the possible existence of a universal $\eta$ distribution in halos, we analyse the radio profiles of a sample of well studied, flat spectrum GHs.
As explained in \S\ref{sec:DataPreparation}, we select A665, A2163, and A2744, where detailed spectral maps are available.
The profiles of $\eta$ and $\alpha$ in these clusters are presented in Figs. \ref{fig:ProfilesA2163}--\ref{fig:ProfilesA2744HaloNW} below.

Avoiding the assumption of spherical symmetry when possible, we compute the radio profiles along two perpendicular directions in each cluster, without performing an azimuthal average.
We choose these two axes such that they intersect at the X-ray peak of the cluster (denoted as $r=0$), and one of them crosses the relic (in A2163 and A2744) or shock (in A665) found in the cluster.
(The analysis of these relics and shocks is deferred to \S\ref{sec:ModelApplications}.)
The $\eta$ profiles are shown in the bottom panels of Figs. \ref{fig:ProfilesA2163}--\ref{fig:ProfilesA2744HaloNW}; the coincident profiles of the
spectral index $\alpha_{0.3}^{1.4}$ are shown in the upper panels.
The orientation examined in each figure is specified in the upper right corner box.

We compute $\eta$ by combining $1.4\GHz$ or $1.5\GHz$ radio maps with X-ray data from ROSAT (the resulting $\eta(r)$ profile is shown in each figure as a blue solid curve) and, when possible, also from \emph{Chandra} (black short-dashed curves, with orange shaded band showing the $10\%$ X-ray uncertainty estimated by \citet{MillionAllen09}).
In addition, we derive somewhat model-dependent $\eta$ profiles, using published $\beta$-models to estimate $F_X$ (red long-dashed curves, with pink shaded band showing the $n_0^2$ uncertainty).
Model parameters are take from the ASCA-based analysis of \citet{FukazawaEtAl04}.

In order to test the flat, $\eta\propto n^0$ model (second model in \S\ref{sec:etaModels}), we plot the central GH fit of {\KL},
\begin{equation} \label{eq:eta0_KL}
\eta_0=10^{-4.4\pm 0.2} \coma
\end{equation}
as a horizontal dotted line with yellow shaded region showing the $1\sigma$ dispersion.
Although $\eta$ lies within this range in the very central parts of A665 and A2163, it exceeds it in their outer parts, and throughout A2744.
Figs. \ref{fig:ProfilesA2163}--\ref{fig:ProfilesA665B} show clear evidence for rising $\eta(r)$ in A665 and A2163.
As the figures show, this rise is stronger than the $\eta\propto n^{-1/3}$ behaviour anticipated from adiabatic {\CRI} evolution with no diffusion (first model in \S\ref{sec:etaModels}).

The data shown in \Figs~\ref{fig:ProfilesA2163}--\ref{fig:LineProfileCompareA2163A2744} reveal a rapidly radially rising $\eta(r)$ profiles in A665 and A2163, and a striking radio similarity between A2744 (where the X-ray morphology is highly irregular) and A2163 (see \S\ref{sec:A2744}).
This suggests that among the three $\eta(r)$ models of \S\ref{sec:etaModels}, observations agree best with the $\eta_j\propto n^{-1}$ distribution resulting from homogeneous {\CRIs} (first model in \S\ref{sec:etaModels}; saturated diffusion).
This can be tested by comparing the $\eta(r)$ profiles computed from the $\beta$-models with the hypothesis
\begin{equation} \label{eq:beta_eta_beta}
\eta=\eta_0(n/n_0)^{-1}=\eta_0(1+r^2/r_c^2)^{-1} \coma
\end{equation}
shown as red long-dotted curves, where we use the measured values of $\eta_0=\eta(r=0)$ for normalisation.

The model can also be tested more directly, using the observed $F_X$, by utilising the $\eta$--$F_X$ relation in \eq{\ref{eq:EtaIxBetaModel}}.
The three clusters at hand are consistent with $\beta=2/3$ \citep[][]{FukazawaEtAl04}, so we may use \eq{\ref{eq:EtaIxBeta23Model}}, whereby the $\eta_j\propto n^{-1}$ scaling leads to $\eta\propto F_X^{-2/3}$.
The resulting hypothesis,
\begin{equation} \label{eq:beta_eta_IX}
\eta(r)=\eta_0[F_X(r)/F_X(0)]^{-2/3} \coma
\end{equation}
with $F_X$ from ROSAT, is shown in each figure as a blue dotted curve.
Note that although this result uses the $\beta$-model scaling of \eq{\ref{eq:EtaIxBetaModel}}, it is much more realistic than the pure $\beta$-model estimate \EqO~(\ref{eq:beta_eta_beta}), because it incorporates the measured $F_X$, allowing for asymmetric and non-monotonic density profiles, and is independent of $r_c$.

Equivalently, we plot, as a function of $r$, the putative central $\eta$ value, $\eta_0$, one would estimate under the $\eta\propto n^{-1}$ scaling.
The corresponding estimate,
\begin{equation} \label{eq:eta0_beta}
\eta_0(r)=\eta(r)[F_X(r)/F_X(0)]^{2/3} \coma
\end{equation}
based on the ROSAT $\eta(r)$ and $F_X(r)$ profiles, is shown as a cyan dot-dashed curve in each figure.

The data is consistent with the $\eta_j\propto n^{-1}$ model if the $\eta(r)\propto n^{-1}$ approximations derived in \eqs{\ref{eq:beta_eta_beta}} and (\ref{eq:beta_eta_IX}) match the respective $\eta(r)$ profiles observed (\ie where there is agreement between the blue solid and dotted curves, or between the red solid and dotted curves), and if the reconstructed $\eta_0(r)$ profile \eq{\ref{eq:eta0_beta}} is flat.

At large radii, where the magnetic field decays below $B_{cmb}$, we expect the $\eta\propto n^{-1}$ profile to gradually flatten (see discussion in {\KL}).
Therefore, in profiles that show an $\eta\propto n^{-1}$ behaviour at small radii, we plot an additional (green dashed) ROSAT-based curve incorporating the effect of the finite magnetic field.
We assume that the magnetic energy density is a fixed fraction of the thermal energy density, $B^2\propto n$, as inferred from recent studies of cluster magnetic fields \citep[\eg][]{MurgiaEtAl09, BonafedeEtAl10}.
We also assume, for simplicity, that the spectrum of {\CRE} injection is flat.
Combining this with the finite-$B$ formula for $\eta_j$ in \eq{\ref{eq:eta_prop}} and with the $n\propto F_X^{2/3}$ scaling of \eq{\ref{eq:beta_n_Ix}}, we obtain
\begin{eqnarray}\label{eq:eta_magnetic_ROSAT}
\eta(r; B_0) & = & \eta_0\left[\frac{F_X(r)}{F_X(0)}\right]^{-2/3}\frac{B(r)^2}{B(r)^2+B_{cmb}^2}  \nonumber \\
& = & \frac{\eta_0}{\left[\frac{F_X(r)}{F_X(0)}\right]^{2/3}+\frac{B_{cmb}^2}{B_0^2}} \fin
\end{eqnarray}
This coincides with \eq{\ref{eq:beta_eta_IX}} in $B\gg B_{cmb}$ regions.

Below we fit the observed $\eta(r)$ profiles using \eq{\ref{eq:eta_magnetic_ROSAT}}, thus obtaining estimates of the central magnetic fields $B_0$.
This is possible in A665 and A2163, in which strong, $B\gtrsim 30\muG$ magnetic fields provide the best fit.
However, such estimates are uncertain, mainly because they rely heavily on the $B^2\propto n$ scaling.
In particular, in the central, $\eta\propto n^{-1}$ regions, the data only require that $B>B_{cmb}$; the $B^2\propto n$ scaling could be replaced by saturation, for example, to any $B_0>B_{cmb}$ value.
The $B^2\propto n$ scaling probably does not hold in clusters with asymmetric $\eta(\vectwo{r})$ profiles, as it would lead to contradicting $B_0$ estimates.
An additional source of confusion is substructure, which can flatten $\eta(r)$ and thus mimic a magnetic cutoff; this would leads to an underestimated magnetic field.

\subsubsection{A2163}
\label{sec:A2163}

We consider A2163 first, because the GH it harbours shows the most regular morphology.
The centre of A2163 is in a state of violent motion, but the temperature map is too complicated to admit a dynamical reconstruction \citep{MarkevitchVikhlinin01}.
\citet{MaurogordatoEtAl08} use multi band observations and simulations to suggest a recent ($\sim 0.5\Gyr$) merger along the NE-SW (or E-W) axis.
They identify several subclusters approximately along the line connecting the halo's peak with the relic.

\begin{figure}
\centerline{\epsfxsize=10cm \epsfbox{\myfarfig{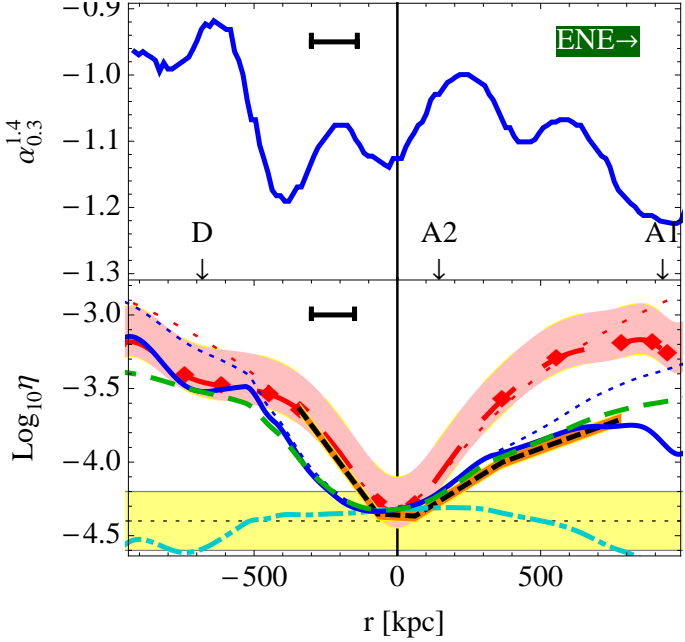}}}
\caption{
Profiles of the A2163 halo along the WSW-ENE line $\mathcal{L}=\{16\mbox{h}15'20'',-6^\circ 11'30''\}$--$\{16\mbox{h}16'23'',-6^\circ 5'12''\}$,
which connects the X-ray peak ($r=0$) with the $1.4\GHz$ brightness peak of the ENE relic (at $r\simeq 1300\kpc$).
Here we show only the GH; the entire halo-relic region is presented later in \Fig~\ref{fig:ProfilesA2163C}. \newline
The $\alpha_{0.3}^{1.4}$ profile (solid curve in upper panel) is extracted from \citet{FerettiEtAl04a}.
We compute $\eta$ by combining $1.5\GHz$ VLA data \citep{FerettiEtAl01} with X-ray data from ROSAT PSPC \citep[solid blue curve; data from][]{ElbazEtAl95} and from \emph{Chandra} \citep[black short-dashed curve with orange shading showing the X-ray uncertainty; data from][]{MillionAllen09}, and with the ASCA-based isothermal $\beta$-model of \citet[][long-dashed red curve; shaded pink region shows the $n_0^2$ uncertainty]{FukazawaEtAl04}.
The radio data is (second order spline) interpolated; data positions are marked by red diamonds.
The sizes of the radio beams are shown as horizontal error bars.  \newline
The $\eta$ profile corresponding to a homogeneous {\CRI} distribution is shown based on the ROSAT data (dotted blue curve; see \eq{\ref{eq:beta_eta_IX}}) and on the $\beta$-model (long dotted red; see \eq{\ref{eq:beta_eta_beta}}).
These curves are in fairly good agreement with the respective $\eta$ profiles measured at $|r|\lesssim 500\kpc$.
Equivalently, the $\eta_0$ value inferred from the ROSAT data at distance $r$ by assuming that $N_i(\vecthree{r})\sim \constant$ (dot-dashed cyan; see \EqO~(\ref{eq:eta0_beta})), is nearly constant and consistent with the central GH average of {\KL} (\eq{\ref{eq:eta0_KL}}; dotted horizontal line, with $1\sigma$ uncertainty shaded yellow).
A halo model for $\eta(r<0)$, invoking a homogeneous {\CRI} distribution and a finite magnetic field $B=B_0(n/n_0)^{1/2}$, is shown (according to \eq{\ref{eq:eta_magnetic_ROSAT}}; dashed green curve) for a central magnetic field $B_0=50\muG$.
Local deviations in $\eta$ and $\alpha$ are probably associated with optically detected clumps A1, A2 and D, that lie along $\mathcal{L}$ \citep[labeled arrows mark their approximate central positions; we use the notations of the $R<21$ R-band magnitude analysis of ][their table 2 and figures 5c,6]{MaurogordatoEtAl08}.
\label{fig:ProfilesA2163}}
\end{figure}

\begin{figure}
\centerline{\epsfxsize=10cm \epsfbox{\myfarfig{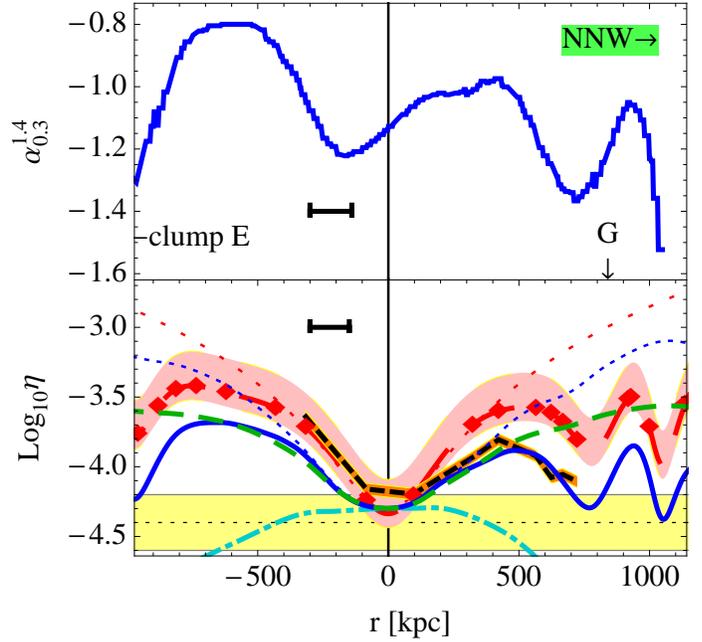}}}
\caption{
Profiles of the A2163 halo along $\mathcal{L}=\{16\mbox{h}15'54'',-6^\circ 17'38''\}$--$\{16\mbox{h}15'37'',-6^\circ 0'17''\}$,
which crosses the X-ray peak ($r=0$) in an SSE--NNW direction, perpendicular to the line examined in \Fig~\ref{fig:ProfilesA2163}.
Symbols and notations are defined in the caption of Fig~\ref{fig:ProfilesA2163}.
Local deviations in $\eta$ and $\alpha$ are probably associated with optically detected clumps E (centred at $r=-1260\kpc$) and G \citep[defined as the clump between A2 and B, see figure 5h of][]{MaurogordatoEtAl08}.
The halo model is shown for $B_0=30\muG$.
\label{fig:ProfilesA2163B}}
\end{figure}

The profiles of $\eta$ and $\alpha$ are shown along the WSW--ENE main elongation of the GH, towards the ENE relic, in \Fig~\ref{fig:ProfilesA2163}, and along the perpendicular NNW-SSE direction in \Fig~\ref{fig:ProfilesA2163B}.

In the central, $r\lesssim 350\kpc$ regions (where $n\gtrsim n_0/4$), the measured $\eta$ profiles agree with the $\eta\propto n^{-1}$ model, in all four rays probed.
Indeed, the measured $\eta(r)$ profiles (solid and dashed curves) coincide with the $\eta\propto n^{-1}$ models (dotted curves), using both $\beta$-model (red curves) and ROSAT-data (blue curves).
In addition, the $\eta_0(r)$ profiles (cyan dot-dashed curves) lie entirely within the {\KL} estimate.
Some asymmetry is evident, for example in the better resolved \emph{Chandra}-based curves, and from the more extended $\eta\propto n^{-1}$ behaviour towards the West.

More quantitatively, we may attempt to fit $\eta$ as a power law in $F_X$, along the four rays depicted in Figs. \ref{fig:ProfilesA2163} and \ref{fig:ProfilesA2163B}.
The results depend on the scales examined, due to the apparent gradual flattening of $\eta(r)$.
Out to $r_{max}=2r_c$ (the minimal distance with sufficient data), we obtain
\begin{equation} \label{eq:A2163FitR2}
\eta(|r|<2r_c)\propto F_X^{-0.60\pm0.02} \coma
\end{equation}
which corresponds according to \eq{\ref{eq:EtaIxBeta23Model}} to $\gamma=-0.90\pm0.03$.
The gradual flattening can be seen by fitting the data out to larger scales, $\eta(|r|<2.5r_c)\propto F_X^{-0.56\pm0.04}$ and $\eta(|r|<3r_c)\propto F_X^{-0.51\pm0.06}$.
These results are consistent with $\eta_j\propto n^{-1}$ ($\gamma=-1$) at small, $r\lesssim 2r_c$ distances from the centre.
The linear regression, employed here and in the halos below, uses the radio data points and the interpolated X-ray data from ROSAT; measurement errors are not propagated.

The deviations from  $\eta\propto n^{-1}$ at large radii are probably caused by the gradual weakening of the magnetic field, falling below $B_{cmb}$ at large distances.
To illustrate this, we plot in Figs. \ref{fig:ProfilesA2163} and \ref{fig:ProfilesA2163B} an additional (dashed green) curve, showing the ROSAT-based $\eta\propto n^{-1}$ model modified by incorporating a finite, $B^2\propto n$ magnetic field, according to \eq{\ref{eq:eta_magnetic_ROSAT}}.

A central magnetic field $B_0\simeq 7B_{cmb}(z)\simeq 30\muG$ provides a good fit, as shown in \Fig~\ref{fig:ProfilesA2163B}.
The Western profile is better fit by a stronger magnetic field; $B_0=50\muG$ is used in \Fig~\ref{fig:ProfilesA2163}.
These values are consistent with the $B_0>20\muG$ lower limit derived for this cluster by {\KL} using the azimuthally averaged profile.

At very large radii where $B\ll B_{cmb}$, the $\eta(r)$ profile of \EqO~(\ref{eq:eta_magnetic_ROSAT}) flattens to a constant.
This leads to a gradual flattening of $\eta(r)$, from $\eta\propto n^{-1}$ near the centre to $\eta(r)\simeq \constant$ at large distances.
Such a behaviour is seen in the azimuthally averaged profile of the cluster in \Fig~\ref{fig:HalosEtaN}, where gradual flattening sets at $r\gtrsim 350\kpc$.
The flattening is somewhat hidden in Figs. \ref{fig:ProfilesA2163} and \ref{fig:ProfilesA2163B} by substructure; optically-identified clumps marked in the figures coincide with local drops in $\eta$ and $\alpha$.

In summary, the GH in A2163 is consistent with homogeneously distributed {\CRIs} and a strong, $B_0\gtrsim 30\muG$ central magnetic field.
The nonsymmetric radial decline of the magnetic field probably gives the halo its East--West elongated structure, and causes the $\eta\propto n^{-1}$ profile to flatten at large, $r\gtrsim 350\kpc$ radii.

\subsubsection{A665}
\label{sec:A665}

\begin{figure}
\centerline{\epsfxsize=10cm \epsfbox{\myfarfig{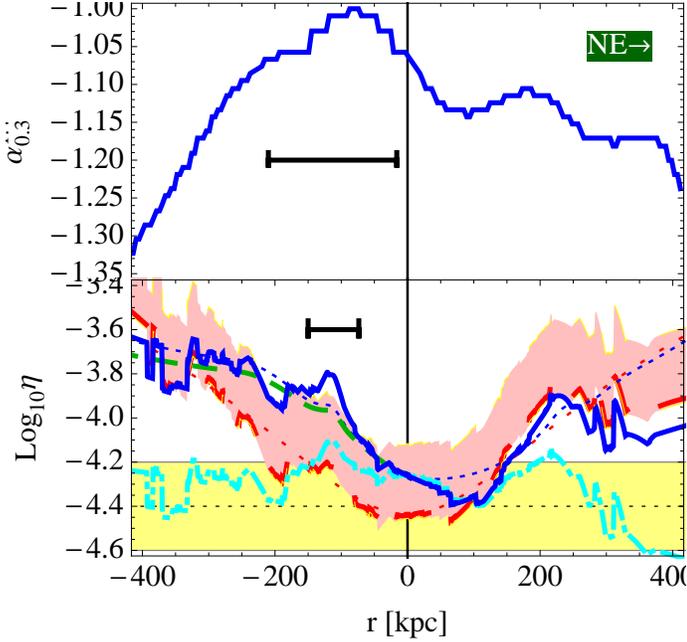}}}
\caption{
Profiles of $\alpha_{0.3}^{1.4}$ (upper panel) and $\eta$ (bottom panel) in A665, along the line with coordinates $\mathcal{L}[\{\mbox{RA,dec}\}(J2000)]=\{8\mbox{h}30'37'',65^\circ 49'5''\}$--$\{8\mbox{h}31'17'',65^\circ 51'23''\}$.
This line crosses the X-ray peak \citep[at $0.8-4\keV$;][]{GovoniEtAl04}, denoted as $r=0$, in an SW--NE orientation, perpendicular to the main halo elongation, with $r$ increasing towards the NE (labeled arrow).
\newline
The $\alpha_{0.3}^{1.4}$ profile (solid curve in upper panel) is extracted from \citet{FerettiEtAl04a}.
We compute $\eta$ by combining $1.4\GHz$ VLA data \citep[][for a $25''\times 25''$ beam]{VaccaEtAl10} with X-ray data from ROSAT HRI \citep[][yielding the solid blue curve]{GomezEtAl00} or from the ASCA-based isothermal $\beta$-model of \citet[][long-dashed red curve; shaded pink region shows the $n_0^2$ uncertainty]{FukazawaEtAl04}.
The sizes of the radio beams are shown as horizontal error bars. \newline
The $\eta$ profile corresponding to a homogeneous {\CRI} distribution is shown based on the ROSAT data (dotted blue curve; see \eq{\ref{eq:beta_eta_IX}}) and on the $\beta$-model (long dotted red; see \eq{\ref{eq:beta_eta_beta}}).
These curves are in fairly good agreement with the respective $\eta$ profiles measured at $r<0$ and at $0<r\lesssim 250\kpc$.
Equivalently, the $\eta_0$ value inferred from the ROSAT data at distance $r$ by assuming that $N_i(\vecthree{r})\sim \constant$ (dot-dashed cyan; see \EqO~(\ref{eq:eta0_beta})), is nearly constant and consistent with the central GH average of {\KL} (\eq{\ref{eq:eta0_KL}}; dotted horizontal line, with $1\sigma$ uncertainty shaded yellow). \newline
A halo model for $\eta(r<0)$, invoking a homogeneous {\CRI} distribution and a finite magnetic field $B=B_0(n/n_0)^{1/2}$, is shown (according to \eq{\ref{eq:eta_magnetic_ROSAT}}; dashed green curve) for a central magnetic field $B_0=30\muG$.
\label{fig:ProfilesA665}}
\end{figure}

\begin{figure}
\centerline{\epsfxsize=10cm \epsfbox{\myfarfig{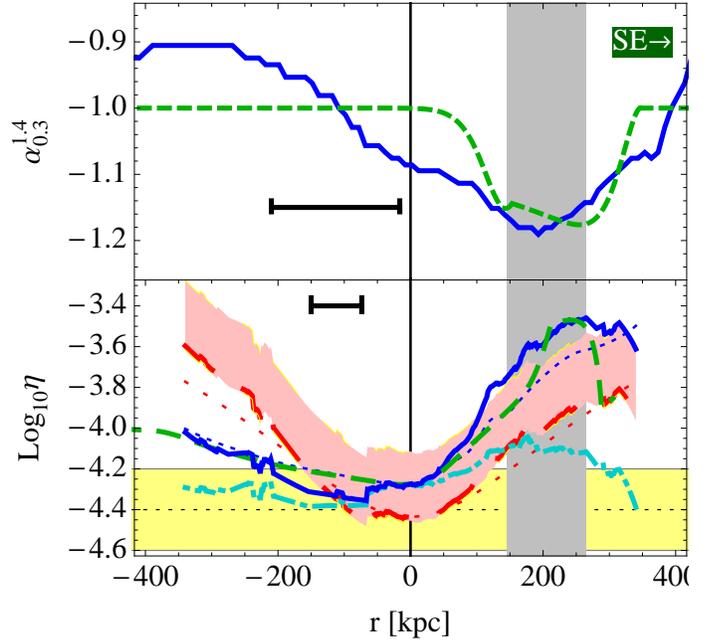}}}
\caption{
Profiles in A665 along $\mathcal{L}=\{8\mbox{h}30'35'',65^\circ 55'35''\}$--$\{8\mbox{h}31'7'',65^\circ 48'17''\}$.
This line crosses the X-ray peak and is oriented in the SE-NW direction, along the main halo elongation, perpendicular to the line shown in \Fig~ \ref{fig:ProfilesA665}.
It partly overlaps the green line examined by \citet[][their figure 3, top panel]{FerettiEtAl04a}.
Most symbols and notations are defined in \Fig~\ref{fig:ProfilesA665}.
The vertical, gray shaded region corresponds to the $k_BT\geq 13\keV$ temperature peak \citep{GovoniEtAl04}, identified as a shock by \citet{MarkevitchVikhlinin01}.
Our model (dashed green) assumes $B_0=30\muG$ and the shock parameters in Table \ref{tab:ModelParameters} (see \S\ref{sec:ModelApplications}).
\label{fig:ProfilesA665B}}
\end{figure}

Next consider A665, which harbours a flat GH.
The core of A665 appears to be moving South, preceded by a shock of Mach number $\mach>1.8$; relativistic {\CRE} accelerated at the shock were suggested as the origin for the brightest halo region \citep{MarkevitchVikhlinin01}.
The $\eta$ and $\alpha$ profiles of A665 are shown along an SW--NE axis in \Fig~\ref{fig:ProfilesA665}, and along the perpendicular, NW--SE direction in \Fig~\ref{fig:ProfilesA665B}.
The latter is aligned with the main elongation axis of the halo, and crosses a Southeast shock.

The azimuthally averaged $\eta(r)$ profile of this cluster, shown in \Fig~\ref{fig:HalosEtaN}, if flat.
However, inspection of the non-averaged profiles shows good agreement with the $\eta\propto n^{-1}$ hypothesis, towards the NW and the SW, out to the maximal $r\simeq 400\kpc$ distances probed.
Indeed, the reconstructed $\eta_0(r<0)$ profiles in both figures lie within the {\KL} fit (assuming that $\eta\propto n^{-1}$; cyan dot-dashed curves).
The $\eta$ profile towards the East shows a different behaviour, but this is dominated by the $r\simeq 200\kpc$ SE shock \citep{MarkevitchVikhlinin01, GovoniEtAl04} and a local feature, possibly due to substructure, $r\simeq 250\kpc$ towards the NE.
Therefore, Figs. \ref{fig:ProfilesA665} and \ref{fig:ProfilesA665B}, and in particular the SW ray, are consistent with $\eta\propto n^{-1}$ in relaxed regions.

More quantitatively, as in A2163, we may attempt to fit $\eta$ as a power law in $F_X$, along the four rays depicted in Figs. \ref{fig:ProfilesA665} and \ref{fig:ProfilesA665B}.
The results again depend on the scales examined, due to the apparent gradual flattening of $\eta(r)$.
Out to $r_{max}=2r_c$ (the maximal distance with available data), we obtain $\eta(|r|<2r_c)\propto F_X^{-0.5\pm0.1}$.
This result is affected by the local features to the East.
If we consider only the two Western rays, we obtain $\eta(-2r_c<r<0)\propto F_X^{-0.56\pm0.08}$, which corresponds according to \eq{\ref{eq:EtaIxBeta23Model}} to $\gamma=-0.84\pm0.13$.
The gradual flattening can be seen by fitting the data out to smaller scales, for example $\eta(|r|<1.5r_c)\propto F_X^{-0.69\pm0.05}$ and
\begin{equation} \label{eq:A665FitR15}
\eta(-1.5r_c<r<0)\propto F_X^{-0.64\pm0.02} \coma
\end{equation}
which corresponds to $\gamma=-0.95\pm0.04$.
These results are again consistent with $\eta_j\propto n^{-1}$ at small, $r\lesssim 2r_c$ distances from the centre.

As the figures show, the X-ray morphology is asymmetric and poorly fit by the $\beta$-model, and the radio and X-ray morphologies differ.
Combined with the shock and possible substructure to the East, this leads to asymmetric $\eta$ profiles and an off-centred $\eta$ minimum in \Fig~ \ref{fig:ProfilesA665}.
Consequently, the flatness of the azimuthally averaged $\eta$ profile is misleading, and does not represent the local distribution.
It certainly does not imply a uniform $\eta(\vectwo{r})$ profile, which is inconsistent with the detailed data.

The $B^2\propto n$ model fits the Southwest profile, provided that $B_0\gtrsim 30 \muG$ (\Fig~\ref{fig:ProfilesA665} shows the $B_0=40\muG$ case).
This is consistent with the $B>17\muG$ lower limit of {\KL}, but greatly exceeds the best fit $B_0=1.3\muG$ of \citet{VaccaEtAl10}; both estimates are inaccurate as they rely on the azimuthally averaged profile.
Moreover, the latter estimate was derived under the assumption of primary {\CREs} at equipartition with the magnetic field.
The shortcomings of such a model were discussed in {\KL}; note that a strong magnetic field that saturates $j_\nu$ would provide a better fit for the radial profile in \citet[][figure 2, at small radii]{VaccaEtAl10}.

To conclude, A665, like A2163, is consistent with homogeneous {\CRIs} and strong, $B_0\gtrsim 30\muG$ magnetic fields, but the corresponding $\eta\propto n^{-1}$ is observed only in the more relaxed regions towards the West.
The $\eta$ profile is dominated in the East by a shock and possible substructure, leading to a misleadingly flat azimuthally averaged $\eta(r)$ profile.
Discussion of the shock-associated steepening and elevated $\eta$ shown in \Fig~\ref{fig:ProfilesA665B} are deferred to \S\ref{sec:ModelApplications}.

\subsubsection{A2744}
\label{sec:A2744}

The rich, hot, and X-ray luminous cluster A2744 appears to be going through a major, $\sim 3:1$ merger in the north-south direction nearly along the line of sight, probably just finishing the first core passage \citep[][whose analysis does not include the Northeast relic region]{BoschinEtAl06}.

\begin{figure}
\centerline{\epsfxsize=10cm \epsfbox{\myfarfig{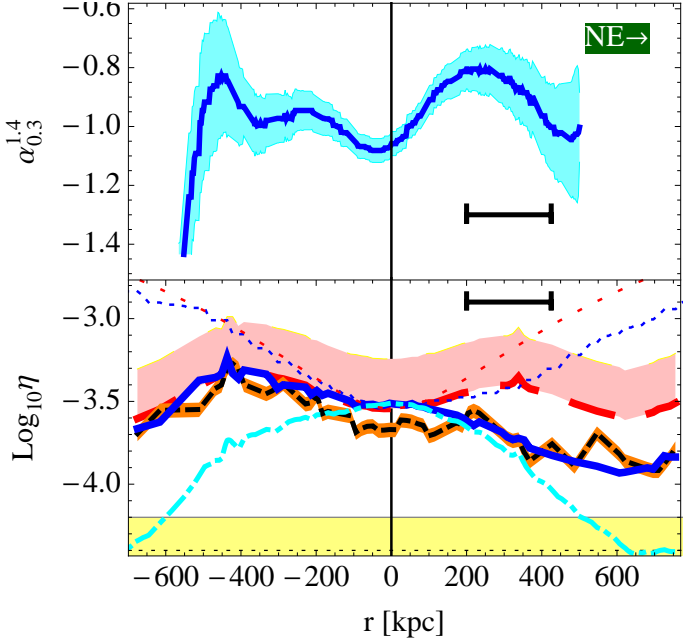}}}
\caption{
Profiles of the A2744 halo along an SW--NE line $\mathcal{L}=\{00\mbox{h}14'9'',-30^\circ 26'3''\}$--$\{00\mbox{h}14'44'',-30^\circ 17'58''\}$, which connects the peak $1.4\GHz$ brightness of the halo ($r=0$) and of the ($r\simeq 1600\kpc$, northeast) relic.
Here we focus on the halo.
Most symbols and notations are defined in Figs. \ref{fig:ProfilesA2163} and \ref{fig:ProfilesA665}.
The radio data is from \citet{OrruEtAl07}.
The $\alpha$ uncertainty is shown as a cyan shaded region.
The ROSAT data is from the PSPC image, background subtracted and smoothed on $30''$, \ie $135\kpc$ scales.
\label{fig:ProfilesA2744HaloNE}}
\end{figure}

\begin{figure}
\centerline{\epsfxsize=10cm \epsfbox{\myfarfig{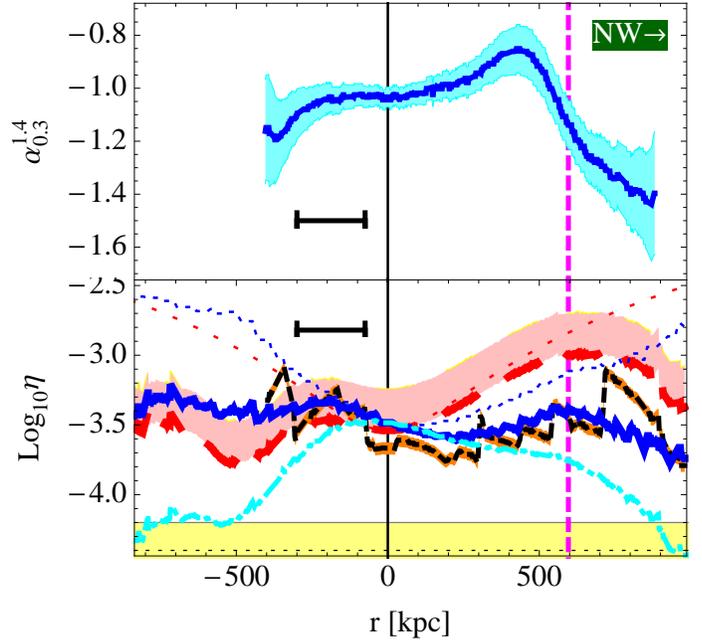}}}
\caption{
Profiles of the A2744 halo along an SE--NW line $\mathcal{L}=\{00\mbox{h}14'29'',-30^\circ 26'12''\}$--$\{00\mbox{h}14'10'',-30^\circ 20'49''\}$, perpendicular to the line examined in \Fig~\ref{fig:ProfilesA2744HaloNE}.
Symbols and notations are defined in Figs. \ref{fig:ProfilesA2163}, \ref{fig:ProfilesA665}, and \ref{fig:ProfilesA2744HaloNE}.
The approximate position of the northwest clump \citep{KempnerDavid04} is shown as a vertical, magenta dashed line.
\label{fig:ProfilesA2744HaloNW}}
\end{figure}

The $\eta$ and $\alpha$ profiles of A2744 are shown along an SW--NE axis in \Fig~\ref{fig:ProfilesA2744HaloNE}, and along the perpendicular, SE--NW axis in \Fig~\ref{fig:ProfilesA2744HaloNW}.
The former crosses the radio relic, found $r\simeq 1600\kpc$ towards the NE.

A linear $I_\nu$--$F_X$ correlation has been reported in this cluster \citep{GovoniEtAl01}.
In X-rays, the cluster shows unusual structure, with ridges extending in four directions (ridges A--D) out to 270--540 kpc distances, and two cool cores.
The combined radio, optical and X-ray data suggests a major merger of mass ratio $\sim 1$, currently at closest approach \citep{KempnerDavid04}.

In accordance, the $\eta$ profiles shown in Figs. \ref{fig:ProfilesA2744HaloNE} and \ref{fig:ProfilesA2744HaloNW} are quite different than found in A665 and A2163.
The central values of $\eta$ are much higher --- by almost an order of magnitude --- than found in the GH sample of {\KL}.
The $\eta$ profiles based on data from ROSAT and \emph{Chandra} are highly asymmetric, initially rising towards the South but declining towards the North.

This peculiar $\eta$ distribution mostly reflects the irregular X-ray morphology, rather than an unusual radio signal, and so does not indicate an irregular {\CRI} distribution.
Indeed, the radio morphology is regular, and strongly resembles other GHs such as A2163.
This can be seen for example by examining the $\eta$ profiles computed using the $\beta$-model, as they are less sensitive to X-ray irregularities.
These profiles are symmetric and radially rising out to $\sim 400\kpc$, except near the Southeast cool core.
Note that the NW profile agrees well with the $\eta\propto n^{-1}$ model, out to the NW clump reported by \citet{KempnerDavid04}.

\begin{figure}
\centerline{\epsfxsize=8cm \epsfbox{\myfarfig{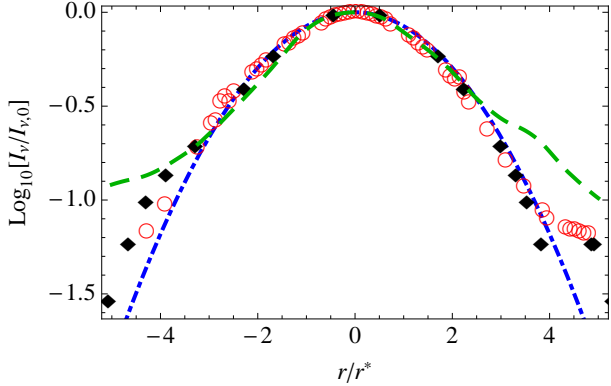}}}
\caption{
Normalised $1.4\GHz$ radio profiles of A2163 ($r$ increasing from South to North; $\mathcal{L}$ and references as in \Fig~\ref{fig:ProfilesA2163B}; black diamonds) and A2744 (Southwest to Northeast; as in \Fig~\ref{fig:ProfilesA2744HaloNE}; red circles).
In A2163, the distance $r$ from the X-ray centre is normalised to $r^*=r_c$.
In A2744, we use $r^*=157\kpc\simeq 0.6r_c$, in order to highlight the similarity to A2163.
(The $r_c$ values are taken from the ASCA analysis of \citet{FukazawaEtAl04}, but in A2744 the fit is problematic; see \citet{KempnerDavid04}.)
The ROSAT-based model for A2163 shown in \Fig~\ref{fig:ProfilesA2163B} (with $B_0=30\muG$), is reproduced here (dashed green curve; deviations beyond $3r_c$ are associated with bright X-ray clumps).
The rescaled profiles resemble each other, and are well fit by the Gaussian $I_\nu=I_{\nu_0}\exp[-0.17(r/r_c)^2]$ (dot-dashed).
The Gaussian behaviour is not due to beam smoothing ($0.8r_c$ FWHM in A2163); \eg the perpendicular profiles are not well fit by a Gaussian.
\label{fig:LineProfileCompareA2163A2744}}
\end{figure}

To demonstrate the similarity between the radio morphology of A2744 and other GHs, we compare in \Fig~\ref{fig:LineProfileCompareA2163A2744} the normalised radio profiles extracted from the SW--NE cut in A2744 ($\mathcal{L}$ as in \Fig~\ref{fig:ProfilesA2744HaloNE}) and from the South-North cut in A2163 ($\mathcal{L}$ as in \Fig~\ref{fig:ProfilesA2163B}).
In order to highlight the similarity between the profiles, in A2163 we normalise the radius $r$ by $r_c$, but in A2744 we normalise it by $0.6r_c$.
The rescaled profiles are in good agreement with each other. Note that $I_\nu$ is well fit by a Gaussian.

Overall, the rising $\beta$-model $\eta(r)$ profiles, the $\eta\sim n^{-1}$ behaviour towards the NW, and the resemblance of the radio morphology to A2163, in which the more relaxed X-ray emission reveals an $\eta\propto n^{-1}$ behaviour, suggest that the {\CRI} distribution in A2744 is homogeneous, too.
However, due to the irregular X-ray emission, the measured $\eta$ profile is effectively flat on intermediate ($\gtrsim 200\kpc$) scales.
This is consistent with the huge, $r_c\simeq 640\kpc$ core found by \citet{GovoniEtAl01B}, largely due to the X-ray ridges \citep{KempnerDavid04}.
As in A665, azimuthal averaging in A2744 hides important features of the $\eta$ profile, and can be misleading.

The combination of regular radio but irregular X-ray emission is consistent with homogeneous {\CRIs}, whereby $j_\nu\propto n^1$, because the X-ray emissivity $j_X\propto n^2$ is more sensitive to inhomogeneities.

Although the radio morphology is regular and resembles other GHs, it is brighter than average.
This suggests a {\CRI} fraction much higher than in other GHs, about an order of magnitude above the estimate of {\KL}; some of the elevated $\eta_0$ may arise, however, from a diminished X-ray peak due to the ongoing merger.

\subsubsection{Conclusion: homogeneous {\CRI} distribution in halos}
\label{sec:HomogeneousCRIsInHalos}

To conclude, direct and indirect observational evidence indicates that in a fair fraction of halos, the radial decline of $I_\nu$ is slower than the radial decline of $F_X$, such that the azimuthally averaged $\eta(r)$ profile is rising.

In other clusters, a constant $\eta(r)$ profile was reported.
However, a detailed morphological analysis of the three halos for which good data is available --- A2163 (where $\eta(r)$ is monotonically rising), A665 and A2744 (a flat $\eta$ was reported in both) --- reveals a rising $\eta(r)$ profile, away from shocks and substructure, in all cases.

In their central $\sim 400\kpc$, the data in all three clusters are consistent with $j_\nu\propto n$, such that $\eta_j\propto n^{-1}$ ($\gamma=-1$).
This behaviour manifests as an $\eta\propto F_X^{-2/3}$ scaling (\cf \EqO~(\ref{eq:EtaIxBetaModel})--(\ref{eq:beta_n_Ix})), or equivalently $I_\nu \propto F_X^{1/3}$, in the relaxed regions where the X-ray emission is regular.
At large radii, the measured $\eta$ profiles gradually flatten with increasing $r$.

More quantitatively, we may attempt to fit $\eta$ as a power law in $F_X$, along the different rays in A665 and A2163 depicted in Figs. \ref{fig:ProfilesA2163}--\ref{fig:ProfilesA665B}.
The results depend weakly on the scales examined, due to the gradual $\eta(r)$ flattening.
Combining the data out to $r_{max}=2r_c$, we obtain
\begin{equation} \label{eq:BestFitGamma}
\eta(|r|<2r_c) \propto F_X^{-0.54\pm0.04} \fin
\end{equation}
This translates to $\gamma=-0.8\pm 0.1$, and is consistent with $\eta_j\propto n^{-1}$ at smaller radii, in particular in regular X-ray regions; see \eqs{\ref{eq:A2163FitR2}} and (\ref{eq:A665FitR15}).
The linear regression leading to \eq{\ref{eq:BestFitGamma}} utilises four rays in each cluster, using the radio data points and the interpolated X-ray data from ROSAT.
We have not included A2744 in the fit because, due to its irregular X-ray morphology, $\eta(r)$ is not symmetric about $r=0$ and is not monotonic in $r$, so the slope strongly depends on $r_{max}$.
Note that we have not included the observational uncertainties in the regression, so the confidence intervals are underestimated.

The $j_\nu\propto n$, or $\gamma=-1$ behaviour we find at small radii is expected if \emph{(i)} diffusion is saturated, such that the {\CRI} distribution is homogeneous, \ie $N_i(\vecthree{r})\simeq \constant$; and \emph{(ii)} the central magnetic field is strong, $B\gtrsim B_{cmb}$.
The two alternative models outlined in \S\ref{sec:etaModels}, where the diffusion is less effective and $\gamma=0$ or $-1/3$, can be ruled out in the three  halos examined above for all plausible magnetic field configurations.

The flattening of $\eta(r)$ at large radii, observed in A665, A2163, and possibly also in A2744, can be explained as arising from a gradual decline in $B$, leading to a transition from the strongly to the weakly magnetised regime $B\lesssim B_{cmb}$ at large radii.
In particular, an $\eta_j\sim N_i B^2/n\simeq \constant$ profile arises in low magnetised regions if the magnetic energy density is proportional to that of the plasma, $B^2\propto n$, as inferred from observations in some clusters \citep[\eg][]{BonafedeEtAl10}.

Our findings suggest that the distribution of {\CRIs} is universally homogeneous in GHs, although a larger sample is required in order to test the ubiquity of this behaviour.
As the flattening of $\eta(r)$ can be explained by magnetic decline, and local drops seen in $\eta$ are associated with clumps, we do not identify any clear radial cutoff in the {\CRI} distribution.

How extended is this distribution? Does it reach the outskirts of clusters? And if so, does it play a role in the peripheral radio emission?
These issues are addressed in \S\ref{sec:eta_in_relics}.

\subsubsection{Reconciling the central $I_\nu\propto F_X^{1/3}$ behaviour with averaged, linear or mildly sublinear correlations}
\label{sec:ReconcilingHomogeneousCRIsInHalos}

How can the $j_\nu\propto n$ emissivity we infer from the data, manifesting as an $I_\nu\propto F_X^{1/3}$ correlation, be reconciled with the linear or mildly sublinear radio--X-ray correlations reported previously, for several clusters?
In such studies, azimuthal averaging or binning the cluster's maps onto a grid led to correlations in the range $I_\nu\propto F_X^{\range{0.6}{1.0}}$ \citep[][and {\KL}]{GovoniEtAl01, FerettiEtAl01}.

This includes, in particular, all three halos in which a close inspection of the radio and X-ray maps indicates that $j_\nu\propto n$ in the central regions, as shown above.
Indeed, binning the data of these clusters, radially or onto a grid, leads to a linear $I_\nu$--$F_X$ correlation \citep[$I_\nu\propto F_X^{0.99\pm0.05}$ in A2744; see][]{GovoniEtAl01}, a mildly sublinear correlation \citep[$I_\nu\propto F_X^{0.64\pm0.05}$ in A2163; see][]{FerettiEtAl01}, or an $\eta(r)\simeq \constant$ profile (in A665; see {\KL} and \Fig~\ref{fig:HalosEtaN}).

In these three clusters, we find that the $I_\nu\propto F_X^{1/3}$ (or equivalently $\eta\propto F_X^{-2/3}$) behaviour near the centre is hidden from such averaging schemes, by a combination of substructure, an irregular X-ray morphology, asymmetry, contaminations, shocks, and weak magnetisation outside the core.
Due to these effects, averaging the data, in particular when cutting the map into equal area bins, smears out the central $I_\nu\propto F_X^{1/3}$ profile.

The averaging process tends to flatten the radial decline of $F_X$ (\eg due to substructure in A2744) and steepen the radial decline of the radio signal (\eg due to radio asymmetry in A665).
Inside the core, the X-ray brightness varies slowly with $r$, so the distinction between an $F_X^{1/3}$ profile and an $F_X^{1}$ profile is small.
Outside the core, due to the decline in $B$ below $B_{cmb}$, the radio profile steepens, leading to a linear $I_\nu\propto F_X$ correlation.
The averaging process typically weighs the data by area, thus emphasising the peripheral regions, where the radio--X-ray relation is indeed linear.

A Combination of these effects effectively masks the $I_\nu\propto F_X^{1/3}$ behaviour in the centre.
It can still be identified by examining the $\eta$ profile along the symmetry axis of the halo, in particular in large, regular halos such as A2163, avoiding substructure, point sources and other contaminations, and refraining from any averaging process.

We predict that when analysed carefully, along these lines, most GHs would reveal an $I_\nu\propto F_X^{1/3}$ profile in their centre, indicative of a homogeneous {\CRI} distribution. Note that in a subset of GHs, in particular small or faint halos, a homogeneous {\CRI} distribution can produce a linear $I_\nu\propto F_X$ profile even in the centre.
This is expected if the magnetic field is weak such that $B\lesssim B_{cmb}$ at small radii.

\subsection{Universal $\eta$ profile among halos and relics: the same, homogeneous {\CRI} origin}
\label{sec:eta_in_relics}

\begin{figure}
\centerline{\epsfxsize=9cm \epsfbox{\myfarfig{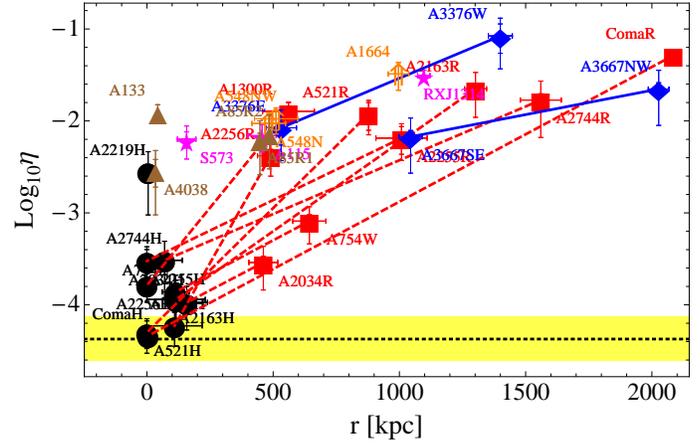}}}
\caption{
Ratio $\eta$ between peak radio and coincident X-ray brightness in relics and halos, as a function of their distance $r$ from the centre of the cluster.
Source symbols are defined in \Fig~\ref{fig:SourcesInuIx}.
Different sources in the same cluster are connected by lines (solid blue in double relic systems; dashed red in halo-relic systems).
Error bars are shown both with and without propagating the $n_0^2$ (but not $r_c$ or $\beta$) uncertainty of the $\beta$-model.
The fit of {\KL} for a sample of GHs is also shown, for reference (\eq{\ref{eq:eta0_KL}}; horizontal dotted line with yellow band showing the $1\sigma$ dispersion).
\label{fig:SourcesEtaR}
\vspace{2mm}}
\end{figure}

\begin{figure*}
\centerline{\epsfxsize=16cm \epsfbox{\myfarfig{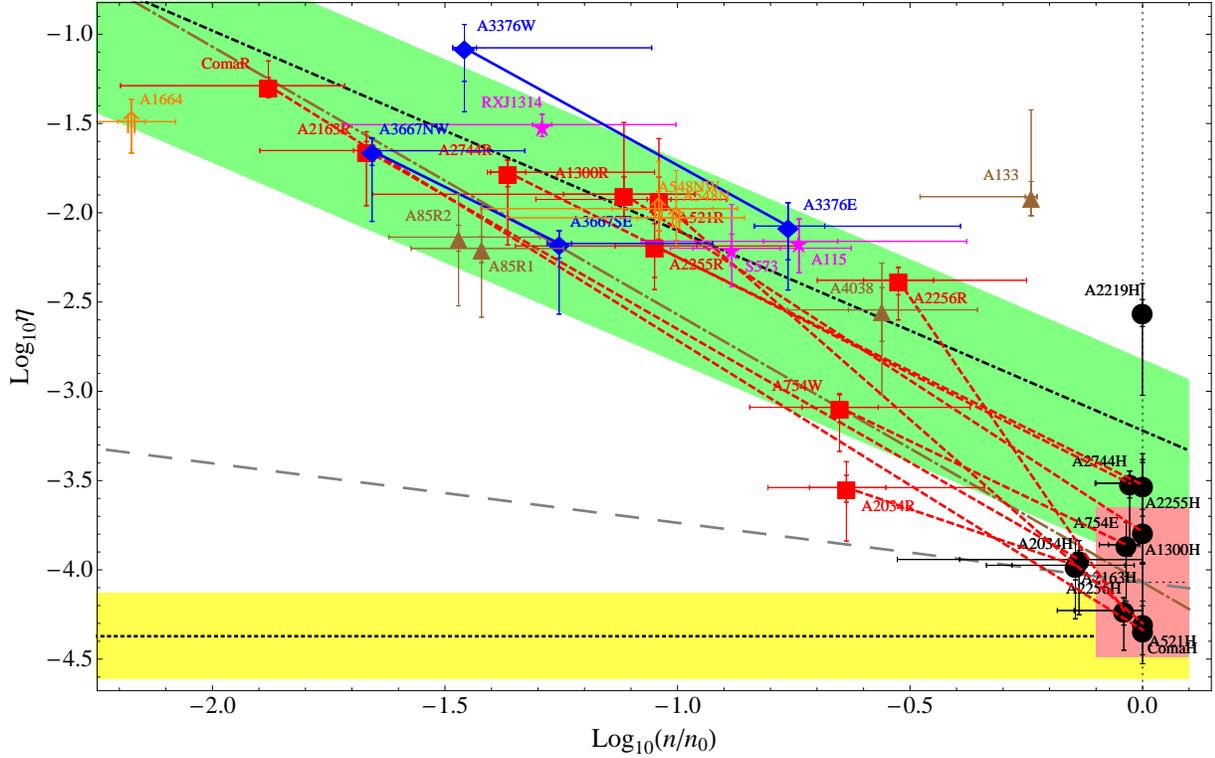}}}
\caption{
Radio peak to X-ray brightness ratio $\eta$ of halos and relics, as a function of the $\beta$-model ratio between local projected and central density, $n/n_0=(1+r^2/r_c^2)^{-3\beta/2}$.
Symbols are defined in \Figs~\ref{fig:SourcesInuIx} and \ref{fig:SourcesEtaR}.
Linear fits are shown for the halos (\eq{\ref{eq:eta0_new_GHs}}; pink rectangle) and for the relics (excluding circular peripheral relics, relics near the first-rank galaxy, and uncertain relics; \eq{\ref{eq:eta_relics_only}}; dot-dashed line with green shaded band).
Also shown, for illustration, are a line of constant radio brightness (\eq{\ref{eq:eta_constant_Inu}}; long dot-dashed, brown) and an $\eta\propto n^{-1/3}$ line corresponding to adiabatic compression (long dashed, gray).
Double error bars are shown for each source, with and without the $\beta$-model uncertainties (see \S\ref{sec:DataPreparation}).
\label{fig:SourcesEtaN}
\vspace{2mm}}
\end{figure*}

We have reviewed several connections between halos and relics in \S\ref{sec:ModelProblems}, and argued in \S\ref{sec:EtaDiagnostic} that the radio--X-ray brightness ratio $\eta\equiv \nu I_\nu/F_X$ provides a useful diagnostic of the nonthermal plasma components.
We now attempt a unified exposition of halos and relics, and plot their $\eta$ values in \Figs~\ref{fig:SourcesEtaR} and \ref{fig:SourcesEtaN}.

The figures show the $\eta$ value of each source, at its peak radio brightness, as explained in \S\ref{sec:DataPreparation}.
Choosing the radio peaks is motivated in part by their possible association with magnetic field maxima.
When the magnetic field amplitude exceeds $B_{cmb}$, $\eta\propto B^2/(B^2+B_{cmb}^2)$ depends weakly on $B$, and so may be used to measure the {\CREs}.

We compute $\eta$ by combining published radio contour maps with $\beta$-models of the X-ray emission, as explained in \S\ref{sec:DataPreparation}.
In \Fig~\ref{fig:SourcesEtaR} we plot $\eta$ against the distance $r$ from the cluster's centre.
In \Fig~\ref{fig:SourcesEtaN}, like in \Fig~\ref{fig:HalosEtaN}, we plot it against the fractional density drop from the centre, $n/n_0=(1+r^2/r_c^2)^{-3\beta/2}$.
In both figures, sources found in the same cluster are connected by lines.
The GH fit of {\KL}, \eq{\ref{eq:eta0_KL}}, is shown for reference (horizontal dotted line surrounded by a $1\sigma$ yellow shaded band).

\subsubsection{Measuring $\eta\propto(n/n_0)^\gamma$ profiles}

Comparing \Figs~\ref{fig:SourcesEtaR} and \ref{fig:SourcesEtaN} suggests that in relics, $\eta$ may be better described as a function of $n/n_0$, rather than as a function of $r$.
Particulary striking is the alignment of the lines (connecting sources found in the same cluster) in \Fig~\ref{fig:SourcesEtaN}, indicating that a power law $\eta\propto (n/n_0)^\gamma$ may roughly agree with all sources.
Therefore, let us focus on the phase space of $\eta$ and $(n/n_0)$, shown in \Fig~\ref{fig:SourcesEtaN}.

Consider first the GHs (black disks in the two figures).
The very high value of $\eta$ in A2219 is due to confusion with a blend of radio sources at the centre of the cluster \citep{OrruEtAl07}, and should be disregarded here.
The remaining nine halos show highly clustered $\eta$ values, as found by {\KL}.
The present sample shows $\eta$ somewhat higher than found in the {\KL} sample, and is best fit by $\eta_0=10^{-3.9\pm0.4}$.
The dispersion is dominated by intrinsic scatter among the halos, needed to obtain an acceptable fit (defined here as $\chi^2/N=1$ with $N$ being the number of degrees of freedom); measurement errors are not propagated (henceforth).
If we include the four additional GHs in {\KL} (A665, A773, A2218 and A2319), the combined fit is slightly lowered,
\begin{equation} \label{eq:eta0_new_GHs}
\eta_0=10^{-4.1\pm0.4} \coma
\end{equation}
but is still somewhat higher than the {\KL} estimate.
This range is shown in \Fig~\ref{fig:SourcesEtaN} as a pink shaded rectangle.

The relics show values of $\eta$ much higher than found in the halos.
As noted in the beginning of \S\ref{sec:HaloAndRelicEta}, the relics tend to have a peak radio brightness roughly similar to that of the halos.
To illustrate this, \Fig~\ref{fig:SourcesEtaN} shows an $I_\nu=\constant$ curve,
\begin{equation} \label{eq:eta_constant_Inu}
\eta(I_\nu=\rm{const.})=\eta_0(n/n_0)^{-3/2}
\end{equation}
(long dot-dashed gray), computed for a characteristic $\beta=2/3$ cluster, with $\eta_0$ fixed using \EqO~(\ref{eq:eta0_new_GHs}).

The relic distribution in \Fig~\ref{fig:SourcesEtaN} shows a correlation, albeit with considerable scatter, between $\eta$ and $n/n_0$: more peripheral relics, found in low density regions, tend to have higher $\eta$.
The best fit for all classical relics, \ie excluding the circular peripheral relics (suspected of projection), relics near the first rank galaxy (suspected of contamination), and the uncertain relics in A754 and A2034, is
\begin{equation} \label{eq:eta_relics_only}
\eta=10^{-3.2_{-0.7}^{+0.4}}(n/n_0)^{-1.1_{-0.6}^{+0.3}} \fin
\end{equation}
This is shown in \Fig~\ref{fig:SourcesEtaN} as a dot-dashed line, surrounded by a shaded green band marking the $1\sigma$ dispersion.
If we include the circular peripheral relics in the fit, it becomes
$\eta=10^{-2.9_{-0.3}^{+0.2}}(n/n_0)^{-0.8\pm0.2}$.
Overall, these results show that the relic data are consistent with $\gamma=1$, although their scatter is substantial and source selection effects can distort the slope.

More than half of the clusters shown in \Figs~\ref{fig:SourcesEtaR} and \ref{fig:SourcesEtaN} harbour two radio sources, in most cases a relic and a halo.
We highlight these clusters by connecting such pairs of sources with a line (solid blue between relics; dashed red between a halo and a relic).
These lines are seen to be more or less parallel to each other in the $\log\eta$--$\log (n/n_0)$ plane, strongly suggesting that the $\eta$ distribution within each cluster --- and not only within halos, say --- is universal.

The slope of these lines provides a direct estimate of $\gamma=d\log\eta/d\log n$.
Such an estimate is more constraining than the profiles deduced by combining sources from different clusters, such as the relics fit of \eq{\ref{eq:eta_relics_only}}, because any scatter that arises from normalisation factors which depend on global cluster parameters, cancels out.

The best fit for the slope of all lines shown in \Fig~\ref{fig:SourcesEtaN}, combined, is somewhat steeper than found among relics,
\begin{equation} \label{eq:eta_lines}
\gamma_{\rm{lines}}=-1.7\pm  0.2 \fin
\end{equation}
This results is dominated by the more numerous (nine) halo--relic pairs, for which
\begin{equation} \label{eq:eta_halo_relic1}
\gamma_{\rm{halo-relic}}=-1.7\pm  0.3 \fin
\end{equation}
The slope is more accurately measured for distant relics.
If we consider only the six halo--relic pairs in which the relic is peripheral, residing where $n<0.1 n_0$, and thus exclude A2034 and A754 (where the relic identification is uncertain anyway) and A2256, we obtain
\begin{equation} \label{eq:eta_halo_relic2}
\gamma_{\rm{halo-relic}}(n<0.1n_0)=-1.6\pm0.2 \fin
\end{equation}
If we consider only the (two) pairs of relics, we obtain a shallower slope,
\begin{equation} \label{eq:eta_relic_relic}
\gamma_{\rm{relic-relic}}=-1.4\pm 0.1 \coma
\end{equation}
which agrees with the slope of the relics fit in \eq{\ref{eq:eta_relics_only}}.
We have not included the observational uncertainties in fitting the line slopes, so the confidence intervals in \eqs{\ref{eq:eta_lines}}--(\ref{eq:eta_relic_relic}) are somewhat underestimated.

\subsubsection{Universal $\eta(n/n_0)$ profile}
\label{sec:UniversalProfile}

The above analysis reveals an apparently universal, monotonically radially rising $\eta(n/n_0)$ profile, involving the radio peaks of different types of diffuse radio sources in different types of clusters.
This is seen in the alignment of the different halos and relics shown in \Fig~\ref{fig:SourcesEtaN}, and in particular in the parallel lines connecting multiple sources found in the same cluster.
This universal $\eta$ profile can be thought of as an (unexpected) extension of the $\eta\sim n^{-1}$ profile derived within halos (in \S\ref{sec:HaloRisingEta}) to large scales, connecting the centres of halo clusters to the peripheral relic peaks.

Approximately the same $\eta(n/n_0)$ profile is found in all clusters with diffuse radio emission, whether a halo, a relic, or both.
In particular, the halos lie close - although not exactly - along the pure power-law $\eta\propto (n/n_0)^\gamma$ profile extrapolated from relics alone, as seen for example by comparing \eqs{\ref{eq:eta_relics_only}} and (\ref{eq:eta0_new_GHs}) (green band and pink rectangle in \Fig~\ref{fig:SourcesEtaN}).
Different types of relics share the same $\eta$ profile, including some peripheral spherical relics (in particular if projection effects are possible) and possibly some relics found near the first rank galaxy.

Inspection of the $\eta(n/n_0)$ distribution of relics in \eqs{\ref{eq:eta_relics_only}} and (\ref{eq:eta_relic_relic}), and the slope of $\eta$ between relics and halos in \eqs{\ref{eq:eta_halo_relic1}}--(\ref{eq:eta_halo_relic2}), suggests that a universal $\eta$ profile would not be a pure power-law, appearing somewhat steeper between halos and relics.
Using the parametrisation $\eta\propto (n/n_0)^\gamma$, this corresponds to a nonconstant $\gamma$, which varies in the range $\gamma=-(1.0$--$1.7)$ among the different types of sources.
This variation is nevertheless quite small, considering that the sources span more than two orders of magnitude in density (and more than that in $\eta$).

Moreover, the steepening of $\eta$ between halos and relics may be due in part to systematic errors caused by our reliance on $\beta$-models to estimate the X-ray brightness.
Recall that the $\beta$-model tends to underestimate the X-ray emission from relics, partly because it does not account for the local X-ray enhancement associated with shock compression, and partly because relics tend to lie along the more elongated X-ray axis (at least in A2163 and A2744).
In reality, then, the $\eta$ values of relics are probably somewhat lower than they appear in \Figs~\ref{fig:SourcesEtaR} and \ref{fig:SourcesEtaN}.
Such a bias does not occur in halos, so improved X-ray modeling should show somewhat better agreement of all sources with a pure power law, $\gamma\simeq -1$.

The universal $\eta(n/n_0)$ profile of radio peaks is found in both rich and poor (S0573) clusters, and in both cool core (see {\KL}) and merger states.
The properties of the {\CREs} and magnetic fields giving rise to the synchrotron radiation must therefore be robust.
They should either be closely related to the thermal X-ray emission, or be saturated as to have little effect on the observed signal.
This is the case in halos, where the {\CREs} arise from a {\CRI} population which has a universal energy fraction with respect to the thermal plasma with little scatter among different clusters, and the magnetic fields are sufficiently strong to saturate the cooling \citep[see][and {\KL}]{KushnirEtAl09}.
But what sets the universal profile of relics?

Moreover, as mentioned above, the $\eta$ profiles among halos, among relics, and within halos appears to be continuous, and therefore, tightly related.
If relic emission arises from primary {\CREs}, accelerated or reaccelerated in a shock, why does the extrapolated $\eta$ profile of relics agree so well with halos, in which {\CREs} are thought to be secondaries?\footnote{Primary {\CRE} halo models would encounter the same problem here, as they invoke a different {\CRE} acceleration mechanism.}
This provides yet another connection between halos and relics, to be added to the list of such connections outlined in \S\ref{sec:HaloRelicConnection}; in present relic models, these are mere coincidences.

\subsubsection{Secondary {\CREs} and strong magnetic fields in relics}
\label{sec:SecondaryCREsInRelics}

A clue to the origin of the relic {\CREs} stems from the halo model of \S\ref{sec:HaloRisingEta}, and by noting that the $\eta$ profile inferred in relics is close --- in both slope ($\gamma\simeq -1$, so $n\eta \simeq \constant$) and normalisation --- to the $\eta$ profile extrapolated from the emission within GHs.
In GHs, the $\gamma=-1$ slope inferred from the data is a natural consequence of saturated {\CRI} diffusion.
This leads to a homogeneous, $N_i\propto \constant$ profile, such that the injection rate of {\CREs} produced in hadronic collisions between the {\CRIs} and the ambient gas scales as $\dot{N}_e\propto n$.
As long as the magnetic field is sufficiently strong to saturate {\CRE} cooling, $B\gtrsim B_{cmb}$, this ensures that $n\eta_j\propto \dot{N}_e/n \simeq \constant$.

It is natural to assume, if diffusion is indeed saturated in halos, and as we find no evidence for a deviation from $N_i\simeq \constant$ out to $r\simeq 400\kpc$ radii in the GHs examined in \S\ref{sec:HaloRisingEta}, that the homogeneous {\CRI} distribution extends out to $r\gtrsim \Mpc$ distances.
Extrapolating the $n\eta$ value of halos to large distances provides a good, although somewhat low, fit to the relics as well.
Therefore, \emph{we propose that relics are synchrotron emission from secondary {\CREs}, produced by the same homogeneous {\CRI} population that gives rise to halos.}
The universal {\CRI} fraction in the centres of halos then ensures that $\dot{N}_e/n$ be universal among both halos and relics.

This, however, does not yet guarantee the $n\eta\simeq \constant$ behaviour observed.
The additional requirement --- strong, $B\gtrsim B_{cmb}$ magnetic fields --- may seem less natural at the large, in some cases $r>2\Mpc$ radii of relics.
Here, the combination of strong fields and low plasma densities entails a small plasma $\beta$ parameter, or equivalently a high magnetic energy fraction,
\begin{equation} \label{eq:epsilon_B}
\epsilon_B\equiv \frac{u_B}{u_{th}} = \frac{B^2/8\pi}{(3/2)\mu^{-1}n k_B T} \simeq 0.10 \frac{b^2}{n_{-3}T_{10}} \coma
\end{equation}
where we defined $b\equiv B/B_{cmb}$ and $n_{-3}\equiv n/10^{-3}$.

Nevertheless, $B>B_{cmb}$ fields are locally plausible downstream of the merger shocks associated with relics.
Even a weak shock amplifies the pressure by a factor of a few, as witnessed by the enhanced X-ray emission and a temperature rise in the vicinity of relics \citep[\eg][]{FinoguenovEtAl10}.
This leads to an elevated $nT$ denominator in \EqO~(\ref{eq:epsilon_B}), effectively lowering $\epsilon_B$.
In addition, as explained in \S\ref{sec:DataPreparation}, our data pertains to the most radio bright point along each relic, presumably corresponding to the most strongly magnetised spot along the shock, so a high value of $\epsilon_B$ is required only locally.

For example, a magnetic field $B>3\muG$ was inferred in the $r\simeq 2\Mpc$, Northwest relic in A3667, both from the Faraday rotation measure \citep{Johnston-Hollitt04} and from an upper limit on Compton emission \citep{FinoguenovEtAl10}.
This translates to $b>0.8$ at the redshift $z=0.055$ and so, according to the $\beta$-model, to $\epsilon_B>0.17$.
The X-ray data suggests a shock of Mach number $\mach\simeq 2$ near the edge of the relic \citep{FinoguenovEtAl10}.
Such a shock raises the pressure by a factor of $\myrp\equiv p_d/p_u=(5\mach^2-1)/4\simeq 5$ (assuming an ideal gas with adiabatic index $\Gamma=5/3$).
If the $\beta$-model distribution is representative of the pre-shock plasma, the observed $B>3\muG$ limit translates to only $\epsilon_B>3.5\%$ downstream, which seems quite plausible.

\subsubsection{A unified, secondary {\CREs} model for relics and halos resolves some previous model discrepancies}

Our model, attributing both relics and halos to secondary {\CREs} produced from the same {\CRI} population, resolves several of the discrepancies of the present, primary {\CRE} relic models, pointed out in \S\ref{sec:ModelProblems}.

The spectrum of secondary {\CREs} injected by hadronic collisions closely follows the spectrum of the primary {\CRI} population.
The {\CRI} distribution is homogeneous and spectrally flat (nearly constant energy per logarithmic interval in particle energy), so the {\CRE} injection spectrum is similarly flat, as inferred from halo observations.
This explains why all relic edges show such a similar, flat, $\alpha\simeq -1$ spectrum, which has no special significance in DSA models, as argued in \S\ref{sec:PeculiarRelics}.
This also explains why the centres of halos and the peaks of relics show, for the most part, a rather similar spectrum.
This is no longer a coincidence, as it was in primary {\CRE} models.

The common source of {\CREs} in halos and in relics explains most of the connections found between them, outlined in \S\ref{sec:HaloRelicConnection}.
The radio bridges observed in several clusters to extend between a halo and a relic simply reflect, in our model, a strongly magnetised region, probably tracing the propagation path of the relic shock.
The strong fields illuminate the pervasive secondary {\CREs} in radio waves; no fine tuning between different halo and relic models is necessary.
Likewise, the similar $P_\nu$--$L_X$ correlations found in halos \citep[][and references therein]{BrunettiEtAl07} and in relics \citep{GiovanniniFeretti04} have the same origin, and are no longer coincidental.

Likewise, the lack of a clear bimodal distinction between the properties of halos and relics, manifest for example in the presence of exceptional halos and relics as described in \S\ref{sec:HaloRelicConnection}, simply reflects, in our model, the similar nature of the two phenomena.
Thus, the varying morphologies, degrees of polarisation, shock associations, etc., among halos and relics, all arise from different magnetic configurations in the cluster.

Specifically, the magnetic fields near the centre of a cluster are thought to be strong and randomly oriented after a merger, and the $B\simeq B_{cmb}$ contour marking the edge of the halo typically being centrally-centred and more or less spherical.
The magnetic field behind a peripheral merger shock, in contrast, is locally enhanced, and thought to have a preferred orientation parallel to the shock.
These properties lead to the regular, central, and unpolarised nature of most halos, and the irregular, peripheral, and polarised nature of most relics.
However, exceptions are to be expected at various stages of the magnetic evolution.
For example, a young merger shock that is magnetising the cluster's centre may lead to an irregular, polarised halo.

Our model alleviates the need to invoke particle acceleration or reacceleration in weak shocks, a process which is neither understood nor observationally constrained, as an explanation for radio relics.
The model does not assume particle acceleration or reacceleration in turbulence, which is similarly unconstrained.
On the contrary, the model places upper limits on the efficiencies of particle acceleration in weak shocks and turbulence, as discussed in \S\ref{sec:Discussion}.

\subsubsection{Challenges for a unified, secondary {\CRE} model}

We have seen that the data supports a unified model which invokes secondary {\CREs}, produced from a homogeneous primary {\CRI} distribution, as the origin of both halos and relics (as well as halo--relic bridges).
This model resolves some of the problems encountered by previous models, and sheds light on the observed connections between halos and relics.
Nevertheless, the model, as described above, is incomplete, and fails to explain some aspects of the halo and relic phenomenology.

First, the data shows that halos and relics do not perfectly align along a pure $\eta\propto (n/n_0)^{-1}$ power law.
The relics have a somewhat higher $\eta$ than anticipated from a simple-minded extrapolation of the $\eta\propto (n/n_0)^{-1}$ profile derived within halos in \S\ref{sec:HaloRisingEta}.
Comparing $\eta$ in different relics, in the same or in different clusters, shows that the $\eta$ profile among relics is somewhat steeper than $\gamma=-1$.
Note that in a diffusion model, one would expect $\gamma$ to flatten --- not steepen --- sufficiently far from the centre.
Moreover, even with the steeper $\gamma$ of relics, extrapolating the relic fit to the cluster's centre, $n/n_0\to 1$, still yields higher $\eta_0$ than observed in halos.
While these discrepancies may be partly relieved by more accurate X-ray data, devoid of the $\beta$-model bias mentioned in \S\ref{sec:UniversalProfile}, they suggest that we have not yet addressed some of the physical processes underlying the radio emission.

Second, the model as stated above cannot account for substantial spectral deviations, away from the flat $\alpha\simeq -1$ observed in the centres of most halos and at the outer edges of relics.
This includes the steep spectrum halos, discussed in \S\ref{sec:SteepHalos}, and the gradual steepening often observed towards the edges of halos and behind (inward of) relics.
Some halo steepening can be explained by the energy dependence of the inelastic cross section for pion production, which governs the injection of {\CREs}, if the magnetic field is sufficiently strong (see {\KL}).
However this cannot explain the very steep spectra of some halos, or the steepening inward of $B\lesssim B_{cmb}$ relics.
In addition, the above model cannot explain the coincidence between the presence of a steep halo and a relic in a cluster, pointed out in \S\ref{sec:HaloRelicConnection}.

Finally, the secondary {\CRE} model of halos, and the unified relic--halo model advocated above, assume steady state {\CRE} injection and static magnetic fields.
More precisely, it is assumed that changes in the magnetic configuration and in particle injection are slow with respect to the cooling time of the {\CREs}.
However, this assumption breaks down in the vicinity of shocks, and probably also at the edges of halos.

In \S\ref{sec:TimeDependentTheory}, we rectify this by deriving the synchrotron signature of time-dependent injection and magnetic fields, focusing in particular on shocks and turbulence.
As we show in \S\ref{sec:ModelApplications}, the remaining model discrepancies outlined here are resolved in the generalised, time-dependent model.


\section{Time-dependent distributions of cosmic-rays and magnetic fields}
\label{sec:TimeDependentTheory}

The analysis of time-dependent {\CRE} injection and magnetic fields is important for understanding key features of radio halos and relics, as we argue in \S\ref{sec:TemporalEvolutionCannotBeNeglected} below.
One type of temporal evolution, discussed in \S\ref{sec:CRAmplification}, involves the compression and acceleration of the ambient plasma and {\CRs} --- both primary {\CRIs} and secondary {\CREs} --- by weak shocks.
Next, we study the evolution of the secondary {\CRE} population injected into time-dependent magnetic fields, in \S\ref{sec:CRE}, and the properties of the resulting  synchrotron emission, in \S\ref{sec:SynchrotronVariability}.
Combining the results derived in \S\ref{sec:TemporalEvolutionCannotBeNeglected}--\ref{sec:SynchrotronVariability}, we discuss the synchrotron signature of a shock in \S\ref{sec:SynchrotronImprintOfShock},
and the signature of {\CRE} diffusion through a turbulent magnetic field in \S\ref{sec:SynchrotronEvolvingB}.

\subsection{Temporal evolution cannot be neglected in relics and in all halos}
\label{sec:TemporalEvolutionCannotBeNeglected}

Diffuse radio emission from galaxy clusters arises from {\CREs}, injected locally into the plasma, losing most of their energy to inverse-Compton and radio synchrotron radiation over a (cosmologically) short timescale.
The cooling time of a {\CRE} that at redshift $z$ emits synchrotron radiation received today with characteristic frequency $\nu=1.4\nu_{1.4}\GHz$ is
\begin{equation} \label{eq:CRE_Cooling0}
t_{cool} \simeq 0.11 \left[\frac{4(b\sqrt{3})^{-\frac{3}{2}}}{1+b^{-2}}\right] \nu_{1.4}^{-\frac{1}{2}}(1+z)^{-\frac{7}{2}} \Gyr \coma \,\,
\end{equation}
where $b\equiv B/B_{cmb}$ is the normalised magnetic field, assumed constant, and an average pitch angle $\myPalpha=\pi/4$ is adopted (see \S\ref{sec:SynchrotronVariability}).
The term in square brackets peaks at unity when $b=1/\sqrt{3}$.

In most parts of a cluster, the plasma properties change slowly, over the $\sim 1 \Gyr$ dynamical timescale, so {\CRE} injection and magnetic fields can be approximated as time-independent (as assumed for the magnetic field in \eq{\ref{eq:CRE_Cooling0}}).
This approximation substantially simplifies the analysis of synchrotron radiation from halos and relics, and has been adopted in virtually all previous models of diffuse radio emission from clusters.

However, diffuse radio emission is also observed from dynamical regions, where fast density and magnetic changes occur over timescales much shorter than $t_{cool}$.
The most radical examples are the weak, Mach number $\mach\sim 2$--$3$ shocks believed to energise the peripheral radio relics, and the weak shocks observed at the edges of halos.
Here, the plasma is rapidly compressed, by factors of $\sim 2$--$3$, and the magnetic field amplitude $B$ may be similarly amplified.
Plasma compression takes place over the shock transition crossing time, $t_{sh}\sim L/v_s \sim 2(\lambda_{sh}/100)(\mach c_s/3000\km\se^{-1})^{-1}(n/10^{-3}\cm^{-3})^{-1/2}\,\rm{ms}$, which is dramatically shorter --- by some $18$ orders of magnitude --- than $t_{cool}$.
Here, $v_s$ and $c_s$ are the shock and sound velocities, and we parameterised the shock width as $L=\lambda_{sh} l_{sd}$, where $l_{sd}$ is the proton skin depth and $\lambda_{sh}$ is an unknown number of order 100.
Note that {\CRE} injection, being proportional to $n$, changes abruptly at the shock, within $\sim t_{sh}$.
It continues to evolve, due to changes in the {\CRI} distribution, over longer length- and time-scales.

Shock acceleration or reacceleration of {\CRs}, and further magnetic field amplification, may proceed on timescales $t_{rel}\sim D/v_s^2$ much longer than $t_{sh}$, but still much shorter than $t_{cool}$.
Here, $D$ is the CR diffusion coefficient, and $t_{rel}$ is defined as the time during which a Lagrangian upstream fluid element sees a substantial change in its relativistic particle population and the associated magnetic field.
It is much shorter than $t_{cool}$ even for inefficient diffusion, typically estimated away from the shock (for $10\GeV$ {\CRs} in $B\sim 3\mu$G fields) as $D \lesssim 10^{31}\cm^2\se^{-1}$ \citep[\eg][]{VolkEtAl96}, such that $t_{rel}\lesssim 5(D/10^{31}\cm^2\se^{-1})(\mach c_s/3000\km\se^{-1})^{-2} \Myr$.

Moreover, even weak shocks travel a considerable distance during $t_{cool}$,
\begin{equation}
v_s t_{cool} \simeq 370 \mach_2 T_{10}^{1/2} \left( \frac{t_{cool}}{0.11\Gyr} \right) \kpc \coma
\end{equation}
where $T_{10}\equiv k_B T/10\keV$ and $\mach_2\equiv \mach/2$.
Hence, the temporal evolution of emission behind the shock can be resolved and modeled.
This has been attempted for example in A521 \citep{GiacintucciEtAl08}, where the spectral steepening behind the relic was interpreted as gradual cooling of a primary {\CRE} distribution injected at the outer relic edge.

Fast temporal evolution of the magnetic field is also thought to take place away from shocks, for example near the centres of young halos, due to strong turbulence triggered by a merger event.
Particularly fast, exponential field evolution is expected when the halo is born \citep[\eg][]{SubramanianEtAl06}, as the ICM becomes strongly magnetised by merger-induced shocks or turbulence.
Similarly fast field evolution may take place at the edges of halos, in particular in MHs where the magnetic structures associated with CFs are thin \citep{KeshetEtAl10}.

The characteristic coherence scale of magnetic fields in the ICM is estimated to be on the order of $\lambda_B\lesssim 20\kpc$ \citep{CarilliTaylor02, Clarke04}.
Substantial changes in the magnetic field can be expected over the corresponding crossing time of fast hydromagnetic modes,
\begin{equation}
t_B \sim \lambda_B/c_s \lesssim 10 T_{10}^{-1/2} (\lambda_B/20\kpc)\Myr \fin
\end{equation}
Alv\'{e}n modes may similarly modify the field, over a time scale $\sim 5 (n_{-3}T_{10})^{1/2} (B/5\muG)^{-1}t_B$, typically longer than $t_B$, but still shorter than $t_{cool}$ in highly magnetised regions.

In cases where the global parameters of the magnetic field evolve slowly, and many, uncorrelated turbulent eddies are integrated over the radio beam, one may effectively assume that the field is stationary ({\KL}); this does not apply to relics, young halos, and halo edges.
A time-dependent approach is needed if only a few eddies are included, if correlation between eddies are important, or if eddy oscillations are resonant with the cooling time.

The preceding arguments indicate that the temporal evolution of {\CRE} injection and magnetic fields cannot in general be neglected in the study of diffuse radio emission from galaxy clusters.
This is particularly true for relics, being associated with shocks, for young halos, where substantial field growth takes or has recently taken place, and for halo edges, where $B\sim B_{cmb}$ may be highly variable.
Indeed, incorporating temporal evolution in relic and halo models is essential for resolving the discrepancies outlined towards the end of \S\ref{sec:HaloAndRelicEta}.

\subsection{Shock compression and {\CR} amplification}
\label{sec:CRAmplification}

Consider an infinite planar shock of Mach number $\mach$, propagating through a homogeneous medium with adiabatic index $\Gamma=5/3$ (see concluding remarks for an arbitrary $\Gamma$).
The low central {\CRI} fraction inferred from halo observations \citep[][and {\KL}]{KushnirEtAl09} indicates that {\CRIs} do not play an important dynamical role in the centres of clusters, except perhaps near local sources.
The analysis of relics in \S\ref{sec:HaloAndRelicEta} indicates that the same holds for relic shocks.
We may therefore use the test-particle approximation to study {\CRs} in weak ICM shocks.
We assume, as usual, that the {\CRs} interact with the ambient plasma by scattering off magnetic irregularities.

According to diffusive shock acceleration (DSA) theory \citep{Krymskii77, AxfordEtAl77, Bell78, BlandfordOstriker78}, relativistic particles near the shock are accelerated by the Fermi mechanism, resulting in a downstream power-law energy spectrum $N(E)\propto E^{s}$ of index
\begin{equation} \label{eq:DSA_s}
s = -\frac{\myrg+2}{\myrg-1} \coma
\end{equation}
where $\myrg$ is the gas compression ratio.
For an adiabatic index $\Gamma=5/3$, $\myrg=4\mach^2/(3+\mach^2)$, so
\begin{equation} \label{eq:DSA_s2}
s = -2 \frac{\mach^2+1}{\mach^2-1} \leq -2 \fin
\end{equation}

Non-cooled {\CRs} (either {\CREs} or {\CRIs}) which already have a flat power-law spectrum ($s=-2$; equal energy per logarithmic particle energy interval) upstream, cannot become flatter in standard DSA, at least not in the test-particle approximation.
It is therefore natural to assume that the flat spectral slope is not altered by the shock, although the normalisation and cutoffs may change.
Under these conditions, it can be shown that the {\CR} energy is shock-amplified by a factor $\mach^2$, and this result does not depend on the details of the interaction with the magnetic irregularities.

To see this, consider the shock frame, oriented such that the shock lies at $\myX=0$ and the plasma flows in the positive $\myX$ direction, with velocity $v=v_s$ upstream ($\myX<0$) and $v_s/r$ downstream.
Parameterising the CR scattering by some diffusion function $D(E,\myX)$, the steady state CR distribution $N(E,\myX)$ satisfies \citep{Krymskii77}
\begin{equation}
\pr_\myX(N v) = \pr_\myX(D\pr_\myX N)+\frac{1}{3}\pr_E (N E)\pr_\myX v \fin
\end{equation}
Integration over $\myX$ yields
\begin{equation} \label{eq:int_CR_PDE}
D\pr_\myX N = N v - N_0 v_s + \frac{(\myrg-1)(s+1)}{3\myrg} v_s\Theta(\myX) N_{sh} \coma \,\,
\end{equation}
where $N_0(E)\equiv N(E,\myX\to-\infty)$ is the CR density far upstream, and the Heaviside step function
\begin{equation}
\Theta(\myX)=
\begin{cases}
0 & \text{if $\myX \leq 0$ ;}
\\
1 & \text{if $\myX > 0$ }
\end{cases}
\end{equation}
is unity downstream and vanishes upstream.
As there is no energy scale in the problem, the spectrum at the shock is a power-law, $N_{sh}(E)\equiv N(E,\myX=0) \propto E^{s}$.

The only bound solution of \eq{\ref{eq:int_CR_PDE}} downstream is uniform, $N_d(E,\myX)=N_{sh}(E)$.
Primary particle acceleration involves no particles far upstream, so by taking $N_0\to 0$ one recovers the standard spectrum \EqO~(\ref{eq:DSA_s}), but cannot determine the normalisation of the distribution, \ie the acceleration efficiency.
However, in the present case of an upstream flat spectrum, we must retain the $N_0$ term, leading to a {\CR} compression ratio
\begin{align}
\label{eq:CR_Compression1} \myrcr \, \equiv \, \frac{N_d}{N_0} & =  \frac{3\myrg}{3+(\myrg-1)(s+1)}  \\
\label{eq:CR_Compression2} & =  \frac{3\myrg}{4-\myrg} \, = \, \mach^2 \coma
\end{align}
where we used $s=-2$ in the last line.
Notice that a larger (smaller) amplification factor would arise if we were to assume that the downstream spectrum is steeper (flatter) than the upstream spectrum in \eq{\ref{eq:DSA_s2}}.

\begin{figure}
\centerline{\epsfxsize=8cm \epsfbox{\myfarfig{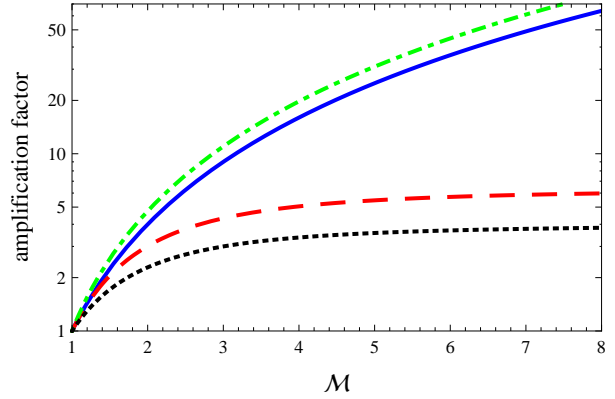}}}
\caption{
The amplification induced by a shock of Mach number $\mach$ in a gas with adiabatic index $\Gamma=5/3$, for
\emph{(i)} {\CRs} with $s=-2$ (\eq{\ref{eq:CR_Compression2}}; solid)
\emph{(ii)} {\CRs} with $s=-2$, assuming adiabatic compression only (dashed);
\emph{(iii)} the ambient gas pressure (dot-dashed);
and
\emph{(iv)} the ambient gas density (dotted).
\label{fig:Reacceleration}
\vspace{2mm}}
\end{figure}

The {\CR} compression ratio $\myrcr(\mach)$ is shown in \Fig~\ref{fig:Reacceleration} as a function of the Mach number.
As the figure demonstrates, the {\CR} amplification we find due to {\CR} interactions with magnetic irregularities is larger than the adiabatic compression ${\myrcr}_{,ad} = \myrg^{(2-s)/3} = \myrg^{4/3}$ usually used in this context \citep[\eg in ][]{EnsslinGopalKrishna01}.
It is bracketed by the ambient plasma's density compression $\myrg$ from below, and by its pressure amplification factor $\myrp$ from above, where \citep{LandauLifshitz60_EM}
\begin{equation}
\myrp\equiv \frac{p_d}{p_u} = \frac{2\Gamma \mach^2}{\Gamma+1} - \frac{\Gamma-1}{\Gamma+1} = \frac{5\mach^2-1}{4} \coma
\end{equation}
and we assumed $\Gamma=5/3$ in the last equality.

In particular, we conclude that a weak shock \emph{lowers} the energy fraction of a flat-spectrum distribution of {\CRs} (either {\CRIs} or {\CREs}),
\begin{align}
\frac{\myrcr}{\myrp}
& = \frac{\Gamma+1}{2\Gamma-(\Gamma-1)\mach^{-2}} \label{eq:DSA_energy_loss1}
\\
& = \frac{4}{5-\mach^{-2}} < 1 \label{eq:DSA_energy_loss2}
\coma
\end{align}
where we neglected logarithmic corrections due to changes in the {\CR} energy cutoffs, and assumed $\Gamma=5/3$ in the second line.
The drop in {\CR} energy fraction due to a shock asymptotes to a constant $2\Gamma/(\Gamma+1)=5/4$ (for $\Gamma=5/3$) as the shock becomes stronger.
This holds up to the strong shock regime, where fresh {\CR} injection can no longer be neglected.

Recall that the radio and X-ray emissivities scale as $j_\nu\propto N_e B^2/(B^2+B_{cmb}^2)$ and $j_X\propto n^2$, respectively, and that {\CREs} are injected according to $\dot{N}_e\sim N_i n$.
We postpone the discussion of magnetic field amplification and its consequences to \S\ref{sec:CRE}.
Without these effects, shock compression and {\CRE} amplification alone initially cause $j_\nu$ to increase by a factor $\myrcr=\mach^2$ behind the {\CR} transition layer, but cause the radio--X-ray ratio $\eta$ to initially \emph{decrease}, by a factor $\myrg^2/\myrcr = 16\mach^2/(3+\mach^2)^2$ (which is larger than unity for $\mach<3$).
However, at the {\CRE} cooling distance $\sim v_s t_{cool}/\myrg$ behind the shock, the shocked {\CREs} have been replaced by newly injected particles.
By this point, the shock-induced jump in $n$ and in $N_i$ cause $j_\nu$ to increase by a larger factor, $\myrcr \myrg=4\mach^4/(3+\mach^2)>1$, and induce an increase of $\eta$ by a factor $\myrcr/\myrg=1+(\mach^2-1)/4>1$.
(These enhancement factors are taken with respect to the upstream, temporarily assuming no change in magnetic field).

If the diffusion function is homogeneous, \ie $D(E,\myX)=D(E)$, then the upstream solution is given by
\begin{equation}
N_u(E,\myX) = N_0 + (N_{sh}-N_0)e^{(v_s/D)\myX} \fin
\end{equation}
The {\CR} transition layer has, therefore, a thickness $\sim D/v_s$.

Notice that we have used the assumption $\Gamma=5/3$ for the ambient gas only in identifying a flat CR distribution with an $s=-2$ spectrum.
Hence, \eq{\ref{eq:CR_Compression1}} (but not \eq{\ref{eq:CR_Compression2}}) holds for any $\Gamma$ and $s$.
In particular, the diverging amplification it implies for an $s=-2$ distribution when $\myrg=4$, persists for any $\Gamma$.

However, $s=-2$ is not the flattest spectrum attainable in \eq{\ref{eq:DSA_s}} for arbitrary $\Gamma$.
If we assume that the spectrum upstream is the flattest spectrum possible in DSA (from previous injection by a strong shock, say), we recover $N_d/N_0=\mach^2$ regardless of $\Gamma$.
Here, \eq{\ref{eq:DSA_energy_loss1}} (but not \eq{\ref{eq:DSA_energy_loss2}}) holds for any $\Gamma$.

Our simplified analysis assumes a pure power-law spectrum, with no cutoffs. 
The results are substantially modified near such cutoffs, and \eq{\ref{eq:CR_Compression2}} becomes an upper limit to the {\CR} amplification.
We neglected {\CR} feedback effects, so the results are not valid in scenarios where the {\CR} pressure is non-negligible with respect to the ambient plasma.
The results are also modified in shocks strong enough to accelerate a flat {\CR} spectrum, \ie when $\mach\gtrsim 5$; here, \eq{\ref{eq:CR_Compression2}} should be regarded as a lower limit.

\subsection{Time-dependent injection and magnetic fields: {\CRE} evolution}
\label{sec:CRE}

\subsubsection{General evolution of the {\CRE} distribution}

A {\CRE} injected into the ICM cools mainly off the ambient photons and magnetic fields, according to
\begin{equation} \label{eq:CoolingODE}
\pr_t E_e = -\mypsi E_e^2 \coma
\end{equation}
where we defined the cooling parameter
\begin{eqnarray}
\mypsi(t,\vecthree{r}) & \equiv & \mypsi_\gamma + \mypsi_B \equiv \frac{4}{3}\sigma_T m_e^{-2}c^{-3}\left(u_{\gamma}+\frac{B^2}{8\pi}\right) \\
& \simeq & 0.83(1+z)^4(1+b^2) \GeV^{-1} \Gyr^{-1} \nonumber \fin
\end{eqnarray}
Here, $\mypsi_\gamma(t,\vecthree{r})$ is the cooling parameter associated with Compton scattering off photons with local energy density $u_\gamma(t,\vecthree{r})$, typically dominated by the cosmic microwave background $u_{cmb}$ as assumed in the second line.

The diffusion of such a {\CRE} through the ICM depends on the poorly constrained magnetic configuration.
A typical estimate of the diffusion function \citep{VolkEtAl96}, which assumes scattering off magnetic regularities with a $P(k)\propto k^{-3/2}$ power spectrum, indicates that the characteristic distance $L$ traversed during $t_{cool}$,
\begin{equation}
L \simeq 20 (1+b^2)^{-1/2} (1+z)^{-19/8} \left(\nu_{1.4} b \right)^{-1/8} \kpc \coma
\end{equation}
is of the order of the magnetic coherence scale when $B\sim B_{cmb}$.

The energy of the {\CRE} evolves according to \EqO~(\ref{eq:CoolingODE}),
\begin{equation} \label{eq:CRE_Cooling}
E(t)^{-1} = E_0^{-1} + \myPsi_{t,t_0} \coma
\end{equation}
where we omitted subscripts $e$ (henceforth), defined $E_0=E(t_0)$, and introduced the integrated cooling parameter
\begin{equation} \label{eq:Def_myPsi}
\myPsi_{t,t_0} \equiv \int_{t_0}^t \mypsi(t')\,dt' \fin
\end{equation}
This implies that all {\CRE} which have energy $\gg E$ at time $t_0$, cool down to $E$ at approximately the same time $t$, determined by $\myPsi_{t,t_0}=E^{-1}$.

Assuming that {\CREs} are injected into the plasma at a rate (number density per unit time) $\dot{\myn}_+(t,E)$, their evolution is described by the diffusion-loss equation for relativistic particles,
\begin{equation} \label{eq:PDE_n}
\pr_t \myn(t,E) = \dot{\myn}_+ - \pr_E\left( \myn \frac{dE}{dt} \right) + \grad(D\grad{\myn})\fin
\end{equation}
In the following, we neglect the last, diffusion term, and reintroduce it later in \S\ref{sec:SynchrotronEvolvingB}.
We also neglect adiabatic losses, $dE/dt=-(1/3)(\grad\cdot \vecthree{v})E$, assuming that the gas evolves slowly compared to $t_{cool}$.
The effects of shocks have been considered separately in \S\ref{sec:CRAmplification}, and will be reintroduced into the analysis at the end of \S\ref{sec:CRE}.

Under these assumptions, the {\CRE} distribution follows the partial differential equation (PDE)
\begin{equation} \label{eq:PDE_n}
\pr_t \myn(t,E) = \dot{\myn}_+ - \pr_E\left( \myn \frac{dE}{dt} \right) \fin
\end{equation}
A flat spectrum $\myn\propto E^{-2}$, with equal energy per logarithmic interval in particle energy, $f(t,E)\equiv E^2\myn=\constant$, is a solution of the homogenous equation for constant cooling ($\psi=\const$).
It is advantageous to introduce the inverse energy $\myx=E^{-1}$ and use $f(t,\myx)\equiv f(t,E=\myx^{-1})$ instead of $\myn(t,E)$.
\EqO~(\ref{eq:PDE_n}) then simplifies to
\begin{equation} \label{eq:PDE_f}
\pr_t f(t,\myx) = Q - \psi \pr_\myx f \coma
\end{equation}
where
\begin{equation} \label{eq:Def_Q}
Q(t,\myx)\equiv E^2\dot{\myn}_+
\end{equation}
is the logarithmic energy injection rate.

The general solution of \EqO~(\ref{eq:PDE_f}) is
\begin{equation} \label{eq:PDE_sol}
f(t,\myx) = f(t_0,\myx-\Psi_{t,t_0}) + \int_{t_0}^t Q(\tau,\myx-\Psi_{t,\tau}) \,d\tau \coma
\end{equation}
where $t_0$ is an arbitrary initial time, and we require $f(t,\myx<0)=0$ and $Q(t,\myx<0)=0$ in order to avoid non-physical results.

For example, consider the simple case where starting from some time $t_0$, {\CREs} are steadily injected with a power-law energy spectrum $Q(t>t_0,\myx)=\myQ_0 \myx^{-q}$ (\ie $\dot{\myn}_+\propto E^q$; we assume $q<1$ throughout this work), and $\psi(t)$ is constant such that $\myPsi_{t,t'}=\mypsi(t-t')$.
Here we recover the known solution
\begin{equation} \label{eq:PDE_sol_start_injection}
f(t,E)
= \frac{\myQ_0 E^{q-1}}{(1-q)\mypsi}\left\{1 - \mbox{Max}\left[0,1-\frac{E}{E_c(t)}\right]^{1-q} \right\} \coma
\end{equation}
where $E_c(t)\equiv [(t-t_0)\mypsi]^{-1}$ is the energy with cooling time $t-t_0$.
At early times, this becomes $f = \myQ_0 E^{q}(t-t_0) + O(t-t_0)^2$, whereas at late times $f\to \myQ_0 E^{q-1}/(1-q)\mypsi$.
The late time limit is a stationary solution of \EqO~(\ref{eq:PDE_f}), as it satisfies
\begin{equation}
\pr_\myx f = Q / \psi \fin
\end{equation}

In general, \EqO~(\ref{eq:PDE_sol}) shows that $f(t,\myx)$ is the sum of all retarded injections $Q(\tau, \myy)$ along a cooling trajectory $\{\tau, \myy\}$, terminating at $\{t, \myx\}$ and determined by the cooling parameter $\psi(t)$ through
\begin{equation} \label{eq:Def_myy}
\myy(\tau) = \myx-\Psi_{t,\tau} \fin
\end{equation}
Assuming sufficiently early $t_0$ or high {\CRE} energy such that $\myx<\Psi_{t,t_0}$, the initial condition $f(t_0)$ can be eliminated from the solution \eq{\ref{eq:PDE_f}}.
The trajectory then begins at $\{\tau,\myy\}=\{t_i,0\}$, where the earliest retarded time $t_i(t,\myx)$ is defined by $\myy(t_i)=\myx-\Psi_{t,t_i}=0$, \ie all the {\CREs} present at time $t_i$ cool to energies $<\myx^{-1}$ by time $t$.
\Eq{\ref{eq:PDE_sol}} then becomes
\begin{equation} \label{eq:PDE_sol2}
f(t,\myx) = \int_{t_i}^t Q(\tau, \myy(\tau)) \,d\tau \fin
\end{equation}

In addition to the standard, instantaneous (or future) estimate of the cooling time,
\begin{equation}
t_{cool}(t,\myx) \equiv \frac{\myx}{\psi(t)} \coma
\end{equation}
it is useful to introduce the retarded cooling time,
\begin{equation} \label{eq:tau_cool_def}
\tau_{cool}(t,\myx) \equiv t-t_i(t,\myx) \coma
\end{equation}
which equals $t_{cool}$ if $\psi(t)=\constant$, but may substantially differ from it when cooling is time-dependent.

\subsubsection{Variable injection or cooling modify the {\CRE} spectrum}
\label{sec:magnetic_variations_modify_CRE}

Consider temporal variations in the cooling parameter $\psi$, for example due to changes in a strong, $B\gtrsim B_{cmb}$ magnetic field, or temporal changes in the injection rate $Q$.
Crudely speaking, the {\CRE} distribution responds to changes in $\psi$ or in $Q$ by evolving towards a new steady state where $f \sim Q / \psi$.
The distribution evolves faster at high energies, where the cooling time is short; it takes gradually longer for the distribution at lower energies to relax into the new steady state.
Due to this gradual response, the spectrum at energy $E$ steepens/flattens, reflecting a temporal growth/decline in $\psi$ or a decrease/increase in $Q$, that occurred roughly a (retarded) cooling time $\tau_{cool}$ earlier.

Quantitatively, the {\CRE} spectrum can be derived from \EqO~(\ref{eq:PDE_sol2}) and written in the form
\begin{eqnarray} \label{eq:CRE_phi}
\phi(t,E) & \equiv & \frac{d\ln f}{d\ln E} = - \frac{d\ln f}{d\ln \myx} \\
& = & \overline{q} -\frac{\myx Q(t_i,0)}{f \mypsi(t_i)} - \frac{1}{f} \int_{t_i}^t d\tau \, \myPsi_{t,\tau} \pr_y Q(\tau, \myy) \nonumber \\
& = & \overline{q} - 1 + \frac{1}{f} \int_{t_i}^t d\tau \, \myPsi_{t,\tau} \pr_\tau \left[ \frac{Q(\tau, \myy)}{\mypsi(\tau)} \right] \nonumber \coma
\end{eqnarray}
where
\begin{equation}
\overline{q} \equiv \frac{\int_{t_i}^t Q(\tau,\myy) q(\tau,\myy)\,d\tau }{\int_{t_i}^t Q(\tau,\myy)\,d\tau }
\end{equation}
is the energy spectral index $q(t,\myx)\equiv -d\ln Q/d\ln \myx$ of {\CRE} injection, averaged over the $\{\tau,\myy\}$ trajectory.
Notice that for injection of a power-law spectrum with a fixed index,
\begin{equation} \label{eq:power_law_injection}
Q(t,\myx)=\myx^{-q_0}\myQ(t)
\end{equation}
with constant $q_0$, we have $\overline{q}=q_0$ regardless of the normalisation time-dependence $\myQ(t)$.

For constant injection and cooling, the integrand in the last term in \EqO~(\ref{eq:CRE_phi}) vanishes and we recover the standard result $\phi=\overline{q}-1$.
For time-dependent injection/cooling, the integrand has the same sign as $\pr_\tau (Q/\mypsi)$ (because $f$ and $\myPsi$ are positive).
This quantifies the steepening (flattening) of the spectrum with respect to $\overline{q}-1$ when cooling strengthens (weakens) or the injection rate decreases (increases) in time.

This effect is most transparent if variations in $Q(t,\myx)$ are small, \ie injection is approximately uniform and flat.
Here, $\overline{q}\simeq 0$, and the {\CRE} spectrum is mostly sensitive to the retarded cooling time $\tau_{cool}$ and to the retarded value of the cooling parameter, $\mypsi_i\equiv \mypsi(t_i)$.
Indeed, for $Q(t,\myx)=\myQ_0=\constant$, the {\CRE} distribution becomes
\begin{equation}
f(t,\myx)=\tau_{cool}\myQ_0 = \frac{\myx \myQ_0}{\mypsi_a} \coma
\end{equation}
so the {\CRE} spectrum is given by
\begin{equation} \label{eq:phi_q0}
\phi(t,E) = -\frac{\myx}{\tau_{cool}\psi_i} = -1-\frac{\mypsi_{a}-\mypsi_i}{\mypsi_i} < 0  \coma
\end{equation}
where we defined a time-averaged cooling parameter $\mypsi_{a}(t,t_i) \equiv \myPsi_{t,t_i}/\tau_{cool}$.
Hence, the spectrum steepens and $\phi<-1$ (flattens and $-1<\phi<0$) if cooling becomes stronger (weaker) in time, in the sense that $\mypsi_{a}>\mypsi_i$ ($\mypsi_{a}<\mypsi_i$).

It is straightforward to generalise this for time-dependent, flat ($q=0$) injection, $Q(t,x)=\myQ_0(t)$, as well as time-dependent cooling.
Here
\begin{equation} \label{eq:f_time_dependent_q0}
f(t,\myx) = \int_{t_i}^{t} \myQ_0(\tau)\, d\tau \equiv Q_{a} \tau_{cool} = \frac{\myx Q_a}{\mypsi_a} \equiv \myQpsi_a \coma
\end{equation}
which defines the average injection $Q_a$, in analogy with the average cooling $\psi_a$.
We define $\myQpsi\equiv \myx Q/\psi$ such that $f=\myQpsi$ is the steady-state distribution corresponding to constant $Q$ and $\psi$.
The spectrum is then given by
\begin{equation} \label{eq:phi_time_dependent_q0}
\phi(t,\myx) = - \frac{Q_i/\psi_i}{Q_a/\psi_a} = -1 - \frac{\myQpsi_i-\myQpsi_a}{\myQpsi_a} <0 \coma
\end{equation}
where $Q_i\equiv \myQ_0(t_i)$ and $\myQpsi_i \equiv Q_i/\psi_i$.

\Eqs{\ref{eq:f_time_dependent_q0}} and (\ref{eq:phi_time_dependent_q0}) imply that
\begin{equation} \label{eq:phi_f_time_dependent_q0}
\phi(t,\myx) f(t,\myx) = (-1) \myQpsi_i \fin
\end{equation}
Thus, for $q=0$ but arbitrary $\myQ_0(t)$ and $\psi(t)$, the product of $\phi$ and $f$, at any given time and energy, equals the product of the steady state $\phi=-1$ and $f$ values that correspond to a putative configuration in which $Q=Q_i$ and $\psi=\psi_i$ are constants, given by their values at the retarded time $t_i=t-\tau_{cool}(t,E)$.

\Eq{\ref{eq:phi_f_time_dependent_q0}} implies that the spectral curvature is given by
\begin{equation} \label{eq:spectral_curvature1}
\widetilde{\phi} \equiv \frac{d\ln \phi}{d\ln E} = -\phi - \frac{d\ln \myQpsi_i}{d\ln \myx} = -\phi - 1 + Q_i^{-1} \frac{d \myQpsi_i}{dt_i} \fin
\end{equation}
Hence, the combination $\phi+\widetilde{\phi}$ measured at any given energy $E$ directly gauges the temporal evolution of {\CRE} injection and magnetic fields at the retarded time $t_i$.
An observed {\CRE} spectrum spanning a wide energy range can thus be used to reconstruct the evolution over an extended period of time.
It is useful to define a measure of the magnetic growth and injection decay at time $t_i$,
\begin{equation} \label{eq:spectral_curvature_kappa}
\mykappa \equiv -Q_i^{-1}d\myQpsi_i/dt_i \fin
\end{equation}
If we can approximate this as a constant $\mykappa=\const\neq -1$, then the solution of \eq{\ref{eq:spectral_curvature1}} becomes
\begin{align} \label{eq:spectral_curvature2}
\phi(E) & = -\frac{1+\mykappa}{1-(E/E_0)^{1+\mykappa}} \, ; \nonumber \\
\widetilde{\phi}(E) & = -\frac{1+\mykappa}{1-(E/E_0)^{-(1+\mykappa)}} \coma
\end{align}
where $E_0$ is an integration constant.
For constant injection, $\mykappa=-dt_{cool}(t_i)/dt_i$, and is positive (negative) for magnetic growth (decay); constant injection and cooling yield $\mykappa=0$.

\subsubsection{Example: sudden change in injection or cooling}
\label{sec:B_jump_CRE}

It is instructive to analyse a simple example of the {\CRE} spectrum resulting from a time-dependent magnetic field or variable injection.
Consider first the case where an otherwise constant magnetic field suddenly changes at time $t_0$, such that the cooling parameter is modified by a factor $\myr$,
\begin{equation} \label{eq:magnetic_jump}
\mypsi(t)=
\begin{cases} \mypsi_1 & \text{if $t \leq t_0$ ;}
\\
\mypsi_2 = \myr\mypsi_1 & \text{if $t > t_0$ .}
\end{cases}
\end{equation}
For simplicity, assume a steady power-law energy injection, $Q=\myQ_0 E^{q}$ with some constant $q<1$.

The corresponding solution, obtained by plugging $\mypsi(t)$ from \EqO~(\ref{eq:magnetic_jump}) into \EqO~(\ref{eq:PDE_sol}) or (\ref{eq:PDE_sol2}), may be written as
\begin{eqnarray} \label{eq:CRE_B_jump}
f(t>t_0,E) & = & \frac{\myQ_0 E^{q-1}}{(1-q)\mypsi_2} \\
& & \times \left\{ 1+(\myr-1)\mbox{Max}\left[0,1-\frac{E}{E_c(t)}\right]^{1-q} \right\} \coma \nonumber
\end{eqnarray}
where
\begin{equation} \label{eq:Ec_def}
E_c(t)\equiv [(t-t_0)\psi_2]^{-1}
\end{equation}
is the energy with (instantaneous) cooling time $t-t_0$.
Notice the similarity between this expression and \EqO~(\ref{eq:PDE_sol_start_injection}), where it was assumed that prior to $t_0$ there was no injection, or equivalently cooling was infinitely fast, corresponding here to the limit $\myr\to 0$.

The spectrum of the {\CREs} can be derived from their distribution \EqO~(\ref{eq:CRE_B_jump}) or directly from \EqO~(\ref{eq:CRE_phi}),
\begin{equation} \label{eq:CRE_B_jump_phi}
\phi(t,E) = q - 1 - (\myr-1)\Phi_{q,\myr}\left( \frac{E}{E_c(t)} \right) \coma
\end{equation}
where we defined
\begin{equation} \label{eq:Phi_def}
\Phi_{q,\myr}(\myzz)\equiv \frac{(1-q)\left(1-\myzz\right)^{-1}\myzz}{\left(1-\myzz\right)^{q-1}+\myr-1} \Theta\left(1-\myzz\right) \fin
\end{equation}
The spectrum depends on $E$ and $t$ only through the combination
\begin{equation}
\myzz \equiv \frac{E}{E_c(t)} \fin
\end{equation}
This self-similar behaviour arises because no time- or energy-scale is associated with the transition in magnetic field, assumed in \eq{\ref{eq:CRE_B_jump_phi}} to be infinitely fast.

By assumption $q<1$, so $\Phi\geq 0$.
\Eqs{\ref{eq:CRE_B_jump_phi}} and (\ref{eq:Phi_def}) thus show that the spectral index $\phi$ decreases (increases) with respect to its steady-state value $\phi_s=q-1$, \ie the spectrum steepens (flattens), if the magnetic field strengthens such that $\myr>1$ (weakens, $\myr<1$).
The deviation of the spectrum from $\phi_s$ is largest at some finite energy $0<E\leq E_c$.
For example, for flat injection ($q=0$), the steady-state spectrum is $\phi_s=-1$, and the extremal spectrum is $\phi_{ext}=-\myr$, which occurs when $E\to E_c$.
As another example, for $q=-1$, $\phi_s=-2$, and $\phi_{ext}=-1-\myr^{1/2}$ occurs when $E/E_c=1-(1+\myr^{1/2})^{-1}$.

The {\CRE} spectrum arising from the magnetic jump \EqO~(\ref{eq:magnetic_jump}) is illustrated in Figs. \ref{fig:BJumpQ0} and \ref{fig:BJumpR}.
These figures show the deviation
\begin{equation} \label{eq:Delta_alpha_def}
\Delta\alpha \equiv \alpha - \alpha_s = \alpha + 1-q/2
\end{equation}
of the radio spectral index $\alpha$ from its steady state value $\alpha_s=(q-2)/2$, in various approximations.
The {\CRE} spectrum can be read from the approximation $\alpha(\nu_s)=[\phi(E)-1]/2$ (shown in dashed curves), where $\nu_s(E)$ is the synchrotron frequency of a {\CRE} with energy $E$.

Although an instantaneous jump in the magnetic field strength is an idealisation of a more gradual magnetic reconfiguration, the corresponding {\CRE} distribution (\EqsO~\ref{eq:CRE_B_jump}, \ref{eq:CRE_B_jump_phi}) provides a good approximation at sufficiently late times or low energies.
The reason is that although the details of the magnetic evolution $B(t)$, spanning some finite timescale $\sim\Delta t$, say, are initially imprinted upon the {\CRE} spectrum at high energies, the corresponding spectral features rapidly diminish and compress as the information gradually propagates to lower energies.
At late times, these fast and small amplitude spectral variations, superimposed on the imprint of the overall magnetic jump, are smeared out in the radio emission, which involves a convolution between the {\CRE} spectrum and the synchrotron function $F_{syn}$, as discussed in \S\ref{sec:SynchrotronVariability} below.

More precisely, while the spectral features associated with the overall change in cooling function span a characteristic energy scale $E_c\simeq [(t-t_0)\mypsi_2]^{-1}$ as seen in \eq{\ref{eq:CRE_B_jump_phi}}, the spectral features corresponding to $\Delta t$ transients lead to spectral variations on an energy scale $E_c \Delta t/(t-t_0)$ which is much smaller at late times.
The amplitude of these spectral variations is of order $\Delta t/(t-t_0)$, an so decreases with time, too.
A more quantitative discussion of the {\CRE} distribution arising from an arbitrary transient magnetic evolution is deferred to Appendix \S\ref{sec:transient_B_evolution}.

It is straightforward to generalise the above discussion to the case where the $\mypsi_1\to \mypsi_2$ change in cooling function is accompanied by a $\myQ_1\to\myQ_2$ change in injection $\myQ$.
If the spectral index $q$ remains constant, \EqsO~(\ref{eq:CRE_B_jump})--(\ref{eq:CRE_B_jump_phi}) are also valid if both injection and cooling jump at $t_0$, with the substitution $\myQ_0\to \myQ_2$ and the revised definition
\begin{equation} \label{eq:myr_general}
\myr \equiv \frac{\myQ_1/\mypsi_1}{\myQ_2/\mypsi_2} = \frac{\myQpsi_1}{\myQpsi_2} \fin
\end{equation}

Note that if at $t_0$ the {\CREs} are compressed by some factor $\myrcr$, for example by a shock, then the above analysis holds with the additional substitution $\myQ_1\to \myQQ_1\equiv \myrcr\myQ_1$.
This is equivalent to an effectively enhanced {\CRE} injection rate prior to $t_0$.

\subsection{Time-dependent injection and magnetic fields: synchrotron emission}
\label{sec:SynchrotronVariability}

\subsubsection{General results}

The synchrotron emissivity of the {\CRE} distribution is given by \citep{RybickiLightman86}
\begin{align} \label{eq:syn_j}
j(t,\nu) & \simeq \sqrt{3}\, e \, r_e \, B(t) \sin(\myPalpha) \int_{E_{min}}^{E_{max}} \myn(t,E) F_{syn} \left( \frac{\nu}{\nu_s(E)} \right) \,dE \nonumber \\
= -& \sqrt{3} \, e \, r_e \, B(t) \sin(\myPalpha) \int_{\myx_{min}}^{\myx_{max}} f(t,\myx) F_{syn}\left( \frac{\nu}{\nu_s(\myx)} \right) \, d\myx \coma
\end{align}
where
\begin{equation} \label{eq:nu_s}
\nu_s \equiv a B E(\myx)^2 \simeq 0.9 b (E/5\GeV)^2 (1+z)^{2} \GHz
\end{equation}
is the emitted synchrotron frequency.
Here, $\myPalpha$ is the pitch angle, we defined
\begin{equation}
a\equiv \frac{3 e \sin\myPalpha}{4\pi m_e^3 c^5} \coma
\end{equation}
$c$ is the speed of light, $m_e$ and $e$ are the electron mass and charge, and $r_e\equiv e^2/m c^2$ is the classical electron radius.

For simplicity, instead of modeling the pitch angle distribution, we adopt a constant value $\myPalpha=\pi/4$, which corresponds here to an isotropic pitch angle distribution.
We shall henceforth approximate the energy range subtended by the power-law spectrum as infinite, $E_{min}\to 0$ and $E_{max}\to \infty$.

The synchrotron source function is given by
\begin{eqnarray}
F_{syn}(\myz) & = & \myz \int_\myz^\infty K_{5/3}(\myz')\,d\myz' \label{eq:FSynFull} \\
& \simeq & c_0 \myz^{c_1} e^{-c_2 \myz} \label{eq:FSynApprox} \coma
\end{eqnarray}
where $K_n(\myz)$ is the modified Bessel function of the second kind.
The approximate form of $F_{syn}$ in \eq{\ref{eq:FSynApprox}}, with $c_0=1.83$, $c_1=0.309$, and $c_2=1.03$, is accurate to within $10\%$.

When the injection is separable in energy and time, $Q(t,\myx)=\myQ(t)\mathbb{Q}(\myx)$, we may compute the synchrotron signal directly from $Q$, for arbitrary cooling $\psi(t)$.
Combining \EqsO~(\ref{eq:PDE_sol2}), (\ref{eq:syn_j}) and (\ref{eq:nu_s}), and switching the order of integration, we obtain
\begin{equation} \label{eq:jnu_approx}
j(t,\nu) \propto B(t) \int_{-\infty}^t d\tau\, \myQ(\tau) G(\nu;t,\tau) \coma
\end{equation}
where
\begin{equation} \label{eq:G_definition}
G(\nu;t,\tau) \equiv \int_0^\infty d\myy\, \mathbb{Q}(\myy) F_{syn}\left[ \frac{\nu}{a B(t)} \left(\myy+\myPsi_{t,\tau}\right)^2\right] \fin
\end{equation}
In \eq{\ref{eq:G_definition}} we took the upper limit $\myy=(m_e c^2)^{-1}-\myPsi_{t,\tau}$ to infinity because $F_{syn}(\myz)$ declines exponentially fast for large $\myz$.
One can compute $G$ analytically for various choices of $\mathbb{Q}(\myx)$. For injection with a flat spectrum, $\mathbb{Q}=\constant$, we obtain
\begin{equation} \label{eq:G_approx}
G(\nu;t,\tau) \propto \sqrt{\frac{B(t)}{\nu}} \, \, \Gamma_{c_1+1/2} \left[ \frac{c_2 \nu }{a B(t)}\myPsi_{t,\tau}^2 \right] \coma
\end{equation}
where $\Gamma_n(z)$ is the incomplete Gamma function, and we have used the approximate form of $F_{syn}$ in \EqO~(\ref{eq:FSynApprox}) for simplicity.

For qualitative estimates, it is sometimes sufficient to approximate the synchrotron function as a Dirac delta function,
\begin{equation} \label{eq:Fsyn_delta}
F_{syn}(\myz)\simeq \delta(\myz-\myz_0)
\end{equation}
with constant $\myz_0\sim 1$, bearing in mind that the convolution in \EqO~(\ref{eq:syn_j}) with the true $F_{syn}$ somewhat smears the spectral features retained by using \eq{\ref{eq:Fsyn_delta}}.
In this approximation
\begin{equation} \label{eq:jFsyn_delta}
j(t,\nu) \propto B(t)^{3/2} \nu^{-1/2} f(t,\myx_\nu)
 \propto  B(t)^{1-\alpha} \nu^{\alpha} \coma
\end{equation}
where the synchrotron spectral index is
\begin{equation} \label{eq:alpha_Fsyn_delta0}
\alpha(t,\nu) \equiv \frac{d\ln j_\nu}{d\ln \nu} = \frac{\phi(t,\myx_\nu)-1}{2} \coma
\end{equation}
and $\myx_\nu$ is defined such that $\nu_s(\myx_\nu)=\nu/\myz_0$,
\begin{equation}
\myx_\nu \equiv \sqrt{a B \myz_0/\nu} \fin
\end{equation}

Typically, the observed radio spectral index $\alpha \lesssim -1$, so according to \eqs{\ref{eq:jFsyn_delta}} and (\ref{eq:alpha_Fsyn_delta0}), $\phi\lesssim -1$ and the radio emissivity scales as at least the square of the magnetic field $B(t)$.
Consequently, regions in which the magnetic field has been growing (decaying) during the past $\tau_{cool}$ show both brighter (weaker) emission and a steeper (flatter) spectrum.
This has two main consequences: \emph{(i)} monotonic magnetic evolution is directly imprinted on the brightness and spectrum profiles; and \emph{(ii)} magnetic fluctuations of period $P\gtrsim 2\tau_{cool}$ induce, on average, a spectral steepening.

\subsubsection{Temporal evolution imprinted on radio emission: qualitative discussion for $q=0$}
\label{sec:Temporal_evoution_qualitative_q0}

Consider temporal variations in the rate of flat ($q=0$) {\CRE} injection, $Q(t,\myx)=\myQ_0(t)$, and in the magnetic field, giving rise to a time-dependent cooling parameter $\mypsi(t)\simeq \mypsi_B(t)+\mypsi_{cmb}$.
Qualitatively, at any given time, the {\CRE} distribution attempts to evolve towards the steady state in which $f\sim \myQpsi \propto Q(t)/\mypsi(t)$.
As mentioned in \S\ref{sec:magnetic_variations_modify_CRE}, this evolution proceeds gradually, with the distribution of higher energy {\CREs} adjusting faster.
Consequently, the {\CRE} distribution at energy $E$ tends to decrease/increase and its spectrum to steepen/flatten, if injection has been diminished/emplified or the magnetic field has strengthened/weakened at a time $\tau_{cool}(t,E)$ earlier.

Synchrotron emission directly reflects these changes in the {\CRE} distribution, although the spectral features are somewhat smeared by the convolution with the synchrotron source function $F_{syn}$.
In addition, the radio emission at time $t$ is proportional to the instantaneous magnetic field energy density $\propto \mypsi_B(t)=\mypsi(t)-\mypsi_{cmb}$.
This slightly complicates the radio response.

For simplicity, consider the approximation $F_{syn}(\myz)\simeq \delta(\myz-\myz_0)$. Using \eqs{\ref{eq:f_time_dependent_q0}} and (\ref{eq:phi_time_dependent_q0}), this yields
\begin{equation} \label{eq:flat_j_t}
\nu j_\nu(t,\nu) \propto \mypsi_B(t)\myQpsi_a
\end{equation}
and
\begin{equation} \label{eq:flat_alpha_t}
\alpha(t,\nu) = -1 - \frac{\myQpsi_{i}-\myQpsi_a}{2\myQpsi_a} \coma
\end{equation}
where the right hand sides of the two equations are to be evaluated at $\myx=\myx_\nu$.
These results qualitatively show how the synchrotron brightness responds immediately and linearly to changes in the magnetic energy density, and in addition responds with a $\Delta t\lesssim \tau_{cool}$ delay to earlier variations in injection and in cooling.

For a more detailed discussion, consider the case where injection is constant, such that $j_\nu\propto \mypsi_B(t)/\mypsi_a$, and \eq{\ref{eq:flat_alpha_t}} becomes
\begin{equation} \label{eq:flat_alpha_t_Qconst}
\alpha(t,\nu) = -1 - \frac{\mypsi_{a}-\mypsi_i}{2\mypsi_i} \fin
\end{equation}
Further assume that the otherwise constant magnetic field instantaneously changes at time $t_0$, leading to a jump by a factor $\myr>1$ ($\myr<1$) in the cooling parameter, from $\mypsi_1$ to $\mypsi_2=\myr\mypsi_1$, as in \EqO~(\ref{eq:magnetic_jump}).
Consequently, $\mypsi(t_i)$ experiences a delayed jump, given by \EqO~(\ref{eq:magnetic_jump}) but with $t_0$ replaced by $t_f=t_0+\tau_{cool}(t_f)$, whereas $\mypsi_{a}$ gradually changes from $\mypsi_1$ to $\mypsi_2$ during the time interval $t_0<t<t_f$.
The synchrotron brightness responds immediately to the magnetic growth (decay) by increasing (decreasing), from $j_{\nu,1}$ to $\myr j_{\nu,1}$.
However, the spectrum is unaffected at $t_0$, so the initially modified, $\myr j_{\nu,1}$ emission is flat ($\alpha=-1$).
The brightness remains elevated (diminished) over $\sim \tau_{cool}$, during which the spectrum gradually steepens (flattens) and the brightness gradually returns to a new steady state, $j_{\nu,2}$, as $\mypsi_a$ gradually evolves from $\mypsi_1$ to $\mypsi_2$.
In the strongly magnetised regime ($\mypsi_B\gtrsim \mypsi_{cmb}$), $j_{\nu,2}\simeq j_{\nu,1}$, otherwise $j_{\nu,2}$ can be substantially higher (lower) than $j_{\nu,2}$.
This steady state is reached at $t_f$, just as the spectral deviation is extremal, $\alpha_{ext}=-(\myr+1)/2$, immediately flattening back to $\alpha=-1$ as $\mypsi_i\to \mypsi_a$.

The synchrotron signature of arbitrary $Q(t)$ and $\mypsi(t)$ can be conceived of as a superposition of multiple such responses.
The above qualitative discussion can thus be generalised for arbitrary $\mypsi(t)$ and $Q(t)$, using \eqs{\ref{eq:flat_j_t}} and (\ref{eq:flat_alpha_t}).
However, in practice, $j_\nu$ and $\alpha$ are smoothed by the convolution with $F_{syn}$, avoided above in the approximation \eq{\ref{eq:Fsyn_delta}}.
In \S\ref{sec:B_jump_syn} we analyse more accurately the radio emission corresponding to an isolated, sudden jump in $Q$ and $\mypsi$, for arbitrary injection power-law $q$.
When the temporal behaviour of $Q$ and $\mypsi$ is erratic, it may be more useful to employ a statistical description of the corresponding radio emission, as shown in \S\ref{sec:fluctuations}.

\subsubsection{Synchrotron signature of an instantaneous change in $Q$ and $B$}
\label{sec:B_jump_syn}

Consider an arbitrary, instantaneous transition in which both the {\CRE} injection rate and the magnetic field change abruptly at $t=0$.
For simplicity, we assume injection of a power-law spectrum with a constant, but not necessarily flat, index $q$, such that $Q(t,E)=\myQ(t)E^q$.
We may thus describe the transition as
\begin{equation} \label{eq:JumpInQandB}
\{\myQ_1,\mypsi_1=\mypsi(B_1)\} \xrightarrow{t=0} \{\myQ_2,\mypsi_2=\mypsi(B_2)\} \fin
\end{equation}

The corresponding {\CRE} distribution $f(t,E)$ and spectrum $\phi(t,E)$ were derived in \S\ref{sec:B_jump_CRE}; see \EqsO~(\ref{eq:CRE_B_jump})-(\ref{eq:Phi_def}) with $\myQ_0$ replaced by $\myQ_2$, and $t_0$ replaced by $0$, so \EqO~(\ref{eq:CRE_B_jump}) becomes
\begin{equation} \label{eq:CRE_QB_jump}
f(t,E) =
\begin{cases}
\frac{\myQ_1 E^{q-1}}{(1-q)\mypsi_1} & \text{if $t<0$ ;}
\\
\frac{\myQ_2 E^{q-1}}{(1-q)\mypsi_2}
\Big\{ 1+ & \!\!\!\!\! (\myr-1) \left[1-\frac{E}{E_c(t)}\right]^{1-q} \Big\}  \\
& \text{if $t>0$ and $E<E_c(t)$ ;}
\\
\frac{\myQ_2 E^{q-1}}{(1-q)\mypsi_2} & \text{if $t>0$ and $E>E_c(t)$ \fin}
\end{cases}
\end{equation}
Here, $\myr \equiv (\myQ_1\mypsi_2)/(\myQ_2 \mypsi_1)=\myQpsi_1/\myQpsi_2$, as defined in \EqO~(\ref{eq:myr_general}).

\begin{figure}
\centerline{\epsfxsize=8cm \epsfbox{\myfarfig{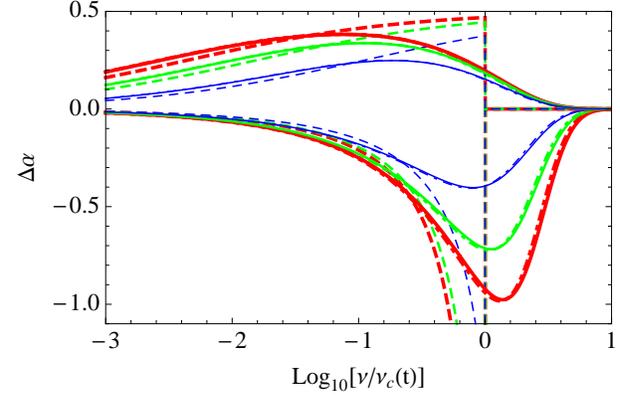}}}
\caption{
Deviation in synchrotron spectral index $\Delta \alpha(\nu)\equiv \alpha-\alpha_s$ from a steady state $\alpha_s$, due to a sudden change in injection and magnetic field.
Here, flat {\CRE} injection ($q=0$) is assumed, and $\myr$ (curves with $\Delta \alpha\leq 0$) or $\myr^{-1}$ (curves with $\Delta\alpha\geq 0$) equals $16$ (red, thick curves), $9$ (green), and $4$ (blue, thin curves), with $|\Delta\alpha|$ monotonically decreasing among these values.
Results are shown based on the exact $F_{syn}$ (\EqO~\ref{eq:FSynFull}; solid curves), on our $F_{syn}$ approximation (\EqO~\ref{eq:FSynApprox}, or equivalently \EqO~\ref{eq:G_approx}; dot-dashed; nearly indistinguishable from the exact results), and on the simple $F_{syn}(\myz)\to \delta(\myz-\myz_0)$ approximation with $\myz_0=1$ (dashed).
The latter, with extremal values $\Delta\alpha_{ext}=(1-\myr)/2$ (for $q=0$), directly reflects the {\CRE} spectrum; see \eq{\ref{eq:phi_s_from_alpha_s}}.
\label{fig:BJumpQ0}
\vspace{2mm}}
\end{figure}

\begin{figure}
\centerline{\epsfxsize=8cm \epsfbox{\myfarfig{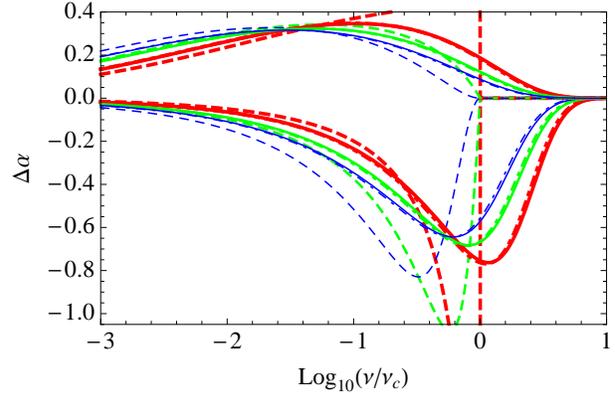}}}
\caption{
Same as \Fig~(\ref{fig:BJumpQ0}), but with $\myr=10$ (curves with $\Delta \alpha\leq 0$) or $\myr=1/10$ (curves with $\Delta \alpha\geq 0$), for power-law injection indices $q=0$ (red, thick curves), $-1$ (green), and $-2$ (blue, thin curves) (large to small effect at high, $\nu>\nu_c$ frequencies).
\label{fig:BJumpR}
\vspace{2mm}}
\end{figure}

In the $\delta$-function approximation of $F_{syn}$ (\eq{\ref{eq:Fsyn_delta}}), the synchrotron spectrum corresponding to \eq{\ref{eq:CRE_QB_jump}} is given by
\begin{equation} \label{eq:alpha_Fsyn_delta}
\alpha(t,\nu) = \frac{q-2}{2} - \frac{\myr-1}{2}\Phi\left( \sqrt{\frac{\nu}{\myz_0 \nu_c(t)}} \right) \coma
\end{equation}
where $\Phi$ is defined in \eq{\ref{eq:Phi_def}}.
This results was discussed qualitatively in \S\ref{sec:Temporal_evoution_qualitative_q0}, and is illustrated as dashed curves in Figs. \ref{fig:BJumpQ0} and \ref{fig:BJumpR}.

In practice, convolving the {\CRE} distribution with the true $F_{syn}$ smears the spectral features of \EqO~(\ref{eq:alpha_Fsyn_delta}).
For example, the maximal deviation for the spectral index from its steady state, $|\alpha_{ext}-\alpha_s|$, is somewhat smaller than derived in \S\ref{sec:Temporal_evoution_qualitative_q0}, and the steady-state spectrum $\alpha=(q-2)/2$ is recovered only gradually, over a timescale $> \tau_{cool}$.
In order to quantify these effects, we must model the synchrotron source function $F_{syn}$ more accurately.

The radio emissivity can be derived by plugging the distribution $f(t,E)$ of \EqO~(\ref{eq:CRE_QB_jump}) into \EqO~(\ref{eq:syn_j}) or, as $Q(t,E)=\myQ E^{q}$ is separable here, by plugging $Q(t,E)$ and $\mypsi(t)$ directly into \EqsO~(\ref{eq:jnu_approx})--(\ref{eq:G_approx}).
Note that in the present case of an instantaneous change in $Q$ and $\mypsi$, we can write \eq{\ref{eq:CRE_QB_jump}} in self-similar form as $f(t,E)=E^{q-1}g(\epsilon=E/E_c(t))$, so
\begin{eqnarray} \label{eq:j_nu_self_similar}
j_\nu(t,\nu) & = & \sqrt{3} \, e \, r_e \sin(\myPalpha_2) B_2 E_c(t)^{q-2} \\
& & \times \int_0^\infty \epsilon^{q-3} g(\epsilon)F_{syn}\left( \epsilon^{-2} \frac{\nu}{\nu_c} \right) \, d\epsilon \fin \nonumber
\end{eqnarray}
This implies that the spectral index $\alpha(t,\nu)$ depends on $t$ and $\nu$ only through the combination $\nu/\nu_c(t)$, where
\begin{equation} \label{eq:nu_c_def}
\nu_c(t) \equiv \nu_s(E=E_c(t)) = a B_2 E_c(t)^2 = \frac{a B_2}{\psi_2^2 t^2}
\end{equation}
is the synchrotron frequency corresponding to the cooling energy $E_c(t)$;
the time-dependent prefactor of the integral in \EqO~(\ref{eq:j_nu_self_similar}) does not modify the spectrum.
This is another manifestation of the lack of a timescale associated with the infinitely fast transition.

The radio emissivity can be computed analytically using the approximate form of $F_{syn}$ in \EqO~(\ref{eq:FSynApprox}).
This yields
\begin{eqnarray} \label{eq:j_nu_q}
j_\nu(t,\nu) & = & A_q a^{-\frac{q}{2}} \myQ_2 \frac{B_2^{2-\frac{q}{2}}\sin^2\myPalpha_2}{B_2^2+B_{cmb}^2} \nu^{-1+\frac{q}{2}} \\
& & \nonumber \times \left[ 1+(\myr-1)J_q\left( \myz \right) \right] \coma
\end{eqnarray}
where the functions $A_q$ and $J_q(\myz)\geq 0$ are defined in \S\ref{sec:jnu_from_jump_detail},
and
\begin{equation} \label{eq:myz_def}
{\myz}(t,\nu) \equiv c_2 \frac{\nu}{\nu_c(t)} = \frac{c_2 \mypsi_2^2 \nu t^2}{a B_2} \fin
\end{equation}
Here, $J_q(\myz\to 0)=1$ and $J_q(\myz\to\infty)=0$, so
\begin{equation} \label{eq:Jq_behviour}
J_q \to
\begin{cases}
1 & \text{as $t \to 0$ ;}
\\
0 & \text{as $t \to \infty$.}
\end{cases}
\end{equation}
This guarantees that $j_\nu(t=0)\propto Q_1/\mypsi_1$, and $j_\nu(t\to\infty)\propto Q_2/\mypsi_2$, as expected.

For flat ($q=0$) injection
\begin{equation} \label{eq:A0Def}
A_0 \equiv \frac{27c_0\sqrt{3}\,\Gamma(1+c_1)}{32\pi c_2^{1+c_1}} \simeq 0.7
\end{equation}
and
\begin{equation} \label{eq:J0Def}
J_0(\myz) \equiv \frac{\Gamma_{1+c_1}({\myz})- \sqrt{{\myz}}\,\Gamma_{\frac{1}{2}+c_1}({\myz})}{\Gamma(1+c_1)} \fin
\end{equation}
For comparison, the well known steady-state ($\myr=1$) result for $q=0$,  $\nu j_\nu= (\myQ_2/2) \sin^2(\myPalpha_2) B_2^2/(B_2^2+B_{cmb}^2)$, would be reproduced if $A_0=0.5$.
The normalisation error in \eq{\ref{eq:A0Def}} arises from the approximation in \eq{\ref{eq:FSynApprox}}; it does not affect the spectrum.

The synchrotron spectrum corresponding to \EqO~(\ref{eq:j_nu_q}) is given by
\begin{equation} \label{eq:alpha_nu_q}
\alpha(t,\nu) = -1 + \frac{q}{2} - \frac{(\myr-1)I_q\left( {\myz} \right)}{ 1+(\myr-1)J_q\left( {\myz} \right) } \coma
\end{equation}
where the function $I_q(\myz)$ is defined in \S\ref{sec:jnu_from_jump_detail}.
For flat injection, \EqO~(\ref{eq:alpha_nu_q}) simplifies to
\begin{equation} \label{eq:alpha_nu_q0}
\alpha(t,\nu) = -1 + \frac{1}{2}\left[ 1 - \frac{\Gamma_{1+c_1}(\myz)+\frac{\Gamma(1+c_1)}{\myr-1}}{\sqrt{\myz}\,\Gamma_{\frac{1}{2}+c1}(\myz)} \right]^{-1} \fin
\end{equation}

\begin{figure}
\centerline{\epsfxsize=8cm \epsfbox{\myfarfig{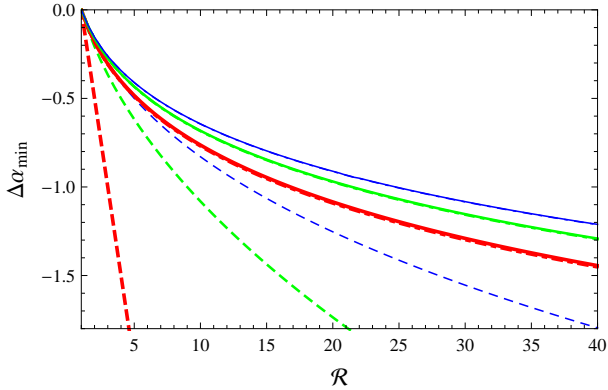}}}
\caption{
Maximal synchrotron steepening shown by plotting the minimal $\Delta\alpha$ as a function of $\myr$, for injection power-law indices $q=0$, $-1$, and $-2$. Line styles are the same as in \Fig~\ref{fig:BJumpR}.
\label{fig:MinDeltaAlpha}
\vspace{2mm}}
\end{figure}

In \Figs~\ref{fig:BJumpQ0}--\ref{fig:MinDeltaAlpha} we present the synchrotron spectrum arising from an instantaneous change in $Q$ and in $B$.
The deviation $\Delta\alpha=\alpha-\alpha_s$ of the spectral index (see \eq{\ref{eq:Delta_alpha_def}}) from its steady-state value $\alpha_s=-1+q/2$ is plotted as a function of frequency, for $q=0$ and different values of $\myr$ in \Fig~\ref{fig:BJumpQ0}, and for different values of $q$ in \Fig~\ref{fig:BJumpR}.
Results are shown for both $\myr>1$ and $\myr<1$; the former arises for example from magnetic amplification or a drop in injection, and leads to $\Delta \alpha\leq 0$ steepening, whereas the latter leads to $\Delta \alpha\geq 0$ flattening.
The strongest steepening $\Delta \alpha_{min}$, typically found when $\nu\simeq \nu_c(t)$, is shown in \Fig~\ref{fig:MinDeltaAlpha} as a function of $\myr>1$, for different values of $q$.

These figures show the outcome of the three expressions we have used for $F_{syn}$: the exact formula \eq{\ref{eq:FSynFull}} (as solid curves), our approximate fit \eq{\ref{eq:FSynApprox}} (which leads to the spectrum in \eqs{\ref{eq:alpha_nu_q}} and (\ref{eq:alpha_nu_q0}); dot-dashed curves), and the $\delta$-function approximation \eq{\ref{eq:Fsyn_delta}} (dashed curves).
The results of our fit are in most cases indistinguishable from the exact solution, indicating that the analytical expressions in \eqs{\ref{eq:j_nu_q}}--(\ref{eq:alpha_nu_q0}) provide good approximations to the actual results.

The $\delta$-function approximation of $F_{syn}$ (nonphysically) avoids smoothing the {\CRE} spectral features, resulting in an inaccurate $\alpha(\nu)$ profile that directly traces the {\CRE} spectrum $\phi(E)$.
The deviation of the {\CRE} spectral index from its steady-state value $\phi_s=-1+q$ can thus be inferred from the figures, through
\begin{equation} \label{eq:phi_s_from_alpha_s}
\Delta\phi\left(\frac{E}{E_c}\right)\equiv \phi-\phi_s=2\Delta\alpha\left(\frac{\nu_s(E)}{\nu_c}\right) \bigg|_{F_{syn}\propto \delta(\myz-\myz_0)} \fin
\end{equation}

Finally, note that $j_\nu$, $\alpha$, and $\Delta\alpha$ do not show a simple symmetry under the transformations $\myr\to 1/\myr$ or $\myr\to 1-\myr$.
What is the effect of spatial or temporal variations in $\myr$, \ie in {\CRE} injection rate and magnetic field strength?

\subsection{Synchrotron imprint of a shock}
\label{sec:SynchrotronImprintOfShock}

\begin{figure}
\centerline{\epsfxsize=8cm \epsfbox{\myfarfig{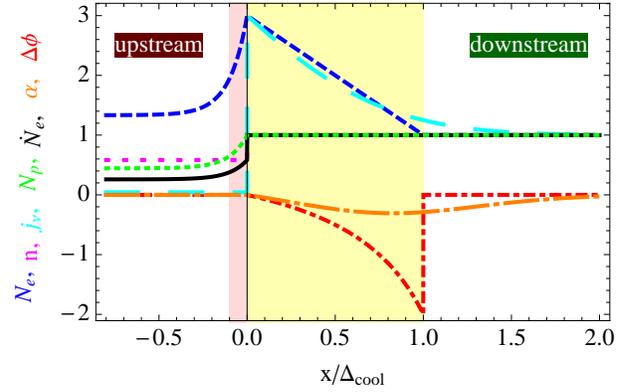}}}
\caption{
Illustration of shock-induced transitions in $N_p$ (dotted), $n$ (long dotted), $\dot{N}_e(E)$ (solid), $N_e(E)$ (dashed), $\phi(E)$ (dot-dashed), $j_\nu[\nu_s(E)]$ (long dashed), and $\alpha[\nu_s(E)]$ (long dot-dashed), plotted in the shock frame, with the shock front at $\myX=0$ and flow in the positive $\myX$ direction.
Nonspectral quantities (\ie all but $\phi$ and $\alpha$) are normalised to unity far downstream.
We assume a weak, $\mach=1.5$ shock with $B_u\ll B_d\simeq 2B_{cmb}$, such that $\myr=3$.
For illustration, we exaggerate the precursor (shaded pink) width $\Dltcr\sim D/v_s$ as $10\%$ of the cooling layer (shaded yellow) width $\Dltcool\sim v_d t_{cool}$, and neglect the width $\Dltg$ of the gas compression layer (horizontal line).
\label{fig:ShockIllustration}
\vspace{2mm}}
\end{figure}

The shock transitions of the components involved in radio emission from clusters span a wide range of length scales.
Consider a typical, weak, $\mach\sim2$ shock in the ICM, with a $B_d\simeq B_{cmb}$ downstream magnetic field.
Here we expect a $\Dltcr\lesssim 10\kpc$ precursor in which {\CREs} and {\CRIs} are amplified by a factor of $\mach^2$ (see \S\ref{sec:CRAmplification}), a negligibly narrow, $\Dltg\sim 10\km$ shock transition in which the ambient plasma is compressed by a factor $r_g=4\mach^2/(3+\mach^2)$ (see \S\ref{sec:TemporalEvolutionCannotBeNeglected}), and a $\Dltcool\sim 200\kpc$ cooling layer in which the downstream {\CRE} distribution relaxes to a steady state (see \S\ref{sec:B_jump_CRE} and \eq{\ref{eq:DeltacoolDef}}).
These transitions are illustrated in \Fig~\ref{fig:ShockIllustration}.

Some level of magnetisation is to be expected in a weak shock, but the details of this process are not well known.
Weak amplification of $B$, by a factor of $\leq r_g$, is expected due to field line compression at the $\Dltg$ shock transition.
Some additional field amplification is expected downstream, due to turbulent fluid motions, but this process in not well constrained.
The energy fraction $\xi_p\propto N_p/n$ of {\CRPs} in clusters varies as $n^{-1}$, according to \S\ref{sec:HaloAndRelicEta}
(and declines faster with radius in previous models).
It can reach several percent of the thermal energy density at peripheral relics, where {\CRIs} may also play a role in the magnetisation process.
For simplicity, below we parameterise the magnetisation as amplification of the magnetic field amplitude by some factor $\myrB\equiv B_d/B_u$, and assume that this is a rapid process, spanning a narrow, $\DltB \ll \Dltcool$ transition.
Consider first the case where no further evolution of the magnetic field takes place downstream; the results are later generalised for slow downstream evolution.

In order to describe the synchrotron signature of a shock, we approximate it as infinite, planar, and stationary.
We work in the shock frame, such that the gas compression layer is stationary at $\myX=0$ and approximated as being infinitely thin.
The flow is assumed to be in the positive $\myX$ direction, such that
\begin{align}
\myX&<-\Dltcr & & \text{upstream (suffix \emph{u});} \nonumber
\\
-\Dltcr<\myX&<0 & & \text{CR precursor;} \nonumber
\\
0<\myX&<\Dltcool & & \text{cooling layer;} \nonumber
\\
\Dltcool<\myX& & & \text{downstream (suffix \emph{d});}  \nonumber
\end{align}
see \Fig~\ref{fig:ShockIllustration}.
We compute the synchrotron emissivity $j_\nu$ and spectrum $\alpha$ as a function of $\myX$, assuming a homogeneous background where $n$ and $N_p$ are uniform; this is generalised for arbitrary initial distributions later.

The synchrotron emissivity rises rapidly across the {\CR} precursor and the shock compression layers, over which both {\CREs} and magnetic fields are amplified.
Consequently,
\begin{equation} \label{eq:ShockBrightnessAmplification}
\frac{j_\nu(\myX=0)}{j_\nu(-\Dltcr)} \simeq \left( \frac{B_d}{B_u} \right)^{2-\frac{q}{2}} \myrcr = \myrB^{2-\frac{q}{2}} \mach^2 \fin
\end{equation}
Note that the {\CRI} amplification does not modify the emission near the shock, at distance much smaller than $\Dltcool$.
Similarly, the magnetic field growth does not modify the spectra of the {\CREs} and of their synchrotron emission this close to the shock.
Therefore, if {\CRE} injection is flat ($q=0$) upstream of the shock, then immediately after the shock, $\phi(x\ll \Dltcool)=-1$ and $\alpha(x\ll \Dltcool)=-1$ are unchanged by the shock transition.

While the width $\Dltcr$ of the {\CR} transition layer is typically well below the resolution of present-day radio telescopes, the $\Dltcool$ cooling layer may in some cases be resolved.
Therefore, we may use the scaling $\Dltg\ll\Dltcr\ll\Dltcool$ to approximate the magnetic and {\CR} amplification layers as infinitely thin.
The results of \S\ref{sec:B_jump_syn} for an instantaneous change in $Q$ and in $B$ at some time $t=0$ then apply.
Bear in mind, however, that we must substitute $\myQ_u\to\myQQ_u$ in order to account for {\CREs} compression at the shock, as mentioned towards the end of \S\ref{sec:B_jump_syn}.

With this substitution, the downstream {\CRE} distribution $f(\myX>0,E)$ evolves approximately according to $f(t>0,E)$ in \eq{\ref{eq:CRE_QB_jump}}.
Here, $f(t,E)$ can be regarded as either the {\CRE} distribution in a Lagrangian fluid element that crossed the shock at time $t=0$, or, equivalently, as the spatial {\CRE} profile downstream of the shock, $f(t(\myX),E)$, where $t(\myX)=\myX/v_d=\myX\myrg/v_s$.
The synchrotron emissivity and spectrum are thus given by \eqs{\ref{eq:j_nu_q}} and (\ref{eq:alpha_nu_q}).
In the limit of flat injection, which is a fairly good approximation in clusters, $j_\nu$ and $\alpha$ are given by \eqs{\ref{eq:J0Def}} and (\ref{eq:alpha_nu_q0}).

\begin{figure}
\centerline{\epsfxsize=8cm \epsfbox{\myfarfig{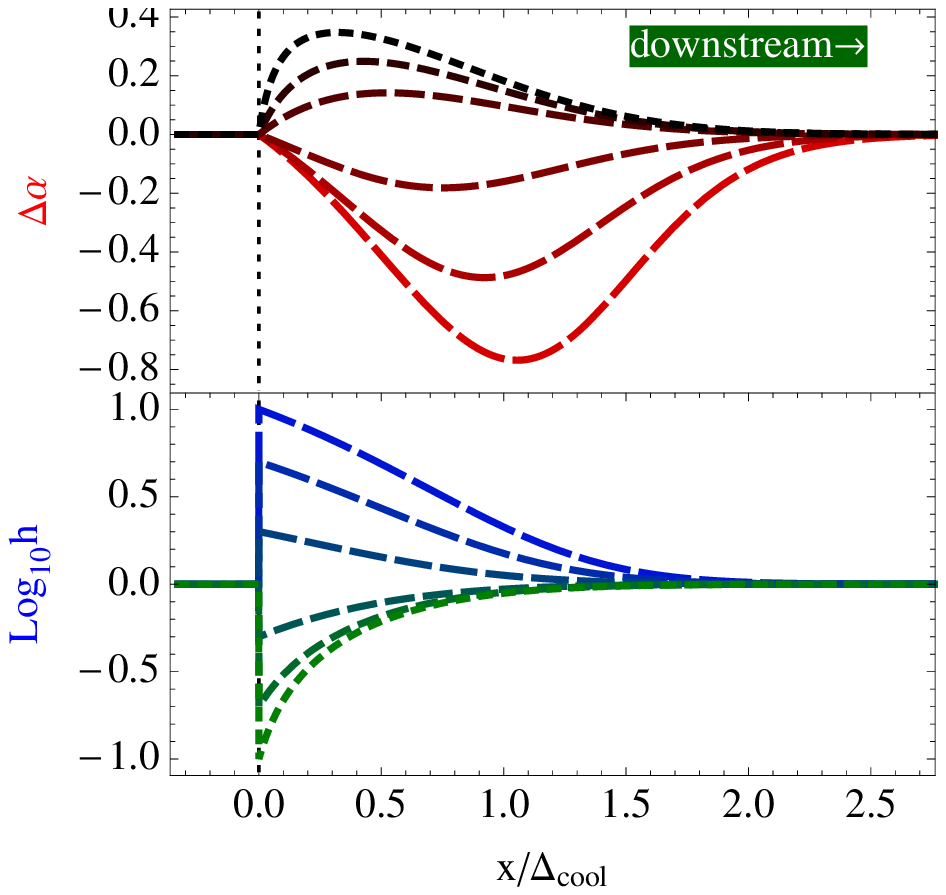}}}
\caption{
Synchrotron signature of a shock at $\myX=0$ (dotted line), assuming flat {\CRE} injection ($q=0$).
The spectral deviation $\Delta\alpha$ (see \eq{\ref{eq:Delta_alpha_def}}; upper panel) and the normalised emissivity $h$ (see \eq{\ref{eq:myhDefGeneral}}; bottom panel) are plotted as a function of the normalised distance $\myX/\Dltcool$ from the shock (with $\myX$ increasing downstream).
Results are shown for $\myr=10$, $5$, $2$, $1/2$, $1/5$, and $1/10$ (long to short dashing).
Equivalently, the abscissa may be regarded as the normalised time $t/t_{cool}$ that elapsed since a fluid element crossed the shock.
\label{fig:ShockSynchQ0}}
\end{figure}

\begin{figure}
\centerline{\epsfxsize=8cm \epsfbox{\myfarfig{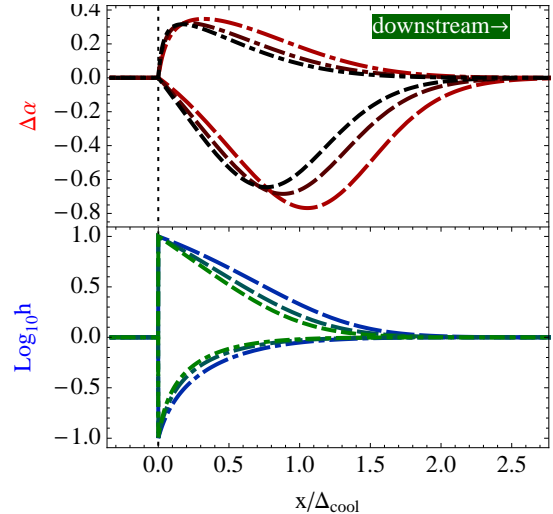}}}
\caption{
Synchrotron signature of a shock for different {\CRE} injection power-laws.
Notations are the same as in \Fig~\ref{fig:ShockSynchQ0}, but here we adopt $q=0$, $-1$, and $-2$ (long to short dashing), shown for fixed $\myr=10$ (dashed) and for $\myr=1/10$ (dot-dashed).
\label{fig:ShockSynchRConst}}
\end{figure}

Figures \ref{fig:ShockSynchQ0} and \ref{fig:ShockSynchRConst} illustrate the synchrotron signature of a shock, for various choices of the injection power-law $q$ and the shock amplification parameters.
As shown in \S\ref{sec:B_jump_CRE}, the {\CRE} distribution depends on the shock amplification of {\CRIs}, {\CREs}, and magnetic fields, only through the combination
\begin{align} \label{eq:myr_Def_Shock}
\myr & \equiv \frac{\myQQpsi_u}{\myQpsi_d} \equiv \frac{\myQQ_u/\mypsi_u}{\myQ_d/\mypsi_d}  \\
& = \frac{\myrcre}{\myrcrp \myrg} \cdot \frac{B_d^2+B_{cmb}^2}{B_u^2+B_{cmb}^2} \nonumber \\
& \simeq \frac{1+b_d^2}{\myrg} \nonumber
\coma
\end{align}
where the compression factors of {\CREs} and {\CRIs} are given by $\myrcre=\myrcrp=\myrcr$, and in the last line we assumed that $B_u\ll B_{cmb}$.
It is therefore convenient to define the synchrotron emissivity normalised by its value far downstream,
\begin{equation} \label{eq:myhDef}
\myh(\myX,\nu) \equiv \frac{j_\nu(\myX,\nu)}{j_\nu(\myX\to\infty,\nu)} =
1+(\myr-1)J_q\left( \myz \right) \coma
\end{equation}
such that $\myh(\myX,\nu)$ and $\Delta \alpha(\myX,\nu)$ depend only on $q$ and $\myr$.

Moreover, \eqs{\ref{eq:myhDef}}, (\ref{eq:alpha_nu_q}) and (\ref{eq:myz_def}) indicate that in the present approximation, where the widths of the transitions in $Q$ and in $B$ are neglected, $\myh$ and $\Delta \alpha$ depend on $t$ and on $\nu$ only through the combination
\begin{equation}
\myz \equiv c_2\frac{\nu}{\nu_c(t)} = c_2\left[ \frac{t}{t_{cool}(\nu)} \right]^2 \coma
\end{equation}
where $t_{cool}(\nu)$ is the cooling time of a {\CRE} with energy $E$ related to $\nu$ through $\nu_s(E)=\nu$,
explicitly given in \eq{\ref{eq:CRE_Cooling0}}.

In the alternative, Eulerian picture, the profiles $\myh(\myX,\nu)$ and $\Delta \alpha(\myX,\nu)$ depend on $\myX$ and on $\nu$ only through the combination
\begin{equation}
\myz = c_2\left[ \frac{\myX}{\Dltcool(\nu)} \right]^2 \coma
\end{equation}
where
\begin{align} \label{eq:DeltacoolDef}
\Dltcool(\nu) & \equiv v_d t_{cool}(\nu) \\
& = 3\sqrt{3\pi} \frac{\sqrt{e m_e c}}{\sigma_T} \frac{\sqrt{B\sin\myPalpha}}{B^2+B_{cmb}^2} \nu^{-1/2} v_d \nonumber \\
& \simeq 160 T_{d,10}^{1/2}(1+z)^{-\frac{7}{2}} \frac{2\sqrt{b_d}}{1+b_d^2} \nu_{1.4}^{-1/2} \sqrt{\frac{\mach^2+3}{5\mach^2-1}} \kpc \nonumber \fin
\end{align}
Here, $T_{d,10}\equiv k_B T_d/10\keV$ is measured downstream, and $\nu_{1.4}$ is the \emph{observed} frequency in units of $1.4\GHz$.

Accordingly, \Figs~\ref{fig:ShockSynchQ0} and \ref{fig:ShockSynchRConst} show the synchrotron signature of a shock using the parametrisation $h_{q,\myr}(x/\Dltcool(\nu))$ and $\alpha_{q,\myr}(x/\Dltcool(\nu))$.
Results for flat injection with different choices of $\myr$ are depicted in \Fig~\ref{fig:ShockSynchQ0}, while different choices of injection index $q$ are shown in \Fig~\ref{fig:ShockSynchRConst}, for fixed $\myr=10$ and for $\myr=1/10$.
In the present definition of $\myh$, these results are valid only downstream;
A generalised version in \eq{\ref{eq:myhDefGeneral}} applies both upstream and downstream.

As the figures show, the spectrum is unchanged as the emissivity jumps at the shock according to \eq{\ref{eq:ShockBrightnessAmplification}}.
Farther downstream, $h$ gradually changes, by a factor of $1/\myr$ as $\myX\to\infty$.
This increase (decrease) in emissivity for $\myr>1$ ($\myr<1$) is faster for steeper injection, and is mostly achieved by $\myX \simeq 2\Dltcool$.
The decline (rise) in $h$ is accompanied by spectral steepening (flattening), which is strongest at $\myX\lesssim \Dltcool$.
Farther downstream, the spectrum gradually recovers its steady state value, roughly by $\myX\sim 2\Dltcool$.

These results can be generalised for an inhomogeneous downstream, in which the initial distributions of gas and {\CRIs} prior to the shock are not uniform.
As long as the downstream magnetic field can be approximated as being constant, our analysis remains valid if we replace $j_\nu$ by its value normalised to the background {\CRE} injection rate $Q\propto N_p n$,
\begin{equation}
j_\nu \to \frac{j_\nu}{N_p n} \propto \eta_j \left( \frac{n}{N_p} \right) \fin
\end{equation}
Note that this is simply proportional to $n\eta_j$ if the {\CRIs} are homogeneously distributed.

Strong variations in the magnetic field complicate the analysis, because they imply a superposition of jumps in $\myr$ which involve significant spectral variations.
However, if the magnetic field changes slowly such that the spectral deviations can be neglected, the results shown above and in \Figs~\ref{fig:ShockSynchQ0} and \ref{fig:ShockSynchRConst} remain approximately valid with the generalised definition
\begin{align} \label{eq:myhDefGeneral}
h(\myX,\nu) & \equiv \left( \frac{j_\nu \mypsi}{\myQ B^{2-q/2}} \right)_{\myX} \bigg/ \left( \frac{j_\nu \mypsi}{\myQ B^{2-q/2}} \right)_{\myX\to\infty} \nonumber \\
& = \left[ \frac{j_\nu (1+b^{-2})}{N_p n b^{-q/2}} \right]_{\myX} \bigg/ \left[ \frac{j_\nu (1+b^{-2})}{N_p n b^{-q/2}} \right]_{\myX\to\infty} \nonumber \\
& \simeq \left( \frac{j_\nu}{N_p n} \right)_{\myX} \bigg/ \left( \frac{j_\nu}{N_p n} \right)_{\myX\to\infty} \coma
\end{align}
where in the last line we assumed that the magnetic field is strong, $B\gg B_{cmb}$.
Here and below, the limit $\myX\to \infty$ can be replaced by any finite $\myX\gg \Dltcool$ position downstream in which the magnetic field does not change appreciably.
For example, in a strongly magnetised cluster where {\CRIs} are homogeneously distributed, \eq{\ref{eq:myhDefGeneral}} becomes simply
\begin{equation}
h(\myX,\nu) \simeq \frac{n \eta_j}{n_0 \eta_{j,0}} \fin
\end{equation}

Note that although the two definition of $\myh$, in \eqs{\ref{eq:myhDef}} and (\ref{eq:myhDefGeneral}), agree downstream if the initial plasma is homogeneous, they differ in general upstream.
\Figs~\ref{fig:ShockSynchQ0} and {\ref{fig:ShockSynchRConst} are valid upstream only using the definition of $\myh$ in \eq{\ref{eq:myhDefGeneral}}.

It is interesting to compare the (normalised) synchrotron emissivity immediately behind (downstream of) the shock with its value far downstream.
This may be written as $\myh(\myX=0_+,\nu)=\myr$, valid for all (constant) $q$, as seen for example from \eq{\ref{eq:myhDef}}, where $J_q(\myz=0)=1$, $J_q(\myz\to\infty)=0$, and $0_+$ denotes the downstream side of the shock front.
This result is easily understood by noting that $\myh(\myX)$ is proportional to the deviation of the {\CRE} distribution $f(\myX)$ from its local, asymptotic steady-state value $\myQpsi(\myX)\propto \myQ/\mypsi$.
In the steady-state far downstream, there is no such deviation.
But immediately behind the shock, $f$ is still given by the upstream (\ie before the precursor) steady state $\myQpsi_u$, enhanced by a factor of $\myrcre$ due to the {\CRE} compression, whereas the local steady-state is $\myQpsi_d$.
Consequently, $\myh(\myX=0_+)=(\myQpsi_u/\myQpsi_d)(\myQQ_u/\myQ_u)=\myr$.

Similarly, consider the jump in synchrotron emission across the shock, \ie in the present approximation, across the {\CR} precursor and the shock compression layer.
The emissivity ratio $r_j$, given by \eq{\ref{eq:ShockBrightnessAmplification}}, translates to a jump by a factor of $\myr$ in the normalised emissivity $h$,
\begin{equation}
\frac{\myh(\myX=0_+)}{\myh(\myX=-\Dltcr)} = r_j \frac{\myQpsi_u}{\myQpsi_d} \myrB^{-2+\frac{q}{2}} = \myr \fin
\end{equation}
This can be understood using the same reasoning applied to the $\myh(\myX=0_+)=\myr$ result above, as the upstream distribution is assumed to be in a steady-state.
We conclude that $\myh$ is unity upstream, jumps by a factor of $\myr$ at the shock, and gradually returns to $\myh=1$ far downstream.

These results are valid, for example, at $\gg \Dltcr$ distances from an outgoing relic shock that crossed the centre of a cluster.
If the magnetic field remains strong ($B\gg B_{cmb}$) behind the shock, at all radii $r<r_{sh}$, then the ratio between the peak radio emissivity in the relic and in the cluster's centre may be written as
\begin{equation}
\frac{\eta_{j}(r_{sh-})}{\eta_{j,0}} \simeq \left( \frac{n}{n_0} \right)^{-1} \myh(r_{sh-}) = \left( \frac{n}{n_0} \right)^{-1} \myr \fin
\end{equation}
The emissivity ratio across the shock may similarly be written as $\myh(r_{sh-})/\myh(r_{sh+})=\myr$.
However, this can translate to different emissivity or $\eta$ ratios, depending on shock magnetisation.
Namely, the emissivity enhancement by the shock (see \eq{\ref{eq:ShockBrightnessAmplification}}) may be written as
\begin{equation}
\frac{\eta_j(r_{sh-})}{\eta_j(r_{sh+})} = \frac{r_j}{r_g^2}=\frac{\myrcr}{\myrg^2}\myrB^{2-\frac{q}{2}} \coma
\end{equation}
which is large in a weak shock only if $B_d\gg B_u$.

In summary, the downstream profiles of synchrotron brightness and spectrum crucially depend on shock magnetisation, through the parameter $\myr>0$.
Three qualitative behaviours can be distinguished (assuming injection with a fixed power-law $q$):
\begin{enumerate}
\item
$\myr>1$ is expected if shock magnetisation is strong in the sense that $\mypsi_d/\mypsi_u > \myrg$.
Here, $\myh(\myX)$ jumps by a factor of $\myr$ at the shock, and gradually returns to unity downstream, accompanied by spectral steepening.
The steepening is maximal $\sim\Dltcool$ downstream of the shock.
More precisely, the maximally steep spectrum
The spectrum flattens back to the steady state farther downstream; $\alpha_s=(q-2)/2$ is roughly recovered by $2\Dltcool$.
For example, for flat ($q=0$) injection and $0<\myr<10$, \eq{\ref{eq:alpha_nu_q0}} implies that the steepest spectrum is well fit by
\begin{equation} \label{eq:FitDltAlpha}
\Delta \alpha \simeq -0.22(r - 1)^{0.91} (1 - 0.23 \ln r)\coma
\end{equation}
reached at
\begin{equation} \label{eq:FitXMaxSteepening}
\myX/\Dltcool \simeq 0.635 r^{0.396} (1 - 0.145 \ln r) \fin
\end{equation}
For $\myr\gtrsim 2$, both a large magnetic amplification factor $\myrB=B_d/B_u$ and a strong, $B_d\gtrsim B_{cmb}$ field downstream are necessary; the emissivity must therefore significantly jump across the shock.

\item
$\myr<1$ is expected if shock magnetisation is weak in the sense that $\mypsi_d/\mypsi_u < \myrg$.
This occurs, for example, if $B_d\ll B_{cmb}$, regardless of $\myrB$.
Here, $\myh(\myX)$ drops by a factor of $\myr^{-1}$ at the shock, and gradually increases back to unity downstream.
This is accompanied by spectral flattening, which is maximal close to the shock front, at $\myX\lesssim\Dltcool/2$.

\item
The special case $\myr=1$ occurs if magnetisation and shock compression are precisely balanced.
The downstream profiles of $\myh(\myX)$ and $\alpha(\myX)$ are uniform and indistinguishable from a steady state.
\end{enumerate}

In all three cases, the emissivity jump at the shock may be large, depending on $\myrB$.
Some additional changes in $j(\myX)$ and $\alpha(\myX)$ may take place downstream, if the magnetic variability is significant.
Incoherent spatial or temporal variations in $B$ typically introduce only a weak, $-0.1<\Delta\alpha<0$ steepening, but {\CRE} diffusion could substantially steepen the spectrum; see \S\ref{sec:SynchrotronEvolvingB}.

Finally note that the above discussion of the emission downstream of a shock applies also for a ``magnetisation front'', in which turbulent motions induce enhanced magnetic fields behind a subsonic, propagating wave.
This scenario corresponds to the $\myr>1$ case because $\myrg\to1$, although $\myr$ is near unity if $B_d\ll B_{cmb}$.
The discussion does not however apply for a tangential discontinuity such as a cold front, where although the magnetic field shows a jump across the discontinuity, the flow is parallel to CF.

\subsection{Irregular, evolving magnetic fields, and {\CRE} diffusion}
\label{sec:SynchrotronEvolvingB}

The above analysis neglected {\CRE} diffusion, and assumed that the magnetic fields are constant away from the near vicinity of shock fronts.
Both assumptions should be relaxed if diffusion is strong and if the magnetic field is turbulent.

Here we present a highly simplified picture of diffusion and turbulence, by examining the synchrotron signature of {\CRE} diffusion across magnetic irregularities in \S\ref{sec:SynchrtronClumpyField}, and of temporal variations in the magnetic field in \S\ref{sec:fluctuations}.
Both effects lead, in general, to radio brightening and spectral steepening.

\subsubsection{{\CRE} Diffusion in an irregular magnetic field}
\label{sec:SynchrtronClumpyField}

In \S\ref{sec:HomogeneousCRIsImplications} we show that the homogeneous distribution of {\CRIs} inferred from observations can be explained by invoking strong diffusion, with coefficient $D>10^{32}\cm^2\se^{-1}$ for $\sim 100\GeV$ {\CRIs}.
This would imply that {\CREs} diffuse similarly, so $1.4\GHz$-emitting {\CREs} could travel a distance of up to $\sim 200\kpc$ before cooling, much farther than the typical coherence length of the magnetic field.
Magnetic irregularities on smaller scales would therefore modify the properties of the radio emission.

The combination of magnetic irregularities and strong diffusion leads, in general, to radio brightening and spectral steepening with respect to the homogeneous case.
To see this, consider the effect of spatial gradients in $B$, neglecting temporal irregularities which are discussed in \S\ref{sec:fluctuations}.
Assume, for simplicity, that {\CRE} injection is homogeneous.
The results can be qualitatively understood by considering the local steady state of $f$ in the absence of spatial or temporal gradients, $\myQpsi\equiv Q/(E\mypsi)$.

As {\CREs} wonder through regions of different magnetisation and cool, their local distribution $\myQpsi_a(\vecthree{r})\sim Q/(E\mypsi_a)$ is regulated by a locally averaged cooling parameter $\mypsi_a(\vecthree{r})\propto B_a^2+B_{cmb}^2$.
Excessive cooling takes place in highly magnetised regions, where the magnetic field amplitude $B_+>B_a$, leading to brighter emission, locally, than would be expected even if the entire ICM was highly magnetised to the same level $B_+$, because $\myQpsi(B_a)>\myQpsi(B_+)$.
The enhanced cooling in these regions leads to local spectral steepening of the {\CREs}; spectral flattening can take place in nearby regions where the magnetic field is weaker.

Integrating the synchrotron and inverse-Compton emission from {\CREs} over a large region recovers the steady-state luminosity, dictated by the injection rate.
However, under the present assumptions, the radio signal is irregular, and the inverse Compton emission does not fall in the radio band.
Hence, the radio luminosity of a source depends on the definition of its volume.
Adopting some noise threshold would highlight the strongly magnetised regions, where the emissivity exceeds the average and the spectrum is steep.

Quantitatively, under the above assumptions one should replace the {\CRE} evolution \eq{\ref{eq:PDE_n}} by the PDE
\begin{equation} \label{eq:PDE_n_diff}
\pr_t \myn(t,E,\vecthree{r}) = \dot{\myn}_+ - \pr_E\left( \myn \frac{dE}{dt} \right) + \grad(D\grad \myn) \coma
\end{equation}
where we assumed that the diffusion is isotropic.
In a steady-state, and assuming that the diffusion is also homogeneous, this equation may be written as
\begin{equation} \label{eq:PDE_f_diff}
Q - \psi \pr_\myx f + D\grad^2f =0 \coma
\end{equation}
generalising \eq{\ref{eq:PDE_f}}.

In order to demonstrate the brightening and spectral steepening of the strongly magnetised regions, consider the case where the diffusion coefficient has a power-law energy dependence $D\propto  E^d$ with $d<1$, and is independent of the magnetic field on the scales affecting $\mypsi$.
Here we may formally solve \eq{\ref{eq:PDE_f_diff}} by expanding in powers of $D/E$,
\begin{align}
f = \myQpsi \bigg[ & 1 + \grad^2\left(\frac{1}{\mypsi}\right) \frac{(D/E)^1}{2-d} \\
& + \grad^2\left(\frac{1}{\mypsi}\grad^2\left(\frac{1}{\mypsi}\right) \right) \frac{(D/E)^2}{(2-d)(3-2d)} +\ldots \bigg] \fin \nonumber
\end{align}
This expansion diverges at low energies.
Consider high energies where the second term in the expansion dominates the correction to the steady-state solution $\myQpsi$.
This term is positive (negative) in magnetic field maxima (minima), inducing radio brightening (dimming) and spectral steepening (flattening), as argued qualitatively above.

An analysis of realistic magnetic configurations is beyond the scope of the present work, so we consider highly idealised magnetic configurations.
Assume, for example, that the magnetic fields configuration is spherically symmetric.
The spectral index of the {\CREs} in the origin is then given by
\begin{equation}
\phi(\vecthree{r}) = -\frac{\myQpsi}{f} \left(1+\frac{D}{Q}\,\frac{\partial^2 f}{\partial r^2} \right) \fin
\end{equation}
For a monotonic $B(r)$ distribution peaked at the centre, $f$ exceeds $\myQpsi$ at $\vecthree{r}=0$, but is still a minimum there, so $\partial^2 f/\partial r^2>0$.
Spectral steepening arises when the curvature term in the parenthesis is larger than $(f-\myQpsi)/\myQpsi$.

As a specific example, consider a configuration where the magnetic field is elevated to a constant amplitude $B$ within a ball, and is negligible outside it.
Solving the steady-state \eq{\ref{eq:PDE_f_diff}} for this configuration leads to a steep spectrum when $B$ is large, for {\CRE} energies near $\sim E_0$, at which the diffusion length is comparable to the diameter of the ball.
The spectrum is shown for various choices of $B$ in \Fig~\ref{fig:B_irregularities}, both in the centre of the ball where the spectrum is steepest, and averaged over the ball, assuming a constant diffusion function.
The spectrum is modified if $D$ is assumed to vary with $E$ or with $B$.

\begin{figure}
\centerline{\epsfxsize=8cm \epsfbox{\myfarfig{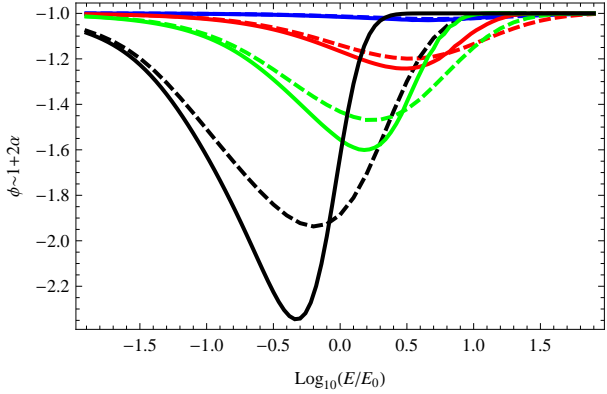}}}
\caption{
{\CRE} logarithmic spectral index $\phi=2+s_p\sim 1+2\alpha$ in the centre of (solid lines), and averaged over (dashed lines), a magnetised ball with $B/B_{cmb}=0.3$, $1$, $2$, and $5$ (top to bottom curves).
Results are computed by numerically solving \eq{\ref{eq:PDE_f_diff}}, assuming constant and flat injection $Q$, a constant diffusion coefficient $D$, and negligible magnetic field outside the ball.
\label{fig:B_irregularities}
\vspace{2mm}}
\end{figure}

\subsubsection{Magnetic variability}
\label{sec:fluctuations}

We have seen that substantial spectral variations may arise from rapid changes in the injection rate and in the magnetic field strength.
Some level of temporal variations in $Q$ and $B$ on the relevant, $\lesssim t_{cool}$ timescales is inevitable in any plausible ICM configuration.
In addition, the combination of {\CRE} diffusion and magnetic inhomogeneities may effectively be regarded as {\CRE} evolution in the presence of time-dependent magnetic fields.
We next consider the spectral signature of {\CREs} steadily injected with a flat spectrum ($Q=\const$) into a dynamically magnetised ICM; the generalisation for steep or variable injection is straightforward.

Coherent variations in $B$ can lead to substantial --- in principle, arbitrarily large --- changes in the {\CRE} spectrum, as in the extreme case of a sudden magnetic jump discussed in \S\ref{sec:B_jump_CRE}.
However, two effects combine to diminish the radio signature of typical magnetic variations: incoherence and synchrotron spectral smoothing.
Consequently, the spectral deviations from steady-state radio emission, expected due to space-time magnetic variability in the ICM, are small away from shocks, as illustrated in \Fig~\ref{fig:BOscillations}.

Consider a time-dependent magnetic field configuration which is statistically in a steady state, in the sense that the magnetic field in the region integrated by the radio beam of frequency $\nu$ does not coherently change over $\lesssim t_{cool}(\myx_\nu)$ timescales.
The time dependence of $B$ in different regions may be decomposed into oscillatory modes of period $T$, $B_T(t)=B_0+\Delta B\sin (\theta+2\pi t/T)$.
Each such mode could have a substantial spectral signature in the radio, provided that the amplitude $\Delta B$ of the oscillation is not small compared to $(B_0^2+B_{cmb}^2)^{1/2}$.
However, the steady state assumption implies that the phases $\theta$ of different such modes are not correlated, so the accumulated spectral effect is suppressed.

\Fig~\ref{fig:BOscillations} shows the deviation $\Delta\alpha(\nu)$ in the radio spectrum introduced by modes of a well defined period $T$, after averaging out the phase $\theta$.
We compute $\Delta \alpha$ for each mode using \eqs{\ref{eq:jnu_approx}} and (\ref{eq:G_approx}), for $B_0=0$ and different values of $\Delta B$, assuming constant and flat {\CRE} injection.
Some spectral signature is seen to survive the phase averaging.
The main effect is weak spectral steepening around the synchrotron frequency of {\CREs} for which the cooling time $t_{cool}(\myx_\nu)\simeq T$.
This steepening is maximised when $\Delta B=3^{1/2}B_{cmb}$.

\begin{figure}
\centerline{\epsfxsize=8cm \epsfbox{\myfarfig{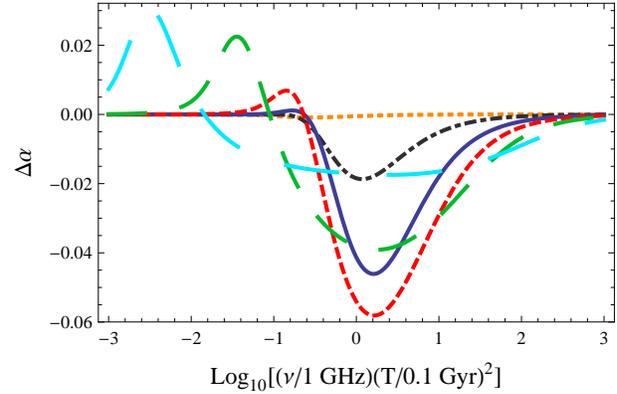}}}
\caption{
Spectral deviation $\Delta\alpha(\nu)$ from a steady state induced by oscillating magnetic modes with period $T$ and amplitude $\Delta B$, after averaging over the oscillation phase.
Results are shown for $\Delta B/B_{cmb}=0.1$ (dotted curve), $0.5$ (dot-dashed), $1$ (solid), $\sqrt{3}$ (maximal steepening; dashed), $4$ (long dashed), and $10$ (very long dashed).
Results are computed using \eqs{\ref{eq:jnu_approx}} and (\ref{eq:G_approx}), assuming constant and flat injection.
\label{fig:BOscillations}
\vspace{2mm}}
\end{figure}

Magnetic modes of period $T$ lead to a {\CRE} spectrum which is quite different from the corresponding radio signature shown in \Fig~\ref{fig:BOscillations}.
In addition to the $t_{cool}(E)\sim T$ steepening, which arises from the last rise in $B$ during the most recent oscillation period, the phase-averaged {\CRE} spectrum $\phi(E)$ shows additional, oscillatory deviations at lower energies.
These low energy spectral features are the accumulated effect of previous $B_T(t)$ oscillation cycles, and can be quite strong.
However, they are effectively erased when convolving the {\CRE} distribution with $F_{syn}$.

The observed radio spectrum involves an integration over all magnetic modes in the beam, weighted by their respective amplitudes.
The outcome is, in general, mild steepening, weaker than shown in \Fig~\ref{fig:BOscillations}.
To estimate the overall effect of an arbitrary magnetic configuration, it is convenient to consider the $\delta$-function approximation of $F_{syn}$ in \eq{\ref{eq:Fsyn_delta}}, keeping in mind that spectral features spanning less than a factor of $\sim 2$ in $\nu$ are erased by the convolution with the true $F_{syn}$.

In this approximation (\cf \eqs{\ref{eq:flat_j_t}}--(\ref{eq:flat_alpha_t_Qconst})),
\begin{align} \label{eq:j_alpha_Fsyn}
& \nu j_\nu(t,\nu) \propto \frac{\mypsi_B}{\mypsi_a} \, ; \\
& \alpha(t,\nu) = -1 - \frac{\mypsi_{a}-\mypsi_i}{2\mypsi_i} \coma \nonumber
\end{align}
where $\mypsi_i$ and $\mypsi_a$ are respectively the values of $\mypsi$ at time $t_i=t-\tau_{cool}(x_\nu)$ and time-averaged between $t_i$ and $t$.
This shows that if $B$ gradually increases (decreases) over a timescale $>\tau_{cool}$, the spectrum steepens (flattens) as the emissivity increases (decreases).
Consequently, brighter emission is on average steeper than faint emission, and the overall effect is that of steepening.

More quantitatively, the observed deviation from a steady-state spectrum may be written as
\begin{equation} \label{eq:alpha_j}
\langle \Delta\alpha(\nu) \rangle \equiv \frac{\langle \alpha j_\nu \rangle}{\langle  j_\nu \rangle} + 1 = \frac{\left\langle \mypsi_B/\mypsi_a\right\rangle - \left\langle \mypsi_B/\mypsi_i\right\rangle}{2\left\langle \mypsi_B/\mypsi_a\right\rangle}  \coma
\end{equation}
where we defined $\langle C(\mypsi) \rangle$ as the average of the functional $C(\mypsi)$ taken over the beam or, if accounting for diffusion, over the histories of {\CREs} ending up in the beam.
The two terms in the numerator are dominated by regions which are presently (at time $t$) radio bright, or were relatively radio dim since $\sim t-t_{cool}$.
Such regions are more likely to have experienced recent magnetic growth.
If the characteristic timescale of such growth exceeds $\sim t_{cool}$, then we may expect $\left\langle \mypsi_B/\mypsi_a\right\rangle<\left\langle \mypsi_B/\mypsi_i\right\rangle$, implying spectral steepening.

\Eq{\ref{eq:alpha_j}} shows how magnetic configurations --- even with constant $\langle B \rangle$ --- can be constructed to yield extreme steepening or even flattening.
However, non-coherent configurations tend to yield a mild, $0.1\lesssim \Delta \alpha <0$ spectral steepening.
This is the case, for example, in magnetic turbulence, corresponding to a power law weighting of the modes shown in \Fig~\ref{fig:BOscillations}.


\section{Time-dependent, secondary {\CRE} model for halos and relics}
\label{sec:ModelApplications}

In \S\ref{sec:HaloAndRelicEta} we presented evidence that the {\CRI} distribution is homogeneous within GHs, and that modeling radio relics and halos as emission from secondary {\CREs} produced by the same {\CRI} population explains several observations, and resolves some of the open questions outlined in \S\ref{sec:ModelProblems}.
However, the steady-state secondary {\CRE} model fails to account for some observations, such as the relatively high radio to X-ray brightness ratio of the relics, the spectral steepening behind some relics, steep spectrum GHs, and evidence for a connection between these steep GHs and the presence of relics.

In order to test the model more precisely and possibly resolve such outstanding issues, we studied in \S\ref{sec:TimeDependentTheory} the temporal evolution of {\CREs} and the resulting synchrotron signature in various scenarios, in particular following weak shocks such as those suspected to energise relics.
For a given {\CRP} spectrum, the profiles of synchrotron brightness and spectrum behind a wave front depend on a single parameter $\myr$, which is sensitive to the wave-induced magnetisation and compression.
In particular, strong magnetisation, where $\myr>1$, leads to enhanced emission near the shock and spectral steepening behind it.

Here we combine the results of the previous sections, and model both relics and halos as emission from secondary {\CREs} produced by a population of homogeneously distributed {\CRIs}, taking into account the time-dependence of the magnetic fields and of the {\CR} distributions.

The phenomenological background used here is summarised above in \S\ref{sec:eta_in_relics}, and the shock model is summarised in \S\ref{sec:SynchrotronImprintOfShock} above.
In \S\ref{sec:model_outline} we present the model and its main predictions.
In \S\ref{sec:ModelTest_StudiedClusters} we test the model against well-studied, individual halo clusters which also harbour a relic or an identified shock.
In \S\ref{sec:ModelTest_Compilation} we apply the model to our sample of relics presented in \S\ref{sec:HaloAndRelicEta}, and use it to estimate the magnetisation efficiency.
In \S\ref{sec:ModelTest_SpectralCurvature} we discuss how the radio spectrum and its curvature gauge the recent magnetic evolution.
Finally, in \S\ref{sec:ModelTest_RelicGHConnection} we examine the connection between steep spectrum GHs and the presence of relics, and argue that it reflects a recent merger.

The derivations of the {\CRI} energy density $u_p$ and spectrum $s_p$ are deferred to \S\ref{sec:Discussion}, where various implications of the model are discussed.

\subsection{Model Description}
\label{sec:model_outline}

\subsubsection{Steady-state radio emission from the ICM: $I_\nu$ is proportional to the column density of magnetised gas}

We have shown in \S\ref{sec:HaloAndRelicEta} that GH observations, in particular the morphologies of flat-spectrum halos, imply a homogeneous distribution of {\CRIs} if the emission arises from secondary {\CREs}.
Moreover, relics can then be explained as emission from secondary {\CREs} produced by the same {\CRI} population, although their radio emission is slightly elevated, suggesting recent magnetisation as described below.

In general, we conclude that for steady state {\CRI} injection and magnetic fields, away from additional radio sources such as radio galaxies and AGNs, the radio emission throughout the ICM is given by
\begin{equation} \label{eq:nuInuPerLambda}
\nu I_\nu(\vectwo{r}) = c_I \lambda_{nB} \equiv c_I \int n \, \frac{B^2}{B^2+B_{cmb}^2} \,d l \coma
\end{equation}
where the coefficient $c_I$ is approximately constant within a cluster, and the integration is performed along the line of sight in the direction $\vectwo{r}$.
Here, a steady state is defined as slow changes with respect to the {\CRE} cooling time in \eq{\ref{eq:CRE_Cooling0}}; deviations in \eq{\ref{eq:nuInuPerLambda}} due to fast temporal changes are discussed below.

We may determine $c_I$ from flat spectrum, regular GHs, which are believed to be highly magnetised regions with steady state {\CRE} injection and fields \citep[][and {\KL}]{KushnirEtAl09}.
Here, the integral of \eq{\ref{eq:nuInuPerLambda}} becomes, approximately, the column density of the gas.
Computing the central column densities using $\beta$ models for the GHs in the present sample and in the sample of {\KL}, we find central $c_I$ values in the range
\begin{align} \label{eq:DefCI}
c_I
& = 10^{-\range{30.9}{31.5}} \erg \se^{-1} \sr^{-1} \\
& = 10^{\range{-0.1}{0.5}} \left(\frac{1\GHz}{10^{22}\cm^{-2}} \right) \muJy \asec^{-2} \nonumber \fin
\end{align}
We used radio observations at frequencies near $1.4\GHz$ to determine $c_I$, but the result is nearly independent of frequency for the flat, $\alpha\simeq -1$ radio spectrum observed.

\subsubsection{Flat spectrum halos: $I_\nu\propto \lambda_n$ becomes $I_\nu\propto F_X$ in weakly magnetised regions}

For the most part, flat-spectrum halos are believed to be in a steady state in which the {\CRE} injection rate $Q$ and the magnetic field amplitude $B$ do not change significantly on $\lesssim t_{cool}$ time scales.
The emission from such halos or parts of halos follows the {\CRI} distribution and so shows a flat, $\alpha\simeq -1$ spectrum.
Weak steepening may be induced by the energy dependence of the cross section for secondary particle production ($|\Delta\alpha| \simeq 0.1$--$0.4$, depending on the {\CRP} spectrum; see {\KL}) and by perturbations about the steady state values of $Q$ and $B$ ($|\Delta\alpha| \lesssim 0.1$; see \S\ref{sec:fluctuations}).
Stronger steepening if {\CRE} diffusion is strong and the magnetic field is spatially irregular; see \S\ref{sec:SynchrotronEvolvingB}.

The morphological properties of steady-state halos --- both GHs and MHs --- were discussed by {\KL}, but the {\CRI} distribution was not well constrained.
The morphology can be described more accurately here, in the context of a homogeneous {\CRI} distribution, using \eqs{\ref{eq:nuInuPerLambda}}--(\ref{eq:DefCI}).

Assuming that the central halo regions are highly magnetised ($b\equiv B/B_{cmb}\gg 1$), the radio brightness is linear in the gas column density, $I_\nu\propto \lambda_n$.
If, in addition, the X-ray data is well fit by a $\beta=2/3$ model, then $I_\nu\propto F_X^{1/3}$.
Indeed, central, regular X-ray regions of select clusters show a radio--X-ray correlation $I_\nu\sim F_X^{1/3}$ (equivalently, $\eta\propto F_X^{-2/3}$), and the radio brightness follows the column density computed from the corresponding $\beta$-model, $I_\nu\sim \lambda_n$.
This includes the relaxed, $r\lesssim 400\kpc$ regions in A665 and in A2163, and the NW sector of A2744.
(The X-ray morphology of A2744 is highly irregular, but it bears great resemblance to A2163 in the radio.)

In weakly magnetised regions, such as in the halo's periphery, the radio emission is diminished.
Adopting the characteristic scaling $B^2\propto n$, in which the magnetic energy density is a fixed fraction of the thermal energy density, yields $I_\nu\propto F_X$ in such regions.
Indeed, the gradual transition from $I_\nu\sim F_X^{1/3}$ to $I_\nu\sim F_X$ behaviour beyond some radius is consistent with the observed profiles shown in \S\ref{sec:EtaInWellStudiedGHs}.

Substructure enhances the X-ray emission, slowing the radial decline of $F_X(\vectwo{r})$, but has no effect on the radio brightness.
The combination of substructure, weak magnetisation outside the core, asymmetry, and contaminations, often masks the underlying $I_\nu\sim F_X^{1/3}$ profile, leading to the linear or mildly sublinear $I_\nu$--$F_X$ relations reported in the literature \citep[][and {\KL}, in clusters including A2163 and A2744, shown in \S\ref{sec:EtaInWellStudiedGHs} to follow $I_\nu\propto F_X^{1/3}$ in relaxed regions]{GovoniEtAl01, FerettiEtAl01}.
This is particularly true if equal area bins, rather than radial bins, are used, thus emphasising the peripheral regions.
Close inspection and removal of irregular X-ray regions is needed in order to recover the underlying scaling.

\subsubsection{Relic observations imply temporal magnetic growth in the fluid frame}

The identification of halos and relics with strong, $B\gtrsim B_{cmb}$ magnetic fields, and the association between relics and weak shocks, are discussed in \S\ref{sec:HaloAndRelicEta} above.
Consider first the case of peripheral relics, where the model is challenged by the weak, $B<B_{cmb}$ steady-state (upstream) magnetic fields expected due to the low ambient density.
In order to reconcile the above observations, the magnetic field must be locally elevated in relics.
Two types of relic magnetisation models are possible.

In one model, the elevated magnetic field is approximately constant in time in the fluid frame.
This is the case, for example, in a magnetised clump that retains its integrity as it moves through the ICM.
This is essentially the simple model studied in \S\ref{sec:HaloAndRelicEta}, failing to reproduce the spectral steepening behind relics, and their association with steep halos.
It also fails to account for the elevated radio brightness, which saturates as $\sim B^2/(B^2+B_{cmb}^2)$ with an increasing magnetic field strength.
Therefore, this simple model could only apply for a subset of relics, and may require a local enhancement of {\CRs} in order to explain the bright radio emission.

The second model, presented here, identifies a relic with a recently amplified (or still rising) magnetic field in the fluid frame.
This occurs most probably in a propagating magnetisation wave, presumably --- but not necessarily --- a shock wave.
In this model, the fluid frame sees a rapid rise in magnetic field, on a timescale shorter than the {\CRE} cooling time $t_{cool}$, so the analysis of \S\ref{sec:SynchrotronImprintOfShock} applies.
In particular, strong magnetisation with $b_d = B_d/B_{cmb}\gtrsim 1$ corresponds to the case $\myr\simeq (1+b_d^2)/\myrg>1$.
Here, the emission is strongly enhanced at the shock, but declines downstream over a timescale $\sim 2 t_{cool}$, accompanied by spectral steepening culminating around $\sim t_{cool}$.

\subsubsection{Dynamic Magnetisation model: main properties}

In this wave model, a shock associated with the relic would have to be weak.
This is partly based on observational constraints, which support the presence of $\mach < 3$ shocks at the edges of relics (although no such a peripheral shock has been confirmed thus far).
In addition, the correlations between relic and halo properties discussed in \S\ref{sec:ModelProblems} and \S\ref{sec:HaloAndRelicEta} indicate that relic shocks cannot be sufficiently strong to appreciably accelerate primary {\CREs}.
Note, however, that the model applies equally well to ingoing (infall, say) or outgoing (merger, say) weak shocks.

The synchrotron emissivity and spectrum behind the shock were derived in \S\ref{sec:SynchrotronVariability} and in \S\ref{sec:SynchrotronImprintOfShock}.
They are given by \eqs{\ref{eq:j_nu_q}} and (\ref{eq:alpha_nu_q}), with the supplementing definitions in \eqs{\ref{eq:myz_def}}--(\ref{eq:Jq_behviour}) and in \S\ref{sec:jnu_from_jump_detail}.
In the limit of flat injection, which is a fairly good approximation in clusters, $j_\nu$ and $\alpha$ are given by \eqs{\ref{eq:j_nu_q}}, (\ref{eq:A0Def})--(\ref{eq:J0Def}), and (\ref{eq:alpha_nu_q0}).
In this limit, maximal spectral steepening is approximately given by \eqs{\ref{eq:FitDltAlpha}}--(\ref{eq:FitXMaxSteepening}).
The shock-induced brightening and subsequent steepening are illustrated in \Figs~\ref{fig:ShockSynchQ0} and \ref{fig:ShockSynchRConst} (curves with $\myr>1$).

Qualitatively, this synchrotron signature can be simply understood as the evolution of the {\CRE} distribution from a high- to a low-density state.
Upstream, the {\CRE} density is in a steady state regulated by cooling off (the CMB and) the weak upstream magnetic field, $\myQpsi_u\propto \mypsi_u^{-1}\propto (1+b_u^2)^{-1}\simeq 1$.
This {\CRE} density is higher or much higher than in the downstream steady state, $\myQpsi_d\propto (1+b_d^2)^{-1}<1$.
The strong downstream field causes the {\CREs} to synchrotron radiate profusely, until the new steady-state is reached at a distance $\sim 2\Dltcool$ downstream.
This transition from a high to a low steady-state density depends on the {\CRE} energy and so occurs gradually, with high energy {\CREs} responding faster.
This leads to the spectral steepening, which is maximal $\sim \Dltcool$ downstream of the wave and can be substantial if $b_d$ is large.
Note that both $t_{cool}$ and $\Dltcool$ are frequency dependent; see \eqs{\ref{eq:CRE_Cooling0}} and (\ref{eq:DeltacoolDef}).

\subsubsection{The model explains the brightness and spectrum profiles of relics}

Taking cognizance of the radio signature of a wave resolves several of the outstanding discrepancies outlined in \S\ref{sec:ModelProblems} and \S\ref{sec:HaloAndRelicEta}, as we show qualitatively here and confirm more quantitatively in \S\ref{sec:ModelTest_StudiedClusters} and \S\ref{sec:ModelTest_Compilation} below.

In particular:
\begin{enumerate}
\vspace{-2mm}
\item
\emph{The elevated radio brightness of relics pointed out in \S\ref{sec:eta_in_relics} is naturally explained by the wave model.}

We showed that extrapolating the radio brightness of GHs to the peripheral relics yields a relic brightness which broadly agrees with observations but is consistently too low, by a factor of $\sim2\text{--}3$.
This discrepancy was discovered by examining relics and halos found both in the same cluster and in a sample of different clusters, focusing on the most radio-bright points along relics, and assuming that the homogeneous distribution of {\CRIs} derived in GHs persists out to large scales.
Note that the discrepancy would be worse if the {\CRI} distribution was to decline at large radii, for example in a model where {\CRIs} are injected at small or large radii and subsequently evolve adiabatically.

In the present, wave model, the extrapolated relic brightness must be modified, by multiplying the expected peak brightness by a factor of $\myr$.
Peak magnetic fields $B_d\gtrsim B_{cmb}$ would suffice to match the theory with observations.
Such magnetic fields are indeed plausible in the downstream of the weak peripheral shocks associated with relics, as argued in \S\ref{sec:SecondaryCREsInRelics}.

\item
\emph{The flat spectrum and downstream steepening observed in some relics (\eg in A521, A3667, A1240, A2256, and A2345) naturally occur in the wave model.}

Regardless of $\myr$, the {\CRE} spectrum just behind the shock has the steady state value, \ie $\alpha=-1$ if the {\CRP} spectrum is flat, because the {\CREs} had no time to cool.
However, the {\CRE} density just behind the shock is elevated (suppressed) with respect to its downstream steady-state, which is regulated by a strong magnetic field, if $\myr>1$ ($\myr<1$).
The subsequent transition to the downstream steady state is accompanied by spectral steepening (flattening), in both frequency and distance from the relic.
The effect is stronger for larger $|\myr-1|$, and vanishes completely for $\myr=1$, for which the spectrum is uniform.
If $\myr>1$ and the emission far downstream is not detectable, the synchrotron profile downstream is indistinguishable from that of primary {\CRE} models with (unrealistically) flat injection.

Previous studies have attempted to model relics, with limited success, using primary {\CRE} models.
In the present, secondary {\CRE} wave model, the predicted emission resembles that of such primary {\CRE} models, with a few notable exceptions:
\begin{enumerate}
\vspace{-2mm}
\item
in the present model, the spectrum at peak brightness is flat, whereas a steep spectrum is always expected in primary {\CRE} models due to the weakness of the shocks, as shown in \S\ref{sec:ModelProblems}.
\item
In the present model, the degree of steepening should vary among relics according to their different $\myr$ values; strong steepening is expected for $\myr>1$, while very little to no steepening is expected behind $\myr\simeq 1$ relics.
In extreme cases where compression is stronger than magnetisation such that $\myr<1$, downstream flattening is expected.
\item
In the present model, radio emission downstream of a shock does not necessarily vanish, as {\CREs} are continuously injected into the plasma.
In some cases, depending on $\myr$, $\lambda_n$ and the magnetic profile, this can lead to a radio trail behind the relic, to a halo--relic bridge, or to a radial halo protrusion, as discussed below.
\end{enumerate}
This is precisely the observational situation: most relics show a spectrum much flatter than anticipated in primary models (see \S\ref{sec:ModelProblems}); different relics show different degrees of steepening, ranging for strong (\eg in A521) to weak or none (\eg A1240, Coma); and some relics show an inward radial trail, in some cases forming a continuous halo--relic bridge (\eg in Coma, A2255, A2744).

Note that some relics show outward, rather than inward steepening.
This is the case, for example, in the Western relic in A2345 (A2345-1), found towards the possibly merging group X1 \citep{BonafedeEtAl09Relics}.
Such a relic may arise from an ingoing shock, for example in front of an infalling clump, or from a projected, oblique, outgoing shock.
It could also arise from an outgoing shock with $\myr<1$, but this would not produce a sharp inward cutoff as seen in A2345-1.

\end{enumerate}

The brightness and spectrum profiles near relics and halo shocks are examined quantitatively in well studied halo clusters in \S\ref{sec:ModelTest_StudiedClusters}, and in our sample of relic clusters in \S\ref{sec:ModelTest_Compilation}.
Such tests gauge $\myr$, and thus provide a measure of the magnetic field amplification in the shocks.

\subsubsection{The model explains the halo--relic connections observed}

The time-dependent, secondary {\CRE} model explains the various connections between halos and relics outlined in \S\ref{sec:ModelProblems}, which are largely unexplained by alternative models.
These connections are natural in the model, because relics and halos are essentially modeled by the same process, namely secondary {\CREs} produced from the same {\CRI} population, synchrotron emitting in strong magnetic fields, although relics require a minor time-dependent correction.

This includes the exceptional halos and relics, which do not follow the bimodal, standard classification between halos and relics according to their morphologies and polarisations.
Thus, although most halos are morphologically regular and unpolarised, some of them show an irregular, clumpy or filamentary morphology, or a strong polarisation.
One example is the irregular, strongly polarised halo in A2255.

In the present model, {\CRE} injection is similar in halos and in relics, so the main difference between them is the magnetic configuration, which determines the radio morphology and polarisation.
Thus, a relaxed magnetic field configuration corresponds to regular, unpolarised emission classified as a halo, whereas a disturbed magnetic field would lead to an irregular profile, which would be polarised in regions where the magnetic field is ordered, for example in the wake of a shock where it would be classified as a relic.
An ICM harbouring merger shocks would therefore be classified as a halo once the overall field is sufficiently amplified, featuring a smooth transition from a central relic or several relics to a halo.
Interestingly, it was recently suggested that A2255 should not be classified as a halo, but rather as an ensemble of relics \citep{PizzoEtAl10}.

Similarly, the model explains why relics qualitatively show a correlation between radio and X-ray emission similar to the correlation found in GHs, as reported by \citet{GiovanniniFeretti04}.
Note that a similar correlation is also found in the brightness and coincident luminosity of MHs ({\KL}).
These correlations are similar in our model because the radio emission arises from the same {\CRI} population, scattering off the X-ray gas.

As mentioned above, the model explains the flat spectrum and inward steepening observed in relics.
It also explains why steady-state halos and relic edges show the same spectrum, as they directly reflect the same {\CRP} spectrum.
Moreover, is also explains why some halos show spectral steepening with increasing radius, similar to that found downstream of relics.
Indeed, if the magnetic field at the periphery of a halo is growing in time, the resulting temporal {\CRE} evolution would be analogous to that expected downstream of a relic.

Finally, as mentioned above, our model predicts extended, faint radio emission downstream of a shock, which may manifest as the observed halo--relic bridges or radial halo protrusions, as shown below.
As discussed in \S\ref{sec:ModelProblems}, such radio emission bridging a halo and a shock is unnatural in previous models, which attribute halos and relic to different {\CRE} populations, as fine-tuning is required to explain the smooth brightness and spectrum profiles.

\subsubsection{Emission between a halo and a shock: radio bridges}

The properties of radio emission in the region between a halo and a shock depend on the magnetic field profile, the distance of the relic, and the value of the parameter $\myr$.

If the magnetic field decays rapidly behind the shock, on a timescale shorter than the shock crossing time, then radio emission will be appreciable only immediately behind the shock, and the emission will be classified as a halo and a relic, well separated from each other.

However, it is more natural to assume that the elevated magnetic field behind the shock remains high over a dynamical time, longer than the shock crossing time.
In such a case, the halo and the shock would be connected by a strongly magnetised filament.
It is plausible that the magnetic field would decline radially along the filament due to the radial decline in ambient density, however this is uncertain because merger shocks tend to strengthen with increasing radius \citep[\eg][]{VazzaEtAl10} and because shock magnetisation is not well understood.

In the part of the filament where the magnetic field is strong, $B\gtrsim B_{cmb}$, radio emission would be linearly proportional to the column density, as in \eq{\ref{eq:nuInuPerLambda}}.
It is likely that the surface brightness would decline radially due the column density profile, even if the magnetic field slightly increases with radius.

The radio profile near the shock front depends on $\myr$.
For $\myr>1$, emission from the shock region would be brighter than the emission farther downstream, at least when normalised by the local density, and the emission would be classified as a relic connected to the halo by a radio bridge.
For $\myr<1$, however, the shock region would be fainter than the filament leading to it, and the entire emission would be classified as a halo with a (radial) protrusion.
In either case, the shock region may not be observable if the column density is too low.

In most cases, observing a relic implies that $\myr>1$, \ie shock magnetisation is stronger than shock compression.
In our model, a filament is always present, although it may fall beneath the radio threshold.
For a given observed relic, a halo--relic bridge would be more likely to be observable if $\myr$ is close to unity.

\subsubsection{Shocks in halos}

Some halos or at least parts of halos are clearly not in a steady state.
The most striking examples are the weak shocks observed in some cases at the edges of halos.
This includes confirmed shocks in 1E 0657--56 \citep[the bullet cluster;][]{MarkevitchEtAl02}, A520 \citep{MarkevitchEtAl05} and A754 \citep{KrivonosEtAl03}, and suspected shocks in A665 \citep{MarkevitchVikhlinin01}, A2219 \citep{MillionAllen09}, and Coma \citep{BrownRudnick10}.

The coincidence of these shocks with the edges of their respective halos provides strong evidence that the shocks play an important role in magnetising the ICM, raising the magnetic field at least to the level of $B_{cmb}$.
Incidentally, as mentioned in \S\ref{sec:HaloRelicConnection}, these shock--halo coincidences are inconsistent with primary {\CRE} models: strong turbulence is not expected in the shock region \citep{GovoniEtAl04}, and even if it was, turbulent acceleration would not be effective immediately behind the shock.

In cases where shocks are embedded in or found at the edge of a halo, time-dependent modeling of the {\CRE} population is essential.
Our analysis predicts a downstream synchrotron signature analogous to the relic signal described above.
Namely, enhance brightness (with respect to the $I_\nu\propto \lambda_n$ profile) with a flat spectrum at the edge of the shock, with gradual decline in brightness back to its steady-state $\sim\Dltcool$ inward of the shock, accompanied by spectral steepening.
These predictions are tested quantitatively in select clusters with suspected shocks, in \S\ref{sec:ModelTest_StudiedClusters} below.
Such modeling would provide a definitive test of our model once the shocks are confirmed.

\subsubsection{Steep spectrum halos: recent magnetic growth}

The spectral steepening induced directly by shock magnetisation is limited to a $\sim 2t_{cool}$ wide region behind the shock.
A single shock can make the integrated spectrum of a halo steep, provided that the shock front is wide and the shock has passed in the recent $\sim t_{cool}$ near the central, brightest part of the halo.

However, steepening is not limited to shocks, or to their direct impact on the plasma.
The results of \S\ref{sec:SynchrotronVariability} indicate that the spectrum of the steep GHs may be induced by any recent ($\sim t_{cool}$), modest ($r_B\sim 2$) magnetic growth, regardless of the physical mechanism driving it.
Note that spectral steepening immediately downstream of a shock is somewhat impeded by the shock-induced gas compression, introducing a factor $\myrg^{-1}$ in $\myr$ (see \eq{\ref{eq:myr_Def_Shock}}).
Magnetisation without gas compression is thus more efficient in producing a steep spectrum.

In particular, the $\Delta\alpha\simeq -0.5$ steepening sufficient to explain all steep GHs (defined as having an average spectrum $\alpha<-1.5$, see \S\ref{sec:SteepHalos}) requires $\myr\simeq5.2$.
If gas compression can be neglected, we may use
\begin{equation}
\myr \simeq \frac{1+b_d^2}{1+b_u^2} \fin
\end{equation}
The spectrum of steep GHs can thus be reproduced if the magnetic field is amplified from $b_1\ll 1$ to $b_2\simeq 2$.
Alternatively, if the halo was already magnetised to $b_1>1$ levels, $\myr=5.2$ requires magnetic amplification by a factor of $\myrB=b_2/b_1\simeq 2.3$ within $\sim t_{cool}$.

In \S\ref{sec:ModelTest_RelicGHConnection} we use the coincidence between steep halos and relics to show that steep GHs are a transient phenomenon, persisting only $\sim 0.3$--$1\Gyr$ after a merger event.
This adolescence should occur in $\sim 15\%$ of all GHs, in agreement with present observations.
Producing such a long-lived, $>t_{cool}$ phenomenon requires more than the impulsive magnetisation induced by a single shock.
We conclude that steep halos are produced by a combination of shocks and a rising level of turbulence, maintaining the steep average spectrum over a $\sim$Gyr timescale.
This results in a high, $B\gg B_{cmb}$ central magnetic field, as found in several clusters by {\KL} and in \S\ref{sec:HaloAndRelicEta}.


\subsection{Illustrating the model in well-studied clusters}
\label{sec:ModelTest_StudiedClusters}

We may test the time-dependent, secondary {\CRE} model of \S\ref{sec:TimeDependentTheory} in individual, well studied clusters, that harbour a relic or a shock.
We use the detailed radio spectral maps published for A665, A2163, and A2744, which harbour the halos analysed in \S\ref{sec:EtaInWellStudiedGHs}.
In addition, we study the spectral map of A2219 \citep{OrruEtAl07}, which was not analysed in \S\ref{sec:EtaInWellStudiedGHs} because the central $\sim 300\kpc$ part of the halo is contaminated by a blend of radio galaxies.

Each of these clusters harbours, in addition to the central GH, either a peripheral relic (in A2163 and A2744) or a suspected shock embedded within the halo (in A665 and A2219).
For each such source, we extract the profiles of radio brightness $I_\nu(r)$ and spectral index $\alpha(r)$ along a line connecting it with the centre of the cluster (defined as the X-ray peak).
In order to estimate the gas column density, we also extract the coincident profile of the X-ray brightness, $F_X$, using ROSAT or \emph{Chandra} data, and in addition use an ASCA-based $\beta$-model of the cluster.
We label the cluster's centre as $r=0$, and orient the axis such that the relic or shock is as $r>0$.

The resulting profiles are analysed in the same method applied to the halos in \S\ref{sec:EtaInWellStudiedGHs}.
The results are displayed in \Figs~\ref{fig:ProfilesA665B} and \ref{fig:ProfilesA2163C}--\ref{fig:ProfilesA2744C}, with the same notations used in \Figs~\ref{fig:ProfilesA2163}--\ref{fig:ProfilesA2744HaloNW}.
In short, each figure shows the $\alpha(r)$ profile in the upper panel, and various estimates of the radio--X-ray ratio $\eta=\nu I_\nu/F_X$ in the bottom panel, based on ROSAT (solid blue curves), \emph{Chandra} (short-dashed black) and the $\beta$-model (long-dashed red).
Dotted curves show our $\eta$ model for the steady-state secondary {\CRE} model with homogeneous {\CRIs}, based on column densities estimated from ROSAT (dotted blue) or from the $\beta$-model (long-dotted red).
The central value $\eta_0$ of the radio--X-ray ratio is also shown (dot dashed cyan), as extrapolated at any distance $r$ using the local radio and X-ray brightness.

For each cluster, we plot (as dashed green curves) the spectrum and brightness profiles computed from the time-dependent {\CRE} model of \S\ref{sec:SynchrotronImprintOfShock}, assuming that each relic reflect an underlying, outgoing, weak shock.
In principle, the model for downstream emission has only two independent parameters, which can be chosen as the Mach number $\mach$ of the shock, and the (normalised to $B_{cmb}(z)$) downstream magnetic field $b_d$.
However, the radio resolution is poor, with the beam width being larger than the characteristic scale $\Dltcool$ for downstream {\CRE} evolution.
Consequently, upstream emission affects the downstream signal, and much be modeled as well.
Approximating the upstream magnetic field as constant, this can be done by introducing one addition degree of freedom, parameterised here as the emissivity ratio across the shock, $r_j$.

\begin{figure}
\centerline{\epsfxsize=10cm \epsfbox{\myfarfig{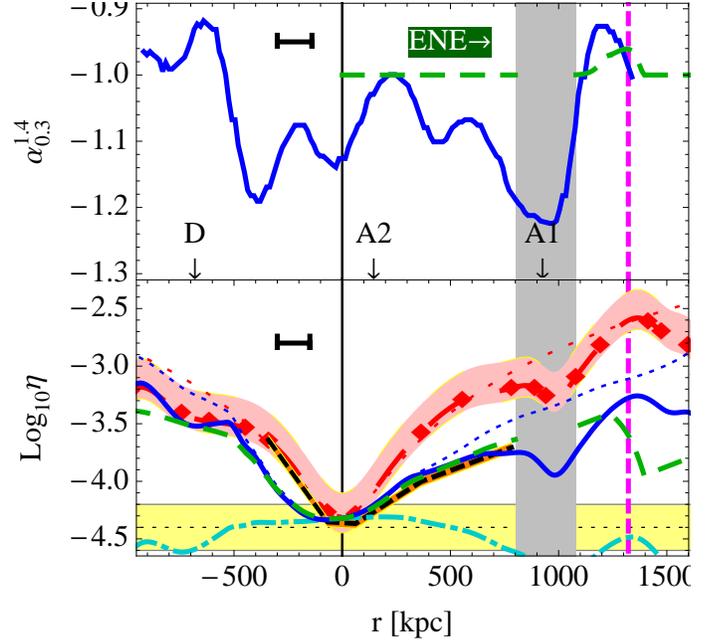}}}
\caption{
Profiles of radio brightness and spectrum in A2163, along the line connecting the X-ray peak with the brightness peak of the ENE relic (at $r\simeq 1300\kpc$; vertical dashed line).
This is a zoomed out version of the same line examined in \Fig~\ref{fig:ProfilesA2163}, and uses the same notations.
The model (dashed green) is discontinuous near the clump A2 (gray shaded vertical band).
\label{fig:ProfilesA2163C}}
\end{figure}

\begin{figure}
\centerline{\epsfxsize=10cm \epsfbox{\myfarfig{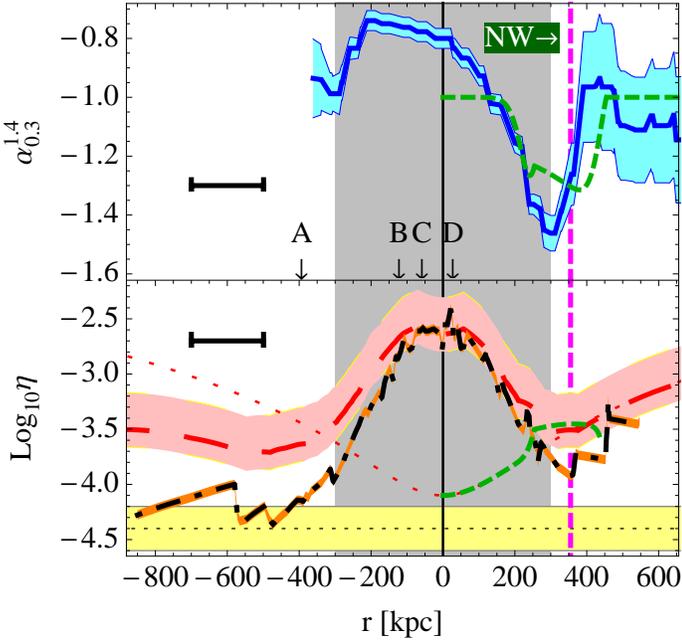}}}
\caption{
Profiles in A2219 along $\mathcal{L}=\{16\mbox{h}40'38'',46^\circ 40'1''\}$--$\{16\mbox{h}40'9'',46^\circ 45'2''\}$, which crosses the $1.4\GHz$ brightness peak ($r=0$) and is oriented in the Southeast ($r<0$)-Northwest direction, along the halo elongation.
Symbols and notations are defined in Figures \ref{fig:ProfilesA2163}, \ref{fig:ProfilesA665}, and \ref{fig:ProfilesA2744HaloNE}.
A filament of clumps (arrows labeled A--D) lies along $\mathcal{L}$ in an oblique orientation with respect to the line of sight \citep{BoschinEtAl04}.
The radio data, from \citet{OrruEtAl07}, is dominated in the central $300\kpc$ (gray shaded vertical band) by a blend of radio galaxies.
A possible shock reported by \citet{MillionAllen09} lies at $r\simeq 350\kpc$ (dashed vertical magenta line).
\label{fig:ProfilesA2219}}
\end{figure}

\begin{figure}
\centerline{\epsfxsize=10cm \epsfbox{\myfarfig{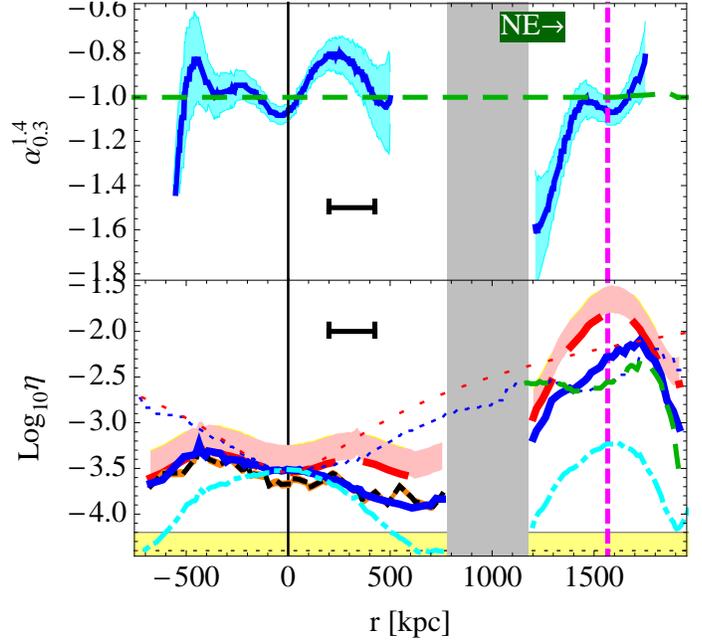}}}
\caption{
Profiles in A2744 along the line connecting the centre of the cluster with the $r\simeq 1600\kpc$, NE relic.
This is a zoomed out version of the line examined in \Fig~\ref{fig:ProfilesA2744HaloNE}, and uses the same notations.
The radio signal, from \citet{OrruEtAl07}, is not detected at $800\lesssim r\lesssim 1200$ (gray-shaded vertical band).
\label{fig:ProfilesA2744C}}
\end{figure}

The three model parameters are summarised, for each of the four clusters, in Table \ref{tab:ModelParameters}.
In addition, the table lists the corresponding values of $\myr$ and $\Dltcool$.
Note that due to the wide beam, the maximal spectral steepening, for example, depends not only on $\myr$, but also on $\Dltcool$ and $r_j$.

Also shown is the energy density fraction $\epsilon_B$ of the downstream magnetic field, with respect to the thermal plasma computed from the isothermal $\beta$-model (see \eq{\ref{eq:epsilon_B}}).
The enhanced pressure immediately downstream of the shock can exceed the local estimate of the $\beta$-model, rendering our $\epsilon_B$ values higher than in reality.

For simplicity, the models assume flat {\CRE} injection ($q=0$), and approximate the beam profile as a top-hat.
Analytic approximations for the brightness and spectrum measured with a finite beam and a finite frequency range, according to the time-dependent, secondary {\CRE} model, are provided in \S\ref{sec:FiniteSpectrumAndBeam}.

\begin{table*}
{
\begin{tabular}{|cc|cc|ccc|ccc|}
\hline
Cluster & $B_0$ & Type & $r$ & $\mach$ & $b_d$ & $r_j$ & $\myr$ & $\Dltcool$ & $\epsilon_B$ \\
\hline
A665 & $30$ & shock & $250$ & $1.2$ & $2$ & $10$ & $3.9$ & $36$ & $4\%$ \\
A2163 & $30$ & relic & $1300$ & $3$ & $1$ & $3$ & $0.67$ & $49$ & $11\%$ \\
A2219 & --- & shock & $350$ & $1.1$ & $3$ & $100$ & $8.7$ & $30$ & $13\%$ ($8\%$) \\
A2744 & --- & relic & $1800$ & $2$ & $1$ & $10$ & $0.9$ & $35$ & $20\%$ \\
\hline
\end{tabular}
\caption{\label{tab:ModelParameters} Parameters of the models used in \Figs~\ref{fig:ProfilesA2163}--\ref{fig:ProfilesA665B} and \ref{fig:ProfilesA2163C}--\ref{fig:ProfilesA2744C}.}
\begin{flushleft}
{\bf Columns}:
(1) cluster name;
(2) central magnetic field $B_0$ (in $\mu$G);
(3) discontinuity type (relic or suspected shock);
(4) discontinuity distance $r$ from the centre of the cluster (in kpc);
(5) shock Mach number $\mach$;
(6) downstream magnetic field $b_d$, in units of $B_{cmb}(z)$;
(7) fractional emissivity jump $r_j$ at the shock;
(8) asymptotic {\CRE} ratio $\myr$ (see \eq{\ref{eq:myr_Def_Shock}});
(9) cooling length downstream $\Dltcool$ (in kpc);
(10) downstream magnetic energy fraction $\epsilon_B$ with respect to the thermal energy computed from the isothermal $\beta$-model (and using the measured downstream temperature).
\end{flushleft}
}
\vspace{0.5cm}
\end{table*}

In addition to the above model parameters, one must specify the spatial distributions of {\CRIs} and gas in order to compute the synchrotron signal.
This is done using the $I_\nu\propto N_p \lambda_n\propto F_X^{1/3}$ relation derived in \S\ref{sec:HaloAndRelicEta}.
We use the $F_X$ profile extracted from ROSAT data, and normalise the $\eta=\nu I_\nu/F_X\propto F_X^{-2/3}$ relation to its measured central value, $\eta_0$.
The centre of A2219 is contaminated by radio galaxies, so here we use the GH average $\eta_0\simeq 10^{-4.1}$ found in \eq{\ref{eq:eta0_new_GHs}}, and derive the $F_X(r)$ profile from the $\beta$-model.

Remarks regarding individual clusters:
\begin{enumerate}
\vspace{-2mm}
\item
In A665, the radio emission along a line connecting the centre of the cluster with a possible shock identified $\sim 200\kpc$ Southeast of the cluster's centre \citep{MarkevitchVikhlinin01, GovoniEtAl04} is modeled in \Fig~\ref{fig:ProfilesA665B}.
The model incorporates the $B^2\propto n$ magnetic radial decline along the Southwest ray (denoted as $r<0$) but not at $r>0$, assuming strong, recent shock magnetisation.
For the model parameters used, magnetisation is strong compared to the gas compression so $\myr>1$, resulting in an enhanced emission and steepening in the near downstream.
\item
In A2163, a model for the line connecting the halo and the relic is shown in \Fig~\ref{fig:ProfilesA2163C}.
The emission is disrupted in the range $800<r/\mbox{kpc}<1100$ by substructure \citep[clump A1 in][]{MaurogordatoEtAl08}, so the model we show is discontinuous; the $B^2\propto n$ scaling is included for $r<800\kpc$.
For the parameters used, magnetisation is weak compared to the gas compression so $\myr<1$, resulting in flattening, rather than steepening behind the shock.
Local X-ray enhancement near the position of the relic is evident in the ROSAT maps \citep[][figure 1]{ElbazEtAl95}, consistent with a shock interpretation.
\item
A2219 appears to be in an advanced stage of merger, involving multiple clumps in a Southeast-Northwest filament oblique to the line of sight \citep{BoschinEtAl04}.
The line connecting the centre of the cluster with the possible shock identified $\sim 350\kpc$ Northwest of the cluster's centre \citep{MillionAllen09} is modeled in \Fig~\ref{fig:ProfilesA2219}.
A blend of radio galaxies in the centre of the cluster, as well as optically identified substructure \citep{BoschinEtAl04}, preclude a reliable model of the central, $|r|<300\kpc$ region.
\item
In A2744, a model for emission along the line $\mathcal{L}$ connecting the halo and the relic is shown in \Fig~\ref{fig:ProfilesA2744C}.
Local X-ray enhancement near the position of the relic is evident in the ROSAT maps, consistent with a shock.
Radio emission along $\mathcal{L}$ drops below the detection threshold at $800\lesssim r/\mbox{kpc} \lesssim 1200$ \citep{OrruEtAl07}, presumably due to a weak magnetic field.
Stronger emission between the relic and the halo is detected almost continuously along a different, curved path, arching to the South of $\mathcal{L}$ (not discussed here).

\end{enumerate}

Figures \ref{fig:ProfilesA665B} and \ref{fig:ProfilesA2163C}--\ref{fig:ProfilesA2744C} show that the secondary {\CRE} model with a homogeneous {\CRI} distribution is successful in qualitatively reproducing the radio brightness profile throughout a cluster, not only within GHs (as shown in \S\ref{sec:HaloRisingEta}) and among different halos and relic (\S\ref{sec:eta_in_relics}), but also throughout individual clusters, from their centres out to the peripheral, $r>1\Mpc$ relics.
Accounting for the time-dependence of the {\CRE} distribution improves, in some cases (when $\myr$ is far from unity), the agreement between model and observation, as seen for example in \Fig~\ref{fig:ProfilesA665B}.

The figures show that the observed spectral profiles can be qualitatively explained in the framework of the secondary {\CRE} model, with a plausible choice of parameters.
Here it is essential to take into account the time-dependent evolution of {\CREs} in order to reproduce the spectral features observed.
Note that, depending on the value of $\myr$, the spectrum behind the shock either steepens (in A665 and A2219) or flattens (possibly in A2163 and A2744).

Overall, the results support the viability of the model as an explanation for the diffuse emission from clusters, including halos, relics, and the emission between them.
There is therefore no need to invoke the ad-hoc assumptions of primary particle acceleration in weak shocks or in turbulence, processes which are neither well understood nor observationally constrained.
We obtain a reasonable fit to the data assuming flat injection, $q=0$, suggesting that the {\CRP} spectrum is flat.
The model also gauges the magnitude of the magnetic field, which is found to be $b_d\gtrsim 1$ inward of relics and shocks.

Quantitatively, however, the modeled curves are crude and the model parameters are poorly constrained.
First, the models are essentially one-dimensional and do not account for substructure, shock-induced overdensities, projection effects involving shock obliqueness, etc.
In particular, the $N_p\lambda_n$ profiles are based on the measured $F_X$ profiles, which are more sensitive to substructure.
Second, the radio resolution in all four clusters is insufficient, allowing only a marginal test of the model because the beams are larger than $\Dltcool$.
Third, the spectral features associated with the two relics are small, while stronger features are related to the substructure not included in the model.
And fourth, we have made the simplifying assumptions of a top-hat beam and a flat, $q=0$ {\CRE} injection, neglecting for example the spectral variations associated with the cross section for secondary production {\KL} and changes in the underlying {\CRP} spectrum.

In addition, the radio profile depends in a non-trivial way on the parameters $\mach$, $b_d$ and $r_j$, allowing for more than one type of solution.
For example, the values we adopt for the Mach numbers of the shocks embedded within the halos are somewhat lower than the values estimated from the temperature jumps observed, \ie $\mach\gtrsim 1.8$ in A665 \citep{MarkevitchVikhlinin01} and $\mach=1.3$--$1.9$ in A2219 \citep{MillionAllen09}.
Reasonable, but somewhat worse fits can also be obtained for these higher Mach numbers, but this would require a stronger downstream magnetic field.

Note that the limitations listed above are far more severe in studies that attempt to model the integrated properties of a cluster such as the average spectra of the radio sources or their total power.
Considering the highly nonuniform nature of the emission, analysing the surface brightness maps is essential for obtaining a reliable model.
Tightly constrained parameters can potentially be derived using a projected, three-dimensional model of the gas and magnetic distributions, in particular if high resolution radio maps can be obtained with beams narrower than $\Dltcool$.
If $\mach$ could be determined from X-rays and the beams were narrow, the downstream emission would depend on the single parameter: $\myr$, or equivalently (if $b_u\ll 1$) $b_d$.

Finally, notice that the $\beta$-model provides a useful fit to the data, as it is accurate to better than a factor of $\sim 3$ even at $r\sim 2\Mpc$ distances.


\subsection{Relic and halo compilation}
\label{sec:ModelTest_Compilation}

In \S\ref{sec:eta_in_relics} we showed that a universal $\eta(n/n_0)$ distribution joins radio relics and halos, as illustrated in \Fig~\ref{fig:SourcesEtaN}.
This suggests that relics can be modeled as emission from secondary {\CREs} which are produced by the same, homogeneous {\CRI} distribution responsible for halos.

However, we found that relics are brighter than expected in a steady-state {\CRE} model with homogeneous {\CRIs}.
Namely, $\eta$ increases too rapidly with radius (\ie with decreasing $n/n_0$), naively corresponding to a {\CRI} density which increases radially instead of being constant.
Thus, we could not find a good power-law fit $\eta\propto (n/n_0)^{\gamma}$ for both halos and relics, and we obtained a smaller (more negative) $\gamma$ than the $\gamma=-1$ value corresponding to homogeneous {\CRIs}, in particular in the range between halos and relics.
This can be seen qualitatively by noting the similar radio brightness characteristic of halos and relics, as shown in \eq{\ref{eq:eta_constant_Inu}}, and quantitatively in the different formal fits derived among relics, and in particular between relics and halos, in \eqs{\ref{eq:eta_relics_only}}--(\ref{eq:eta_relic_relic}).

Now we consider the time-dependence of the {\CRE} distribution, induced by changes in the magnetic field and in the {\CRI} distribution.
This is essential because, while flat spectrum halos are thought to be approximately in a steady state, in the sense that the {\CRE} injection rate $Q$ and the magnetic field amplitude $B$ vary on a timescale longer than $t_{cool}$, this is not the case for relics.

In \S\ref{sec:TimeDependentTheory} we saw that if the magnetic field rises in the fluid frame, synchrotron emission is significantly enhanced, and can exceeds its final steady-state value during a timescale $\sim t_{cool}$ required for the {\CREs} to adjust to their new, $B$-dependent steady state.
Therefore, the brightness of a relic exceeds its steady-state expectation if it is associated with a weak shock or a magnetisation wave, in direct proportion to the parameter $\myr$ defined in \eq{\ref{eq:myr_general}}.

In order to test the time-dependent model and gauge its parameters, we reanalyse the data of the halo and relic sample presented in \S\ref{sec:DataPreparation}--\S\ref{sec:SourceSelection} in the framework of the time-dependent theory.
As in \S\ref{sec:eta_in_relics}, relics near a first rank galaxy, peripheral circular relics, and the uncertain relics in A754 and A2034, are excluded from the analysis.

First, we simultaneously fit the data of the halos and relics as a power law in $(n/n_0)$, assuming that the relics all have the same enhancement factor $\myr$. This leads to the fit
\begin{equation}
\frac{\eta}{\myr} = 10^{-3.8\pm0.1} \left( \frac{n}{n_0} \right)^{-0.7\pm0.3} \coma
\end{equation}
where for halos $\myr=1$ (by assumption), and for relics $\myr=10^{1.0\pm0.3}$.
The enhancement factor of relics corresponds, if gas compression is negligible, to a downstream field $b_d\simeq 3.2$.
This is an unrealistically strong field, as in distant relics the corresponding magnetic energy density $u_B$ would exceed the local thermal energy density $u_{th}$.

The unrealistically high value obtained for a universal $\myr$ in relics indicates that the different environments of the relics must be taken into account.
Hence, we again fit the halo and relic data as a power law in $(n/n_0)$, but this time we allow different values of $\myr$ in each relic, determined such that the magnetic field constitutes a fixed fraction $\epsilon_B$ of $u_{th}$, where $u_{th}$ is determined from the $\beta$-model of the cluster.
This leads to the best fit
\begin{equation} \label{eq:fit_relics_epsB}
\frac{\eta}{\myr} = 10^{-3.8\pm0.1} \left( \frac{n}{n_0} \right)^{-1.3\pm0.2} \coma
\end{equation}
where for halos $\myr=1$, and for relics
\begin{equation}
\myr=1+\epsilon_B \frac{u_{th}}{u_{cmb}} \coma
\end{equation}
where
\begin{equation} \label{eq:FitSourcesEpsilonB}
\epsilon_B=10^{-0.8\pm0.4}\simeq 20\% \coma
\end{equation}
within a factor of $\sim 2.5$.
This fit is illustrated in \Fig~\ref{fig:SourcesEtaNModel}, where we plot the $\eta/\myr$ value of each source against its $(n/n_0)$ estimate.

\begin{figure}
\centerline{\epsfxsize=9cm \epsfbox{\myfarfig{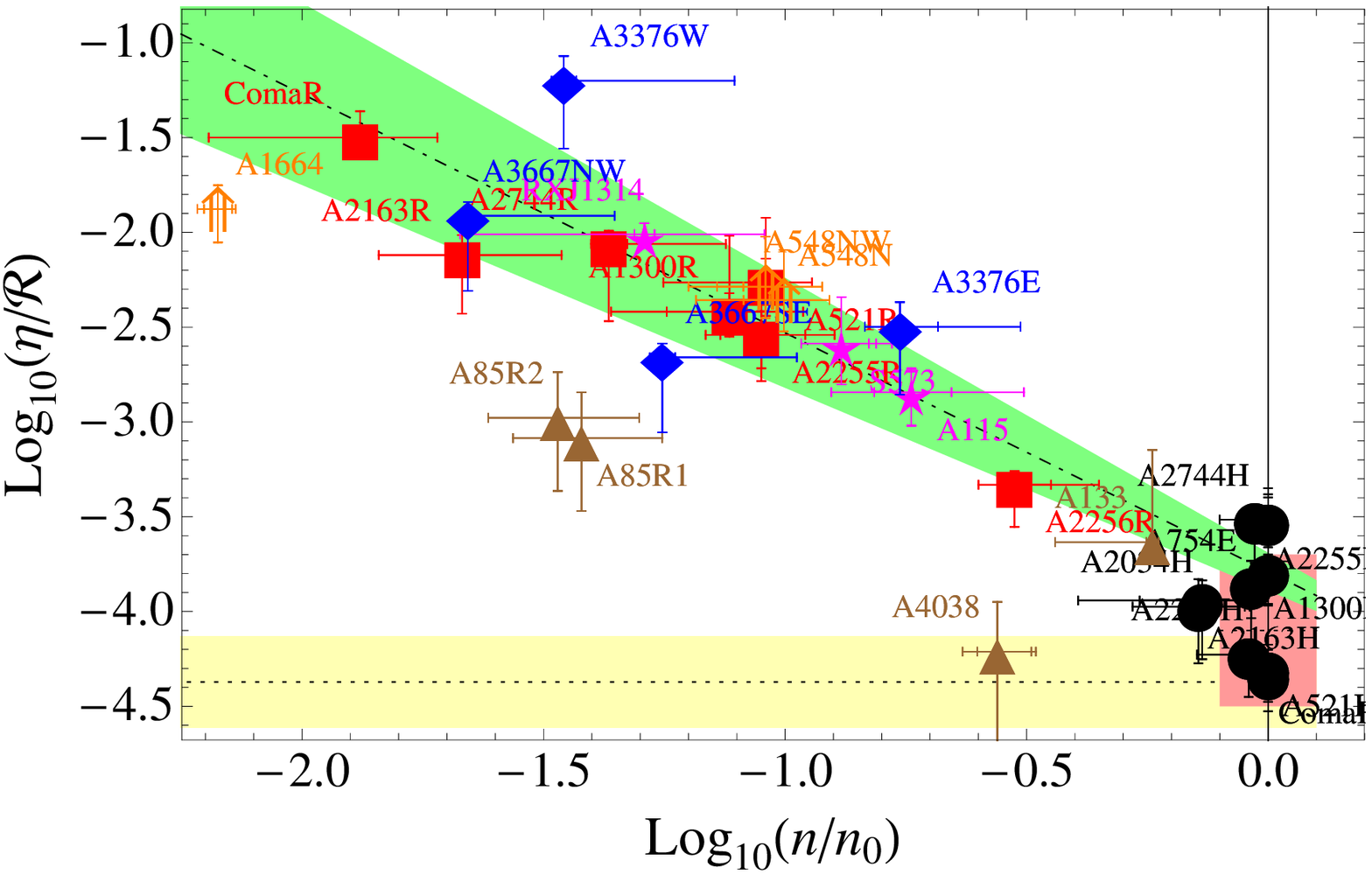}}}
\caption{
The distribution of relics and halos in the $(\eta/\myr)$--$(n/n_0)$ phase space, taking into account the $\myr$ enhancement of relic emission due to magnetic field amplification.
Symbols and notations are defined in \Figs~ \ref{fig:SourcesInuIx} and \ref{fig:SourcesEtaN}.
Here we consider only the magnetisation and neglect gas compression, appropriate for a magnetisation-wave relic model.
The fit of \eqs{\ref{eq:fit_relics_epsB}}--(\ref{eq:FitSourcesEpsilonB}) is shown (dot-dashed line with green shaded band), based on uncontaminated halos and classical relics (see text), with the best fit value $\epsilon_B=20\%$.
Note that we exclude relics near the first rank galaxy (brown triangles) from the fit.
\label{fig:SourcesEtaNModel}
\vspace{2mm}}
\end{figure}

Note that although the $\gamma\simeq -1.4$ obtained here is similar to that found among relics in the steady-state model in \S\ref{sec:eta_in_relics}, here the fit agrees with the halos, too.
The fact that $\gamma<-1$ does not contradict the homogeneous {\CRI} assumption, considering the statistics and the simplifying assumptions of the model.
For example, we have neglected the projection effects of the relics, assuming that they lie in the plane of the cluster's centre.
In practice, the data points should be shifted to somewhat lower values of $(n/n_0)$ (\ie to the left in \Fig~\ref{fig:SourcesEtaNModel}), resulting in a higher $\gamma$.

The best-fit value $\epsilon_B\simeq 20\%$ is larger than typically expected in weak shocks, but the statistical uncertainty is large and we made strong simplifying assumptions.
Note that in the fit procedure we have not constrained the value of $\epsilon_B$ to be smaller than unity, so obtaining $\epsilon_B<1$ is reassuring.

More importantly, we have not included the effect that the relic shock has on the plasma.
If the relic is found downstream of a shock with Mach number $\mach$, we should make the following substitutions:
\begin{align} \label{eq:ShockSubstitutions}
& \frac{n}{n_0} \to \frac{n^*}{n_0}=\frac{n}{n_0}\myrg \, ; \nonumber \\
& u_{th} \to u_{th} \myrp \nonumber \, ; \\
& \myr \to \frac{1+b_d^2}{\myrg} \, ; \\
& F_X \to \myrg^{2} F_X \nonumber \, ; \\
& \eta \to \eta^* \equiv\myrg^{-2}\eta \nonumber \coma
\end{align}
where $\myrp=(5\mach^2-1)/4$ and $\myrg=4\mach^2/(3+\mach^2)$ are the fractional increases in plasma pressure and density, respectively, across the shock.

Incorporating these modification in the model produces a fit similar to that in \eq{\ref{eq:fit_relics_epsB}}, but with a lower, $\mach$-dependent value of $\epsilon_B$.
For example, assuming that all relics lie downstream of a Mach $\mach=2$ shock, we obtain
\begin{equation} \label{eq:fit_relics_epsBM2}
\frac{\eta^*}{\myr}(\mach=2) = 10^{-3.7\pm0.1} \left( \frac{n^*}{n_0} \right)^{-1.3\pm0.2} \coma
\end{equation}
where for halos $\myr=1$, and for relics
\begin{equation}
\myr=\myrg^{-1}\left(1+\epsilon_B \frac{u_{th}}{u_{cmb}} \right) \coma
\end{equation}
with a downstream magnetic fraction
\begin{equation} \label{eq:FitSourcesEpsilonBM2}
\epsilon_B(\mach=2) = 10^{-1.4\pm0.4} \simeq 4\% \coma
\end{equation}
within a factor of $\sim 2.5$.
The corresponding source distribution and best fit are shown in \Fig~\ref{fig:SourcesEtaNModelM2}.

\begin{figure}
\centerline{\epsfxsize=9cm \epsfbox{\myfarfig{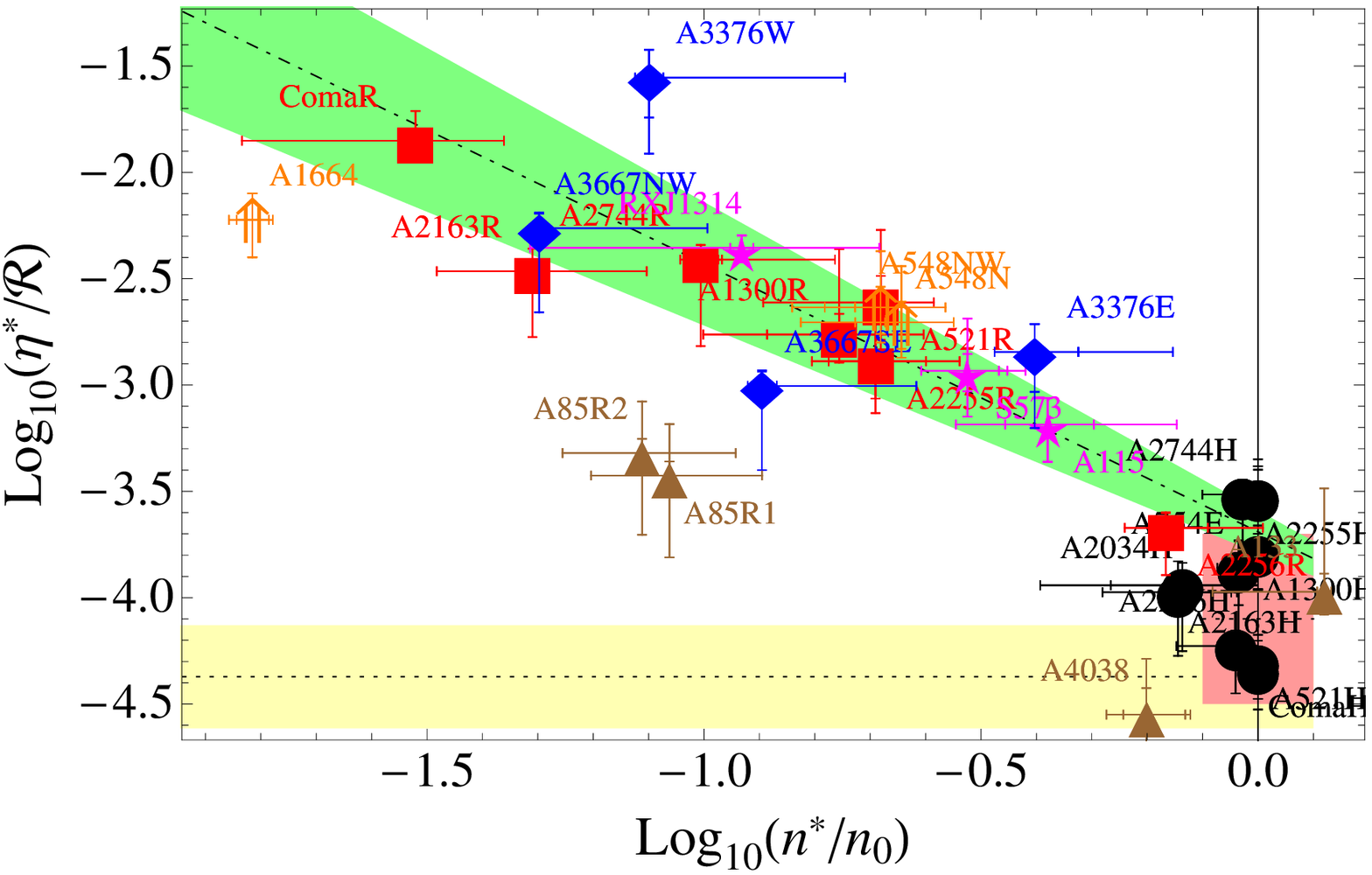}}}
\caption{
The distribution of relics and halos in the $(\eta^*/\myr)$--$(n^*/n_0)$ phase space, taking into account both the $\myr$ radio brightening and the $\myrg$ gas compression, assuming that relics are associated with an $\mach=2$ shock.
Symbols and notations are defined in \Figs~ \ref{fig:SourcesInuIx} and \ref{fig:SourcesEtaN}.
The fit of \eqs{\ref{eq:fit_relics_epsBM2}}--(\ref{eq:FitSourcesEpsilonBM2}) is shown (dot-dashed line with green shaded band), based on uncontaminated halos and classical relics (see text), with the best fit value $\epsilon_B=4\%$.
Note that we exclude relics near the first rank galaxy (brown triangles) from the fit.
\label{fig:SourcesEtaNModelM2}
\vspace{2mm}}
\end{figure}

In summary, relics and halos can be explained as emission from secondary {\CREs} produced by a homogeneous population of {\CRIs}, if the temporal evolution of the {\CRE} population is taken to account.
A Good agreement with observations is found using the $\beta$-model if relics are $\mach\sim 2$ shocks with a typical $\epsilon_B\sim 4\%$ magnetic energy fraction downstream, or strong magnetisation waves with $\epsilon_B\sim 20\%$.


\subsection{Magnetic archeology using the radio spectrum}
\label{sec:ModelTest_SpectralCurvature}

In \S\ref{sec:magnetic_variations_modify_CRE} we showed that the spectrum of the {\CRE} distribution reflect the temporal evolution of the ratio between the {\CRE} injection and the cooling parameter, $\myQpsi\propto Q/(1+b^2)$, with a $\sim t_{cool}$ time delay.
Above we considered the radio steepening or flattening behind shocks, resulting from gas compression and magnetic field amplification.
But even away from shocks, in regions where changes in the density are slow (span timescales $\gg t_{cool}$), the spectrum directly traces the recent evolution of the magnetic field.

In particular, the {\CRE} spectral index $\phi(t,E)$ decreases (increases) with respect to its steady state value $\phi_s=\overline{q} - 1$ if the magnetic field was growing (decaying) at the retarded time $t-\tau_{cool}(E)$; see \eqs{\ref{eq:tau_cool_def}} and (\ref{eq:CRE_phi}).
Similarly, for flat injection, the spectral curvature $\widetilde{\phi}(t,E)$ decreases (increases) with respect to its steady state $\phi_s-\phi$ for magnetic growth (decay) at $t-\tau_{cool}(E)$, corresponding to $\mykappa>0$ ($\mykappa<0$); see \eqs{\ref{eq:spectral_curvature1}}--(\ref{eq:spectral_curvature_kappa}).

The evolution of the magnetic field is similarly imprinted upon the radio spectrum, although it is somewhat smeared out by the convolution with $F_{syn}$, projection along the line of sight, integration over the beam width, and the finite spectral range needed to measure the spectrum.
The spectrum resulting from a rapid change in $B$ is shown in \eq{\ref{eq:alpha_nu_q}}; the spectrum for an arbitrary $B(t)$ evolution is given for flat injection in the $\delta$-function approximation of $F_{syn}$ in \eq{\ref{eq:flat_alpha_t_Qconst}}.

A brightness map at a single frequency also gauges the magnetic evolution, but extracting information about the magnetic field requires a model for the gas distribution.
The homogeneous {\CRI} distribution simplifies such an analysis, as no {\CRI} model is required.

Recall that in a steady-state, the radio spectrum gauges the amplitude of the magnetic field because the {\CRE} spectrum steepens with increasing energy $E_e\propto (\nu/B)^{1/2}$ due to the energy-dependent cross section for secondary production.
This leads to a $\Delta \alpha\simeq 0.1$--$0.4$ steepening, depending on the {\CRP} spectrum, when the magnetic field drops below $B\simeq 10(\nu/700\MHz)\muG$ (see {\KL}).
The above results indicate that similar or larger spectral deviations occur when the magnetic field is not in a steady state, if it evolves on $\sim t_{cool}$ timescales in regions where $B\gtrsim B_{cmb}$.

As an illustration, consider the simple case of flat injection and a magnetic evolution satisfying $\mykappa=\const$, in the $\delta$-function approximation of $F_{syn}$.
Here
\begin{equation} \label{eq:alpha_kappa}
\alpha(\nu) = -\frac{1}{2} \left[ 1+ \frac{1+\mykappa}{ 1 -(\nu/\nu_0)^{\frac{1+\mykappa}{2}} } \right]
\end{equation}
and
\begin{equation} \label{eq:alpha_dot_kappa}
\widetilde{\alpha} \equiv \frac{d\ln\alpha}{d\ln\nu} = -\frac{\left(\alpha+\frac{1}{2}\right)\left(\alpha+1+\frac{\mykappa}{2}\right)}{\alpha}
\coma
\end{equation}
where $\nu_0$ is a constant.
Note that $d\widetilde{\alpha}/d\alpha<0$ as long as $\alpha^2>(2+\mykappa)/4$, so stronger steepening is expected when the spectrum is steeper.

Such a behaviour was indeed reported in relics by \citet{vanWeerenEtAl09}.
We reproduce their data in \Fig~\ref{fig:SpectralCurvature}, showing several relics in the phase space of spectral curvature (defined here as $\alpha_{74}^{610}-\alpha_{610}^{1400}$) and spectral index ($\alpha_{74}^{1400}$).
For comparison, we plot some $\mykappa=\constant$ curves according to \eqs{\ref{eq:alpha_kappa}}--(\ref{eq:alpha_dot_kappa}), with different values of $\mykappa$.

\begin{figure}
\centerline{\epsfxsize=8cm \epsfbox{\myfarfig{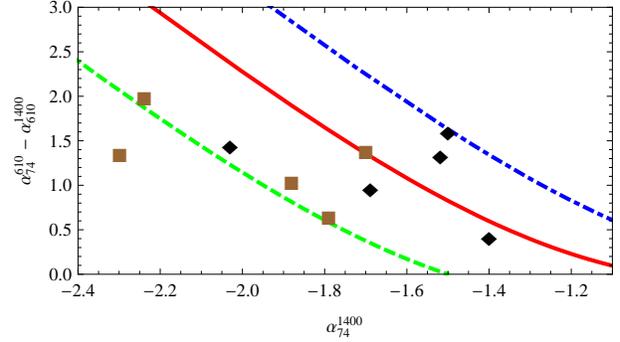}}}
\caption{
Spectral curvature vs. index for radio relics from \citet{vanWeerenEtAl09}, showing relics suspected as merger shocks (diamonds) and as radio phoenixes (squares). The curvature-index relation for flat injection is demonstrated according to \EqsO~(\ref{eq:alpha_kappa})-(\ref{eq:alpha_dot_kappa}) for a constant magnetic field ($\mykappa=0$; solid curve), magnetic growth with $\mykappa=+1$ (dashed), and magnetic decay with $\mykappa=-1$ (dot-dashed).
\label{fig:SpectralCurvature}}
\end{figure}

As the figure shows, most relics suggest recent magnetic growth ($\mykappa>0$).
However, in order to accurately constrain the magnetic evolution, one must take into account the $F_{syn}$ smearing and the integration over the beam, neglected above, and consider a more realistic magnetic evolution.

We have seen evidence for substantial spectral steepening induced by magnetic growth behind relics. Could recent magnetic growth also explain the observed steep-spectrum halos?


\subsection{Steep halos and their association with nearby relics}
\label{sec:ModelTest_RelicGHConnection}

\subsubsection{Steep GHs require extended magnetic growth}

As a steep spectrum is a transient phenomenon in our time-dependent, secondary {\CRE} model, the most plausible stage at which a globally steep halo can be produced is during the initial magnetisation of the cluster to $\gtrsim B_{cmb}$ levels, as the halo is born.
The GH--merger connection indicates that such magnetisation is triggered by a merger event, in part but probably not exclusively in the wake of merger shocks.
The magnetisation process may be driven in part by the onset of turbulence, as irregular flows amplify the preexisting magnetic field.

Note that steady state turbulence probably cannot explain the steep halos observed.
The timescale for an $\myrB\sim 2$ (less than an e-fold) amplification of the magnetic field in an eddy spanning the typical, $\sim 10\kpc$ magnetic coherence length is $\lesssim 5\Myr$, much less than the cooling time.
Naively, as recently magnetised eddies dominate the emission, the overall spectrum steepens even if turbulence is in a steady state.
However, as shown in \ref{sec:fluctuations}, in regions with a constant average magnetic field this results in only mild, $\-0.1<\Delta\alpha<0$ steepening.
{\CRE} diffusion could induce stronger steepening under such circumstances (see \S\ref{sec:SynchrotronEvolvingB}), but may be insufficient to explain the $\alpha<-1.5$ spectra observed.

Therefore, a possible scenario in which a steep GH can form involves a combination of merger shocks and an increasing energy in turbulent motions, induced by a recent merger event, causing a substantial fraction of the mass to see magnetic amplification by a factor $\gtrsim2$ within the recent $\lesssim 2 t_{cool}$.
The sound crossing time of a typical, $\sim 1\Mpc$ halo is of the order of $0.5\Myr$, longer than $t_{cool}$, so in this scenario some parts of the halo would undergo substantial magnetisation, by more than one e-fold.
This is consistent with the high, $\sim 30\muG$ magnetic fields inferred in the centres of some GHs, in \S\ref{sec:HaloAndRelicEta} and in {\KL}.

The sound crossing time is short, however, compared to the estimated age of halos, $t_{halo}\sim 5\Gyr$, corresponding to the time necessary for the magnetic field to decay \citep[see, for example,][]{SubramanianEtAl06}.
The fraction of steep halos is therefore expected to be small, as estimated below.
The details of the magnetisation and magnetic decay processes are not well understood, and not much data is available for the steep halos, so we can only crudely sketch their properties.

Regardless of the mechanism magnetising the steep halos, it is unlikely to operate identically and simultaneously across the cluster, so spectral variations are to be expected between different parts of the halo.
In particular, an outgoing radial mode would induce radial spectral steepening, observable wherever the halo is sufficiently magnetised ($b\gtrsim 1$).
Similarly, as $t_{cool}\sim b^{-3/2}$ for large $b$, {\CREs} in the central, presumably most magnetised regions cool quickly, and the resulting spectrum flattens more rapidly there than in the periphery.
This too, leads to outward spectral steepening.

Such radial steepening was identified in several flat halos (see discussion in {\KL}), but little is known about the spectral distribution within steep halos.
In the well studied case of A521, where $\alpha_{0.3}^{1.4}=-1.86\pm0.08$ \citep{DallacasaEtAl09}, we find a central spectrum $\alpha_{0.2}^{1.4}\simeq -1.4$, suggesting substantial outward steepening.
The spectral steepening observed both with increasing frequency and increasing radius can be used to test the model and extract the magnetic field profile, as discussed in {\KL}.
Moreover, as shown in \S\ref{sec:TimeDependentTheory}, it can be used to gauge the evolution of the magnetic field.

\subsubsection{Using relics to time the evolution of halos}

In the absence of available spectral maps for the steep halos, we use the published, integrated spectrum of each halo in order to constrain the magnetic evolution.
As mentioned in \S\ref{sec:HaloRelicConnection}, all six steep-spectrum halo clusters harbour either an identified relic (in four out of the six), or a relic-like, filamentary feature (in the remaining two).
Assuming that relics are outgoing shocks, we may use them as a proxy for the time that elapsed since the merger.

This is clearly a crude approximation, because multiple merger shocks can form at different times since the --- ill defined --- time of the merger.
For example, in the steep GH cluster A754, in addition to the $r\sim 600$--$700\kpc$ relic, a more distant, $r\simeq 1\Mpc$ shock was identified \citep{KrivonosEtAl03}.
Nevertheless, timing mergers with relics is useful for obtaining a qualitative insight to the nature of the steep halos.

\Fig~\ref{fig:HaloAlphaVsRelicR} thus shows all the clusters reported in the literature as harbouring both a halo with a measured spectrum --- either flat or steep --- and a classical (see \S\ref{sec:SourceSelection}) relic.
For each such system, we plot the average spectral index of the \emph{halo}, measured around $1.4\GHz$, as a function of the distance of the \emph{relic} from the centre of the cluster.
We supplement the figure by the two systems (with a measured halo spectrum) in which a shock is embedded within or at the edge of the halo, because such shocks shall, presumably, become relics at a later stage.

The spectral data used in the figure is summarised in \Tab~\ref{tab:HaloSpectra}.
Relic positions refer to the brightest contours, as described in \S\ref{sec:DataPreparation};
For Coma, we use $\alpha_{0.3}^{1.4}=-1.27\pm0.07$ based on the data summarised in \citet{ThierbachEtAl03}. This is similar to the estimate of \citet{KimEtAl90}, but steeper than the best fit $\alpha\simeq -1.03$ of \citet{ThierbachEtAl03} based on a broken power-law model \citep{Rephaeli79}.
In A754, the spectral uncertainty reflects the difference between the two blobs (labeled 2 and 3) in \citet{KaleDwarakanath09}.

\begin{figure}
\centerline{\epsfxsize=9cm \epsfbox{\myfarfig{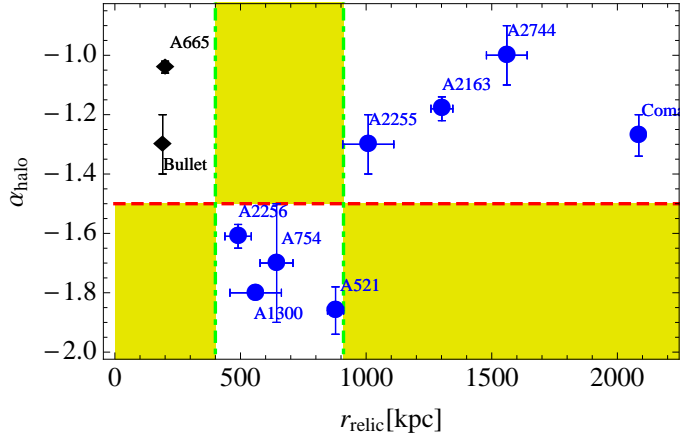}}}
\caption{
Spectral index of halos found in clusters that harbour a relic (disks) or have an embedded shock (diamonds) at a distance $r$ from the centre of the cluster.
The data is summarised in tables \ref{tab:ClusterData} and \ref{tab:HaloSpectra}.
Note that none of the halos is steep (flat) if the neighbouring relic or shock lies outside (inside) the range $400\kpc\lesssim r \lesssim 1\Mpc$, as illustrated by the shaded regions.
\label{fig:HaloAlphaVsRelicR}}
\end{figure}

\begin{table}
{
\begin{tabular}{|ccc|}
\hline
Source Name & Spectrum & Reference \\
\hline
1E 0657--56 & $\alpha_{1.3}^{8.8}=-1.3\pm0.1$ & L00 \\
A521H & $\alpha_{0.30}^{1.40}=-1.86_{-0.08}^{+0.10}$ &  D09 \\
A665 & $\alpha_{0.30}^{1.40}=-1.04\pm0.02$ & F04b \\
A754E & $\alpha_{0.15}^{1.36}=-1.70_{-0.20}^{+0.20}$ &  K09 \\
A1300H & $\alpha_{0.30}^{1.40}=-1.8$ &  G09 \\
A2163H & $\alpha_{0.30}^{1.40}=-1.18_{-0.04}^{+0.00}$ &  F04b \\
A2255H & $\alpha_{0.15}^{1.20}=-1.30_{-0.10}^{+0.10}$ &  P09 \\
A2256H & $\alpha_{0.02}^{2.70}=-1.61_{-0.04}^{+0.00}$ &  B08b \\
A2744H & $\alpha_{0.30}^{1.40}=-1.00_{-0.10}^{+0.10}$ &  O07 \\
ComaH & $\alpha_{0.03}^{1.40}=-1.27\pm{0.07}$ & (this work) \\
\hline
\end{tabular}
\caption{\label{tab:HaloSpectra} Spectrum of halos in clusters which harbour a relic or a halo-embedded shock.
Reference abbreviations are as in Table \ref{tab:ClusterData}, and in addition:
B08b -- \citet{Brentjens08};
K09 -- \citet{KaleDwarakanath09};
L00 -- \citet{LiangEtAl00}.}
}
\vspace{0.5cm}
\end{table}

Although the statistics is poor due to the small number of observed halo--relic systems, the figure reveals an interesting trend.
The steep spectrum halos are found only in clusters in which the relic is less than $\sim 1\Mpc$ away from the cluster's centre.
Clusters with more distant relics all have a flat spectrum.
This is another example of the many connections between halos and relics, outlined in \S\ref{sec:HaloRelicConnection}.

Interpreting relics as outgoing shocks, the figure supports the notion that steep spectrum halos are a transient stage in the halo evolution.
For a typical shock Mach number $\mach\sim 2$ and a downstream temperature of $\sim 10\keV$, a transition from a steep to a flat spectrum at $\sim 1\Mpc$ suggests that steep halos survive over a timescale $t_{steep}\sim 1\Gyr$.
Compared to the estimated lifetime of a halo, $t_{halo}\sim 5\Gyr$, one may expect a fraction $t_{steep}/t_{halo}\simeq 20\%$ of the halos to be steep, consistent with present observations ($\sim 20\%$ of the GHs observed are steep; see \S\ref{sec:SteepHalos}).
That $t_{steep}>t_{cool}$ supports the conclusion that halos are not magnetised simply by a single weak shock, but rather by a combination of multiple shocks and an increasing level of turbulence.

Assuming that the shocks found at $r\simeq 200\kpc$ in the bullet cluster and in A665 shall eventually become relics, once they separate from their currently flat halos, suggests that it takes a considerable time for the halos to develop a steep spectrum.
The steep halos shown in \Fig~\ref{fig:HaloAlphaVsRelicR} are associated with relics found at $400\kpc\lesssim r \lesssim 1\Mpc$.
It would be difficult to distinguish putative relics found at smaller radii from the halos.
Interestingly, the two steep halo clusters with no relic identification do show a filamentary, relic-like structure at a smaller radius.
However, it is difficult to determine their distance because the $r<500\kpc$ filament in A697 \citep{VenturiEtAl08} is nearly radial, and the $300\lesssim r \lesssim 450\kpc$ filament in A1914 is riddled with point sources \citep{BacchiEtAl03}.

Assuming an $r\sim300\kpc$ transition from a flat to a steep spectrum, and adopting $\mach\sim 2$ and $k_B T_d\sim 10\keV$, the evolution of a steep halo spectrum would require $\sim 0.3\Gyr$, comparable to the sound crossing time of the halo.
This slightly lowers the expected incidence rate of steep halos, to $\sim 15\%$, and provides further evidence that their magnetisation involves multiple shocks and turbulence.


\section{Discussion}
\label{sec:Discussion}

Our unified picture of radio emission from galaxy clusters is  briefly reviewed in \S\ref{sec:model_assumptions_and_results}.
We explain how accounting for the temporal evolution of the magnetic field and for {\CRE} diffusion leads to a model that attributes GHs, MHs, relics, and halo--relic bridges to secondary emission from {\CREs} produced by the same, homogeneous distribution of {\CRIs}.

The magnetic field configuration inferred from the data, consistent with an $\epsilon_B=\constant$ magnetic energy fraction, is discussed in \S\ref{sec:magnetic_scaling}.
We point out that some halos and relics locally appear to reach a saturation $\epsilon_B$ level on the order of $10\%$.
In \S\ref{sec:primary_vs_secondary} we show that our results are inconsistent with primary {\CRE} models; in particular, such models require fine tuning as they attribute GHs, MHs, relics and bridges to different acceleration mechanisms, which operate in very different environments.

In order to derive the properties of the {\CRI} distribution, we first determine in \S\ref{sec:KinematicEmissivity} the kinematic emissivity in the centres of GHs, where the magnetic fields are strongest so the gas column density is a good approximation for $\lambda_{nB}$.
Next, in \S\ref{sec:CRP_spectrum} we derive the spectrum of the {\CRIs} from the radio spectra observed at the edges of relics, where the {\CRE} spectrum still retains its steady state, upstream value, unaffected by the shock and the subsequent downstream magnetic evolution.
Combining these results, we compute in \S\ref{sec:CRI_energy_density} the energy density of the {\CRIs}.

Next, we examine the implications of the inferred {\CRI} distribution.
In \S\ref{sec:HomogeneousCRIsImplications} we show that both the virial shock and SNe shocks can account for the inferred {\CRI} energy, provided that {\CRI} escape from the cluster is quenched, and argue that strong diffusion or gas mixing are needed to explain the {\CRI} homogeneity.
In \S\ref{sec:CRIDispersionAmongClusters} we discuss the dispersion of the {\CRI} energy density among different clusters, and demonstrate how combining the results $u_p\propto n_0$ and $\epsilon_B\propto \constant$ can recover the observed $P_\nu\till L_X$ and $P_\nu\till R_\nu$ correlations in GHs, where $R_\nu$ is an average halo radius.
Additional hadronic signals from galaxy clusters, such as $\gamma$-rays produced by $\pi^0$ decay, and hard X-ray to $\gamma$-ray inverse Compton emission, are discussed in \S\ref{sec:Additional_hadronic_signals}.

In \S\ref{sec:limits_on_particle_acceleration_and_injection} we show how the strong magnetic fields inferred in our model impose strict upper limits on particle acceleration in weak shock and in turbulence.
We conclude in \S\ref{sec:model_assumptions_and_uncertainties} by discussing the assumptions and uncertainties of our analysis.

\subsection{Time-dependent secondary emission and implications for radio sources}
\label{sec:model_assumptions_and_results}

\subsubsection{Model overview}

In \S\ref{sec:ModelApplications} we showed how emission from a time-dependent distribution of {\CREs} produced by variable cooling in strong, evolving, and irregular magnetic fields, can explain flat and steep spectrum halos, relic, and halo--relic bridges.
More generally, as the {\CRI} distribution we infer is homogeneous, any strongly magnetised region in the ICM, where $B\gtrsim B_{cmb}$, should show bright radio emission, proportional to the column density $\lambda_{nB}$ of magnetised gas, given by \eqs{\ref{eq:nuInuPerLambda}}--(\ref{eq:DefCI}).

The surface brightness in \eqs{\ref{eq:nuInuPerLambda}}--(\ref{eq:DefCI}) is modified if the plasma properties are variable.
As the {\CRE} distribution evolves over a {\CRE} cooling time (given in \eq{\ref{eq:CRE_Cooling0}}), magnetic evolution during the past $\sim t_{cool}$ is imprinted upon the radio signal.
Recent magnetic growth (decay) thus leads to enhanced (diminished) radio emission followed by spectral steepening (flattening), subsequently returning to the steady state, $\alpha\simeq -1$ spectrum.
High resolution radio spectra could be used to reconstruct the recent evolution of the magnetic field, constraining the dynamical state of the cluster.

Of particular interest is the simple case of a fast, $\Delta \ll t_{cool}$ plasma transition at time $t=0$, involving arbitrary transitions $n_1\to n_2=\myrg n_1$ in density, $b_1\to b_2=\myrB b_1$ in normalised magnetic field amplitude, and $N_{cr,1} \to N_{cr,2}=\myrcr N_{cr,1}$ in {\CR} density.
The radio emission is initially brightened by a factor $\myrB^2\myrcre$ (assuming flat injection), while retaining the initial, $t<0$ spectrum.
This early time emission is related to the late time, steady-state emissivity by a factor
\begin{align} \label{eq:myr_general2}
\myr & = \frac{j_\nu(\Delta t<t\ll t_{cool})}{j_\nu(t\gg t_{cool})} \\
& = \frac{n_1}{n_2} \cdot \frac{B_2^2+B_{cmb}^2}{B_1^2+B_{cmb}^2} = \myrg^{-1} \frac{1+b_2^2}{1+b_1^2} \nonumber \fin
\end{align}
This factor relates, for example, the value of $I_\nu/\lambda_n$ at the edge of a shock-induced relic, to its value in the downstream halo.
These result are not unique to a shock, and would apply for any disturbance propagating in the ICM.

Note that the brightness change in \eq{\ref{eq:myr_general2}} is independent of the {\CR} amplification factor $\myrcr$, because {\CRIs} and {\CREs} at a given energy experience the same amplification.
These amplification factors are important, for example, in relating the emission upstream and downstream of a shock.
We thus showed in \S\ref{sec:CRAmplification} that {\CRs} with a flat spectrum (the flattest spectrum attainable in DSA) are amplified by a factor of $\mach^2$ at weak shocks, independent of the details of the diffusion mechanism or even the equation of state.

We parameterise the {\CRI} distribution as an $N_p\propto n^\sigma$ scaling with the gas density.
Analysing the morphologies of two well studied GHs in \S\ref{sec:HaloAndRelicEta}, we find that $\sigma=0.2\pm0.1$ within $r<2r_c$ (see \eq{\ref{eq:BestFitGamma}}), with evidence for flattening (towards $\sigma=0$) at smaller radii.
In an independent, complementary method, we fit the radio brightness peaks of all relics and halos in relic clusters, as a power-law in column density, accounting for the temporal magnetic evolution in relics by introducing a free parameter $\epsilon_B$.
This yields $\sigma = -0.3\pm0.2$, both if we assume that relics are $\mach=2$ shocks (see \eq{\ref{eq:fit_relics_epsBM2}}, corresponding to $\epsilon_B\sim 4\%$), and if we assume that they are magnetisation fronts (see \eqs{\ref{eq:fit_relics_epsB}}; here $\epsilon_B\sim 20\%$).
These estimates are consistent with a homogeneous, $\sigma=0$ distribution of {\CRIs}, considering the systematic uncertainties, although we cannot rule out small deviations from homogeneity.

\subsubsection{Different radio sources in the model}

In our model, flat spectrum GHs arise from steady state, strong magnetic fields, illuminating the secondary {\CREs} in radio waves.
Relics have a similar origin, but are additionally amplified at their outer edge by a factor $\myr$ due to recent, shock-induced magnetisation.
The region connecting a relic with its downstream halo can be thought of as a halo extension, observable where the column density of magnetised gas is sufficiently high, in the form of a halo--relic bridge or a halo protrusion.

A steep radio spectrum or spectral steepening are interpreted in our model as arising from any combination of recent magnetic growth, {\CRE} diffusion across an irregular magnetic field, and strong magnetic variability about a constant mean.
This naturally explains the downstream steepening behind some relics, the steep halos recently observed, and the steepening towards the edges of some flat halos.
In some of these cases, evidence of recent magnetic growth is independently inferred from observations.

For example, in our picture, relics gauge the time that elapsed since the merger event that presumably magnetised the cluster.
Thus, a relic close to (far from) the centre of the cluster indicates a young (middle-aged) merger, whereas a relaxed halo with no relic corresponds to an old merger.
Among the halo--relic clusters, relics lying within $\sim 1\Mpc$ from the cluster's centre are preferentially found next to a steep halo, as shown in \S\ref{sec:ModelTest_RelicGHConnection}.
This central relic--steep halo association supports the notion that the steep spectrum of the halo reflects a young merger, involving recent magnetic growth, an irregular magnetic morphology, and strong turbulence.

We therefore propose searching for steep GHs in clusters where central relics have been observed.
Conversely, we suggest that irregular features found in steep halos, such as the nearly radial protrusion West of the GH in A697, and the radio filament Southwest of the GH in A1914, are associated with shocks.

The model suggests that MHs in relaxed, cool core clusters would never show a steep, $\alpha<-1.5$ spectrum near their centre, because their regular morphology reveals little recent magnetic evolution in the past $t_{cool}$.
This is consistent with observations, although present knowledge about the spectrum of MHs is limited.

Note that in addition to magnetic evolution and {\CRE} diffusion, some spectral steepening may also arise from the energy-dependence of the cross section for secondary {\CRE} production, in cases where the {\CRP} spectrum is sufficiently steep (see {\KL}).
However, in \S\ref{sec:CRP_spectrum} we show that relics reveal a flat, $s_p\simeq -2.2$ {\CRP} spectrum, so the steepening induced by the cross section is small, of order $\Delta \alpha\simeq 0.1$.

\subsubsection{Halos in relic vs. non-relic clusters}

We noted in \S\ref{sec:eta_in_relics} that the GHs in the present sample are brighter by a factor of a few, on average, than the GHs in the {\KL} sample, and show a larger dispersion in the central radio to X-ray brightness ratio $\eta_0$.
This can be seen for example in the $\eta$ values plotted in \Fig~\ref{fig:SourcesEtaN}, and in the kinematic emissivity $\nu j_\nu/n$ shown in \Fig~\ref{fig:SourcesEmissivityHalos}.

In these figures, central radial bins are used for the GHs in the {\KL} sample, whereas the brightest radio regions are used for the GHs in the present sample, typically located near but not precisely at the centre of the cluster.
Note that the A521, A1300, and A2256 are steep halos, where $\nu I_\nu$ depends strongly on $\nu$, contributing to the large dispersion.

An important difference between the two samples is the association with relics.
The GHs in the present sample are all (except A2219, which is contaminated and excluded from the analysis) found in clusters which also harbour a relic, while only one of the GHs studied in {\KL} harbours a relic --- A2163 --- which also shows the highest $\nu j_\nu /n$ value within its sample.

As mentioned above, timing mergers using relics, we may interpret GH clusters devoid of a relic as old mergers, where the relics have already left the system.
Therefore, the larger dispersion in halo brightness among relic clusters may be due to $I_\nu$ variations induced by recent magnetic evolution.
The lower halo brightness in non-relic clusters could be attributed to more substantial magnetic decay, leading to weaker or more patchy magnetic fields than found in relic clusters.

With the present data, we can only infer from the relic travel time that in such a scenario, magnetic decay would proceed on timescales longer than $\sim 1.5\Gyr$.
Better halo and relic statistics could be used to constrain the details of the magnetic decay process, using relics as merger clocks.

\subsubsection{Relics in non-halo clusters}

Clusters with a GH but without a relic can be explained as old mergers or as newborn halos where the shocks are still embedded in the halo.
But what is the origin of relics found in clusters that do not harbour a GH?
As mentioned in \S\ref{sec:HaloRelicConnection}, in spite of the association of halos and relics with mergers and with each other, we find no evidence for an enhanced incidence rate of halo/relic detection in relic/halo clusters.
Although a quantitative treatment of the selection effects involved is yet to be performed, this observation suggests that relics in non-halo clusters are abundant.

Note that some of these non-halo relics are brighter than relics found in halo clusters at similar radii, for example the relics in A115 and in RXJ1314, and the double relic system in A3376.
A subset of non-halo relics may not be associated with the cluster's {\CRI} distribution but rather with local {\CR} sources, such as AGNs.
Another subset of relics could arise from primary {\CRE} acceleration in strong accretion shocks \citep{EnsslinEtAl98, KeshetEtAl04}, in particular those relics found at large distances from the cluster's centre.

Consider those non-halo relics that are generated by the cluster's homogeneous {\CRI} distribution.
Some of them, in particular the $r\lesssim 1\Mpc$ relics, may lie near an undetected steep spectrum halo which was not selected by high frequency observations.
Alternatively, one could explain the absence of a halo by invoking scenarios in which the central magnetisation is weak or substantially delayed.
The merger involved could be minor, insufficient to magnetise the central parts of the cluster to $B_{cmb}$ levels, but still capable of forming a relic shock.
The absence of a central radio shock trail may arise if the shock is ingoing rather than outgoing, if its trajectory is offset from the cluster's centre, or if it was very weak when passing through the cluster's centre.

\subsection{Magnetic field distribution with $\epsilon_B\sim \const$}
\label{sec:magnetic_scaling}

In \S\ref{sec:HaloAndRelicEta} we presented evidence showing that at small radii $I_\nu \propto \lambda_n$, whereas at large radii $I_\nu\propto F_X$ (away from substructure and contaminations).
We interpret this as a linear dependence of the radio emissivity upon density at small radii (flat region), $j_\nu\propto n$, and a quadratic, $j_\nu\sim n^2$ scaling
at large radii.
For emission from secondary {\CREs} produced by a homogeneous {\CRI} distribution, $j_\nu\propto n B^2/(B^2+B_{cmb}^2)$, so this behaviour naturally arises if the magnetic field declines radially, approximately as $B^2\sim n$.
Thus, the strong, $B\gtrsim B_{cmb}$ fields in the central region saturate the dependence on $B$ such that $j_\nu\propto n$, whereas in weakly magnetised regions $j_\nu\propto n (B/B_{cmb})^2\propto n^2$.
Such radial breaks, marking the transition from strong to weak magnetic fields, were discussed in {\KL}.

Extrapolating the magnetic field from the radial break towards the centre of the cluster by assuming $B^2\propto n$, \ie that the magnetic energy density $u_B$ constitutes some fixed fraction $\epsilon_B$ of the thermal energy density $u_{th}$, yields strong central magnetic fields, as indicated in {\KL}.
Here, for example, we obtain $B_0\sim 30\muG$ in A665 and A2163.
Note, however, that (i) the large radii scaling $I_\nu\propto F_X$ is not well constrained by our analysis, although it was more carefully demonstrated in other studies mentioned below; and (ii) the observations could be explained equally well by saturation to any constant $B$ in the flat region, but this is less natural than assuming a monotonic decline in $B$.

In the framework of a putative, \emph{primary} {\CRE} model, the same emissivity scaling arises if the {\CRE} energy density $u_e$ is assumed to be some fixed fraction of $u_{th}$, provided that the magnetic field behaves as described above in the context of a secondary {\CRE} model: $B\gtrsim B_{cmb}$ (or simply constant) at small radii, and $B^2\sim n$ at large radii.
Therefore, $u_p\sim \constant$ secondary {\CRE} models and $u_e\sim n$ primary {\CRE} models require precisely the same magnetic scaling.

Indeed, studies assuming primary {\CREs} have reported a $B^2\propto n$ scaling at large radii \citep[see, for example,][]{MurgiaEtAl09, VaccaEtAl10}.
Note, however, that primary models that assume equipartition between {\CREs} and the magnetic field, $u_e=u_B$ \citep[and not just $u_e\propto u_B$; see \eg][]{VaccaEtAl10}, fail to reproduce the observations at small radii, because equipartition typically implies that $B< B_{cmb}$ even in the central region.

We find that significant magnetic amplification, on the order of $\epsilon_B\simeq 10\%$, is needed in order to explain relics using the secondary {\CRE} model.
However, the inferred value of $\epsilon_B$ is somewhat degenerate with the shock strength.
Thus, for a typical relic shock Mach number $\mach\simeq 2$, our relic sample yields $\epsilon_B\simeq \range{4}{10}\%$ (see \Figs~\ref{fig:SourcesEtaNModelM2} and \ref{fig:SourcesEmissivityShock}), whereas for $\mach\to 1$ it requires $\epsilon_B\simeq 20\%$ (see \Fig~\ref{fig:SourcesEtaNModel}).
Moreover, these estimates rely on $\beta$-model extrapolations of the column density to large radii, and so are vulnerable to the $\sim$factor 2 uncertainty of the $\beta$ models.

\subsubsection{Apparent saturation at $\epsilon_B\sim 0.1$}

Assuming that the radial profile of the magnetic field declines with increasing radius such that $u_B\propto u_{th}$, one may use the radial break to gauge the magnetic field throughout the cluster.
In the two GHs with a regular X-ray morphology, A665 and A2163, we thus obtained an acceptable morphological fit, indicating that the central magnetic field is approximately $B_0\sim 30\muG$ in both cases.
In A2163, this corresponds to a magnetic fraction $\epsilon_B\simeq 10\%$.
In A665, where a weak shock is identified $\sim 400\kpc$ from the centre, we obtain a higher, $\epsilon_B\simeq 20\%$ magnetic fraction.

Similar magnetic fractions, typically $\epsilon_B\gtrsim 5\%$, are also inferred from lower limits on the central magnetic fields in GH clusters, obtained by {\KL} by assuming that $u_B\propto u_{th}$ and that the GH size is no smaller than the magnetic break radius $R_B$.
Indeed, in \S\ref{sec:CRIDispersionAmongClusters} we show that assuming that $u_p$ and $\epsilon_B$ are universal constants, approximately reproduces the $P_\nu\till L_X$ and $P_\nu\till R_\nu$ correlations observed (see \eqs{\ref{eq:Prop_Pnu_rnu_model}} and (\ref{eq:Prop_Pnu_LX_model})).

Combining these results suggests that the magnetic fields in GH clusters share, approximately, a universal magnetic fraction, on the order of $10\%$ of the thermal energy density.
However, the magnetisation of GH clusters are thought to be magnetised by the turbulence and shocks induced by a merger event, and the merger details and histories differ among different GHs.
Therefore, our results suggest that in both halos and relics, the magnetic field has reached a saturation level of $\epsilon_B\sim 10\%$ (within a factor of a few), above which further magnetic growth is quenched.

Such magnetisation levels are higher than thought to be induced, microscopically, even by a strong shock, where $\epsilon\sim 1\%$ is typically assumed.
Therefore, inferring $\epsilon_B\simeq 10\%$ both behind weak shocks and in turbulent regions in the ICM suggests that in both cases, the plasma is magnetised to a saturation level by the turbulent motions.
For a discussion of turbulent magnetisation arising from shock-induced vorticity, see \citet{SironiGoodman07}.

\subsection{The results disfavour primary {\CRE} models}
\label{sec:primary_vs_secondary}

There are several indications that primary {\CRE} acceleration is unlikely to play an important role in halos.
Some of the evidence was outlined by {\KL}, who showed that neither GHs nor MHs are likely to arise from primary {\CREs}.
However, it was so far thought that radio relics do arise from primary {\CREs}.

The present study shows that modeling relics as primary {\CRE} acceleration or reacceleration leads to unnatural conclusions, whereas relics are easily explained in the framework of a secondary model, if magnetic amplification in the shock or in its wake are taken into account.
This is based on the universal, flat spectrum observed at the edges of relics, the varying degree of inward steepening, the multiple connections between halos and relics such as halo--relic bridges, and the relation $\eta\sim (n/n_0)^{-1}\eta_0\myr$ which is satisfied by both halos and relics.

Primary {\CRE} models attribute GHs and relics to different physical processes, typically electron acceleration or reacceleration in turbulence and in shocks, respectively.
However, the numerous connections between halos and relics, outlined in \S\ref{sec:ModelProblems}, are highly unnatural if the two phenomena arise from {\CRE} populations of such a different origin, and it is necessary to fine tune the acceleration parameters.
In contrast, halos and relics are essentially identical in our secondary {\CRE} model, and the connections between them arise naturally, as shown in \S\ref{sec:model_outline}.

This argument complements and strengthens an analogous relation between GHs and MHs, pointed out by {\KL}.
Namely, the environments of GHs and MHs are very different from each other, with GHs extending over large volumes in the low density ICM of merger clusters, and MHs found in the high density cores of relaxed clusters, in direct association with cold fronts.
In a primary {\CRE} framework, the nature of particle acceleration and the resulting {\CRE} parameters should differ substantially between the two types of sources.
However, the properties of GHs and MHs are very similar (same radio--X-ray relations, spectrum, morphology, and weak polarisation), and some clusters show features of both (\eg cool cores and CFs in some GH clusters, and an apparent transition from a MH to a GH in A2319; see {\KL}).
This occurs naturally in our secondary {\CRE} model, but requires fine tuning and ad-hod assumptions in a primary framework (see {\KL}).

Therefore, primary {\CRE} models are strongly disfavoured by the present analysis, as a viable explanation for either GHs, or MHs, or relics.
As all three types of sources can be explained as secondary emission from the same primary {\CRI} distribution, it would be unlikely for any of them to show a substantial contribution from primary {\CREs}.

Additional evidence favouring secondary {\CRE} models:
\begin{enumerate}
\item
In both primary and secondary {\CRE} models, fitting the GH morphologies implies that the central magnetic field is strong, as shown in \S\ref{sec:magnetic_scaling}.
In a primary model, it is assumed that secondary {\CREs} are negligible, but this would imply that the primary {\CRIs} constitute a fraction $\xi_p\ll 10^{-3}$ of the thermal energy density in the centres of clusters.
Such a small value of $\xi_p$ would be difficult to reconcile with the expected {\CRI} output of SNe and the virial shock;
see \S\ref{sec:HomogeneousCRIsImplications}.
It may also contradict the level of {\CRI} output from the same putative {\CRE} sources.
\item
Weak shocks are observed at the very edges of several GHs, as described in \S\ref{sec:HaloRelicConnection}.
At least in some cases, the spectrum at the edge is flat, $\alpha\simeq -1$, and steepens behind the shock, in resemblance of relic spectra; see for example the shock in A665 shown in \Fig~\ref{fig:ProfilesA665B}.
In the secondary {\CRE} model, this is precisely the expected behaviour, as there is no essential difference between such shocks and relics.
In primary {\CRE} models, however, turbulent acceleration cannot explain the radio emission immediately behind the shock \citep[see for example][]{GovoniEtAl04}, nor the spectral steepening, whereas weak shock acceleration cannot explain the emission far from the shock.
It is possible in principle to invoke a combination of weak shock acceleration near the shock and turbulent acceleration far from it, but fine tuning would be necessary.
\item
Reacceleration of particles can be ruled out in relics that show a pure power law spectrum.
Acceleration of particles according to DSA also provides an unnatural explanation for the observed relic spectra, as argued in \S\ref{sec:ModelProblems}.
We predict a pure power-law spectrum immediately behind shocks at the edges of halos; if confirmed, this would have similar implications for acceleration and reacceleration in halos.
\item
In \S\ref{sec:CRAmplification} we showed that a weak shock propagating into a medium with a preexisting, flat distribution of relativistic particles, can only lower the energy fraction of these particles.
This suggests that weak shock reacceleration of a non-cooled, low energy distribution of {\CRs} is inefficient, and unlikely to play an important role in relics or halos.
\end{enumerate}

\subsection{Kinematic radio emissivity $\myk$}
\label{sec:KinematicEmissivity}

In order to evaluate the energy density $u_p$ of the {\CRIs}, we must first compute the steady-state, kinematic (\ie per unit density) radio emissivity of the diffuse sources.
We thus normalise $\nu I_\nu$ to the gas column density $\lambda_n$, for each sources, and correct for the cosmological dimming.
Assuming $\alpha=-1$, no k-correction is needed for the combination $\nu j_\nu$, so
\begin{equation} \label{eq:SpecificEmissivity}
\myk \equiv \frac{\nu j_\nu}{n} \simeq 4\pi(1+z)^4 \frac{\nu I_\nu}{\lambda_n} \coma
\end{equation}
where $\lambda_n$ is computed using the $\beta$-model of each cluster.

For a homogeneous distribution of {\CRIs}, $\myk$ is constant within a cluster, in all strongly magnetised regions.
Measuring the value of $\myk$ along a single, highly magnetised line of sight, where $\lambda_n\simeq \lambda_{nB}$, thus fixes the value of $u_p$ throughout the cluster.
The most reliable estimate of $u_p$ is based on the centres of GHs, where the approximation $\lambda_n\simeq \lambda_{nB}$ is good.
For relics, one needs to correct for the magnetic field evolution.

\subsubsection{$\myk$ in Halos}

\Eq{\ref{eq:SpecificEmissivity}} provides a good approximation for the kinematic emissivity in the centres of regular, flat spectrum GHs.
Here, it is reasonable to assume that the gas is highly magnetised, temporal evolution is slow, and the integral $\lambda_{nB}$ in \eq{\ref{eq:nuInuPerLambda}} approaches $\lambda_n$.
For sources that are weakly magnetised, are not very extended, or are seen in projection, $\lambda_n$ overestimates the column density $\lambda_{nB}$ of the magnetised gas, so \eq{\ref{eq:SpecificEmissivity}} underestimates the true kinematic emissivity.

The kinematic emissivity estimated in halos using \eq{\ref{eq:SpecificEmissivity}} is shown in \Fig~\ref{fig:SourcesEmissivityHalos}, plotted against the column density.
Results are shown both for the present sample of brightest halo emission in relic-cluster GHs, and for the radial profiles of GHs and MHs studied by {\KL} using radio data from \citet{MurgiaEtAl09}.
Also shown is the radial profile of the GH in A2744, discussed in \S\ref{sec:EtaInWellStudiedGHs}, using radio data from \citet{MurgiaEtAl09}.
Notice that for A2163 and A2744 we show both a radial profile and a data point corresponding to the brightest emission; the differences between them reflect the radial binning and the azimuthal averaging of $I_\nu$.
As in {\KL}, MH data is shown only where the contamination from the central AGN is less than $10\%$.
Note that the estimated $\myk$ values in MHs are systematically less certain than in GHs, because the $\beta$-model provide in general a less accurate fit for CCs.

\begin{figure*}
\centerline{\epsfxsize=16cm \epsfbox{\myfarfig{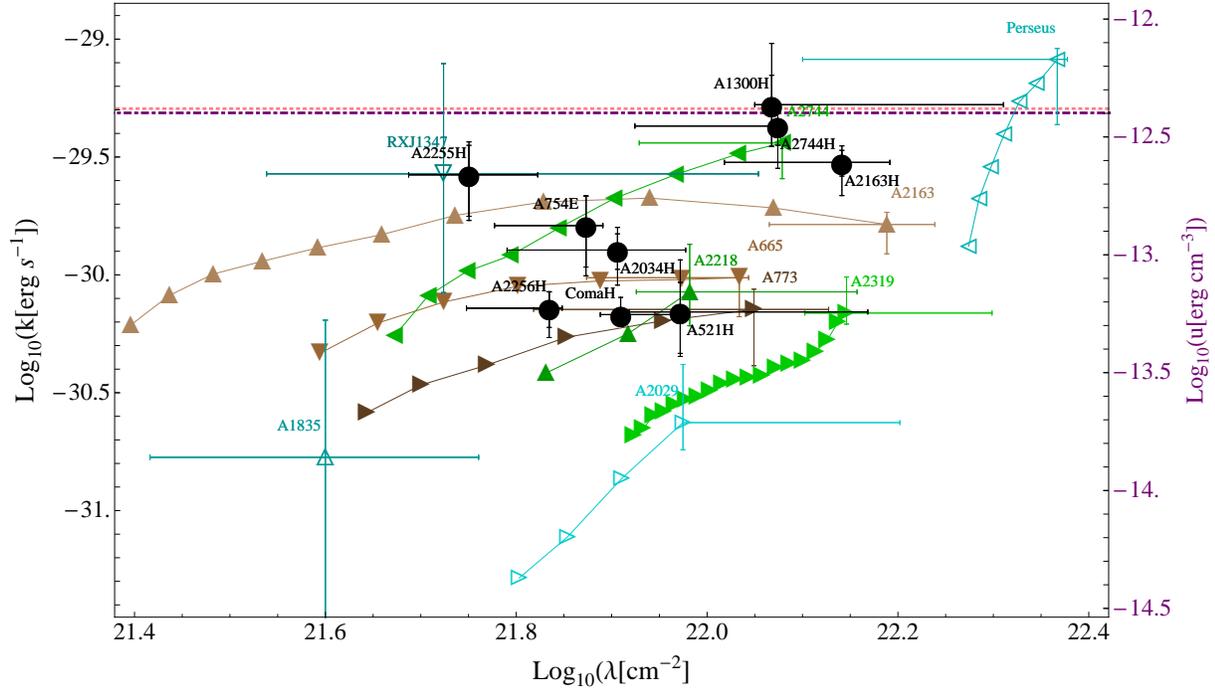}}}
\caption{
Kinematic (per unit density) emissivity (left axis) of GHs (filled symbols) and MHs (empty symbols), and the corresponding {\CRI} energy density between $10\GeV$ and $10^{7}\GeV$ (right axis).
The halo data in the present sample, consisting of the brightest radio emission of GHs found in relic clusters, are shown as black disks.
The radial profiles of the GH in A2744 and in the halos studied in {\KL} are shown as triangles, connected in each halo by lines to guide the eye, using radio data from \citet{MurgiaEtAl09}.
These radial profiles are shown for both GHs (filled green or brown triangles) and MHs (empty cyan triangles), where only the central data points are labeled with the cluster's name, and include error bars.
The {\CRI} energy density of each source (shown using the right axis) is computed assuming an $s_p=-2.2$ {\CRP} spectrum, according to \eq{\ref{eq:up10to16_perK}}.
For reference, we also show (with the right axis) the energy densities of the CMB (dotted pink horizontal line) and of the downstream of a typical virial shock, assuming $n_d=10^{-5}\cm^{-3}$ and $k_B T_d=10\keV$ (dot-dashed purple line).
\label{fig:SourcesEmissivityHalos}
\vspace{2mm}}
\end{figure*}

The radial profiles indicate that the kinematic emissivity is nearly uniform towards the centres of the GHs in A665, A773, and A2163 (brown triangles).
Indeed, a constant kinematic emissivity corresponds to a homogeneous {\CRI} energy density and a strongly magnetised plasma, as discussed below and shown using the right axis of the figure.

However, the radial profiles of the GHs in A2218, A2319, and A2744 (green triangles), and in all the MHs (see radial fits in figure 2 of {\KL}), show a decline in $\myk$ with increasing radius (\ie towards the left in \Fig~\ref{fig:SourcesEmissivityHalos}).
The reasons for such a radial decline were discussed in \S\ref{sec:HaloAndRelicEta} and in {\KL}.
They consist of a combination of weak or patchy magnetisation, asymmetry, substructure, shocks, and contaminations, aggravated by the averaging and binning procedures.
In MHs, the radial decline is also associated with the sharp drop in magnetisation above the CFs.
These effects lead, in general, to an underestimated value of $\myk$, and therefore of $u_p$, at large radii.

In the specific case of A2744, we showed in \S\ref{sec:EtaInWellStudiedGHs} that in spite of the azimuthally averaged radial decline in  $\myk$, the data is consistent with a homogeneous distribution of {\CRIs}, masked by the highly irregular X-ray morphology (see in particular \Fig~\ref{fig:LineProfileCompareA2163A2744}).
As such a homogeneous {\CRI} distribution is inferred in all three halos with good data studied in \S\ref{sec:HaloAndRelicEta}, it is likely that a careful analysis of the GHs in A2218 and A2319 and of the MHs would reveal a similar distribution; however, this is not possible using the data presently available to us.

Assuming a homogeneous {\CRI} distribution in each cluster, our best estimate for its density is based on the most central data point available, where $\lambda_n$ and the magnetic field are maximal, and the artifacts mentioned above are minimised.
The systematic errors due to the finite sizes of the sources, projection effects, weak magnetisation, substructure, and shocks, all tend to lower our estimated kinematic emissivity with respect to its true value.

On the other hand, the above estimates have all assumed steady-state magnetic fields and {\CRE} injection.
Recent magnetic growth or an irregular magnetic field configuration can locally brighten the radio emission.
In such regions, the $\myk$ values computed from \eq{\ref{eq:SpecificEmissivity}} may overestimate its steady-state value.

Our approximation is probably justified in the centres of halos, at least in those that show a regular morphology and a flat spectrum.
In particular, estimates based on the points of maximal radio brightness along the GH, shown in \Fig~\ref{fig:SourcesEmissivityHalos} as black disks, may be somewhat more reliable, as they correspond to optimal projection and magnetisation.
Notice that these data span a factor of $\sim 7$ in $\myk$ (and therefore in $u_p$); similar to the dispersion found in the central radio--X-ray brightness correlation.

The estimated $u_p$ values may be inaccurate in irregular GHs such as A2255, and in steep GHs such as A521 and A2256.
Note that in the latter, a k-correction would lead to a slightly higher kinematic emissivity than shown in the figure, but the results would be frequency dependent.

\subsubsection{Relics: correcting $\myk$ for temporal evolution}

Temporal variations in the magnetic field and in the {\CRE} injection rate can modify the emissivity with respect to its steady state value, which is the quantity we need in order to compute $u_{p}$.
In relics, this effect is strong, and we correct for it using the time dependent model of \S\ref{sec:TimeDependentTheory}.
For this purpose, we identify the brightest relic emission as the near-downstream region of a shock or a magnetisation wave.

Here, the peak radio emissivity exceeds its steady-state downstream value by a factor $\myr$, and the column density is raised by a factor $\myrg$ with respect to the upstream.
Correcting for these effects, we estimate the steady-state kinematic emissivity in relics as
\begin{align} \label{eq:SpecificEmissivityRelics}
\myk^* \equiv \frac{\nu j_\nu^*}{n^*} & \simeq 4\pi(1+z)^4 \frac{\nu I_\nu/\myr}{\lambda_{n,u} \myrg} \\
& \simeq 4\pi(1+z)^4 \frac{\nu I_\nu}{\lambda_{n,u}(1+b_d^2)} \nonumber\coma
\end{align}
where we assumed a weak upstream magnetisation, $b_u\ll 1$.

The main approximation here, $\lambda_{nB}\simeq \lambda_n$, is less justified than it is in the centres of regular GHs, because relic magnetisation is weaker and is confined to one side of the discontinuity.
Indeed, due to the curvature of the discontinuity and its finite area, only a fraction of the line of sight passes through strongly magnetised plasma.
Nevertheless, the approximation is plausible if the discontinuity surface is parallel to the line of sight, because (i) we choose the point with peak radio brightness, corresponding to optimal projection; and (ii) the gas density declines rapidly, roughly as $n\sim r^{-2}$, outside the core.

We assume that the $\lambda_n$ values computed using the $\beta$-model pertain to the upstream, as in some cases we find the X-ray brightness near relics to be elevated with respect to its $\beta$-model estimate (see \S\ref{sec:HaloAndRelicEta}).
The resulting, corrected values of the steady-state kinematic emissivity are plotted against $\lambda$ in \Fig~\ref{fig:SourcesEmissivityShock}, using \eq{\ref{eq:SpecificEmissivity}} for $\myk$ in GHs, and \eq{\ref{eq:SpecificEmissivityRelics}} for $\myk^*$ in relics.
Here we assume that all relics are associated with $\mach=2$ shocks, with a magnetic energy fraction $\epsilon_B=10\%$ downstream.
As in \S\ref{sec:HaloAndRelicEta}, we omit the uncertain relics in A754 and in A2034.

\begin{figure*}
\centerline{\epsfxsize=16cm \epsfbox{\myfarfig{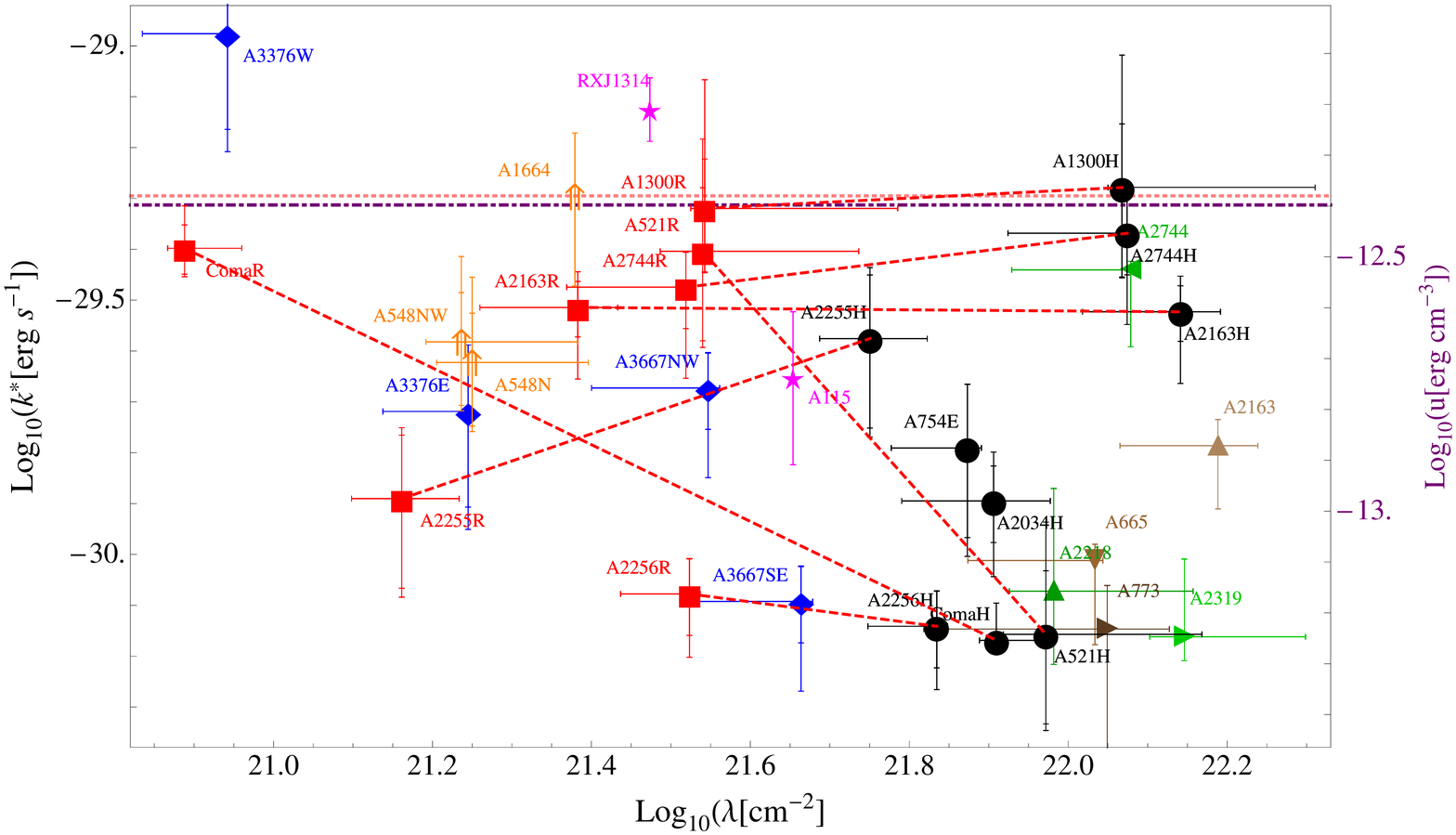}}}
\caption{
The kinematic radio emissivity $\myk$ in GHs (black disks and triangles, computed using \eq{\ref{eq:SpecificEmissivity}}) and the estimated steady-state emissivity $\myk^*$ in relics (other symbols, defined in \Fig~\ref{fig:SourcesInuIx}; computed using \eq{\ref{eq:SpecificEmissivityRelics}} assuming relic shocks with $\mach=2$ and $\epsilon_B=10\%$), plotted (using the left axis) along with the corresponding {\CRI} energy density between $10\GeV$ and $10^{7}\GeV$ (right axis), against the column density.
Notations are identical to those used in \Fig~\ref{fig:SourcesEmissivityHalos}.
To reduce clutter, we show only the innermost data point for each radial GH profile displayed in \Fig~\ref{fig:SourcesEmissivityHalos}.
In clusters that harbour both a halo and a relic, the two sources are connected by a dashed red line.
\label{fig:SourcesEmissivityShock}
\vspace{2mm}}
\end{figure*}

In the figure, dashed red lines connect halos and relics that are found in the same cluster.
The shock parameters $\mach$ and $\epsilon_B$ were fit to produce, on average, lines with $\myk\simeq \myk^*$, corresponding to a homogeneous {\CRI} distribution, in analogy with the fit procedure demonstrated in \Figs~\ref{fig:SourcesEtaNModel} and \ref{fig:SourcesEtaNModelM2}.
Indeed, the $\myk$ and $\myk^*$ values in most halos and relics (and therefore, $u_p$ in their respective clusters) span a factor of $\sim7$, similar to the dispersion in the central $\nu I_\nu\till F_X$ correlation.
Notice that lower values of $\mach$ and $\epsilon_B$ would correspond to a {\CRI} density increasing with radius; significantly larger values of these parameters are unlikely.

The shock parameters $\mach$ and $\epsilon_B$ play a degenerate role in determining the correction factor in \eq{\ref{eq:SpecificEmissivityRelics}}, as $b_d\propto (5\mach^2-1)\epsilon_B$.
For simplicity, we assumed that they have the same values in all relics, but in reality they are likely to vary with cluster parameters and with $r$.
For example, simulations suggest that stronger shocks are found at larger radii \citep[\eg][]{VazzaEtAl10}.
Adopting $\mach$ values that increase with $r$ would lower the values of $\myk^*$ in the more peripheral relics, resulting in a flatter $u_p$ distribution, for example in Coma and in A3376, but not in A2255.

\subsection{{\CRP} spectrum}
\label{sec:CRP_spectrum}

In order to convert the measured kinematic emissivity $\myk$ to the {\CRI} energy density $u_p$, we must first determine the {\CRP} spectrum.
If the spectrum is not flat, this conversion depends on $\nu$ and on $B$, through the $E_e$-dependence of the cross section for {\CRE} production.

One possibility is to compute the {\CRP} spectrum based on the centres of regular, flat spectrum GHs, where steady-state injection and magnetic fields are plausible.
However, the existence of steep spectrum GHs, and the sensitivity of the spectrum to spatial and temporal irregularities, would render such an estimate uncertain.

Moreover, the magnetic field is strong in the centres of GHs, and in the strong field regime $\alpha$ is not sensitive to $s_p$ (see figure 3 of {\KL}).
Thus, depending on the magnetic field value, the radio spectral range $\alpha\simeq -(0.9\text{--}1.1)$ observed in the centres of flat GHs could be explained by {\CRP} spectral indices in the range $s_p=-(2\text{--}3)$.
In addition, one expects deviations from a pure power law near $E_p\sim\GeV$.
For a given $\nu$, a strong field implies lower $E_e$, so the contribution of such low energy {\CRPs} becomes increasingly large.

The radio spectrum is sensitive to $s_p$ when the magnetic field is weak.
For example, for $B=B_{cmb}(z=0)$, we obtain (see Fig. 3 of {\KL}, inset)
\begin{equation} \label{eq:sp_alpha_approx}
s_p\simeq 2.15 \alpha_{1.4} \fin
\end{equation}
Here we used model A of {\KL}, adopting the spectral fits for $e^+$ and $e^{-}$ production in inelastic $p\till p$ scattering according to \citet{KamaeEtAl06}, with corrected parameters and cutoffs (T. Kamae \& H. Lee 2010, private communications), and the $\delta$-function approximation of $F_{syn}$ in \eq{\ref{eq:Fsyn_delta}}, which is accurate when $s_p(E_p)$ varies slowly.
Thus, in a steady-state region known to be marginally magnetised and uncontaminated by substructure or discrete sources, measuring $\alpha_{1.4}=-1.0$ would imply that $s_p\simeq -2.2$, whereas measuring $\alpha_{1.4}=-1.4$ would indicate $s_p\simeq -3.0$.

Several GHs show spectral steepening with increasing frequency, increasing radius, or decreasing temperature \citep[see][and references therein]{FerettiGiovannini08, FerrariEtAl08, GiovanniniEtAl09}.
{\KL} used the radial spectral steepening observed in GHs such as A665, A2163, and A2744, from $\alpha\simeq -1.0$ to $\alpha\simeq -1.4$, to infer a steep {\CRP} spectrum, $s_p\lesssim -2.7$.
However, as shown in \S\ref{sec:TimeDependentTheory}, such spectral steepening can arise from recent magnetic growth, in particular downstream of shocks.
The presence of shocks or relics in the vicinity of the steep spectrum regions in all three clusters mentioned above, suggests that the origin of their spectral steepening is, at least in part, dynamical.

In some clusters, a flat radio spectrum is measured in regions in which we infer magnetic fields of order $B_{cmb}$; see for example the $r\sim 500\kpc$ region of A2163 in the spectral map of \citet{FerettiEtAl04a}, and the slices through this map shown in \Figs~\ref{fig:ProfilesA2163} and \ref{fig:ProfilesA2163B}.
Such observations suggest a flat {\CRP} spectrum, in the range $s_p=-\range{2.0}{2.4}$.
However, these estimates are still sensitive to unconstrained dynamical changes in {\CRE} injection and in magnetisation.

As shown in \S\ref{sec:TimeDependentTheory}, immediately downstream of a shock, at distances $\Delta\ll v_d t_{cool}$, the radio spectrum reflects the {\CRE} distribution upstream.
Upstream of mergers shocks, it is much more plausible to assume steady state, regular magnetisation and injection, than it is in the turbulent, post-merger ICM.
Hence, the radio spectrum measured at the outer edges of shocks, found either in relics or at the edges of GHs, provides a direct diagnostic of the {\CRP} spectrum.

The spectrum at the edges of relics are typically found to be in the range $\alpha=-\range{1.0}{1.1}$, as shown in Table \ref{tab:RelicSpec} for relics at distances $r\simeq \range{0.9}{2}\Mpc$.
A similarly flat spectrum is found behind the shock marking the $r\simeq 400\kpc$ edge of the GH in A665 (see \Fig~\ref{fig:ProfilesA665B}), and behind the suspected shock in A2219 (see \Fig~\ref{fig:ProfilesA2219}).

These observations, showing a similar radio spectrum in low magnetised upstream regions spanning a wide range of radii, suggest a universal {\CRP} spectrum in clusters.
Combining the spectra inferred at the outer edges of the relics summarised in Table \ref{tab:RelicSpec}, and interpreting them as a universal spectrum, yields $\alpha=-1.02\pm0.02$.
Assuming that the magnetic fields in the downstream regions are $B\lesssim 10\muG$, we may use \eq{\ref{eq:sp_alpha_approx}} to obtain
\begin{equation} \label{eq:sp_estiamte}
s_p = -2.20\pm0.05 \fin
\end{equation}

For such a flat {\CRP} spectrum, the spectral steepening of secondaries due to the energy-dependence of their production cross section is minor, of order $|\Delta \alpha|\simeq 0.1$, as shown in \Fig~3 of {\KL}.
The steady state radio spectrum is therefore flat over a wide range of frequencies and magnetic field strengths, as long as \eq{\ref{eq:sp_estiamte}} is valid in the corresponding {\CRP} energies.
Spectral deviations are expected in very low frequencies and strong magnetic fields, where the typical {\CRP} energy responsible for the emission approaches $1\GeV$, as discussed in {\KL}.

Conversely, in our model, assuming that the $\gg \mbox{GeV}$ {\CRP} spectrum is well fit by \eq{\ref{eq:sp_estiamte}}, spectral deviations from $\alpha=-1$ in uncontaminated regions are all associated with temporal changes in magnetic fields and in {\CRE} injection, or with {\CRE} diffusion and an irregular magnetic configuration, except in low frequencies and strongly magnetised regions where $E_e\lesssim 1\GeV$ (see \eq{\ref{eq:nu_s}}).

A correlation was reported between the distance of a relic from the cluster's centre, the size of the relic, and its spectrum: more distant relics tend to be larger and show a flatter spectrum \citep{vanWeerenEtAl09}.
One possible interpretation involves {\CRI} injection near the centre of the cluster, and energy-dependent {\CRI} diffusion outwards, as suggested by {\KL}.
However, the reported correlation involves the spatially averaged relic spectrum, which depends on the threshold and does not directly reflect the spectrum of the {\CRIs}.
For example, spectral steepening behind the relic should be stronger for nearby relics, where the downstream magnetic field can significantly exceed $B_{cmb}$.

\subsection{{\CRI} energy density}
\label{sec:CRI_energy_density}

For a flat, $\alpha=-1$ synchrotron spectrum, the logarithmic energy injection rate $Q$ of {\CREs}, defined in \eq{\ref{eq:Def_Q}}, is related to the synchrotron emissivity through
\begin{equation}
Q = 2(1+b^{-2})\nu j_\nu \fin
\end{equation}
For a homogeneous {\CRI} distribution, $Q$ is proportional to $n$.
If, in addition, the magnetic field is strong, we may approximate the kinematic (per unit density) injection rate as
\begin{equation} \label{eq:Kinematic_Q_vs_myk}
Q/n = 2(1+b^{-2})\myk \simeq 2\myk \fin
\end{equation}
In \S\ref{sec:KinematicEmissivity}, $\myk$ was estimated from observations, so we can now use \eq{\ref{eq:Kinematic_Q_vs_myk}} to determine the {\CRP} energy density $u_p$ by finding its relation to $Q/n$.

We use model A of {\KL} to compute $Q/n$ for an arbitrary {\CRP} spectrum.
Integrating the {\CRP} distribution weighted by the inclusive cross sections $\sigma^{\pm}(E_e,E_p)$ for $e^{\pm}$ production, the kinematic injection of {\CREs} is given by
\begin{equation} \label{eq:Q_Kamae}
\frac{Q^{\pm}(E_e)}{n} = c E_e \int \frac{d\sigma^{\pm}\left(E_e,E_p\right)}{d\ln E_e} \frac{du_p}{d\ln E_p} \frac{d \ln E_p}{E_p} \fin
\end{equation}
Fitting formulae for $d\sigma^{\pm}/d\ln E_e$ are provided by \citet{KamaeEtAl06}, with corrected parameters and cutoffs (T. Kamae \& H. Lee 2010, private communications).
These cross sections take into account diffractive dissociation processes and the Feynman scaling violation, and incorporate the $\Delta$ resonance as well as several hadronic resonances around $1.6\GeV/c^2$.

Assuming that the {\CRP} spectrum is a pure power-law,
\begin{equation} \label{eq:Def_K_p}
\frac{du_p}{d\ln E_p}=\myCp \left( \frac{E_p}{1\GeV}\right)^{2+s_p} \coma
\end{equation}
we may use \eq{\ref{eq:Q_Kamae}} to compute the ratio $\myCp/\myk$ as a function of $s_p$, for a given value of the synchrotron-emitting {\CRE} energy $E_e$.
The ratio $\myCp/\myk$ is not sensitive to $E_e$ when the {\CRP} spectrum is flat, but depends strongly on it when $s_p\lesssim-2.5$.
The typical energy of a {\CRE} emitting at $\nu=1.4\GHz$ in a $B\simeq B_{cmb}$ field at $z=0$ is $E_e\simeq 6\GeV$ (see \eq{\ref{eq:nu_s}}).

For $s_p=-2.2$, which appears to be the relevant case for the {\CRP} distribution in galaxy clusters as shown in \S\ref{sec:CRP_spectrum}, combining \eqs{\ref{eq:Q_Kamae}} and (\ref{eq:Def_K_p}) yields
\begin{equation} \label{eq:K_p_perK_s2.2}
\myCp/\myk \simeq 2.8\times 10^{16} \se \cm^{-3} \coma
\end{equation}
constant to within $<4\%$ as $E_e$ changes by a factor of $3$ from its canonical value $E_e=6\GeV$.
In \Fig~\ref{fig:CRPLogEPerK} we plot the ratio $\myCp/k$ for a range of {\CRP} spectral indices, for {\CRE} energies $E_e/\mbox{GeV}=2$, $6$, and $20$.

\begin{figure}
\centerline{\epsfxsize=8cm \epsfbox{\myfarfig{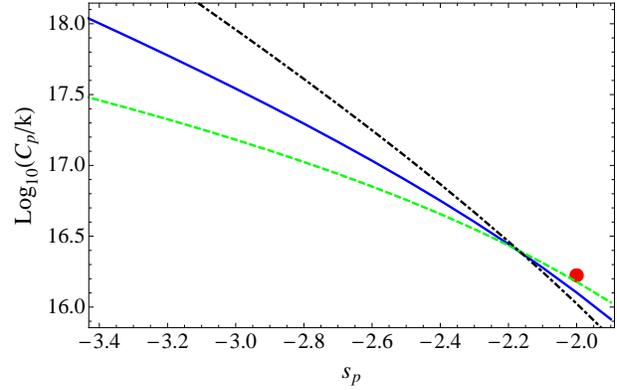}}}
\caption{
The ratio between the logarithmic {\CRP} energy density $du_p/d\ln E_p$ at $E_p=1\GeV$, and the kinematic synchrotron emissivity $\myk$, as a function of the {\CRP} spectral index $s_p$.
We compute $\myCp/\myk$ from \eq{\ref{eq:Q_Kamae}} for $E_e=6\GeV$ (solid curve), $2\GeV$ (dashed), and $20\GeV$ (dot-dashed).
Also shown is the $s_e=s_p=-2$ approximation of \eq{\ref{eq:K_p_to_k_ratio}} (red circle).
\label{fig:CRPLogEPerK}
\vspace{2mm}}
\end{figure}

To demonstrate these results, consider a flat {\CRP} spectrum, $s_p=-2$, and assume for simplicity that the resulting {\CRE} spectrum is also a flat, $s_e=-2$ power-law.
(This approximation is good to $\sim 10\%$ for $E_e>$ a few GeV; see {\KL}.)
Here one may approximate \citep[\eg][]{KushnirEtAl09}
\begin{equation} \label{eq:Approx_sigma_sM2}
E_e^+\frac{d\sigma^+}{d\ln E_{e^+}}+E^- \frac{d\sigma^-}{d\ln E_{e^-}} \simeq \frac{\sigma_{inel}}{10} E_p^2 \delta(E_p-10E_e) \coma
\end{equation}
where $\sigma_{inel}\simeq 40\mb$ is the total cross section for an inelastic $p\till p$ collision.
In this approximation,
\begin{equation} \label{eq:K_p_to_k_ratio}
\frac{\myCp}{\myk} \simeq \frac{2}{c \sigma_{inel}/10} \simeq 1.7 \times 10^{16} \se \cm^{-3} \coma
\end{equation}
shown in \Fig~\ref{fig:CRPLogEPerK} to be in approximate agreement with the numerical results.

Next, we consider the total energy density of {\CRIs}.
For $s_p=-2.2$, integrating the {\CRP} distribution of \eqs{\ref{eq:Def_K_p}} and (\ref{eq:K_p_perK_s2.2}) over the energy range $[10\GeV,10^{7}\GeV]$, we find
\begin{equation} \label{eq:up10to16_perK}
u_{p[10,10^7]} = 8.2\times 10^{-14} \left(\frac{\myk}{10^{-30} \erg\se^{-1}} \right) \erg \cm^{-3} \fin
\end{equation}
The upper limit $10^{7}\GeV$, corresponding to the position of the knee in the Galactic {\CRP} spectrum, is not well justified here, but the results are not sensitive to its value; allowing infinite {\CRP} energies raises the coefficient in \eq{\ref{eq:up10to16_perK}} by $<7\%$.

The values of $u_{p[10,10^7]}$ inferred for the sources shown in \Figs~\ref{fig:SourcesEmissivityHalos} and \ref{fig:SourcesEmissivityShock} are provided by the right axes of these figures, using the normalisation in \eq{\ref{eq:up10to16_perK}}.
The halo centres and the classical relics show $\myk$ values in the range
\begin{equation} \label{eq:myk_range}
\myk = 10^{-\range{29.3}{30.2}}\erg\se^{-1} \fin
\end{equation}
Using \eq{\ref{eq:up10to16_perK}}, we conclude that the {\CRI} energy density in these clusters lies in the range
\begin{equation} \label{eq:up10to16}
u_{p[10,10^7]} \simeq 10^{-\range{12.4}{13.3}} \erg\cm^{-3} \fin
\end{equation}
As the figures show, this is much smaller than the thermal energy density $u_{th}$, except possibly at the outskirts of the cluster.

This {\CRI} energy density is in good agreement, in the centres of clusters, with the estimate of \citet[][based on the $P_\nu\till L_X$ correlation in GHs]{KushnirEtAl09} and with the estimate of {\KL} (based on the $\nu I_\nu\till F_X$ in the centres of halos).
All three estimates are dominated by the central regions of GHs, and show that {\CRIs} hold a fraction $\epsilon_p\sim 10^{-3}$ (with an uncertainty factor of a few) of the thermal energy density in the centres of massive clusters.

However, here we infer a much higher {\CRI} energy density at the peripheries of clusters, because the {\CRI} distribution is shown to be homogeneous, rather than declining radially.
Indeed, interpreting relics as emission from secondary {\CREs} implies that the {\CRI} distribution is homogeneous out to $\gtrsim 2\Mpc$ scales.
Consequently, we find a {\CRI} energy fraction that increases with radius, and reaches $\epsilon_p\sim 0.1\till 1$ behind the virial shock; $1\till2$ orders of magnitude above the \citet{KushnirEtAl09} estimate.
(The peripheral value of $\epsilon_p$ in the {\KL} model is sensitive to the SFR history and {\CRI} diffusion).

\subsection{{\CRI} production and evolution}
\label{sec:HomogeneousCRIsImplications}

The two main sources discussed in the literature as possible explanations for the origin of {\CRIs} in clusters involve particle acceleration in strong shocks, either in supernovae or in the virial shock of the cluster.
In both cases, the shocks are believed to deposit a substantial fraction of the thermal energy, typically estimated in the range $\epsilon_p\simeq 5\%\till50\%$ \citep[see for example][]{KeshetEtAl03}, in a {\CRI} distribution with a flat spectrum compatible with \eq{\ref{eq:sp_estiamte}}.

In order to explain the homogeneous {\CRI} energy density inferred above, one needs to account not only for the production of the {\CRIs}, but also for their
evolution, in particular their escape from the cluster and the mechanism distributing them homogeneously throughout the ICM.

\subsubsection{{\CRI} origin and escape}

Consider first the virial shock.
The thermal energy density behind (immediately downstream of) the shock is $u_{th}\simeq 4\times 10^{-13}n_{-5}T_{10}\erg\cm^{-3}$, where $T_{10}\equiv k_B T/10\keV$ is the downstream temperature of bulk protons and electrons, assumed to have the same temperature.
This value of $u_{th}$, and the comparable CMB energy density $u_{cmb}\simeq 4.2\times 10^{-13}\erg\cm^{-3}$, are shown in \Figs~\ref{fig:SourcesEmissivityHalos} and \ref{fig:SourcesEmissivityShock} as horizontal lines.
The halos and classical relics show lower or similar energy densities $u_p$, ranging from $u_p\sim 0.1u_{th}$ in Coma to $u_p\sim u_{th}$ in A1300, with a median value $\sim 0.4u_{th}$.
Therefore, the measured {\CRI} energy density is consistent with the \emph{instantaneous} {\CRI} injection rate by the virial shock, for acceleration efficiencies in the range $\epsilon_p\sim \range{0.1}{1}$.
This is consistent with typical estimates of $\epsilon_p$ based on observations of SNe in the Galaxy, considering the systematic uncertainties in our analysis.

Next, consider the accumulated {\CRIs} accelerated in the virial shock throughout the life of the cluster.
Computing the resulting energy density requires a model for the shock evolution, for energy losses and gains, in particular by adiabatic compression, and for the escape of {\CRIs} from the cluster, for example through regions where the virial shock is weak or disrupted.
Assume that on average, a fraction $f_{esc}$ of the accelerated particles escape from the cluster, and the remaining {\CRIs} are distributed homogeneously.
Assuming, in addition, that after these {\CRIs} leave the shock they gain, on average, a factor $f_{ad}$ in energy, yields
\begin{align}
u_{p,\text{vir}} & = \epsilon_p(1-f_{esc})f_{ad}  \frac{(3/2)(M/\mu m_p)k_B T_a}{(4/3)\pi R^3} \\
& \simeq 2 \times 10^{-13} (1-f_{esc})f_{ad}\epsilon_{p,0.5}M_{14}T_{a,1}R_{2}^{-3} \erg \cm^{-3} \nonumber \fin
\end{align}
Here, $M=10^{14}M_{14}M_{\odot}$ is the mass of the cluster, $R=2R_{2}\Mpc$ is the radius of the virial shock, $T_a=1\, T_{a,1}\keV$ is the downstream temperature, flux-averaged over the cluster's evolution, and we defined $\epsilon_{p,0.5}\equiv \epsilon_p/0.5$.

This estimate is consistent with the median of the radio sources observed, for the above parameter choice and $(1-f_{esc})f_{ad}T_{a,1}\simeq 1/2$, but this averaged quantity is poorly constrained.
For an estimate of the {\CRI} energy distribution in the negligible diffusion limit, where $f_{esc}=0$ and $f_{ad}$ is large, see \citet{KushnirWaxman09}.

Next, consider the contribution of SNe.
The accumulated {\CRI} output of SNe can be crudely estimated \citep{VolkEtAl96, KushnirEtAl09} if we assume that a fraction $f_{\text{II}}$ of the mass-averaged $Z=0.3Z_{0.3}$ solar metallicity of a cluster is seeded by Type II SNe, which on average produce $0.1M_{\odot}M_{Fe,0.1}$ of iron and deposit a fraction $\eta_p=0.2\eta_{p,0.2}$ of the $E=10^{51}E_{51}\erg$ explosion energy in $E_p>10\GeV$ {\CRIs}.
If a fraction $f_{esc}$ of these {\CRIs} escape, and the rest are homogeneously distributed within the virial shock radius, then the {\CRI} energy density due to SNe becomes
\begin{equation}
u_{p,\text{SNe}} \simeq 10^{-13} \frac{(1-f_{esc})f_{ad}\eta_{p,0.2}f_{\text{II}} Z_{0.3} E_{51} M_{14}}{R_{2}^{3} M_{\text{Fe,0.1}}} \erg \cm^{-3} \fin
\end{equation}
Here, the energy gain factor $f_{ad}$ may be smaller than for virial shock {\CRIs}, because SNe {\CRIs} may lose energy as they travel away from their host galaxies and into the ICM.

The above estimates show that the {\CRI} distribution we infer from radio observations could plausibly be explained as the accumulated {\CRI} output of either the virial shock or SNe shocks (or a combination of both), provided that only a small fraction, $f_{esc}\lesssim 1/2$ of the {\CRIs} escape from the cluster.
If diffusion is negligible, $f_{esc}$ is guaranteed to be small, but this requires efficient gas mixing in order to explain the homogeneous {\CRI} distribution, as discussed below.
Considering the uncertainties folded in the above estimates, we are unable to determine which of the two {\CRI} sources dominates the ICM distribution.

Note that if {\CRI} acceleration in the virial shock and in SNe shocks have similar efficiencies, then one should expect $\eta_p\simeq \epsilon_p/3$.
However, the escape probability is likely to be at least as high in the case of virial shock acceleration, which injects particles into the rarefied cluster's periphery, as it is for SNe {\CRIs}.
Therefore, if the {\CRIs} originate mostly from the virial shock, then SNe make a non-negligible contribution to $u_p$, unless the {\CRIs} they produce experience substantial energy losses when leaving their host galaxies.
If, on the other hand, the {\CRIs} originate mostly from SNe, then the high {\CRI} energy fraction behind the virial shock suggests that these {\CRIs} play a role in the evolution of the shock, and their escape from the cluster could be regulated by it.

The flat spectrum of the {\CRIs}, inferred from relic observations in \eq{\ref{eq:sp_estiamte}}, is consistent with the injection spectrum typical of strong shocks.
This suggests that for the dominant {\CRI} source, the fraction of particles escaping the cluster is indeed small, otherwise the energy-dependence of the diffusion may lead to spectral steepening, as discussed in {\KL}.

One could distinguish between the two {\CRI} source models by testing for a correlation between $u_p$ and the estimated {\CRI} output of each model.
Both models predict a correlation between $u_p$ and the average density $M/R^3$.
In addition, the SNe model entails a correlation with tracers of the accumulated SNe activity, such as the total star formation and metallicity, whereas the virial shock model implies a correlation with tracers of the accreted energy, such as the cluster's temperature.
Preliminary, low significance evidence for correlations between $\eta_0$ and the metallicity and the specific star formation rate (SFR) were pointed out by {\KL}.
A quantitative study is needed to test these correlations, in particular considering the large dispersion in $u_p$ among relic clusters not studied by {\KL} .

\subsubsection{{\CRI} diffusion or gas mixing}
\label{sec:HomogeneousCRIsImplications_Diffusion}

In both models for their origin, The injection of {\CRIs} is local, and far from being homogeneous through the ICM.
Therefore, in both cases, the homogeneous {\CRI} distribution inferred from observations implies that some mechanism acts to homogeneously distribute the {\CRIs} across the cluster, either through {\CRI} diffusion or by gas mixing.

Otherwise, for virial shock acceleration, adiabatic compression would lead to a radio-to-X-ray brightness ratio profile $\eta\propto n^{-1/3}$, which we rule out in \S\ref{sec:HaloAndRelicEta}.
For SNe acceleration, {\CRI} injection would trace the SNe history of the cluster, and in the absence of diffusion, would be further modulated by subsequent adiabatic compression.
Although the radial distribution of SNe in the cluster is not well constrained, the resulting {\CRI} distribution would be centrally peaked, in resemblance of the distribution of metals \citep{SandersonEtAl09}, which are also believed to be injected by SNe.

The homogeneous {\CRI} distribution extends from the centre of the cluster to distances $r\gtrsim 2\Mpc$.
This distance scale is comparable to the typical radius of the virial shock \citep[see for example][]{KeshetEtAl04}.
Therefore, diffusion or mixing must operate across the entire cluster, faster than the timescale for substantial change in the {\CRI} injection rate.
In the virial shock model, this is the timescale over which the shock parameters change appreciably, on the order of a few dynamical times.
In the SNe model, this timescale is comparable to the time that elapsed since the peak in star formation, at $z\sim 1$.
In either case, we must require that {\CRIs} travel at least across distances $r_D\sim 2\Mpc$ over a timescale $t_D\simeq 10\Gyr$.

Consider first the case where diffusion plays the dominant role in distributing the {\CRIs}.
Here, maintaining a homogeneous distribution of {\CRIs} imposes a lower limit on the diffusion coefficient,
\begin{equation} \label{eq:D_estimate}
D \gtrsim r_D^2/t_D \simeq 10^{32}\cm^2 \se^{-1} \fin
\end{equation}
\Eq{\ref{eq:D_estimate}} applies to the {\CRIs} that give rise to synchrotron emission at frequencies discussed in this work, $\nu\simeq 1\GHz$.
Using \eq{\ref{eq:nu_s}} and the approximate relation $E_p\simeq 20E_e$ \citep{GinzburgSyrovatsky61}, this corresponds to $E_p\simeq 200(B/1\muG)^{-1/2}\GeV$.
The lower limit on $D$ in \eq{\ref{eq:D_estimate}} is at least an order of magnitude higher than typically estimated at these energies \citep[see, for example][]{VolkEtAl96}.

Such a strong diffusion wold aggravate the need for a mechanism quenching particle escape beyond the virial shock, as any escaping particles near the shock would be rapidly replenished by {\CRIs} diffusing from downstream.
The ram pressure of the upstream flow could stem the {\CRI} escape, as long as (i) it exceeds the {\CRI} energy density, \ie $\epsilon_p<0.9$; and (ii) upstream diffusion is not too strong, with $D\ll v_u R \simeq 10^{33}R_2(v_u/1000\km\se^{-1})\cm^2\se^{-1}$.
Notice that $D\sim 10^{\range{32}{33}}\cm^2\se^{-1}$ satisfies the necessary requirements both upstream and downstream, although it does not need to be equal on both sides of the shock.

Strong gas mixing could, in principle, explain the homogeneous {\CRI} distribution without invoking {\CRI} diffusion, thus relaxing the {\CRI} escape problem. Such mixing could arise from the erratic gas motions induced by consecutive merger events, and from the {\CRI} populations advected by the infalling substructure.
It is unclear, however, if such mixing would be sufficiently efficient to homogenise the {\CRIs} over $\gtrsim 2\Mpc$ scales, and if the gas mixing would not be ruled out by independent tracers of the gas, such as metallicity.

A one-dimensional treatment of gas mixing in the absence of diffusion \citep{KushnirWaxman09} shows some steepening in $\eta$ with respect to the adiabatic $\eta\propto n^{-1/3}$ profile (\ie flattening of the {\CRI} distribution with respect to $N_i\propto n^{2/3}$), but not at a level sufficient to explain the observations.
Notice that the {\CRI} distribution must be homogeneous \emph{before} the major merger that led to the present radio emission by magnetising the plasma.
Indeed, similar {\CRI} energies are inferred immediately behind merger shocks in relics and at the edges of GHs, presumably gauging the {\CRI} distribution
prior to the most recent, ongoing merger.

To conclude, strong diffusion or gas mixing are needed in order to reconcile the homogeneous {\CRI} distribution with the inhomogeneous distributions of the {\CRI} sources. We consider diffusion to be the more plausible mechanism, because (\emph{i}) gas mixing does not appear to be sufficiently effective according to preliminary estimates; (\emph{ii}) the distribution of metals shows a $Z(r>0.02r_{500})\propto r^{-0.3}$ radial decline, in both cool and non cool core clusters \citep{SandersonEtAl09}, which would have been eroded by efficient gas mixing; (\emph{iii}) without diffusion, adiabatic compression would imply that low frequency, $\alpha\simeq -1$ observations are already probing the $E_p\sim \GeV$ energy range \citep{KushnirEtAl09}, where one expects deviations from a flat {\CRI} spectrum;
and (\emph{iv}) in our model, {\CRI} diffusion plays an important (although not exclusive) role in producing the steep spectrum of young halos.

\subsubsection{Escaping ultra high energy {\CRIs} do not contradict local observations; are comparable at $\sim10^{18}\eV$}

The fraction of {\CRIs} escaping from the cluster, and their distribution, is unknown.
The flat spectrum and large energy density we find within the cluster suggest that the escaping fraction at $E_p\lesssim 100\GeV$ energies is small.
Reconciling significant escape at these energies with the flat spectrum observed would imply an energy-independent escape mechanism.

The virial shock can accelerate {\CRIs}, according to the \citet{Hillas84} criterion, up to energies
\begin{align}
E_p & \simeq \beta e B R \\
& \simeq 2 \times 10^{18}v_{s,3} R_2 \left(\epsilon_{B,-2}n_{-5}T_{10}\right)^{1/2} \eV \nonumber \coma
\end{align}
where $\epsilon_{B,-2}\equiv\epsilon_B/10^{-2}$ and $v_{s,3}\equiv v_s/10^3\km\se^{-1}$.
In SNe, the same criterion yields lower {\CRI} energies, $E_p\sim 10^{15}v_{s,3}\eV$ for $R=10\pc$ and $\epsilon_B nk_BT=10\eV\cm^{-3}$.
Note that here we use $\epsilon_B\sim 1\%$, smaller than we infer downstream of weak shocks, because it is the magnetic field near and upstream of the shock front that limits the acceleration of the particles.

In the SNe model, it is unclear if any of the {\CRIs} appreciably escape from the cluster.
In contrast, in the virial shock model, the highest energy {\CRIs} must be able to escape.
For an $s_p=-2.2$ spectrum, \eq{\ref{eq:up_cos}} below yields a cosmic {\CRI} background of $du_p/d\ln E_p\simeq 3\times 10^{-19}\xi_{p,0.2}T_{\text{cos},0.3}\erg\cm^{-3}$ at $E_p=10^{18}\eV$, where $T_{\text{cos}}=0.3T_{\text{cos},0.3}\keV$ is the mass-averaged temperature of baryons in the Universe \citep[\eg][]{KeshetEtAl03}.
This is similar to the observed {\CRI} distribution at the top of Earth's atmosphere in the same energies, $du_p/d\ln(E_p)\simeq 3.5\times 10^{-19}\erg\cm^{-3}$ \citep{NaganoWatson00}.
Indeed, \citet{MuraseEtAl08} have proposed that the observed {\CRIs} between the second knee and the ankle, \ie in the energy range $10^{\range{17.5}{18.5}}\eV$, arise from the virial shocks of galaxy clusters.
While our estimate is too crude to support such a claim, it indicates that our model is consistent with ultra high energy cosmic-ray observations.

\subsection{{\CRI} energy density in different clusters, and the $P_\nu\till L_X$ correlation}
\label{sec:CRIDispersionAmongClusters}

The dispersion in $u_p$ among the different radio sources shown in \Figs~\ref{fig:SourcesEmissivityHalos} and \ref{fig:SourcesEmissivityShock} is somewhat larger than the measurement uncertainties.
This suggests an inherent dispersion among the host clusters, either in $u_p$ or in the underlying approximations, in particular $\lambda_n\simeq \lambda_{nB}$.

We have searched for correlations between $u_p$ and the bulk properties of the host cluster, including the cluster's total mass $M$, gas mass $M_g$, temperature $T$, central density $n_0$ and central column density $\lambda_{n,0}$ \citep[all based on][]{FukazawaEtAl04}.
We have also examined a possible evolution with redshift.
However, due to the small size of the sample, the similar bulk properties of the host clusters, and the large uncertainty in $u_p$, we are unable to show any statistically significant correlation or evolution.

Indirect estimates of the dependence of $u_p$ upon cluster parameters may be inferred from the reported correlations between the bulk properties of GHs and their host clusters.

Consider first the radio power $P_\nu$ expected according to our homogeneous {\CRI} model, in a flat-spectrum GH approximated as spherical.
The result depends on the strength of the magnetic field with respect to $B_{cmb}$.
The radio power of the strongly magnetised part of the cluster is
\begin{equation} \label{eq:P_nu_Strong_B}
P_{\nu,\text{strong}}\propto \int n u_p \,dV\propto u_p M_{g,B} \coma
\end{equation}
where $M_{g,B}$ is the mass of the highly magnetised gas, and the integration is performed over volume out to the magnetic break radius $R_B$.
The radio power of the weakly magnetised part of the cluster is
\begin{equation}
P_{\nu,\text{weak}}\propto \int n^2 u_p \, dV \propto u_p (L_X-L_{X,B}) \coma
\end{equation}
where we adopted the $u_B\propto u_{th}$ scaling of \S\ref{sec:magnetic_scaling}, defined $L_{X,B}$ as the X-ray luminosity of the highly magnetised gas, and the integration here is performed beyond $R_B$.
We shall henceforth consider strongly magnetised halos, where $R_B>r_c$, and $P_\nu$ and $L_X$ are dominated by the strongly magnetised gas.

These predicted scalings can be compared to the reported GH correlations,
\begin{equation} \label{eq:Prop_Pnu_rnu_obs}
P_{\nu,\text{obs}} \propto R_\nu^{4.2\pm0.7} \coma
\end{equation}
where $R_\nu$ is the size of the radio bright region \citep{CassanoEtAl07},
and
\begin{equation} \label{eq:Prop_Pnu_LX_obs}
P_{\nu,\text{obs}}\propto L_X^{\range{1.7}{2.2}}
\end{equation}
\citep[][]{BrunettiEtAl07,KushnirEtAl09,BrunettiEtAl09}.

It may appear, for example if one assumes that $M_{g,B}\propto R_\nu^3$, that these observations imply that $u_p$ varies with cluster parameters, being larger in more massive, hotter, and X-ray brighter clusters \citep[see for example][]{KushnirEtAl09}.

However, it is important to take into account the scaling of $R_\nu$ with cluster parameters, as emphasised by {\KL}.
In our model, $R_\nu\simeq R_B$ is the radius where the magnetic field declines below $B_{cmb}$, marking the transition from strong to weak fields, and most of the radio emission is produced within $R_B$.

For simplicity, assume that the magnetic energy density in GHs is a universal, fixed fraction of the thermal energy density, such that $B^2\propto nT$ with a constant coefficient.
This is plausible, for example, if the magnetic fields are saturated at $\epsilon_B\sim 0.1$, as suggested in \S\ref{sec:magnetic_scaling}.
In order to obtain power-law scalings which can be compared with \eqs{\ref{eq:Prop_Pnu_rnu_obs}} and (\ref{eq:Prop_Pnu_LX_obs}), we must approximate the $\beta$-model as a power-law.
Assume, therefore, that most of the radio emission arises in a region where the density may be approximated as an isothermal sphere, $n\propto n_0 r^{-2}$.

Under these assumptions, $R_\nu=R_B\propto (n_0 T)^{1/2}$, and \eq{\ref{eq:P_nu_Strong_B}} yields
\begin{equation} \label{eq:Prop_Pnu_rnu_model}
P_\nu/u_p \propto n_0 R_\nu^3 \propto R_\nu^5/T \propto R_\nu^{4.4} \coma
\end{equation}
where in the last proportionality we used the best fit scaling relations $R_\nu\propto R_{vir}^{2.6}$ \citep{CassanoEtAl07} and $R_{vir}\propto T^{0.6}$ \citep{ZhangEtAl08}, with $R_{vir}$ being the virial radius.

For the assumed density distribution, we may approximate $L_X\propto n_0^2$, implying that
\begin{equation} \label{eq:Prop_Pnu_LX_model}
P_\nu/u_p \propto n_0^{5/2}T^{3/2} \propto L_X^{2} \coma
\end{equation}
where in the last proportionality we used the relation $L_X\propto T^2$ \citep{Markevitch98}.
The scalings derived in \eqs{\ref{eq:Prop_Pnu_rnu_model}} and (\ref{eq:Prop_Pnu_LX_model}) for this simplified model agree surprisingly well with the observed relations in \eqs{\ref{eq:Prop_Pnu_rnu_obs}} and (\ref{eq:Prop_Pnu_LX_obs}), considering the inaccurate approximation.

While the arguments leading to these scalings are crude, they serve to show that the correlations observed in the bulk halo properties do not necessarily imply that $u_p$ varies among clusters.

This is not to suggest that the homogeneous {\CRI} distribution in clusters is part of a cosmic {\CRI} background with a well defined energy density $u_{p,\text{cos}}$.
In addition to the large dispersion found in the values of $u_p$ among different clusters,
these values are much too high to characterise a cosmological background.
To see this, assume that every baryon in the Universe deposits a fraction $\epsilon_p$ of its thermal energy in {\CRIs}.
The resulting {\CRI} background would then have an average energy density
\begin{align} \label{eq:up_cos}
u_{p,\text{cos}} & \simeq \epsilon_p \frac{3}{2} \left( \frac{\Omega_b \rho_c}{\mu m_p} \right) k_B T_{\text{cos}}  \\
& \simeq 6\times 10^{-17} \epsilon_{p,0.2} T_{\text{cos},0.3} \erg\cm^{-3} \nonumber \coma
\end{align}
lower by $\sim 3$ orders of magnitude than inferred in clusters.
Here, $\Omega_b \rho_c\simeq 2\times 10^{-7}m_p\cm^{-3}$ is the average baryon mass density in the Universe, and $T_{\text{cos}}=0.3T_{\text{cos},0.3}\keV$ is its mass-averaged temperature \citep[\eg][]{KeshetEtAl03}.

Notice that if $u_p$ does in fact increase with the cluster's size, then the {\CRIs} are more likely to be dominated by the virial shock, rather than by SNe. Indeed, the {\CRI} output of the virial shock is thought to be proportional to the cluster's temperature, $T\sim M^{2/3}$.
In contrast, in the SNe model, the {\CRI} distribution should correlate with the specific SFR, which is not thought to increase with $M$ \citep[see for example][]{Goto05}.

\subsection{Additional hadronic signals}
\label{sec:Additional_hadronic_signals}

For a homogeneous distribution of {\CRIs}, pion-production is proportional to the local density.
Integrated signals proportional to pion production, such as the $\gamma$-ray emission from  $\pi^0$ decay, are therefore proportional to the gas mass of the cluster. This is different from the synchrotron signal, which is proportional to the strongly magnetised gas mass, and therefore is a transient and typically centrally-peaked phenomenon.

The clusters which are optimal targets for detection of the $\pi^0$ signal are the most gas-rich clusters, or more generally, sources with maximal gas mass enclosed within the point spread function of the detector.

Assuming that the energy a {\CRP} deposits in neutral pions is $\sim 3$ times larger than the energy transferred to secondary {\CREs}, the subsequent $\pi^0$ decay would lead to a $\gamma$-ray emissivity
\begin{align}
\nu j_\nu^{(\pi^0)} & \simeq 3 \nu j_\nu^{(syn)}(B\gg B_{cmb}) \\
& \simeq 3 \times 10^{-32} n_{-2} \left( \frac{\myk}{10^{-30}\erg\se^{-1}} \right) \erg \se^{-1}\cm^{-3} \nonumber \coma
\end{align}
and an integrated luminosity
\begin{align} \label{eq:gamma_from_pi0}
\nu L_\nu^{(\pi^0)} & \simeq \nu j_\nu^{(\gamma)} \frac{M}{m_p} \\
& \simeq 4\times 10^{41} \left( \frac{\myk}{10^{-30}\erg\se^{-1}} \right) \left( \frac{M_g}{10^{14}M_\odot} \right) \erg \se^{-1} \nonumber \coma
\end{align}
where $\myk$ here pertains to its strongly magnetised ICM value.

In addition to the different scaling (with $M_g$ rather than with $L_X$), this $\gamma$-ray signal is slightly stronger than anticipated in models that assume $N_p\sim n$ \citep[such that $\nu L_\nu^{(\gamma)}\simeq 3\nu L_\nu^{syn}\simeq 3\times 10^{41} L_X^{1.7}\erg\se^{-1}$;][]{KushnirEtAl09}, because the peripheral regions also contribute.

Nevertheless, detecting the weak signal of \eq{\ref{eq:gamma_from_pi0}} would be challenging even in the most massive clusters, using existing and planned $\gamma$-ray telescopes.
A more promising $\gamma$-ray signal from clusters is the inverse-Compton emission from primary electrons near the virial shock \citep{LoebWaxman00, TotaniKitayama00}, which should be detectable in several clusters using the 5-year Fermi data \citep{KeshetEtAl03}.

The inverse-Compton signal produced as secondary {\CREs} scatter off the CMB is complementary to the synchrotron emission, in the sense that it is proportional to the mass of the \emph{weakly} magnetised gas.
Assuming slow changes in magnetic fields and in {\CRE} injection, the expected surface brightness is
\begin{align}
\nu I_\nu^{(iC)}(\vectwo{r}) \simeq \frac{10^{-\range{8.9}{9.5}}}{10^{22}\cm^{-2}} \int \frac{n\,d l}{1+b^2} \erg \se^{-1} \cm^{-2} \sr^{-1} \coma
\end{align}
and the luminosity is
\begin{align}
\nu L_\nu^{(iC)} \simeq \frac{10^{\range{40.9}{41.8}}}{10^{14}M_\odot} \int \frac{\rho\,d V}{1+b^2} \erg \se^{-1} \coma
\end{align}
where $\rho=m_p n$ is the mass density.

This signal is too weak to account for the hard X-ray emission observed from some clusters, which is more likely to arise from primary {\CRE} acceleration at the virial shock; see \citet{KushnirWaxman10}.

\subsection{Upper limit on particle acceleration in turbulence and in weak shocks}
\label{sec:limits_on_particle_acceleration_and_injection}

Primary acceleration or reacceleration (for short: acceleration) of electrons in turbulence and in weak shocks have been proposed as possible explanations for the injection of {\CREs} in radio relics and halos.
These primary processes are not well understood, and various ad hoc prescriptions were adopted in computations and simulations reported in the literature.
In \S\ref{sec:primary_vs_secondary} we reviewed various evidence indicating that primary {\CRE} acceleration cannot play an important role in the diffuse radio emission from the ICM.
More importantly, as direct and indirect evidence implies that the magnetic field is strong in halos and in relics, the faint radio emission observed imposes restrictive limits on the efficiency of primary particle acceleration in these environments.

Consider first the case of relics.
The regions near the outer edge of a relic, showing the brightest emission with the flattest spectrum, are interpreted as the near downstream of a shock, where accelerated particles had little time to cool.
Arguably, these regions are strongly magnetised, as confirmed independently for example in A3667.
We may use the brightness of these regions to compute the efficiency of relativistic electron injection.

As discussed in \S\ref{sec:ModelProblems}, the spectrum at the edges of relics is almost precisely $\edalpha=-1$ in all cases where good data is available, corresponding to a {\CRE} spectrum $s_e\simeq -3$.
Although this is a peculiar injection spectrum in a DSA model, it is exactly the spectrum expected in our secondary {\CRE} model, reflecting the upstream steady state of secondary {\CREs} injected with a flat, $s_e\simeq -2$ spectrum and subsequently cooling.

We may therefore infer the necessary {\CRE} energy density in a primary model, directly from our results for the secondary model, provided that we retain the (logarithmic) electron distribution function $f$ instead of replacing it by its steady state value $f=Q/(E\psi)$ in \eq{\ref{eq:Kinematic_Q_vs_myk}}.

Assuming that the emission is dominated by secondary {\CREs}, the efficiency of any primary acceleration is bound by the energy fraction
\begin{align} \label{eq:epsilon_e}
\epsilon_e \equiv \frac{u_e}{u_{th}} & < \frac{E f \Lambda_e}{(3/2)\mu^{-1}n k_B T} = \frac{2(1+b^{-2})k\Lambda_e}{\psi(3/2)\mu^{-1}k_B T} \\
& \simeq 2\times 10^{-4}\frac{\Lambda_e}{T_{10}} \left( \frac{B}{1\muG} \right)^{-2} \left( \frac{\myk}{10^{-29}\erg\se^{-1}} \right) \nonumber \fin
\end{align}
Here, $\Lambda_e\equiv u_e/f=1\Lambda_{e,1}$ relates the total {\CRE} energy density to its logarithmic contribution at the energy given by $\nu_s(E)\simeq \nu$, where $\nu$ is the frequency at which $\myk$ is measured.
For a flat spectrum, $\Lambda$ is large (\eg $\Lambda_e\simeq 16$ for $E_{e,max}\simeq 10^{4}\GeV$), but for the $s_e\simeq -3$ spectrum of primary acceleration, $\Lambda_e\simeq 1$.

Projection and other systematic effects may have caused us to underestimate the values of $\myk$ in \Fig~\ref{fig:SourcesEmissivityShock}.
In addition, the figure shows the values of $\myk*=\myk/(1+b_d^2)$ in relics, were the division by $\myr r_g=(1+b_d^2)\sim 2$ is justified in our model, but not for primary {\CRE} acceleration.
Therefore, we consider $\myk\simeq \text{a few}\times10^{-29}\erg\se^{-1}$ as an upper limit to the kinematic emissivity in relics, implying that
\begin{equation}
\epsilon_e\lesssim 10^{-3}(B/1\muG)^{-2} \fin
\end{equation}

We infer strong, $B\gtrsim B_{cmb}>3\muG$ magnetic fields in relics.
Such a strong field was independently inferred in the Northwest relic in A3667, both from the Faraday rotation measure \citep{Johnston-Hollitt04} and from an upper limit on Compton emission \citep{FinoguenovEtAl10}.
Therefore, the above constraints imply a negligible primary acceleration efficiency, at the level of $\epsilon_e\lesssim 10^{-4}$.

This constraint is much more restrictive than present, $\epsilon_e<10\%$ limits on particle acceleration, inferred for example from the jump conditions across resolved shocks \citep{NakarEtAl08}.
We find similar or stronger constraints for weak shocks at the edges of halos, which show similar $\myk$ but stronger magnetic fields, for example in A665.

Next, consider particle acceleration in turbulence.
Here we may directly use \eq{\ref{eq:Kinematic_Q_vs_myk}}, with $Q$ being the injection rate of primary, rather than secondary, {\CREs}.
This yields an upper limit to primary injection,
\begin{equation}
Q < 2(1+b^{-2}) n k = 2(1+b^{-2}) \nu j_\nu \fin
\end{equation}
In the centres of GHs, this corresponds to $Q\lesssim 10^{-31}\erg\se^{-1}\cm^{-3}$.
Averaged over a GH, assumed to be strongly magnetised, we may use the average emissivities reported by \citet{MurgiaEtAl09} to find $Q\lesssim 10^{-33}\erg\se^{-1}\cm^{-3}$.
These estimates are much lower than the injection rate typically invoked in primary models, usually estimated assuming weak fields.

\subsection{Model assumptions and uncertainties}
\label{sec:model_assumptions_and_uncertainties}

The assumptions made in our analysis include:
\begin{enumerate}
\item
Most of our relic analysis approximates the gas distribution within clusters using isothermal $\beta$ models, derived mainly from ASCA X-ray data \citep{FukazawaEtAl04}.
At relics or other weak shocks, we assume that the $\beta$ model pertains to the upstream, as in some cases we find the X-ray brightness near relics to be elevated with respect to its $\beta$-model estimate (see \S\ref{sec:HaloAndRelicEta}).
We assume that a good approximation for the confidence interval of any $\beta$ model estimate is obtained by adopting the largest propagated uncertainty of any one of the three parameters of the $\beta$ model (see \S\ref{sec:DataPreparation}).
\item
We assume that the bulk plasma can be approximated as thermal, with equal electron and proton temperatures.
A constant adiabatic index $\Gamma=5/3$ is assumed.
Metallicity is approximated as homogeneous within each cluster, as given by \citet{FukazawaEtAl04}.
Ionisation is approximated as homogeneous and constant, with an average particle mass $\mu m_p\simeq 0.6m_p$.
\item
Elongated relics are assumed to be magnetisation fronts or weak shocks of Mach numbers around $\mach\sim 2$.
This is based on observational evidence for a strong magnetic field or an enhanced X-ray emission in some relics.
\item
The synchrotron signals are computed, for the most part, using the $\sim 10\%$ accuracy approximation for the synchrotron function in \eq{\ref{eq:FSynApprox}}.
\item
For inelastic collisions between the {\CRIs} and the gas, only the pion decay channels of individual protons are considered.
The assumptions underlying the computations of {\CRE} production from a {\CRP} are reviewed in \citet{KamaeEtAl06}.
\item
A flat, concordance $\Lambda$CDM model is adopted, with dark matter fraction $\Omega_M=0.26$, baryon fraction $\Omega_b=0.04$, and a Hubble constant $H=70\km \se^{-1}\Mpc^{-1}$.
\end{enumerate}

Our estimate of {\CRI} homogeneity is based on a combination of two independent analyses, of the morphologies of two well studied GHs ($\sigma=0.2\pm0.1$), and on the peak brightness of relics and halos ($\sigma=-0.3\pm0.2$).
The morphological estimate is based on direct radio and X-ray data, out to $r\sim 500\kpc$ radii.
The relic-based estimate is based on $\beta$-model extrapolations of the X-ray data, so we cannot rule out small deviations from homogeneity at large radii.
The $\myk$ values we infer from relics are uncertain, both because of the $\beta$-model uncertainties, and due to the $\myr$ correction.
Thus, the $\myk$ values we infer in relics are somewhat degenerate with the $\epsilon_B$ and $\mach$ values we assume or fit.
Note that a {\CRI} distribution that declines with $r$ would require stronger shocks or stronger magnetisation than we used.
There is not much freedom in the choice of these parameters, as the shocks are known to be weak and the magnetisation we find is already strong.
Conversely, the estimated $\epsilon_B\sim 0.1$ that gives the best fit to the relic data (for $\mach\sim 2$) is uncertain, as it is based on the $\beta$-models at large radii.

We estimate $u_p$ using the centres of GHs, in order to reduce the uncertainties due to the $\beta$-model, projection, weak or irregular magnetisation, etc.
Similarly, we estimate $s_p$ using the spectrum of relics, rather than GHs, in order to reduce the uncertainties due to magnetic evolution, {\CRI} diffusion, deviations from a power law spectrum at low, $E_p\sim$GeV energies, and the low energy saturation of the inclusive cross section for {\CRE} production.
Nevertheless, $u_p$ could be inaccurate if the gas or the magnetic field are clumpy.
Here, the $\beta$-model may overestimate the column density $\lambda_n$, and $\lambda_n$ may underestimate the magnetised column density $\lambda_{nB}$.
An error in our estimate of $s_p$ could arise if the {\CRP} spectrum is not homogeneous.

\subsubsection{Projection effects}

For extended sources such as GHs, one must integrate the emission along the line of sight, well outside the plane of the cluster (defined as the plane perpendicular to the line of sight).
This is the motivation for analysing $\eta(\vectwo{r})$, in \S\ref{sec:HaloAndRelicEta}, with respect to the local projected density $n(\vectwo{r})$, defined as the density in the plane of the cluster, as explained in \S\ref{sec:EtaDiagnostic}.
Equivalently, this is why $I_\nu(\vectwo{r})$ scales with the column density $\lambda_n(\vectwo{r})$ (for homogeneous {\CRIs}), rather then with $n(\vectwo{r})$.

The elongated morphologies and high polarisation levels of classical relics suggest an extended shock front observed edge-on, propagating nearly perpendicular to the line of sight.
This motivates an integration along the line of sight in relics, too, as implicitly assumed in \S\ref{sec:HaloAndRelicEta}, where we show that the $\eta(\vectwo{r})$ values of relics scale nearly as $n(\vectwo{r})^{-1}$.
This is explicit in \S\ref{sec:ModelApplications}, where we estimate $I_\nu(\vectwo{r})$ as being nearly linearly proportional to $\lambda_n(\vectwo{r})$, in both halos and relics.
In practice, the line of sight integration may lead to overestimated values of the parameters $n$, $\lambda_n$ and $F_X$, that are actually associated with the relic emission.
One should account for the finite size of the shock front, its oblique angle with respect to the line of sight, and its distance out of the cluster's plane.

Accounting for these projection effects would lead to smaller $n$, $\lambda_n$, and in particular $F_X$ values coincident with the relic.
Consequently, the relic data points shown in the $\eta\text{--}n$ phase space in \Figs~\ref{fig:SourcesEtaN}, \ref{fig:SourcesEtaNModel}, and \ref{fig:SourcesEtaNModelM2}, and in the $\myk\till\lambda$ phase space in \Figs~\ref{fig:SourcesEmissivityHalos} and \ref{fig:SourcesEmissivityShock}, would move upwards and towards the left.
Our model would then require somewhat higher values of the relic shock parameters $\mach$ and $\epsilon_B$, in order to explain the observations.

Nevertheless, the agreement between the present model --- uncorrected for projection --- with observations, the reasonable values of $\mach$ and $\epsilon_B$ obtained above, and the morphology and polarisation of relics, suggest that these projection effects are not substantial.
Furthermore, our choice of identifying each source with the position showing the highest radio surface brightness, presumably selects for optimal projection conditions, and reduces the impact of these spurious effects.

If classical relics do indeed correspond to shocks observed nearly edge-on, then diffuse radio emission from clusters is far more complicated than it presently appears.
As the brightness of an observed relic is similar and often higher than in its halo counterpart,
we should assume that halo observations are often contaminated by unidentified emission from projected relics, leading to spurious brightening and spectral steepening.
A careful analysis of the observations carried out by next generation radio telescopes, such as MWA, LOFAR, and SKA, may be able to disentangle the different sources of radio emission and provide strong constraints on the cluster's dynamics.


\section{Summary}
\label{sec:Summary}

Present models for the diffuse radio emission from galaxy clusters, attributing relics to primary {\CREs}, and halos to primary or to steady-state secondary {\CREs}, are challenged and in some cases inconsistent with observations.
These challenges include (see \S\ref{sec:ModelProblems}):
(\emph{i}) the multiple connections between halos and relics, such as radio bridges sometimes observed to connect a halo and a relic; (\emph{ii}) the remarkably similar radio spectra $\alpha\simeq -1$ at the edges of all well studied relics, and the selective appearance of downstream steepening; and (\emph{iii}) the steep spectrum of a small, $\lesssim 20\%$ fraction of GHs.
Primary halo models face several additional challenges, such as explaining the shocks found at the edges of several halos; see \S\ref{sec:primary_vs_secondary} and {\KL}.

We extract radio and X-ray data from the literature, for all the known relic clusters (see \S\ref{sec:HaloAndRelicEta}), supplemented by a sample of non-relic clusters harbouring GHs or MHs studied in {\KL}.
We study the morphological, spectral and energetic properties of the radio sources, and analyse them using the properties of the thermal plasma, inferred from X-ray data or from X-ray based $\beta$-models for the gas distribution.
The assumptions we make are summarised in \S\ref{sec:model_assumptions_and_uncertainties}.

Taking into account the effects of irregular and time-dependent magnetic fields (see \S\ref{sec:TimeDependentTheory}), we find that diffuse radio emission from the ICM, in its different forms, can be explained by a simple model (see \S\ref{sec:ModelApplications}), as synchrotron emission from secondary {\CREs} produced from a {\CRI} distribution which is spectrally flat and is homogeneous on cluster scales (see \S\ref{sec:ModelApplications}), radiating in strong, and in some cases evolving, magnetic fields.

The main attributes of our model are:
\begin{enumerate}
\vspace{-2mm}
\item
{\CRIs} are homogeneously distributed within each cluster, at least out to the $\sim 2\Mpc$ scales where relics are observed.
Parameterising the {\CRI} distribution as an $N_i\propto n^\sigma$ scaling with the local gas density $n$, we infer $\sigma = +0.2\pm0.1$ from the morphologies of the central, $r<2r_c$ regions of two well studied GHs (see \eq{\ref{eq:BestFitGamma}}), and $\sigma = -0.3\pm0.2$ from a sample of halo and relic peaks (see \eqs{\ref{eq:fit_relics_epsB}} and (\ref{eq:fit_relics_epsBM2})).
These estimates are consistent with a homogeneous, $\sigma=0$ distribution of {\CRIs}, considering the systematic uncertainties.

\item
The logarithmic energy density of the {\CRIs} is
\begin{equation}
\frac{du_p}{d\ln E_p} \simeq 10^{-\range{12.9}{13.8}}\left(\frac{E_p}{\text{GeV}} \right)^{-0.20\pm0.05} \erg\cm^{-3}
\end{equation}
(see \eqs{\ref{eq:sp_estiamte}}, (\ref{eq:Def_K_p}), (\ref{eq:K_p_perK_s2.2}), and (\ref{eq:myk_range})).
The {\CRI} energy density, $u_{p[10,10^7]} \simeq 10^{-\range{12.4}{13.3}} \erg\cm^{-3}$, is a small fraction, $\epsilon_p\sim 10^{-3}$ of the thermal energy density in the centres of clusters, in agreement with previous studies \citep[][{\KL}]{KushnirEtAl09}, but constitutes a fair fraction of the thermal energy at the cluster's periphery, $1\till2$ orders of magnitude higher than previously thought.

\item
Inelastic p--p collisions of these {\CRIs} with the ambient plasma lead to the injection of {\CREs} at a logarithmic rate
\begin{align}
Q \simeq 10^{-\range{31}{32}} n_{-2} \left(\frac{E_e}{\text{GeV}} \right)^{-0.06\pm0.02} \erg \se^{-1}\cm^{-3} \fin
\end{align}

\item
In strongly magnetised, $B\gtrsim B_{cmb}$ regions in the ICM, these {\CREs} emit most of their energy as synchrotron radiation, giving rise to the diffusive radio signals observed.
The synchrotron emissivity $j_\nu$ is proportional to the gas density $n$, so the surface brightness $I_\nu$ is proportional to the column density $\lambda_B$ of magnetised gas, and the luminosity $L_\nu$ is proportional to the magnetised mass.
Strongly magnetised regions arising from merger-induced turbulence or sloshing motions thus lead to GHs and MHs, respectively.

\item
More precisely, in regions where the magnetic field configuration and the injection rate of {\CREs} evolve on time scales longer than the cooling time $t_{cool}$ of the {\CREs} (see \eq{\ref{eq:CRE_Cooling0}}), the {\CRIs} reach a steady state distribution
\begin{equation} \label{eq:summary_f_steady_state}
f(E) \equiv E^2 N_e \simeq \frac{Q}{E\psi} \propto \frac{Q}{E(B^2+B_{cmb}^2)} \fin
\end{equation}
The resulting radio signal follows
\begin{equation}
\nu j_\nu \simeq 10^{-\range{31.3}{32.2}} \frac{B^2n_{-2}}{B^2+B_{cmb}^2} \erg\se^{-1} \cm^{-3} \coma
\end{equation}
\begin{align}
\nu I_\nu(\vectwo{r}) \simeq \frac{10^{\range{-0.1}{0.5}}}{10^{22}\cm^{-2}} \int \frac{B^2n\,d l}{B^2+B_{cmb}^2} \GHz \muJy \asec^{-2} \coma
\end{align}
and
\begin{align}
\nu L_\nu \simeq \frac{10^{\range{40.9}{41.8}}}{10^{14}M_\odot} \int \frac{B^2}{B^2+B_{cmb}^2}  \rho\,d V\erg \se^{-1} \fin
\end{align}

\item
In regions where magnetic fields or {\CRE} injection evolve faster than $t_{cool}$ (see \eq{\ref{eq:CRE_Cooling0}} for the relevant frequency range), the {\CRE} distribution continuously evolves towards the steady state given by \eq{\ref{eq:summary_f_steady_state}}.
At higher energies, $f$ adjusts faster to the evolution.
Consequently, in regions where the magnetic field recently grew stronger (weaker), the radio signal is initially amplified (suppressed), followed by a gradual decline (increase) in brightness accompanied by spectral steepening (flattening).
As the brightness reaches a steady state, the flat steady state spectrum $\alpha\simeq -1$ is recovered.
(These effects are analysed in \S\ref{sec:TimeDependentTheory}.)
One may therefore use radio spectral maps to reconstruct the recent magnetic evolution.

\item
In particular, a weak shock initially (\ie in the near downstream) amplifies the radio brightness by a factor $\myrB^{2-(q/2)}\myrcre\simeq (\mach b_d/b_u)^2$, while the spectrum remains initially flat, $\alpha=-1$.
This is precisely the signature of relics, unexplained by previous models.

Farther downstream, assuming no additional temporal evolution in $B$ and $Q$, the kinematic (per unit density) radio emissivity $\myk$ declines by a factor $\myr\simeq r_g^{-1}(1+b_d^2)/(1+b_u^2)$, which is typically larger than unity.
Here, the downstream decline in $\myk$ is accompanied first by spectral steepening, and later by flattening back to $\alpha\simeq -1$, as inferred inward of some relics.

In the downstream, $\myk$ is elevated with respect to the unshocked ICM, leading to the observation of relic tails, halo--relic bridges, or halo protrusions, depending on the column density, the detection threshold, and the value of $\myr$ (see \S\ref{sec:ModelApplications}).

\item
The turbulence induced by a recent merger event, recognised for example by the presence of merger shock relics near the centre of the cluster, can gradually magnetise the ICM, appreciably raising the magnetic field over timescales $\lesssim t_{cool}$.
The resulting radio halo would show spectral steepening in regions where the recent fractional growth in magnetic energy is large, or where the field is irregular, most notably at the edges of the halo where $B\simeq B_{cmb}$.
This explains the steep spectrum of a subset of halos.
In particular, we show that these are the halos associated with nearby, $r\lesssim 1\Mpc$ relics (see \Fig~\ref{fig:HaloAlphaVsRelicR}).
\item
Such recent magnetic growth also provides an alternative explanation for the radial spectral steepening observed in the edges of flat spectrum halos that harbour more distant relics, interpreted in {\KL} as evidence for a steep {\CRP} spectrum.

\end{enumerate}

Additional conclusions and implications of our model:
\begin{enumerate}
\vspace{-2mm}

\item
The values of $\myr$ needed to reconcile halos and relics imply that a fraction $\epsilon_B\sim 0.1$ of the thermal energy density $u_{th}$ is deposited in magnetic fields (see \S\ref{sec:magnetic_scaling}).
For example, assuming that all relics are $\mach=2$ shocks would imply that $\epsilon_B=0.04_{-0.02}^{+0.06}$; higher (lower) magnetic fractions are needed if the shocks are weaker (stronger).

\item
Studying the radio maps of the well-studies clusters A665, A2163, and A2744, we identify an $I_\nu\propto \lambda_n \sim F_X^{1/3}$ scaling at small radii, and a transition to $I_\nu\propto F_X$ at large radii.
We show how this behaviour is often masked by substructure, asymmetry, weak fields, contaminations, and shocks, in particular if one uses azimuthal averaging or bins the data onto a grid (see \S\ref{sec:ReconcilingHomogeneousCRIsInHalos}).
The morphology is explained by a $B^2\propto n$ magnetic scaling, in which $\epsilon_B$ is uniform in the plasma (see \S\ref{sec:magnetic_scaling}).
At large radii, where $B$ declines below $B_{cmb}$, this scaling leads to the $I_\nu\propto F_X$ profile; here, studies assuming primary {\CRIs} infer the same magnetic behaviour.

\item
We use the morphological break observed in A665 and A2163 to infer a central magnetic field $B_0\simeq 30\muG$, corresponding to a homogeneous $\epsilon_B\simeq 0.1$ (see \S\ref{sec:EtaInWellStudiedGHs}).
These results suggest that both halos and relics reach magnetic saturation at $\epsilon_B\sim 0.1$ levels (accurate to within a factor of $2$ or so; see \S\ref{sec:magnetic_scaling}).

\item
Assuming that $\epsilon_B$ and $u_p$ are universal constants, we crudely derive the scalings $P_\nu\sim L_X^{2}$ and $P_\nu\propto R_\nu^{4.4}$, which are in good agreement with the observed correlations (see \S\ref{sec:CRIDispersionAmongClusters}).

\item
The strong magnetic fields we infer in halos and behind relics imply that particle acceleration in weak shocks and in turbulence is very inefficient (see \S\ref{sec:limits_on_particle_acceleration_and_injection}).
Using relics, we find that the $\mach\lesssim 2$ shocks involved cannot deposit more than a small, $\epsilon_e<10^{-4}$ fraction of the downstream thermal energy in relativistic, $>10\GeV$ electrons.
Using halos, we find that the ICM turbulence cannot significantly inject primary {\CREs}, with an upper limit for $E_e>10\GeV$ {\CREs} of
$Q\lesssim 10^{-31}\erg\se^{-1}\cm^{-3}$ in the centres of GHs, and $Q\lesssim 10^{-33}\erg\se^{-1}\cm^{-3}$ averaged over the GH.

\item
The homogeneous {\CRI} distribution inferred from observations implies that {\CRI} diffusion, or some other mixing mechanism, must be strong (see \S\ref{sec:HomogeneousCRIsImplications}).
Along with additional evidence supporting strong diffusion, this imposes a lower limit $D \gtrsim 10^{32}\cm^2 \se^{-1}$ on the $\sim 100\GeV$ {\CRI} diffusion coefficient.

\item
The observed {\CRI} distribution could be generated by SNe, by the virial shock of the cluster, or by a combination of both, provided that the fraction of $E_p\lesssim 100\GeV$ {\CRIs} escaping the cluster is small, $\lesssim 50\%$ (see \S\ref{sec:HomogeneousCRIsImplications}).

\item
The {\CRI} energy density integrated over $E_p$ is $u_p \simeq 10^{-\range{12.4}{13.3}} \erg\cm^{-3}$.
In the outskirts of the cluster, $u_p$ reaches a few $10\%$ of the thermal energy density downstream of the virial shock.
Here, {\CRIs} could play an important role and may modify the shock properties; {\CRIs} escaping beyond the virial shock could magnetise the inflow.

\item
Weak shocks amplify a maximally flat distribution of relativistic particles by a factor $\leq \mach^2$, regardless of the details of the diffusion mechanism or the equation of state (see \S\ref{sec:CRAmplification}).
This corresponds to a decrease in the fractional energy of the relativistic particles, by a factor between $1$ and $5/4$ (for $\Gamma=5/3$).

\item
{\CRE} diffusion across an irregular magnetic field can lead to significant spectral steepening, because a locally strong field leads to enhanced brightness and excessive cooling (see \S\ref{sec:SynchrotronEvolvingB}).
Magnetic oscillations about a constant or a slowly changing average lead to a mild, $-0.1<\Delta \alpha<0$ spectral steepening, which arises because brighter emission is correlated with recent magnetic growth and therefore with spectral steepening.

\item
Hadronic signals such as $\gamma$-ray emission from neutral pion decay, or the sum of synchrotron and inverse Compton emission from the secondary {\CRIs}, scale with the projected gas mass; specific predictions are given in \S\ref{sec:Additional_hadronic_signals}.

\end{enumerate}

Future high resolution data and a more detailed analysis would allow us to better test the model, calibrate its parameters, and utilise it in the study of clusters:
\begin{enumerate}
\vspace{-2mm}
\item
Detailed brightness and spectral maps could be used to test the time-dependent model, and to gauge the recent magnetic evolution throughout the cluster. In particular, spectral maps in multiple frequencies, or maps of the spectral curvature, would strongly constrain the evolution.
\item
The strong magnetic fields we find in halos, relics, and halo--relic bridges, could be tested with Faraday rotation measures (see {\KL} for a possible preliminary indication for GH magnetisation).
\item
X-ray observations at large distances from the cluster's centre could be used to confirm the relic--weak shock association, and determine $\mach$.
With a measured $F_X$, our relic model would be left with a single free parameter, $b_d$ (assuming that $b_u\ll 1$).
\item
Deviations from the homogeneous, flat spectrum {\CRI} distribution could be constrained by better modeling relics, which probe the cluster's periphery.
If found, they could identify the {\CRI} sources, and would provide a measure of the {\CRI} diffusion, instead of the present lower limit on $D$.
\item
Better data could be used to identify a redshift evolution of $u_p$, and its correlations with various properties of the cluster; we failed to detect a significant signal with the present uncertainties (see \S\ref{sec:CRIDispersionAmongClusters}).
\item
Correlations between $u_p$ and SNe tracers, such as metallicity and star formation, could be used to identify the contribution of SNe to the cluster's {\CRI} population.
Preliminary, low significance evidence for correlations between $\eta_0$ and the metallicity and the specific SFR were pointed out by {\KL}.
\item
Correlations between $u_p$ and tracers of the {\CRI} output of the virial shock, such as its downstream temperature, could be used to identify the contribution of the virial shock to the cluster's {\CRI} population.
\item
The weak signal arising from primary {\CREs} accelerated by the virial shock could be identified in radio \citep{KeshetEtAl04}, X-ray \citep{KushnirWaxman10}, and $\gamma$-ray \citep{KeshetEtAl03} observations, directly or through a correlation with cluster tracers.
This would constrain the magnetisation and {\CRI} output of the virial shock.
\item
We have neglected projection effects, effectively approximating for example relics as infinite, planar shock-fronts parallel to the line of sight. A more realistic, three-dimensional model would better constrain the model and the dynamical state of the cluster (see \S\ref{sec:model_assumptions_and_uncertainties}).
\end{enumerate}


\begin{table*}
 \caption{Definition of main parameters used in the paper. }
 \label{tab:Parameters}
 \begin{tabular}{@{}|cccc|}
  \hline
  Name & Meaning & Definition & Reference \\
  \hline
    $\bar{m}=\mu m_p$    & average particle mass & $\mu m_p\simeq0.6m_p$  & \S\ref{sec:Introduction} \\
    $\vectwo{r}$    & two-vector in the plane of the cluster & \nosymb & \S\ref{sec:DataPreparation} \\
    $\vecthree{r}$  & three-vector in the cluster frame & \nosymb & \S\ref{sec:DataPreparation} \\
    $\Xi_{0}$   & (arbitrary) quantity $\Xi$ measured at $r=0$ & \nosymb & \S\ref{sec:Introduction} \\
    $\Xi_{u,d}$   & (arbitrary) quantity $\Xi$ measured upstream, downstream & \nosymb & \nosymb \\
    $I_\nu$ & radio surface brightness & \nosymb & \S\ref{sec:Introduction} \\
    $F_X$   & X-ray surface brightness between $0.1$ and $2.4\keV$ & \nosymb & \S\ref{sec:Introduction} \\
    $\eta$  & radio to X-ray brightness ratio & $\nu I_\nu/F_X$ & \S\ref{sec:Introduction} \\
    $j_\nu$ & synchrotron emissivity & \nosymb & \S\ref{sec:EtaDiagnostic} \\
    $j_X$   & X-ray emissivity between $0.1$ and $2.4\keV$ & \nosymb & \S\ref{sec:EtaDiagnostic} \\
    $\eta_j$  & radio--X-ray emissivity ratio & $\nu j_\nu/j_X$ & \S\ref{sec:EtaDiagnostic} \\
    $\eta_L$  & radio--X-ray luminosity ratio & $\nu P_\nu/L_X$ & \S\ref{sec:etaModels} \\
    $\myetaL(r)$  & radio--X-ray luminosity ratio within $r$ a& $\nu P_\nu(<r)/L_X(<r)$ & \S\ref{sec:etaModels} \\
    $n$     & bulk electron number density & \nosymb & \S\ref{sec:Introduction} \\
    $\rho$  & mass density & $n m_p$ & \S\ref{sec:HomogeneousCRIsImplications} \\
    $N_p,N=N_e$     & {\CRP}, {\CRE} number density & \nosymb & \S\ref{sec:Introduction} \\
    $B$     & magnetic field amplitude  & \nosymb & \S\ref{sec:Introduction} \\
    $B_{cmb}$  & magnetic field at equipartition with the CMB & $(8\pi u_{cmb})^{1/2}$ & \S\ref{sec:Introduction} \\
    $b$     & magnetic field normalised to $B_{cmb}$ & $B/B_{cmb}$ & \S\ref{sec:SecondaryCREsInRelics} \\
    $R_\nu$     & radius of observed radio halo & \nosymb & \S\ref{sec:etaModels} \\
    $R_B$     & radius where $B=B_{cmb}$ & \nosymb & \S\ref{sec:etaModels} \\
    $\beta$,$r_c$,$n_0$ & $\beta$-model parameters & \nosymb & \S\ref{sec:DataPreparation} \\
    $u_{th}$    & thermal energy density & $(3/2)\mu^{-1}nk_BT$ & \S\ref{sec:Introduction} \\
    $u_p,u_e$ & {\CRP}, {\CRE} energy density & \nosymb & \S\ref{sec:Introduction} \\
    $u_{B}$    & magnetic energy density & $B^2/8\pi$ & \S\ref{sec:Introduction} \\
    $\epsilon_B$  & magnetic energy fraction & $u_B/u_{th}$ & \SEq{\ref{eq:epsilon_B}} \\
    $\epsilon_p$, $\epsilon_e$  & {\CRP}, {\CRE} energy fraction & $u_p/u_{th}$, $u_e/u_{th}$ & \S\ref{sec:HomogeneousCRIsImplications}, \SEq{\ref{eq:epsilon_e}} \\
    $\mysigma$  & $N_p(n)$ power-law index & $d\ln N_p/d\ln n$ & \S\ref{sec:EtaDiagnostic} \\
    $\gamma$  & $\eta_j(n)$ power-law index & $d\ln \eta_j/d\ln n$ & \S\ref{sec:EtaDiagnostic} \\
    $\alpha$ & radio spectral index & $d\ln (I_\nu)/d\ln(\nu)$ & \S\ref{sec:Introduction} \\
    $\alpha_{\nu_1}^{\nu_2}$ & $\alpha$ measured between $\nu_1$ and $\nu_2$ & $\ln (I_{\nu_2}/I_{\nu_1})/\ln(\nu_2/\nu_1)$ & \S\ref{sec:SteepHalos} \\
    $\avalpha$ & average $\alpha$ in a relic & \nosymb & \S\ref{sec:PeculiarRelics} \\
    $\edalpha$ & $\alpha$ in the outer edge of a relic & \nosymb & \S\ref{sec:PeculiarRelics} \\
    $E_p$, $E=E_e$ & energy of a {\CRP}, {\CRE} & \nosymb & \S\ref{sec:SteepHalos}, \S\ref{sec:PeculiarRelics} \\
    $f=f_e$     & logarithmic distribution function & $E^{2}N$ & \S\ref{sec:CRE} \\
    $\phi=\phi_e$  & spectral index of $f$ & $d\ln f/d\ln E$ & \SEq{\ref{eq:CRE_phi}} \\
    $Q=Q_e$     & logarithmic {\CRE} injection & $E^2\dot{\myn}_+$ & \SEq{\ref{eq:Def_Q}} \\
    $\psi$ & cooling parameter & $-E^{-2}dE/dt$ & \S\ref{sec:EtaDiagnostic} \\
    $\myQpsi$ & asymptotic value of $f$ & $Q/(\psi E)$ & \S\ref{sec:magnetic_variations_modify_CRE} \\
    $\myx$  & inverse {\CRE} energy & $E^{-1}$ & \S\ref{sec:CRE} \\
    $\myPsi_{t,t_0}$    & integrated cooling parameter & $\int_{t_0}^t \mypsi\,dt'$ & \SEq{\ref{eq:Def_myPsi}} \\
    $\myy(\tau)$  & retarded inverse {\CRE} energy & $\myx-\Psi_{t,\tau}$ & \SEq{\ref{eq:Def_myy}} \\
    $t_{cool}$ & instantaneous cooling time & $(E\mypsi)^{-1}$ & \SEq{\ref{eq:CRE_Cooling0}} \\
    $\tau_{cool}$ & retarded cooling time & $t-t_i$ & \SEq{\ref{eq:tau_cool_def}} \\
    $s_p$ & {\CRP} spectral index & $d\ln (N_p)/d\ln (E_p)$ & \S\ref{sec:SteepHalos} \\
    $s_e$ & {\CRE} spectral index & $d\ln (N_e)/d\ln (E_e)$ & \S\ref{sec:PeculiarRelics} \\
    $\myh$  & $j_\nu$ normalised to its far downstream value & $j_\nu/j_\nu(\myX\to\infty)$ & \SEq{\ref{eq:myhDef}} \\
    $\myk$ & kinematic radio emissivity & $\nu j_\nu/n$ & \SEq{\ref{eq:SpecificEmissivity}} \\
    $\lambda_n$ & column density of gas & $\int n \,dl$ & \SEq{\ref{eq:beta_n_Ix}} \\
    $\lambda_{nB}$ & column density of magnetised gas & $\int n B^2\, dl/(B^2+B_{cmb}^2)$ & \SEq{\ref{eq:nuInuPerLambda}} \\
    $\lambda_B$ & magnetic coherence scale & \nosymb & \S\ref{sec:TemporalEvolutionCannotBeNeglected} \\
    $v_s$   & shock velocity & \nosymb & \S\ref{sec:TemporalEvolutionCannotBeNeglected} \\
    $c_s$   & sound velocity & $(\Gamma k_B T/ \mu m_p)^{/2}$ & \S\ref{sec:TemporalEvolutionCannotBeNeglected} \\
    $\Gamma$ & adiabatic index & $5/3$ & \S\ref{sec:PeculiarRelics} \\
    $\mach$ & shock Mach number & $v_s/c_s$ & \S\ref{sec:PeculiarRelics} \\
    $\myrg$ & gas compression factor & $n_d/n_u$ & \SEq{\ref{eq:DSA_s}} \\
    $\myrcr$ & CR compression factor & $N_d/N_u$ & \SEq{\ref{eq:CR_Compression1}} \\
    $\myrB$  & magnetic compression factor & $B_d/B_u$ & \S\ref{sec:SynchrotronImprintOfShock} \\
    $\Dltg$  & thickness of shock compression layer & \nosymb & \S\ref{sec:SynchrotronImprintOfShock} \\
    $\Dltcr$  & thickness of {\CR} amplification layer & \nosymb & \S\ref{sec:SynchrotronImprintOfShock} \\
    $\DltB$  & thickness of magnetic amplification layer & \nosymb & \S\ref{sec:SynchrotronImprintOfShock} \\
    $\Dltcool$  & thickness of cooling layer & \nosymb & \S\ref{sec:SynchrotronImprintOfShock} \\
    $\Lambda_e$ & Coulomb logarithm for {\CRE} energy & $u_e/f$ & \SEq{\ref{eq:Approx_sigma_sM2}} \\
  \hline
 \end{tabular} 

 \medskip
    Symbols are ordered contextually.
\end{table*}



\section{Acknowledgments}
It is a pleasure to thank A. Loeb, D. Kushnir, E. Waxman, B. Katz, M. Markevitch, E. Nakar, G. Brunetti, C-K. Chan, A. Vikhlinin, E. Million, and P. Johnson for valuable discussions.
This work was supported by NASA through
Einstein Postdoctoral Fellowship grant number PF8-90059 awarded by the
\emph{Chandra} X-ray Center, which is operated by the Smithsonian
Astrophysical Observatory for NASA under contract NAS8-03060.

\appendix
\onecolumn

\section{Arbitrary transient magnetic evolution}
\label{sec:transient_B_evolution}

Consider magnetic evolution similar to that in \EqO~(\ref{eq:magnetic_jump}), but with arbitrary transient behaviour $\mypsi_e(t)$,
\begin{equation} \label{eq:magnetic_jump2}
\mypsi(t)=
\begin{cases}
\mypsi_1 & \text{if $t \leq t_0$ ;}
\\
\mypsi_e(t) & \text{if $t_0<t\leq t_0 + \Delta t$ ;}
\\
\mypsi_2 = \myr\mypsi_1 & \text{if $t_0 + \Delta t < t$ .}
\end{cases}
\end{equation}
Plugging this into \EqO~(\ref{eq:PDE_sol}) and assuming steady, power-law injection $Q=\myQ_0 E^q$, yields the {\CRE} distribution
\begin{equation}
f(t>t_0+\Delta t,E) = \frac{\myQ_0 E^{q-1}}{(1-q)\mypsi_2} \times
\begin{cases}
\left[ 1+(\myr-1)\left(1-\frac{E}{E_c}\right)^{1-q}+\Delta_1 + \Delta_2  \right] & \text{if $E \leq E_c$ ;}
\\
\left( 1+\Delta_3 \right) & \text{if $E_c<E\leq \epsilon E_c$ ;}
\\
1 & \text{if $\epsilon E_c<E$ .}
\end{cases}
\end{equation}
where we defined $E_c(t)\equiv [(t-t_0-\delta t)\mypsi_2]^{-1}$, $\delta t\equiv \Delta t -\mypsi_2^{-1} \myPsi_{t_0+\Delta t,t_0}$, and
\begin{equation}
\epsilon \equiv \frac{t-t_0-\delta t}{t-t_0-\Delta t} = 1 + \frac{\Delta t}{t-t_0}\left( 1-\frac{\delta t}{\Delta t} \right) + O\left( \frac{\Delta t}{t-t_0}\right)^2 \fin
\end{equation}
The quantities
\begin{equation}
\Delta_1 \equiv \left(1-\frac{E}{E_c}\right)^{1-q}-\left(1-\frac{E}{\epsilon E_c}\right)^{1-q}
= (1-q) \frac{E}{E_c}\left(1-\frac{E}{E_c}\right)^{-q} \frac{\delta t-\Delta t}{t-t_0} + O \left( \frac{\delta t-\Delta t}{t-t_0} \right)^2 \coma
\end{equation}
\begin{equation}
\Delta_2 \equiv \frac{(1-q)\mypsi_2}{\myQ_0 E^{q-1}} \int_{t_0}^{t_0+\Delta t}Q(\tau, y) \,d\tau \leq (1-q)\mypsi_2 E \left(1-\frac{E}{\epsilon E_c}\right)^{-q} \Delta t \coma
\end{equation}
and
\begin{equation}
\Delta_3 \equiv \frac{(1-q)\mypsi_2}{\myQ_0 E^{q-1}}  \int_{t_i}^{t_0+\Delta t}Q(\tau, y) \,d\tau \leq \Delta_2
\end{equation}
(where $t_0<t_i<t_0+\Delta t$) are all small, $O[\Delta t/(t-t_0)]$ corrections at late times.

\section{Power-law cooling evolution}
\label{sec:power_law_B_evolution}

Consider magnetic evolution that gives rise to a power-law temporal behaviour of the cooling parameter,
\begin{equation}
\mypsi(t) = \begin{cases} \mypsi_1  &\text{    if $t\leq t_1$\,;}
\\
\mypsi_1 (t/t_1)^{\beta} &\text{    if $t_1<t<t_2$\,;}
\\
\mypsi_2\equiv \mypsi_1 (t_2/t_1)^{\beta} &\text{    if $t_2\leq t$\coma}
\end{cases}
\end{equation}
where $t_1$, $t_2$ ($t_2>t_1$), $\mypsi_1$ ,$\mypsi_2$, and $\beta\neq -1$ are constants.
For simplicity, consider steady {\CRE} injection with a flat spectrum, $Q(E,t)=\constant$.

At early times $t<t_1$ and very late times $t\gg t_2$, the solution asymptotes to $f\to f_{1,2} \equiv Q/(E\mypsi_{1,2})$, with a spectral index $\phi=-1$.
At intermediate times
\begin{equation}
f(t_1<t<t_2,E) = Q t \times
\begin{cases} \left[1 + \frac{1}{\mypsi_1 E t} -\frac{\mypsi/\mypsi_1+\beta t_1/t}{1+\beta}  \right] &\text{    if $E\leq E_1$ ;} \\
\left[1-\left(1-\frac{1+\beta}{\mypsi E t} \right)^{\frac{1}{1+\beta}} \right]
&\text{    if $E>E_1$\coma}
\end{cases}
\end{equation}
where we defined a characteristic {\CRE} cooling energy
\begin{equation}
E_1(t) \equiv \frac{1+\beta}{\mypsi t - \mypsi_1 t_1} \coma
\end{equation}
above which memory of the $t< t_1$ distribution is lost.
The energy power-law index of $f$ is, accordingly,
\begin{equation}
\phi(t_1<t<t_2,E) \equiv \frac{d\ln f}{d\ln E} = \begin{cases}
-\left[1+\mypsi_1 E (t-t_1)-E/E_1\right]^{-1}
&\text{    if $E\leq E_1$\,;} \\
-\left\{1+\beta-\mypsi E t \left[ 1-\left( 1-\frac{1+\beta}{\mypsi E t}\right)^{\frac{\beta}{1+\beta}}  \right]\right\}^{-1}
&\text{    if $E>E_1$\fin}
\end{cases}
\end{equation}
For $\beta>0$ ($\beta<0$), the low energy distribution gradually softens (hardens) until reaching $E_1$, above which the distribution hardens (softens) back.
At $E=E_1$, we find $f=Q(t-t_i)$ and a spectral index $\phi=-[(t/t_1)^{1+\beta}-1]/[(1+\beta)(t/t_1-1)]$, which corresponds to a very steep (flat) spectrum for $t\gg t_1$ if $\beta>0$ ($\beta<0$).
Note that this problem has a characteristic timescale $\Delta t=t_2-t_1$, so the spectrum is not self similar.

At late times
\begin{equation}
f(t>t_2,E) = \begin{cases}
\frac{Q}{\psi_1}\left[ E^{-1} +\beta E_1(t_2)^{-1}-t(\psi_2-\psi_1) \right]
&\text{    if $E\leq E_2$ ;} \\
Q\left\{ t-t_2\left[ 1-(1+\beta)\left( 1+\frac{1}{E\psi_2 t_2}-\frac{t}{t_2} \right) \right]^{1/(1+\beta)} \right\}
&\text{    if $E_2<E\leq E_3$ ;} \\
f_2\equiv \frac{Q}{E\psi_2}
&\text{    if $E_3\leq E$ \coma }
\end{cases}
\end{equation}
where $E_2(t)\equiv \left[E_1(t_2)^{-1}+E_3(t)^{-1}\right]^{-1}$ and $E_3(t)\equiv [\psi_2(t-t_2)]^{-1}$ are cooling energies in different regimes.

\section{Synchrotron signature of a jump in {\CRE} injection or magnetic field strength}
\label{sec:jnu_from_jump_detail}

Consider power-law {\CRE} injection with an arbitrary index $q$, and an instantaneous jump in $\myQ$ and $\mypsi$ as in \eq{\ref{eq:JumpInQandB}} at $t_0$.
Then, using \EqO~(\ref{eq:FSynApprox}) to approximate $F_{syn}$, we obtain at $t>t_0$
\begin{equation}
j_\nu(t,\nu) = A_q\, a^{-\frac{q}{2}} \myQ_0 \frac{B^{2-\frac{q}{2}}\sin^2\myPalpha}{B^2+B_{cmb}^2} \nu^{-1+\frac{q}{2}} \left[1+(\myr-1)J_q\left(\myz\right) \right] \coma
\end{equation}
where
\begin{equation}
A_q\equiv \frac{27c_0\sqrt{3}\,\Gamma(1+c_q)}{32\pi c_2^{1+c_{q}}(1-q)}
\end{equation}
is a dimensionless coefficient, $c_{q}\equiv c_1-q/2$, and $\myz$ is defined in \eq{\ref{eq:myz_def}}.
Here we defined
\begin{eqnarray}
J_q({\myz}) & \equiv &
_qf_p\left(\left\{\frac{q-1}{2},\frac{q}{2}\right\};\left\{\frac{1}{2},-c_{q}\right\}; -\myz \right) \\ & & \nonumber
-\frac{\pi^{3/2}\sqrt{\myz}}{2\Gamma(1+c_q)}
\left[
\frac{(1-q) H\left( \frac{1+q}{2},\frac{q}{2};\frac{3}{2},\frac{1}{2}-c_{q} \right)}{\cos(c_{q}\pi)}
+\frac{2 \left(\frac{{\myz}}{4}\right)^{\frac{1}{2}+c_q} \Gamma(2-q) H\left( \frac{1}{2}+c_1,1+c_1;\frac{3}{2}+c_{q},2+c_{q} \right)}{\sin(2c_{q}\pi)\Gamma(-2c_1)}
\right]
\coma
\end{eqnarray}
where $_qf_p$ is a generalised hypergeometric function, and
\begin{equation}
H(a_1,a_2;b_1,b_2) \equiv \frac{_qf_p(\{a_1,a_2\};\{b_1,b_2\};-{\myz})}{\Gamma(b_1)\Gamma(b_2)}
\end{equation}
is a regularised generalised hypergeometric function.

The radio spectrum is then given by
\begin{equation}
\alpha(t,\nu) = -1 + \frac{q}{2} - (\myr-1)\frac{I_q\left({\myz}\right)}{1+(\myr-1)J_q\left({\myz}\right)} \coma
\end{equation}
where
\begin{eqnarray}
&& I_q({\myz}) \equiv \nonumber
\frac{ \pi^{3/2}q(1-q){\myz}}{4\Gamma(1+c_q)}
\Bigg\{
\frac{H\left(\frac{1+q}{2},\frac{q}{2},\frac{3}{2},\frac{1}{2}-c_{q}\right)}{q{\myz}}
-\frac{\sqrt{{\myz}}}{\cos(c_{q}\pi)} \left[
\frac{(1+q)H\left(\frac{3+q}{2},1+\frac{q}{2},\frac{5}{2},\frac{3}{2}-c_{q}\right)}{2}
\right]
-\frac{H\left(\frac{1+q}{2},1+\frac{q}{2},\frac{3}{2},1-c_{q}\right)}{\sin(c_{q}\pi)}
\\ &&
\,\,\,\, -\frac{(1+c_{q})(\frac{{\myz}}{4})^{c_{q}}\Gamma(-q)}{\sin(2c_{q}\pi)\Gamma(-1-2c_1)}
\left[
\frac{(1+c_1){\myz}H\left(\frac{3}{2}+c_1,2+c_1,\frac{5}{2}+c_{q},3+c_{q}\right)}{1+c_{q}}
-\frac{2H\left(\frac{1}{2}+c_1,1+c_1,c_{q}+\frac{3}{2},2+c_{q}\right)}{1+2c_1}
\right] \Bigg\} \fin
\end{eqnarray}

\section{Finite spectral range and finite beam}
\label{sec:FiniteSpectrumAndBeam}

The spectrum is often measured by comparing observation performed at substantially different frequencies.
In such cases, the spectral features are in general smeared, and one should consider the two-frequency spectral measure
\begin{equation}
\alpha_{\nu_1}^{\nu_2} \equiv \frac{\ln(j_{\nu_2}/j_{\nu_2})}{\ln(\nu_2/\nu_1)}
= \frac{\int_{\nu_1}^{\nu_2} \alpha(\nu)  \frac{d\nu}{\nu}}{\ln(\nu_2/\nu_1)} \fin
\end{equation}
For flat injection,
\begin{equation}
\alpha_{\nu_1}^{\nu_2}(t) = -1
+ \frac{1}{\ln(\nu_2/\nu_1)}
\ln \left\{ \frac{\Gamma(1+c_1)-(\myr-1)\left[\sqrt{\myz_2}\,\Gamma_{\frac{1}{2}+c1}(\myz_2)-\Gamma_{1+c_1}(\myz_2)\right]} {\Gamma(1+c_1)-(\myr-1)\left[\sqrt{\myz_1}\,\Gamma_{\frac{1}{2}+c1}(\myz_1)-\Gamma_{1+c_1}(\myz_1)\right]} \right\}  \coma
\end{equation}
where $\myz_{1,2}(t)\equiv \myz(t,\nu_{1,2})$.

Similarly, the size of the beam is often comparable to the {\CRE} cooling length, so the signal must be convolved with the beam shape.
For a top-hat beam of full width corresponding to a distance range $2\Delta$, the integration of \eq{\ref{eq:j_nu_q}} can be carried out analytically.
In the flat injection limit, this yields
\begin{align}
j_\nu(t,\nu;\Delta) & = A_0 \myQ_d \frac{B_2^{2}\sin^2\myPalpha_2}{B_2^2+B_{cmb}^2} \nu^{-1}
\times \begin{cases}
1+\frac{\myr-1}{\Delta}\left[ K\left(\frac{\myX+\Delta}{\Dltcool}\right)-K\left(\frac{\myX-\Delta}{\Dltcool}\right) \right] & \mbox{if }\myX>\Delta \,;\\
1+\frac{\Delta-\myX}{2\Delta} \left(\frac{r}{r_j}-1\right) + \frac{r-1}{\Delta/\Dltcool}\left[K\left(\frac{\myX+\Delta}{\Dltcool}\right)+ \frac{\Gamma(c_1+3/2)}{4\Gamma(1+c_1)} \right] & \mbox{if }\myX<\Delta \coma
\end{cases}
\end{align}
where we defined
\begin{equation}
K(\chi)\equiv \frac{2\chi\Gamma_{1+c_1}(\chi^2)-\chi^2\Gamma_{\frac{1}{2}+c_1}(\chi^2)-\Gamma_{\frac{3}{2}+c_1}(\chi^2)}{4\Gamma(1+c_1)}\fin
\end{equation}

\end{document}